%% file: paper.tex
\documentclass[twocolumn,times]{aastex61}
\usepackage{graphicx}
\usepackage{epstopdf}
\usepackage{hyperref}

\newcommand{\simlt}{\mathrel{\hbox{\rlap{\hbox{\lower4pt\hbox{$\sim$}}}\hbox{$<$}}}}
\newcommand{\simgt}{\mathrel{\hbox{\rlap{\hbox{\lower4pt\hbox{$\sim$}}}\hbox{$>$}}}}

\begin{document}

\title[PTF SLSN-I]{Spectra of Hydrogen-Poor Superluminous Supernovae from the Palomar Transient Factory}

\author[0000-0001-9171-5236]{Robert M. Quimby}
\altaffiliation{email: rquimby@mail.sdsu.edu}
\affiliation{Department of Astronomy / Mount Laguna Observatory, San Diego State University, 5500 Campanile Drive, San Diego, CA 92812-1221, USA}
\affiliation{
  Kavli Institute for the Physics and Mathematics of the Universe (WPI),
  The University of Tokyo Institutes for Advanced Study,
  The University of Tokyo,
  Kashiwa, Chiba 277-8583, Japan
}

\author{Annalisa De Cia}
\affiliation{European Southern Observatory, Karl-Schwarzschild-Str 2, D-85748 Garching bei M{\"u}nchen, Germany}
\affiliation{Department of Particle Physics and Astrophysics, Weizmann Institute of Science, Rehovot 7610001, Israel}

\author{Avishay Gal-Yam}
\affiliation{Department of Particle Physics and Astrophysics, Weizmann Institute of Science, Rehovot 7610001, Israel}

\author[0000-0002-8597-0756]{Giorgos Leloudas}
\affiliation{Department of Particle Physics and Astrophysics, Weizmann Institute of Science, Rehovot 7610001, Israel}
\affiliation{Dark Cosmology Centre, Niels Bohr Institute, University of Copenhagen, Juliane Maries Vej 30, DK-2100 K{\o}benhavn {\O}, Denmark}

\author[0000-0001-9454-4639]{Ragnhild Lunnan}
\affiliation{The Oskar Klein Centre \& Department of Astronomy, Stockholm University, AlbaNova, SE-106 91 Stockholm, Sweden}
\affiliation{Cahill Center for Astrophysics, California Institute of Technology, Pasadena, CA 91125, USA}

\author[0000-0001-8472-1996]{Daniel A. Perley}
\affiliation{Astrophysics Research Institute, Liverpool John Moores University, IC2, Liverpool Science Park, 146 Brownlow Hill, Liverpool L3 5RF, UK}
\affiliation{Dark Cosmology Centre, Niels Bohr Institute, University of Copenhagen, Juliane Maries Vej 30, DK-2100 K{\o}benhavn {\O}, Denmark}

\author{Paul M. Vreeswijk}
\affiliation{Department of Particle Physics and Astrophysics, Faculty of Physics, Weizmann Institute of Science, Rehovot 76100, Israel}

\author{Lin Yan}
\affiliation{Cahill Center for Astrophysics, California Institute of Technology, Pasadena, CA 91125, USA}
\affiliation{Infrared Processing and Analysis Center, California Institute of Technology, Pasadena, CA 91125, USA}

\author{Joshua S. Bloom}
\affiliation{Department of Astronomy, University of California, Berkeley, CA 94720-3411, USA}

\author[0000-0003-1673-970X]{S. Bradley Cenko}
\affiliation{Astrophysics Science Division, NASA Goddard Space Flight Center, 8800 Greenbelt Road, Greenbelt, MD 20771, USA}
\affiliation{Joint Space-Science Institute, University of Maryland, College Park, MD 20742, USA}

\author{Jeff Cooke}
\affiliation{Centre for Astrophysics and Supercomputing, Swinburne University of Technology, PO Box 218, H30, Hawthorn, Victoria 3122, Australia}

\author[0000-0001-7782-7071]{Richard Ellis}
\affiliation{Department of Physics and Astronomy, University College London, Gower Street, London, WC1E 6BT, UK}

\author{Alexei V. Filippenko}
\affiliation{Department of Astronomy, University of California, Berkeley, CA 94720-3411, USA}
\affiliation{Miller Senior Fellow, Miller Institute for Basic Research in Science, University of California, Berkeley, CA 94720, USA}

\author[0000-0002-5619-4938]{Mansi M. Kasliwal}
\affiliation{Cahill Center for Astrophysics, California Institute of Technology, Pasadena, CA 91125, USA}

\author{Io K. W. Kleiser}
\affiliation{Cahill Center for Astrophysics, California Institute of Technology, Pasadena, CA 91125, USA}

\author[0000-0001-5390-8563]{Shrinivas R. Kulkarni}
\affiliation{Cahill Center for Astrophysics, California Institute of Technology, Pasadena, CA 91125, USA}

\author[0000-0001-6685-0479]{Thomas Matheson}
\affiliation{National Optical Astronomy Observatory, 950 North Cherry Avenue, Tucson, AZ 85719-4933, USA}

\author[0000-0002-3389-0586]{Peter E. Nugent}
\affiliation{Lawrence Berkeley National Laboratory, Berkeley, California 94720, USA}
\affiliation{Department of Astronomy, University of California, Berkeley, CA 94720-3411, USA}

\author{Yen-Chen Pan}
\affiliation{Department of Astronomy and Astrophysics, University of California, Santa Cruz, CA 95064, USA}

\author{Jeffrey M. Silverman}
\affiliation{Department of Astronomy, University of California, Berkeley, CA 94720-3411, USA}
\affiliation{Department of Astronomy, University of Texas, 2515 Speedway, Austin, TX, USA}

\author{Assaf Sternberg}
\affiliation{Observatoire Astronomique de l'Universit{\'e} de Gen{\'e}ve, Chemin des Maillettes 51, CH-1290, Versoix, Switzerland }

\author{Mark Sullivan}
\affiliation{Department of Physics and Astronomy, University of Southampton, Southampton, Hampshire SO17 1BJ, UK}

\author{Ofer Yaron}
\affiliation{Department of Particle Physics and Astrophysics, Faculty of Physics, Weizmann Institute of Science, Rehovot 76100, Israel}

\begin{abstract}

Most Type I superluminous supernovae (SLSNe-I) 
reported to date have been identified by their high peak
luminosities and spectra lacking obvious signs of hydrogen. We
demonstrate that these events can be distinguished from 
normal-luminosity SNe (including Type~Ic events) solely from their
spectra over a wide range of light-curve phases. We use this
distinction to select 19 SLSNe-I and 4 possible SLSNe-I from the Palomar
Transient Factory archive (including 7 previously published
objects). We present 127 new spectra of these objects and combine
these with 39 previously published spectra, and we use these to
discuss the average spectral properties of SLSNe-I at different
spectral phases. We find that \ion{Mn}{2} most probably contributes to
the ultraviolet spectral features after maximum light, and we give a detailed
study of the \ion{O}{2} features that often characterize the early-time
optical spectra of SLSNe-I. We discuss the velocity distribution of
\ion{O}{2}, finding that for some SLSNe-I this can be confined to a
narrow range compared to relatively large systematic velocity shifts.
\ion{Mg}{2} and \ion{Fe}{2} favor higher velocities than \ion{O}{2}
and \ion{C}{2}, and we briefly discuss how this may constrain 
power-source models.  We tentatively group objects by how well they match
either SN\,2011ke or PTF12dam and discuss the possibility that
physically distinct events may have been previously grouped together 
under the SLSN-I label.
  
\end{abstract}


\section{Introduction}\label{intro}

The highest luminosity supernovae (SNe), often called ``superluminous
supernovae'' (SLSNe; \citealt{galyam2012}), are of special interest
because they mark the upper extremum of stellar explosions, can be
detected out to very high redshifts ($z$), and may be used to probe the
early universe. SLSNe may derive their luminosity from unique power
sources or from processes that make only minor contributors to normal
SNe \citep{kasen_bildsten2010, woosley2010,
  chevalier_irwin2011}, and some may offer important constraints on
the final stages of stellar evolution \citep{woosley2017}. SLSNe may
further have applications in cosmology \citep{inserra_smart2014,
  scovacricchi2016}, studies of the stellar initial mass function
\citep{tanaka2012}, and probing the metal content of early star-forming
regions \citep{berger2012, vreeswijk2014}; moreover, they may provide a
means to directly study the first stars \citep{whalen2013}. But to
confidently use SLSNe as probes of the high-redshift universe, we must
build a better physical understanding of these stellar explosions and
identify what it is exactly that distinguishes these events from
normal-luminosity SNe so that we can account for any evolution
with redshift.

\begin{figure}
\begin{center}
 \includegraphics[width=\linewidth]{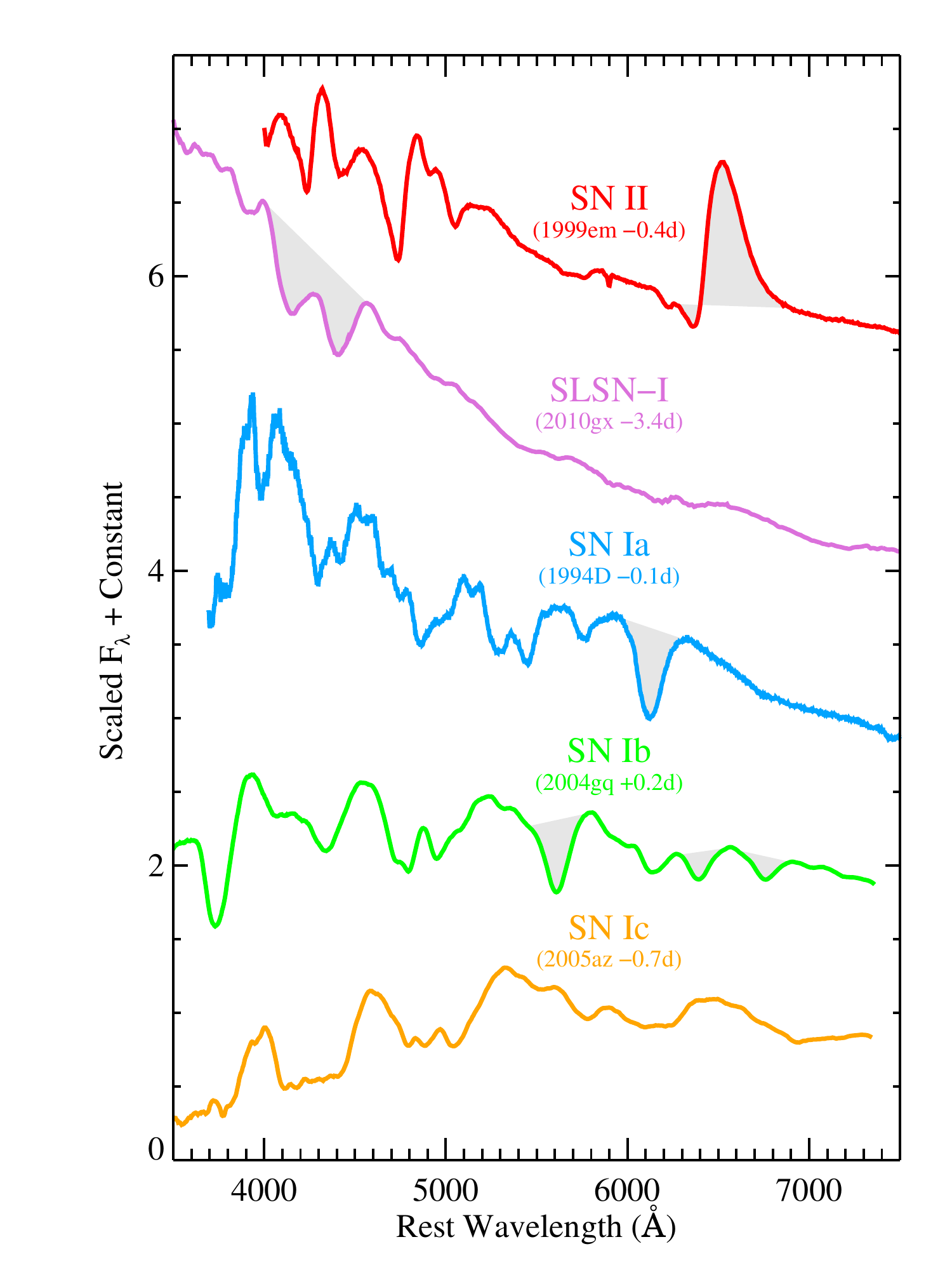}
 \caption{Spectra near peak optical brightness of a SLSN-I compared to
   normal-luminosity SNe. Some key features are shaded in gray for
   emphasis: \ion{H}{1} in the SN~II, \ion{O}{2} in the SLSN-I,
   \ion{Si}{2} in the SN~Ia, and \ion{He}{1} in the SN~Ib. Except for
   SN\,1994D, the data have been smoothed for clarity.  }
   \label{fig:compare_spec}
\end{center}
\end{figure}

Observationally, SNe are sorted into a number of different
types primarily by their spectra \citep{filippenko1997, galyam2016}. A
supernova (SN) is classified as Type~II if it exhibits obvious hydrogen
features in spectra taken near maximum light, Type~Ia if hydrogen is
lacking but \ion{Si}{2} is strong, Type~Ib if hydrogen is lacking,
\ion{Si}{2} is weak, and helium lines are well detected, and finally
Type~Ic if none of these classifications hold (we will use SN~II, SN~Ia,
SN~Ib, and SN~Ic to refer to these spectral types, respectively; see
Fig.~\ref{fig:compare_spec}). There are further refinements of this
classification scheme for objects with relatively narrow emission features 
(SN~IIn and SN~Ibn), transitional objects (e.g., SN~IIb), and sometimes 
objects are subclassified by their light-curve properties (e.g., SN~II-P 
and SN~II-L).

Normal SNe typically have optical luminosities in the $-14 < M < -20$\,mag
range \citep{li2011a, richardson2014}. The SLSN label has
traditionally been assigned to events with peak absolute magnitudes
brighter than about $M<-21$ in the optical \citep{galyam2012}.
Many papers have been published on specific SLSN events
(e.g., \citealt{hatano2001, smith2007, quimby2007c, galyam2009,
  barbary2009, quimby2011, chomiuk2011, rest2011, leloudas2012,
  howell2013, nicholl2014, benetti2014}), and there is a growing
number of papers exploring the diversity of the population
(e.g., \citealt{inserra2013, nicholl2015, lunnan2017, decia2017}). The
SLSN group may now include over 100 distinct events\footnote{For example, 
see the  Open Supernova Catalog listing at
  https://sne.space/?claimedtype=SLSN.} \citep{guillochon2017}, but
this is just 0.26\% of all reported SNe --- a testament to the low
volumetric rates at which SLSNe are produced \citep{quimby2013,
  prajs2017}.

SLSNe with obvious spectroscopic evidence for hydrogen near maximum
light have been classified as SNe~II (or SLSNe-II to highlight their
extreme luminosities), while others lack the defining features noted
above and fall into the default SN~Ic (or SLSN-I) category. Some
initially hydrogen-poor SLSNe develop hydrogen features in their later-time
spectra \citep{yan2015, yan2017b}, although these are usually classified as
SLSNe-I. Additionally, a SLSN-R class has been introduced
\citep{galyam2012}, but this may not be distinct enough from SLSNe-I to
warrant a separate class \citep{decia2017}.

Since the first examples were published a decade ago, the physical
nature of these objects has been debated. Models developed to explain
normal-luminosity events ($M>-20$\,mag) cannot easily be stretched to
account for the immense energies released by SLSNe
(the radiation budgets alone can exceed $10^{51}$\,erg;
e.g., \citealt{chatzopoulos2011}), so new power sources have been
sought.

Among the first models to be considered were the pair-instability
explosion models that had been developed to predict the deaths of the
first stars \citep{fowler_hoyle1964, barkat1967}. These models
initially assumed zero-metallicity progenitors, but they have
been compared to explosions in the $z\approx0$ universe
(e.g., \citealt{smith2007, galyam2009}). Not all agree that these
models explain the data, however (e.g., \citealt{dessart2012,
  jerkstrand2017}). Nonetheless, recent developments in stellar
evolution theory incorporating rotation have pointed to possible
avenues for stars with the required, extremely-massive cores to exist
in the modern universe \citep{yoon_langer2005, woosley_heger2006,
  yusof2013}, and this is supported by observations
\citep{crowther2016}, so this progenitor model continues to be
explored (e.g., \citealt{whalen2014, kozyreva2017}).

Most of the hydrogen-rich SLSNe-II show time-variable, narrow emission
lines in their spectra that indicate a relatively slow-moving wind is
being overtaken by fast-moving ejecta
(e.g., \citealt{fransson1996}). This interaction potentially offers an
additional source of power that may help to explain their high
luminosities (e.g., \citealt{smith2007, chatzopoulos2011,
  chevalier_irwin2011}). SLSNe-I do not exhibit these tell-tale spectral
features (e.g., \citealt{pastorello2010, quimby2011,
  inserra2013}). However, SLSNe-I may yet be powered by interaction if
the wind is very extended and moving at a high velocity
\citep{chatzopoulos_wheeler2012}, or if the wind is depleted of
hydrogen and not photoionized by the SN (see, for example,
the helium or carbon-oxygen wind models of
\citealt{tolstov2017}). SLSNe-I also show somewhat weak spectral
features from relatively cool ions that may be diluted by a hot
continuum produced by an underlying interaction \citep{chen2017}. But
the lack of obvious interaction signatures opens the possibility that
different SLSNe are powered primarily through different means (or that
the signs of interaction in most SLSNe-II are a red herring).

An attractive explanation for the unusually high energies
of SLSNe-I is that additional energy is deposited in the ejecta over
time as a nascent magnetar spins down \citep{kasen_bildsten2010,
  woosley2010}. Although the details are lacking on how this spindown
energy is injected into the ejecta, the bolometric evolution of
several SLSNe-I has been fit with this model yielding a plausible range
of initial spin periods, magnetic fields, and ejecta masses
\citep{inserra2013, liu2017, nicholl2017}. Although the light curves
of some SLSNe-I can be fit with the magnetar model, other power sources
have been shown to fit certain events as well or better
\citep{chatzopoulos2013}, and thus photometry alone has not ended the
debate over what powers SLSNe-I.

However, there are other potential tests to discriminate between
SNe powered internally from magnetar spindown or centrally
located $^{56}$Ni, externally from the outer ejecta interacting with
slower-moving material, or from a combination of magnetar, $^{56}$Ni,
and interaction power sources. The wealth of information provided
through spectroscopy may hold the key to separating these models. For
example, if SLSNe are powered primarily by an exceptionally large
yield of $^{56}$Ni, the spectra may show evidence for an unusually
large amount of iron-peak elements \citep{dessart2012}. Moreover,
if energy is added from within, then this could change the velocity
structure of the ejecta by accelerating the slowest-moving material in
the interior and forming a bubble of evacuated space, similar to a
pulsar wind nebula \citep{metzger2014}. The interaction model can
similarly result in a shell-like structure, but in this case the
interior velocity structure should retain the homologous expansion
velocities rendered from the explosion. Thus, the velocity evolution
and the final velocity distribution at late times may serve to
distinguish the magnetar model from the interaction model.

To determine the velocity structure of the ejecta, the ions responsible
for the spectroscopic features must be properly
identified. Significant work has been done on identifying the features
in normal-luminosity SN spectra, but the application of this
work to SLSNe-I is complicated by two issues: (1) the strength of
spectral features tends to be much lower in SLSNe-I than in SNe~Ic
(e.g., see Fig.~\ref{fig:compare_spec}), and (2) owing to ionization and
possibly composition differences, there are likely features in the
SLSN-I spectra that are not present in normal-luminosity SNe,
and these may blend with or totally dominate the normal features. The
latter is certainly true for the \ion{O}{2} ion, which dominates the
optical spectra of most young SLSNe-I \citep{quimby2011} but which is
not typically seen in lower-luminosity events; two notable exceptions,
SN 2008D \citep{soderberg2008,mazzali2008,modjaz2009} and OGLE-2012-SN-006
\citep{pastorello2015}, are discussed below. These features offer
the only means to extract velocity information from the spectra in
some cases --- but, unfortunately, these features are the product of many
blended lines \citep{mazzali2016}, and a simple method for extracting
velocity information from them has yet to be developed.

The connection between normal-luminosity SNe~Ic and SLSNe-I may also
help constrain the source of power. The nascent magnetars may
serve a wide range of power to the SN ejecta depending largely
on the initial spin period and magnetic-field strength
\citep{kasen_bildsten2010, woosley2010}. Many of these combinations
may result in relatively low spindown luminosities. Because SLSNe have
traditionally been selected by their luminosities
(e.g., \citealt{galyam2012}), there may be an artificial division
between the fainter SNe that come from these conditions and
the high-luminosity objects that result from more optimal initial
conditions, even though the two are physically related. However, if
such a power source is present and significant, then the velocity and
perhaps ionization evolution may still be detectable in the
lower-luminosity events. Yet to discover this connection, a method to
identify SNe physically similar to SLSNe-I but with lower
luminosities must be developed.

In this paper, we discuss a method to select objects spectroscopically
similar to the published sample of SLSNe-I independent of their light
curves. We show in \S\ref{selection} that although at certain
light-curve phases SLSNe-I have spectral features that are similar to
those of normal-luminosity SNe~Ic at other light-curve phases, the spectra of
SLSNe-I tend to be better matched to the spectra of other SLSNe-I (and
not SNe~Ic) over a wide range of phases. The two groups can thus be
spectroscopically divided. In \S\ref{obs} we apply this classification
scheme to the spectra of 1815 SNe discovered by the Palomar
Transient Factory and present 19 objects that are best classified as
SLSNe-I and an additional 4 objects that are possibly SLSNe-I. In
\S\ref{sequence} we consider 133 spectra of SLSNe-I taken over a
variety of phases. These spectra are organized into a sequence and we
assign spectral phases, $\phi$, to SLSN-I and SN~Ic spectra by matching
these data to our SLSN-I sequence. We also discuss in this section the
clustering of objects as more similar to PTF12dam or SN\,2011ke. In
\S\ref{compare_spec} we compare the mean spectral properties of
SN\,2011ke-like and PTF12dam-like events at four different spectral
phases and note some potential differences. Individual spectra are
examined in \S\ref{lineids} to identify line features. We pay
particular attention to \ion{O}{2}, identify \ion{Mn}{2} with high
probability for the first time, and note the presence of obvious
hydrogen and helium lines in PTF10aagc and PTF10hgi,
respectively. With secure line identifications in hand, we present the
\ion{O}{2} and \ion{Fe}{2} velocity evolution of PTF12dam in
\S\ref{velocities}. We discuss our findings and provide a summary of
our conclusions in \S\ref{conclusions}.

\section{Spectroscopic Selection of SLSNe-I}\label{selection}

SNe are usually classified by matching their spectral features to
objects of known types.  There are three different techniques used for
spectral matching of SNe: $\chi^2$ minimization, cross correlation,
and ``feature'' matching in a series of wavelength bins (e.g.,
\citealt{riess1997, harutyunyan2008}). We choose to use the $\chi^2$
minimization routine, {\tt superfit} \citep{howell2006}, for our main
analysis and we test our findings using a custom cross-correlation
code based on {\tt SNID} \citep{blondin2007}. Briefly, {\tt superfit}
compares an input spectrum to a library of template SN spectra. Each
template is reddened (or dereddened) using a Cardelli extinction law
\citep{cardelli1989}, a galaxy template is added to account for any
host-galaxy contamination, and the $\chi^2$ value of the model fit is
determined. The templates are then rank ordered by their $\chi^2$
values. We describe below how we build our comparison spectral library,
account for the varying rest-frame wavelength coverage of these
templates, and then interpret the match results to derive a final
spectral classification for an input object.

\subsection{Spectral Template Library}

The library of template spectra used for the fitting is naturally an
important concern.  In principle, it would be best to have theoretical
models with uniform wavelength and phase coverage for all possible
classes of SNe. However, there are very few models available
today that are sufficiently realistic for our needs. We are thus
forced to base our spectral templates on observations.
 
For our analysis, we constructed a new library of spectral templates
based on observations. For the SNe~Ia, we use the CfA
Supernova Archive, which includes 1924 unique observations of 221 SNe~Ia
with spectroscopic subtypes identified (\citealt{blondin2012}; see
also the Berkeley Supernova Ia Program, \citealt{silverman2012}). We
similarly used the CfA Supernova Archive's 480 spectra of 44 
stripped-envelope core-collapse SNe including Types Ib, Ic, and the more
rare Ic-bl, IIb, and Ibn, with light-curve phase information
\citep{modjaz2014}. To these we added 107 well-observed spectra of
stripped-envelope core-collapse SNe, as well as 373 spectra of 22 SNe~II 
and SNe~IIn, from a
number of sources (see Table~\ref{table:libraries}). These data were
downloaded from WISeREP\footnote{https://wiserep.weizmann.ac.il/}
\citep{yaron_galyam2012}. In total, our library contains 2884 spectra
of normal-luminosity SNe.

\input{library_refs.tex} 
\input{ref.SLSN-I.table.tex} 

Next, we need a library of SLSN spectral templates. For these we use 92
publicly available spectra of 20 objects published before early 2015
as SLSNe-I or SLSNe-II. We further include 31 PTF spectra for three of
these objects, which we publish here for the first time (see \S\ref{obs}). 
Our SLSN template library is thus based on events that
have been classified as SLSNe from their luminosities.

For all of our spectral libraries, we must remove features in the data
that do not relate to the SNe, including cosmic rays, telluric
features, and narrow spectral lines from gas in the host galaxy. We
first remove narrow lines (only those unrelated to the SNe) by
fitting Gaussian profiles at the expected locations of typical 
host-galaxy lines. We perform this fit section by section with all lines in
a given section constrained to have the same full width at half-maximum
intensity (FWHM) in the fit. We then
remove only lines that were significantly detected according to the
fit. We also remove telluric features using a high-resolution telluric
spectrum degraded to the approximate resolution of the spectra. Given
changes in atmospheric conditions this fitting is imperfect, but it can
greatly reduce the effects of telluric features in spectral templates
for which the reducers have not already removed these features (this
practice is unfortunately not standard). We then smooth the templates
using a generalized Savitzky-Golay filter as described in
Appendix~\ref{sgfilter} (this smoothing step was not done for the SN~Ia
spectra, but a large fraction of these data are high signal-to-noise
(S/N) ratio observations that would not benefit from smoothing; e.g., see
Fig.~\ref{fig:compare_spec}). For templates lacking error spectra, we
estimate the error spectra by fitting out the broad SN features
with an iterative B-spline fit. The error is then estimated from the
standard deviation of the B-spline-subtracted spectrum in several
intervals and then interpolated to the entire spectral range.

\subsection{Wavelength Range for Spectral Matching}

A significant problem in SN template libraries is that the
libraries are constructed from observations and thus do not uniformly
cover all of the desired parameter space. First, the libraries include
objects at different redshifts and they are observed with different
instruments. Thus, the rest-frame wavelength coverage varies
significantly from spectrum to spectrum. Second, we have an order of
magnitude more SN~Ia templates than the other object types. Likely our
templates do not account for the full diversity of each class of
object (except possibly for the SNe~Ia). As discussed below, this may
result in false matches with the wrong object type. Third, the
templates are by no means evenly distributed with respect to light-curve 
phase. Gaps in the temporal coverage of one type of object may
again lead to matches skewed to another object type with better
coverage at the relevant light-curve phase. A final problem of note is
that some templates have more noise than others, and some may be
contaminated by host-galaxy lines, telluric features, cosmic rays, or
other artifacts. Below we discuss how we account for these potential
problems.

\begin{figure}
\begin{center}
 \includegraphics[width=\linewidth]{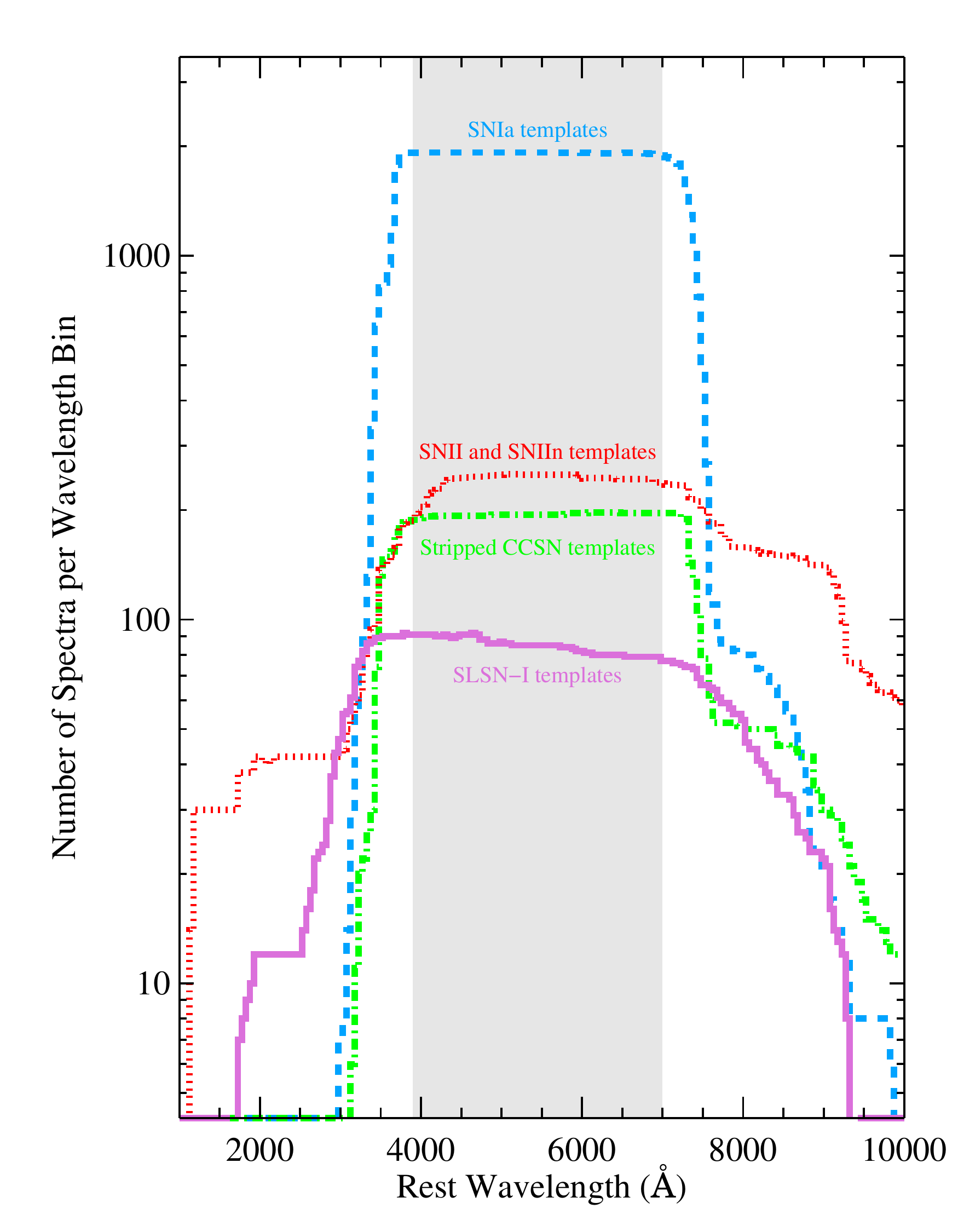}
 \caption{Wavelength coverage of our spectral template libraries
   (number of templates per wavelength bin). }
   \label{fig:coverage}
\end{center}
\end{figure}

Figure~\ref{fig:coverage} shows the number of spectral templates in
our libraries for different types of SNe as a function of
wavelength. Most of the normal-luminosity SNe are low-redshift objects
with only ground-based optical coverage; a notable exception is
SN~1987A, which has a number of ultraviolet (UV) spectra.
Consequently, there are few templates of normal-luminosity SNe with
rest wavelengths below about 3500\,\AA. Spectroscopic coverage of
these objects is limited to $\lambda < 7500$\,\AA\ as well, since this
is the effective limit of the FAST spectrograph \citep{fabricant1998}
at the Fred Lawrence Whipple Observatory, which supplied most of the
observations.

In contrast, SLSNe-I are typically found at significantly higher redshifts 
(e.g., $z \approx 0.3$); hence, ground-based follow-up observations naturally
cover shorter rest-frame wavelengths. The SLSNe-I in the template
sample were also more frequently observed with dual-channel
instruments offering superior blue and red wavelength coverage. A
greater fraction of SLSNe-I thus have coverage below 3000\,\AA\ and above
8000\,\AA\ than the SN~Ia and stripped-envelope core-collapse 
samples from the CfA archive.

Based on the wavelength coverage of our spectral templates shown in
Figure~\ref{fig:coverage} and the spectral features of various
SNe shown in Figure~\ref{fig:compare_spec}, we choose to input
only the rest-frame 3900--7000\,\AA\ range of our test spectra to our
template matching codes. This way, templates with better spectral
coverage will not be artificially favored. If we did not do this, then
a SLSN-I spectrum might be much more likely to match other SLSN-I
templates simply because many of the other templates would be rejected
owing to insufficient wavelength coverage.

\subsection{Automated Template Matching}

Typically, the final classification of SN spectra is left to
human judgment. Template matching is performed using an automated
code, which returns a rank-ordered list of possible matches. However,
the top match from the template libraries is not guaranteed to have
the same spectral classification as the input spectrum. The input
spectrum may be for a subtype or taken at a phase that is missing from
the libraries. More generally, the libraries may not fully account for
the diversity of SN spectra. Because of this and possible
systematic errors in the libraries and input spectra (e.g., imperfect
calibration, telluric removal, or the presence of artifacts), 
template-matching codes serve as imperfect tools that require human oversight.

A disadvantage of human interaction in spectral classification is that
it can be subjective. Different groups may disagree with the
interpretation of the output from spectral matching codes (for
example, see the difference in opinion on the classification of
PS1-10afx; \citealt{quimby2013,chornock2013}). In this work, we favor
a classification scheme that is (mostly) free of human interpretation
and should thus be readily reproducible by others using the same
technique. We show below that although the top match output by a
spectral matching code may not always belong to the same class as the
input object, it is the case that the top matches belonging to the
correct class tend to be systematically higher in the rankings than
they would be for input objects of different types. We can thus
compare how highly ranked the top matches of each class are and
compare the differences in these average scores to a training set to
determine the true classification (with quantifiable
uncertainty). With a sufficiently large set of library spectra and careful
consideration of sample bias, spectral classification can be
automated.

With the smoothed spectral template libraries in hand, we can now
check if SLSN-I spectra at various phases are equally well matched by
SLSN-I and SN~Ic templates or if the spectra alone indicate separate
populations. To do this test, we input each spectral template
individually into {\tt superfit} and determine how well these spectra
match each of the smoothed spectra in our template libraries. For each
spectrum we exclude matches to any templates of the same object in our
libraries and create a rank-ordered list of the best-matching
templates. For this test we artificially redshift the input spectra to
$z=0.1$ for convenience. This choice ensures that matches to the SN
templates, which include objects spanning a wide range of velocities,
will always result in a positive redshift. For the SNe~Ia, we use only
the elliptical galaxy template to account for host-galaxy
contamination (most of the templates have little or no host-galaxy
contamination), we set the allowed range of $A_V$ from $-2$ to 2\,mag,
and we fix the redshift search range to $0.07 \le z \le 0.13$ in steps
of $\delta z=0.01$. (Note that the $A_V$ parameter is normally
intended to account for reddening by the host galaxy, but it can also
be used to adjust for intrinsic color differences of SNe themselves;
thus, negative values are acceptable.)  For all other SNe, we use the
Sc galaxy template, allow $A_V$ to vary from $-4.5$ to 2\,mag, and fix
the redshift search range to $0.03 \le z \le 0.17$ (the larger $A_V$
range helps to account for the greater intrinsic color range of
this group; see \S\ref{slsnIvsIc}). These values were selected
considering the range of expansion velocities and reddening in our
libraries and after a number of {\tt superfit} trials. As noted
earlier, the rest-wavelength range for the input templates is fixed to
at most 3900--7000\,\AA, and we only include templates that cover 95\%
or more of this wavelength range. All other {\tt superfit} parameters
are left at their default settings.

\subsection{Testing the Automated Spectral Classification}

\begin{figure}
\begin{center}
 \includegraphics[width=\linewidth]{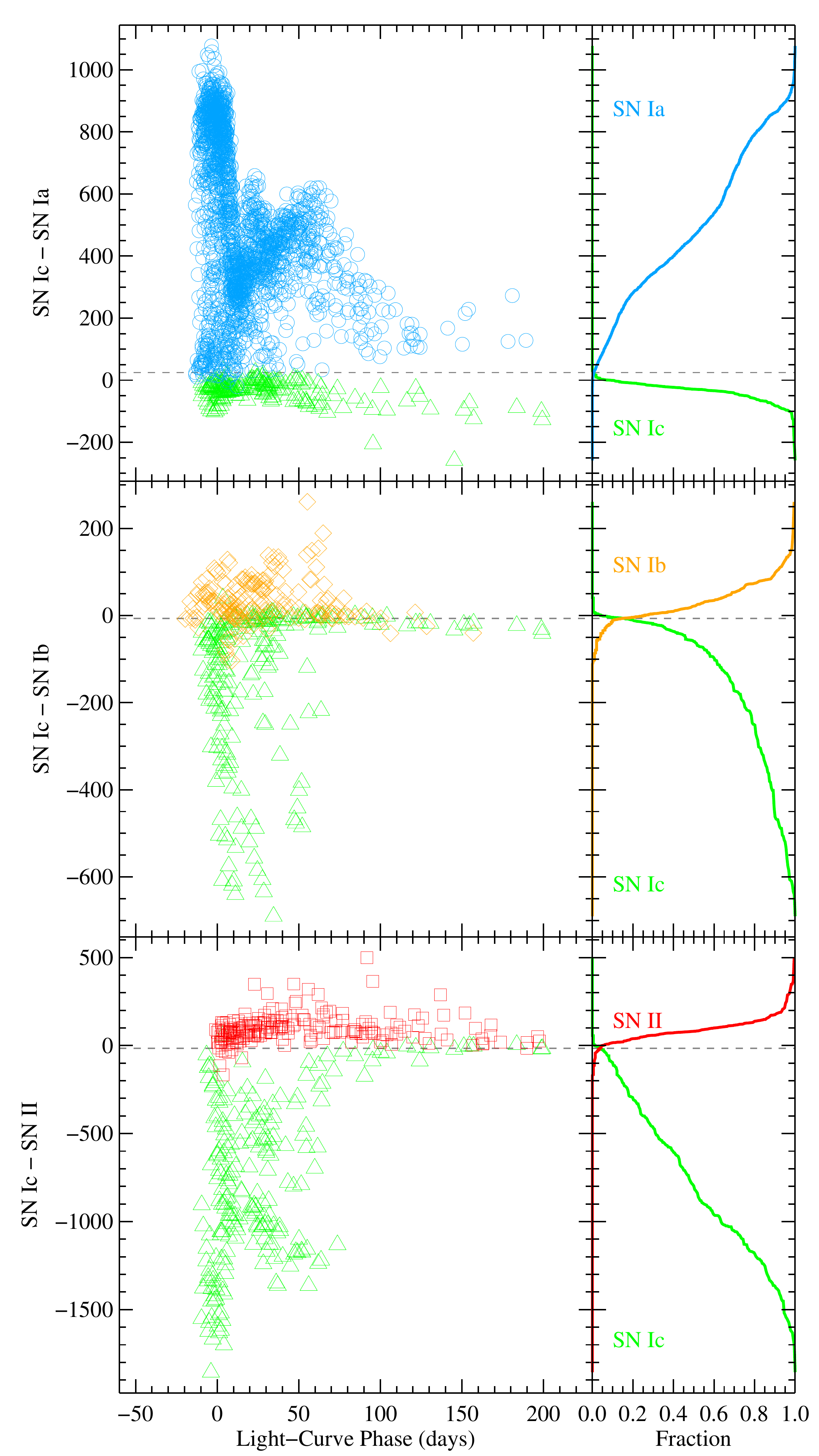}
 \caption{The population of SNe~Ic can be separated from SNe~Ia, SNe~Ib, 
   and SNe~II using the difference in the average {\tt superfit} index of
   the best-matching templates from our spectral libraries. In each
   panel, the ordinate is computed by taking the indices of the top 5
   matches from the rank-ordered list output by {\tt superfit} and
   subtracting the average of these for a given SN type from
   the average of the top 5 SN~Ic indices. Plots in the left column
   show this index difference as a function of light-curve phase,
   which is days from explosion for SNe~II and days from maximum light
   for all other types. The plots in the right-hand column show the
   cumulative fractions of observations with index differences greater
   than (for the SNe~Ic) or less than (for the other types) a given
   value.  }
   \label{fig:stack_plot_SNIc}
\end{center}
\end{figure}

In Figure~\ref{fig:stack_plot_SNIc}, we show that with these settings
and our template libraries, we can distinguish the classes of
normal-luminosity SNe. The figure shows $\Delta I_{\rm Ic - X} =
\langle I_{\rm Ic} \rangle - \langle I_{\rm X} \rangle$, the
difference between the average index of the top 5 SN~Ic template
matches found by {\tt superfit}, $\langle I_{\rm Ic} \rangle$, minus
the average index, $\langle I_{\rm X} \rangle$, for SN~Ia, SN~Ib, or
SN~II templates. Here ``index'' just means the ranking of the template
as determined by {\tt superfit}, with larger indices indicating worse
matches. For example, if the top 5 matches found by {\tt superfit} are
all SN~Ia templates, then the average index will be $\langle I_{\rm
  Ic} \rangle = 2.0$ (the indices are zero-indexed; that is, the index
of the best-matching template is zero.). Lower values for $\Delta
I_{\rm Ic - X}$ indicate that the SN~Ic templates tend to be more
highly ranked. In each case it is apparent that the SNe~Ic can be
distinguished from the other populations; this is readily confirmed by
a formal Kolmogorov-Smirnov (KS) test.

We plot the $\Delta I_{\rm Ic - X}$ values as a function of light-curve 
phase. This allows us to demonstrate some of the degeneracy
between different SN types. For example, SNe~Ia are more easily
distinguished from SNe~Ic at maximum light than they are two weeks later
when the Si~II line at 6150\,\AA\ weakens and the overall appearance becomes 
more SN~Ic-like. The division between SNe~Ic and SNe~Ib is less distinct
especially at later phases, when the helium lines have weakened and the
available spectra are limited. Obviously the distinction
between classes can be most strongly made in this analysis at phases
where there are larger numbers of library templates in each group.

Marginalizing over light-curve phase, we can determine the fraction of
spectra with $\Delta I_{\rm Ic - X}$ values less than a given
value. The figure shows this for the SNe~Ic and the fraction with
$\Delta I_{\rm Ic - X}$ greater than the given value. We can then
determine the cutoff value where the fraction of SNe~Ic above the cutoff
equals the fraction of the SNe~Ia, SNe~Ib, or SNe~II below the cutoff. This
is shown with the horizontal dashed lines in the figure. For objects
with multiple spectra, we can average the results to determine the
overall classification. The contamination rate is estimated from the
fraction of the populations that are incorrectly classified by this
cutoff. For example, for SNe~Ic the contamination rates of SNe~Ia, SNe~Ib,
and SNe~II are 0.8\%, 15.1\%, and 3.7\%, respectively. We use this
cutoff value to classify each object as more SN~Ic-like or more like a
SN~Ia, SN~Ib, or SN~II.

The final result of this procedure is a calibrated system to classify
SN spectra. If we use the same procedures with the same
template libraries to find the best matches to a new SN, we can
determine the $\Delta I_{\rm Ia - X}$, $\Delta I_{\rm Ib - X}$ (etc.)
scores and compare these to our calibrated cutoff scores to find if
this new object is most similar to a SN~Ia, SN~Ib (etc.). The population
overlap from the template libraries then gives an indication of how
trustworthy this classification is. For example,
Figure~\ref{fig:stack_plot_SNIc} shows that a $\Delta I_{\rm Ic - Ib}$
score of $-300$ would signify that the object is definitely more
SN~Ic-like than SN~Ib-like (none of the library SN~Ib templates scores
this low), while a score of $-10$ would indicate an ambiguous
classification (11\% of the library SNe~Ib have $\Delta I_{\rm Ic - Ib}$
scores this low or lower).

We can test this classification system using the template
libraries. Of the 302 objects tested (we exclude the SLSNe-II
SN\,2006gy and SN\,2008am), we find 295 (98\%) are classified in
agreement with the published types by the process described above. Of
the 7 that are classified differently, 4 are SNe~Ia that are
incorrectly classified as SLSNe-I. As discussed below, this is to be
expected given the large SN~Ia sample and the slight overlap between
the populations, but such interlopers can be removed based on other
metrics. Another difference is that the lone spectrum of the
SN~1991bg-like (peculiar SN~Ia; e.g., \citealt{filippenko1992})
SN\,2006em \citep{gonzalez-gaitan2011} is found to be marginally more
consistent with a SN~Ic strictly following our method above (the top
{\tt superfit} matches for this object are all SNe~Ia-CL, but there
are a few matches to the SN~Ic 2004aw relatively high in the ranking
that throw off the classification).  Additionally, the Type~Ic SNe
2004dn and 2005kl are classified as SNe~Ib through our method. For the
latter, at least \citet{modjaz2014} find the spectral
classification to be ambiguous because data were not taken in the
phase when the helium lines are readily visible, so the SN~Ib
classification may be accurate. However, SN\,2004dn was firmly
classified as a SN~Ic by \citet{sun_galyam2017}.

\subsection{SLSN-I Spectra Differ from Normal-Luminosity Events}\label{slsnIvsIc}

Having demonstrated that normal-luminosity SNe can be
accurately classified using their $\Delta I_{\rm X-Y}$ scores, we turn
to SLSNe. We use the procedure outlined above
and test the $\Delta I_{\rm SLSN-I-X}$ values for each of the
normal-luminosity types. Given the importance on the number of
template library spectra demonstrated above, we deem the 22 SLSN-II
spectra insufficient for this procedure. Thus, we cannot automatically
distinguish between SLSNe-II and SLSNe-I (or any other type). Lacking
this ability, we can still determine if an object is best matched by
SLSN-I templates and then visually check for the obvious presence of
hydrogen in the spectra to determine if it should properly be
classified as a SLSN-II.

\begin{figure}
\begin{center}
 \includegraphics[width=\linewidth]{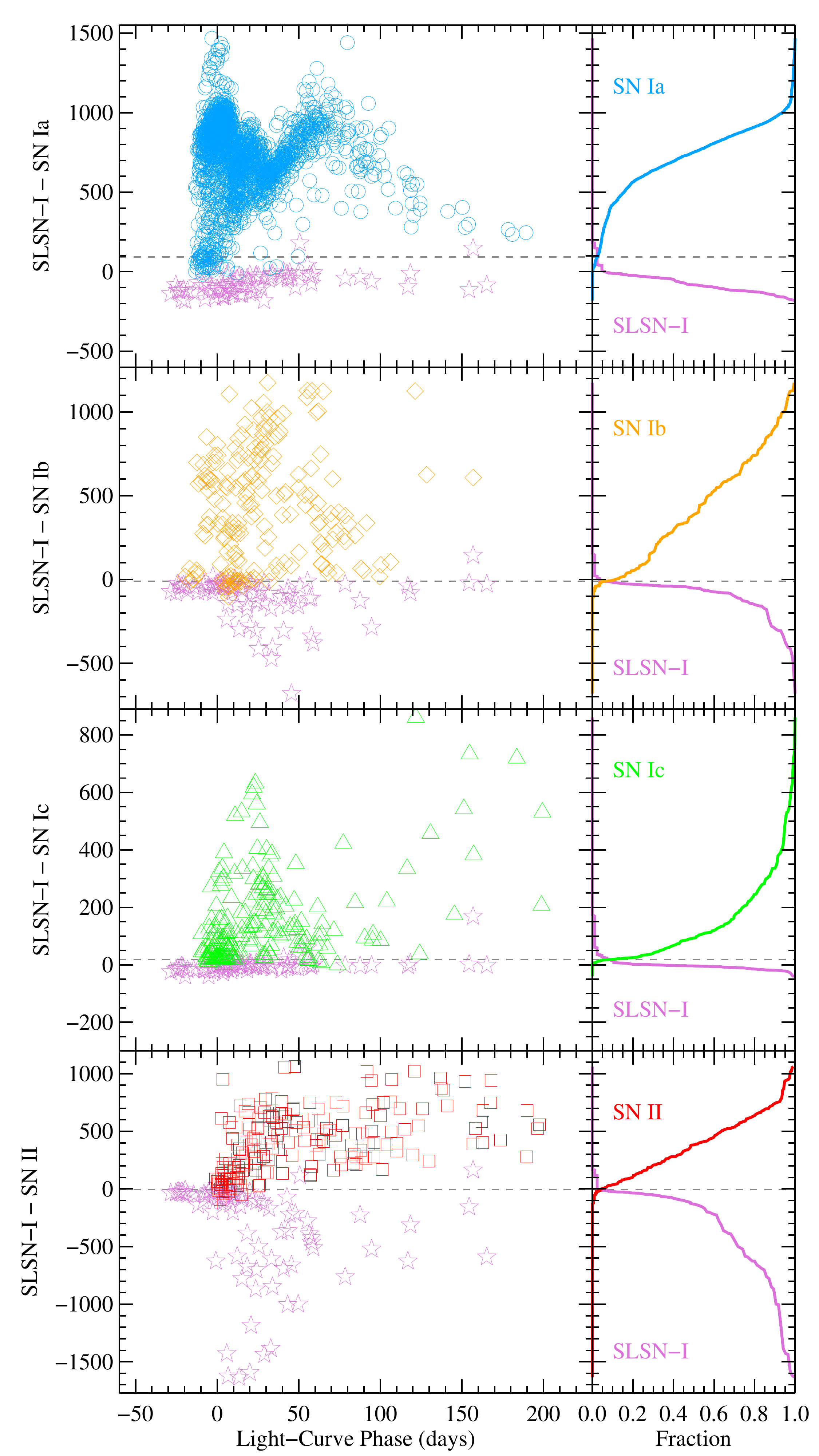}
 \caption{Similar to Figure~\ref{fig:stack_plot_SNIc} but for SLSNe-I
   compared to other types of SNe.  }
   \label{fig:stack_plot_SLSN-I}
\end{center}
\end{figure}

Figure~\ref{fig:stack_plot_SLSN-I} shows the $\Delta I_{{\rm
    SLSN-I}-{\rm X}}$ values for the 118 spectra of confirmed SLSNe-I
along with the same index difference for SNe~Ia, SNe~Ib, SNe~Ic, and
SNe~II. In each case we find that the SLSN-I population is clearly
offset from the normal-luminosity events. Perhaps of most interest,
the SLSN-I group can, in fact, be distinguished from normal-luminosity
SNe~Ic based only on their spectra (see \S\ref{compare_spec} for a discussion
of line features that contribute to this division).
The contamination rate is only 8.5\%
and the $p$-value from a formal KS test is $10^{-43}$, which strongly
rejects the null hypothesis that SN~Ic and SLSN-I spectra are drawn
from the same parent population. This implies at a minimum that the
spectra of SNe~Ic in the classical sense (lacking hydrogen, strong
silicon, and strong helium lines) carry information about
the luminosity of the SN. It is no surprise that the SLSNe-I
observed at early light-curve phases, which are dominated by
\ion{O}{2} lines in the wavelength range considered, can be
distinguished from ordinary SNe~Ic, which never show \ion{O}{2}
features. However, the spectroscopic distinction persists to later
phases when the \ion{O}{2} lines vanish and the spectra of SLSNe-I have
been shown to be similar to those of lower-luminosity SNe~Ic
\citep{pastorello2010}. In particular, at later phases SNe~Ic like
SNe~1994I, 2004aw, and 2002ap appear to prefer matches to other SNe~Ic and 
SNe~Ib more than to SLSNe-I (see Appendix~\ref{latematch}).

Most SNe~Ib can be trivially delineated from SLSNe-I with the notable
exceptions of SN\,2005la and SN\,2006jc, which carry the formal
classification of Type~Ibn as their spectra exhibit narrow lines
(interpreted to be the consequence of interaction with slowly moving
CSM; SN\,2005la is sometimes given the transitional label
Type~Ibn/IIn; \citealt{pastroello2008}). Similarly, most SNe~Ia are
well separated from the SLSN-I population except for the SN~Ia-SS
subtype (the SS, CN, BL, and CL subtypes of SNe~Ia are defined by
\citealt{blondin2012}). From maximum light down to about 10 days
before, these objects have muted spectral features including weak
Si~II $\lambda$6150, which are not so dissimilar to SLSNe-I. However,
shortly after maximum light SN~Ia-SS objects become clearly more
SN~Ia-like than SLSN-I-like. For completeness, we note that a few
SN~Ia-CL templates have low $\Delta I_{\rm SLSN-I-SNIa}$ values near
maximum light, but higher values at other phases. The SLSN-I and SN~Ia
populations are still clearly offset across all phases shown in
Figure~\ref{fig:stack_plot_SLSN-I}, but the simple, phase-independent
cutoff will consequently yield greater contamination from SNe~Ia in
the SLSN-I category.

\begin{figure}
\begin{center}
 \includegraphics[width=\linewidth]{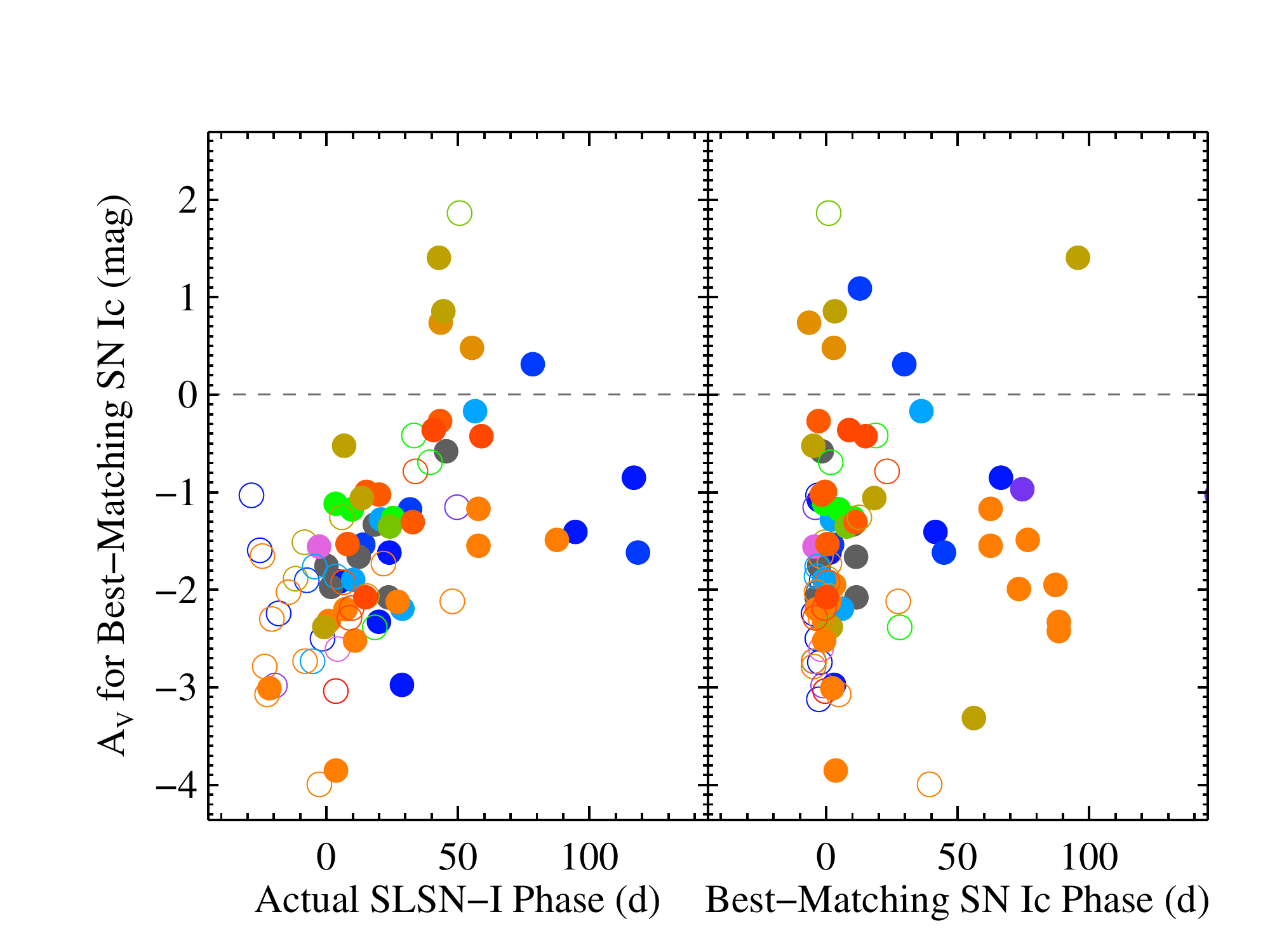}
 \caption{ Average $A_V$ values found by {\tt superfit} for the top three
   SN~Ic templates that best match SLSNe-I as a function of the actual
   light-curve phase (left panel) and the light-curve phase of the
   best-matching SN~Ic template (right panel). Filled circles mark
   points where the best-matching SN~Ic templates are among the top 10
   matches returned by {\tt superfit} on average; open circles mark
   spectra that were less well fit by SN~Ic templates. The symbols are
   color coded by object as in other figures.}
   \label{fig:av_plot}
\end{center}
\end{figure}

Of all the normal-luminosity SN types, SNe~Ic provide the best matches
to SLSN-I spectra. However, there is a qualitative difference in these
spectra: SLSNe-I typically have much bluer continuum slopes. This can
be quantified using the $A_V$ parameter output by {\tt superfit}. In
Figure \ref{fig:av_plot} we show the best-fit $A_V$ values for the
best-matching SN~Ic templates. For the continuum slope, the SN~Ic
templates must be corrected by $A_V \approx -2$\,mag to match SLSNe-I
out to about 1 month past maximum light. After SLSNe-I have evolved
1--2 months past maximum light the mean $A_V$ values approach zero,
although there is considerable scatter.

Based on the spectral template matching, we also find that SLSNe-I are
more likely related to SNe~Ic-bl than to ordinary SNe~Ic (see also
\citealt{liu2016}). Given only the choice of SN~Ic or SN~Ic-bl
templates --- that is, judging only by the $\Delta I_{\rm SNIc-SNIc-bl}$
scores --- we find that 10 of the 15 SLSNe-I in our reference sample are
better sorted into the SN~Ic-bl group by the method described above.

\section{Observations}\label{obs}

Now that we have a method to classify objects as SLSNe-I based only on
spectra, we turn to the full sample of SN spectra gathered by
the Palomar Transient Factory. We will use the techniques described in
\S\ref{selection} to identify which PTF objects are
spectroscopically similar to SLSNe-I.

\subsection{PTF Survey and Follow-Up Observations}

The Palomar Transient Factory is a wide-field photometric survey for
time-variable objects \citep{rau2009,law2009}. Objects were detected
using the 1.2\,m Oschin Schmidt Telescope at Palomar Observatory from
early 2009 through 2012 (the survey was extended into 2017 as the
Intermediate Palomar Transient Factory to bridge the time gap before
the Zwicky Transient Facility). A reengineered version of the
CFHT-12k camera (renamed the PTF camera) surveyed 7.2 square degrees
of sky at $1''$ pixel$^{-1}$ sampling per exposure. In 60~s, typical SDSS-$g$ 
and Mould-$R$ limits were 21\,mag and 20.5\,mag, respectively. The result
of the large field-of-view and depth of the PTF survey, combined with
over a thousand nights of observations, was the discovery of tens of
thousands of candidate transient sources (time-variable, astrophysical
sources that are not associated with point sources brighter than the
survey limit).

To help assess and classify these candidates, the PTF survey was
coupled with an ambitious follow-up program. The neighboring
Palomar~1.5\,m telescope was used for photometric monitoring of
transient sources. Decoupled from the main survey, the Palomar~1.5\,m
was able to reach greater depths, employ more filters, and target
objects at times when the 1.2\,m Oschin was surveying other
fields. Several nights a month were also reserved for spectroscopic
classification of candidates with the Palomar 5.1\,m Hale telescope
and the twin Keck~10\,m telescopes. Additional photometric and
spectroscopic follow-up observations were undertaken with a variety of
telescopes including the twin Gemini~8\,m telescopes, the KPNO~4\,m
Mayall, the WHT~4.2\,m, the NOT~2.5\,m, and the Lick~3.0\,m Shane.

Even with all of these resources, it was impossible to observe every
transient source identified by the PTF survey. During the survey,
various selection processes were used in order to tailor the survey to
specific targets and to maximize the scientific return from the
limited follow-up observations. In general, candidate targets were
first identified from the positive residuals remaining after a
point-spread-function (PSF) matched image template was subtracted from new
observations. Initial, automatic filtering reduced this list to
candidates that were found in at least two images at the same sky
location within the night and that were found to be roughly consistent
with the appearance of a point source. The best candidates were then
vetted by humans who ultimately judged if the sources were likely to
be astrophysical transients. The list of transient candidates was then
passed on to the spectroscopic observes who selected targets to
observe. There was not a consistent rubric for this selection. For
most of the PTF survey preference for spectroscopic observations was
given to targets that appeared to be on the rise, but the final
selection was up to the observer at the telescope, who could
potentially have a bias to a specific type of SN (e.g., some
members of the collaboration were interested in observing SNe~Ia while
others preferred more rare events). The PTF selection function is thus
quite complex but is mostly dictated by the apparent brightness of the
targets: brighter targets are easier to observe so more of them can be
observed on a given night with a given instrument.

After a spectrum of a candidate was obtained, the data were quickly
reduced and posted to a central repository, the ``PTF Marshal,'' where
all collaboration members could access the data and provide
comments. Typically an initial classification was provided within a
day. Further observations could then be requested for interesting
objects or those for which the initial classification proved
ambiguous. As the spectra were obtained under multiple observing
programs on multiple telescopes, there were naturally multiple
pipelines used for the final reduction of the data. Some of these
final reductions were posted to the Marshal shortly after the data
were taken, but often final reductions were not available until an
object was selected for publication. See \citet{galyam2011} for
further details on the PTF candidate vetting process.

\subsection{Spectral Extraction Procedures}

Most of the spectra presented here were extracted using a custom
pipeline implemented in IRAF\footnote{IRAF is distributed by the
  National Optical Astronomy Observatory, which is operated by the
  Association of Universities for Research in Astronomy (AURA), Inc., 
  under a cooperative agreement with the U.S. National Science Foundation (NSF).},
Python, and IDL. The data were typically processed using calibration
frames taken on the same night. The overscan is subtracted from each raw
image and the data are trimmed down to the active window size. The
images are then divided by uniformly illuminated exposures (flat
fields) obtained from either internal lamps or illuminated dome
screens. Next, the background sky including emission lines is fitted
and removed from the corrected two-dimensional (2-D) frames using the 
IDL routine {\tt bspline\_iterfit.pro} following the procedure described by
\citet{kelson2003}. The 1-D spectra are then extracted in the optimal
manner. An initial wavelength solution is found using calibration-lamp
observations. This solution is adjusted based on night-sky lines to
account for flexure (the best fit 1-D sky model above is extracted and
propagated with the target spectra for this purpose). Observations of
standard stars are used to determine the total system throughput
as a function of wavelength, and this is used to produce final, 
flux-calibrated spectra. Observations with Keck/LRIS employed the 
Atmospheric Dispersion Corrector; observations with other instruments
were typically obtained with the slit rotated to the parallactic angle to
minimize differential slit losses \citep{filippenko1982}. This pipeline was used
to extract 1-D spectra from the Keck/LRIS and P200/DBSP observations
obtained under Caltech time. The reduction of other datasets was
typically done through a similar processes.

We also obtained UV spectra of select events with the {\it Hubble Space
Telescope (HST)}. These data were taken as part of the programs
GO-12223 and GO-12524. STIS/MAMA observations from GO-12223 were
extracted using the usual Space Telescope Science Institute (STScI) 
procedures. WFC3 UV grism
observations from GO-12524 were extracted using the {\tt aXe}
package from STScI. For these extractions we use {\tt LACOSMIC} to
identify cosmic rays in individual exposures and then stack the data
with these pixels masked to produce clean images. Source positions
were identified using SExtractor on F200LP and F300X exposures and
then input into {\tt axecore} for extraction. We retain only the
dominant spectral order for our analysis.

\subsection{Spectroscopic Selection of PTF SLSNe-I}

During the survey, several SLSNe-I were identified and monitored.
These include some of the first objects to define the SLSN-I class
(e.g., \citealt{quimby2011}). However, considering our limited
understanding of SLSNe-I in the early days of the survey, there is no
guarantee that all SLSNe-I were correctly identified in time to obtain
further follow-up observations.

With the spectroscopic classification scheme described in
\S\ref{selection}, we can now search through the spectra in the
PTF Marshal to provide a consistent classification of SNe
discovered (or identified) by the survey. We begin with the 3432
spectra of 1815 objects in the PTF sample (2009--2012) that were
initially classified as SNe and for which redshift estimates
are available. For each spectrum we use the {\tt superfit} output and
the cutoff scores from our libraries to determine a classification. We
first assess if the spectrum is more like a SN~Ia or a SN~Ib. We then
determine if the spectrum is more like whichever of these
provides a better match or a SN~Ic. We repeat this process for SNe~II,
SNe~IIb, SNe~IIn, and finally SLSNe-I (as noted above, our libraries
contain too few SLSNe-II for definitive comparison).

Through this process, we initially identify 100 possible SLSNe-I in the
PTF spectral archive. However, we can expect that many of these are
simply SNe~Ia (and SNe~Ia-SS in particular) observed near maximum light. As
noted in \S\ref{selection}, our simple, phase-independent cut
allows a fraction of SNe~Ia (especially SNe~Ia-SS) to spill into the
SLSN-I category. Given the 1783 SN~Ia spectra in the PTF archive and a
contamination rate of 2.3\%, we would expect around 42 SNe~Ia to falsely
be classified as SLSNe-I with our scheme (note that many of the SNe~Ia in
our template library were selected by targeted surveys and the sample
may therefore have a biased distribution of SN~Ia subtypes when
compared to the PTF sample; the PTF sample likely contains a higher
fraction of SNe~Ia-SS). Because the distributions of S/N ratios
and host-galaxy contamination factors likely differ from those of
the training set, which is preferentially populated with well-observed
events, some additional contamination is possible in our first
selection of SLSNe-I.

To remove contaminants from the SLSN-I spectroscopic sample, we
performed several additional checks. Many of the objects in the
initial list had two or more spectra available. In cases where the
individual spectra favor different classifications, we gave weight to
observations of higher quality. Many objects in the initial list had
at least one spectrum with very low S/N ratio that resulted in
ambiguous classifications. As these were often reobserved under
better conditions or with longer exposure times, the later spectra
could be used to more reliably set the classification. We further
removed objects that had $\Delta I$ scores only marginally more
consistent with SLSNe-I than another type and which showed the
characteristic spectral features of the other type. For example,
objects that showed a clear \ion{Si}{2} $\lambda$6150 feature were
reclassified as SNe~Ia. We removed all such marginal SLSN-I candidates
that had SN~Ia (and not SN~Ic or SN~Ib) as the second-best
classification.

After careful consideration of the initial 100 possible SLSNe-I, we
find that 19 of these objects are spectroscopically consistent with
SLSNe-I while another 4 objects are possibly consistent but a
definitive statement cannot be made. In Table~\ref{table:sample} we
list these objects and label them as ``SLSN-I'' and ``possible
SLSN-I'' candidates, respectively. The remaining SLSN-I candidates are
mostly SNe~Ia that passed the automatic cut as described above, but we
also removed several objects with poor-quality spectra. Note that to
this point we have made no conscious use of the photometry or 
host-galaxy properties in identifying our sample; we have simply
constructed a sample of objects that have spectra that are more
consistent with SLSNe-I than with any other type.

The SLSN-I reference sample does, however, contain several PTF objects
(PTF09atu, PTF09cnd, PTF09cwl, SN\,2010gx, PTF10hgi, SN\,2011ke,
PTF11rks, and PTF12dam). As noted above, we always exclude any
spectral matches to templates in our libraries that belong to the
object being classified, but nonetheless the knowledge that these
objects have previously been classified as SLSNe-I could bias our
vetting process. For the 11 new SLSNe-I in the spectroscopic sample, we
did not knowingly use any information other than the spectra for the
final classification, but it should be noted that some of these
objects had also been initially classified as SLSNe-I, and their 
light curve and host-galaxy properties were not formally blinded during the
final vetting process. It should also be noted that there are
additional objects in the PTF archive that have spectra but which were
not classified as SNe or for which redshifts have not been
determined. These spectra are typically of inferior quality and it is
uncertain if they alone can provide useful constraints on the
classification (e.g., the initial quick-look analysis failed to yield a
classification). There may further be human errors, such as spectra
assigned to the wrong object, that can disrupt the classification
process (e.g., see PTF10gvb below).

\subsection{PTF SLSN-I Sample}\label{ptfsample}

Below we give the details for the 19 SLSN-I and 4 possible SLSN-I
spectroscopic samples. Unless otherwise specified, host-galaxy
information is taken from \citet{perley2016} and light-curve peak
dates and absolute magnitudes are adopted from
\citet{decia2017}. Spectral observations are listed in
Table~\ref{table:speclog} and shown in Figures 
\ref{fig:spec01}--\ref{fig:spec11} in the Appendix.

\input{ptf.SLSN-I.table.tex} 

{\bf PTF09q} was found at the beginning of the PTF survey while the
system was still being commissioned. It was actually discovered before
the PTF naming convention had been settled, so it was first announced
as PTF-OT4 \citep{kasliwal2009_atel1} and later given the IAU name
SN\,2009bh \citep{kasliwal2009_cbet1}; for convenience we adopt the
final PTF name (PTF09q) in this work. The object was initially
classified as a SN~Ic. In our reanalysis of this classification, we
find that SN~Ic templates, such as SN\,2005az at $t=-6$\,d, do provide
reasonable matches to the single spectrum obtained, but SLSNe-I, such
as SN\,2011ke at $t=+24$\,d, are possible matches as well (template
phases are taken from the original source; most SNe~Ia, SNe~Ic, and
SLSNe-I phases are relative to $B$-band, $V$-band, and SDSS-$g$
maximum light, respectively). We thus place PTF09q in the possible
SLSN-I sample. The object appears to be hosted by a massive galaxy,
which would be unusual for a SLSN-I (e.g., \citealt{perley2016}), but
there are exceptions to this (e.g., \citealt{nicholl2017b,bose2017}),
and the host galaxy could be star-forming. The photometric peak of
PTF09q is well below the standard $M<-21$\,mag traditionally required
of SLSNe \citep{galyam2012}.

{\bf PTF09as} was identified by the PTF survey on 2009 March 27 (UT
dates are used throughout this paper) and a single spectrum was
obtained on 2009 March 31. A transient source matching the location of
this object was independently reported by CRTS as
CSS090319:125916+271641 \citep{drake2009_atel1} and was later given
the IAU name SN\,2009cb \citep{drake2009_cbet1}. The source was
actually first classified as a SN~Ia by PTF \citep{quimby2009_atel1,
  quimby2009_cbet1}, but our reanalysis casts doubt on this initial
classification. With the large libraries of SN templates now
available, we find that the spectrum of PTF09as is far better matched
to SLSN-I templates, such as SN\,2011ke at $t=+53$\,d. The SN is
apparently hosted by a dwarf galaxy.

{\bf PTF09atu}, {\bf PTF09cwl} (= SN\,2009jh), and {\bf PTF09cnd} have
previously been published as SLSNe-I \citep{quimby2011}. We confirm
that the spectra of these objects are better matches to other SLSNe-I
than to any other SN type, and we present multiple new
spectroscopic observations of each target here for the first time. We
also provide significantly improved extractions of the previously
published spectra for these objects (and for SN\,2010gx below).

{\bf PTF10bfz} was identified by PTF as an optical transient on 2010
Feb. 1, but its spectral classification was not immediately
obvious. Later that year after several spectra had been taken it was
concluded to be a SN~Ic-BL event \citep{arcavi2010}. However, our
reanalysis strongly favors classification as a SLSN-I. The fourth
spectrum in particular is best matched to SLSN-I templates including
PTF11rks at $t=+7$\,d. This classification is supported by an apparent
dwarf host galaxy and a peak brighter than $-21$\,mag.

{\bf PTF10bjp} was identified as a transient candidate on 2010 Feb. 21. Two
spectra were obtained that, although relatively noisy, showed broad
features consistent with those of SNe and of SLSNe-I in
particular. Our reanalysis confirms this source to be a SLSN-I. The
first spectrum is well matched by PTF12dam at $t=+49$\,d. The apparent
host galaxy is a dwarf. The peak absolute magnitude recorded is
$-20.5$, which would be below the traditional cutoff for SLSNe.

{\bf SN\,2010gx} (= PTF10cwr) and {\bf PTF10hgi} have previously been
published as SLSNe-I (e.g., \citealt{pastorello2010,
  quimby2011,inserra2013, chen2013}). We confirm that these sources
have spectra that are best matched by SLSN-I templates through our
automated process. However, PTF10hgi is peculiar in the sense that it
has obvious hydrogen Balmer and \ion{He}{1} lines in contrast to other
SLSNe-I, and thus it may be better classified as a SLSN-IIb. We discuss 
this further in \S\ref{hhe}.

{\bf PTF10gvb} was first identified as a possible SN on 2010
May 6, and it was spectroscopically vetted with LRIS on Keck-I later
that same night. This first spectrum is mostly featureless except for
two broad dips around 5400\,\AA\ and 6100\,\AA\ in the host-galaxy
rest frame. This spectrum is roughly similar to early-phase spectra of
PTF09cnd and PTF12dam, but the PTF10gvb spectra lack the characteristic
\ion{O}{2} features around 4200 to 4500\,\AA. A second spectrum taken
on 2010 May 15 is well matched by the peculiar SN~Ib SN\,2009er near
maximum light and also to various SNe~Ic-bl. A final spectrum taken on
2010 is similar to the SN~Ic-bl SN\,2003jd around three weeks after
maximum light (including data outside the 3900--7000\,\AA\ range, the
phase of the best matches increases to roughly $+50$ days). Based on
this final spectrum it was internally classified as a SN~Ic-bl. However,
the first two spectra are roughly consistent with a SLSN-I, so we
consider this object a possible SLSN-I. We note that when we first
applied our spectroscopic selection process described above, this
object was not identified as a possible SLSN-I because spectra of
another object had been mistakenly included\footnote{We thank Maryam
  Modjaz (private comm.) for pointing out this object as a potential SLSN-I.}. 
The bolometric light curve of PTF10gvb was studied
by \citet{prentice2016}. The host galaxy is blue, of intermediate
mass, and star-forming (Taggart et al., in prep.).

{\bf PTF10nmn} has been presented by \citet{galyam2012} and will be
further discussed by Yaron et al. (in prep.). We confirm that this
object is most spectroscopically similar to SLSNe-I (we do not
use the SLSN-R classification in this work). In particular, the
first spectra are similar to those of PTF12dam at 2--3 months after 
maximum brightness.

{\bf PTF10uhf} first showed a mostly featureless spectrum on 2010
Sep. 8, which was two days after the target was first identified by
PTF. Later spectra exhibit several SN-like features, and we find a
reasonable match to SN\,2011ke at $t=+26$\,d for the second
spectrum. As noted by \citet{perley2016}, the apparent host galaxy of
this target is atypically luminous for a SLSN-I. PTF10uhf reached an
absolute magnitude of $-20.7$, which is below the traditional SLSN
threshold but which would be extremely luminous for a typical SN~Ic.

{\bf PTF10vqv} was announced as a possible SLSN-I similar to PTF09cnd
by \citet{quimby2010_atel2}. Seven spectra were obtained, although
most have relatively low S/N ratios. We can confirm,
however, that the spectra are best matched by SLSN-I templates if the
lowest-quality spectra are ignored. The second spectrum, which is of
reasonable quality, is well matched to PTF09cnd at $t=+28$\,d. The
peak absolute magnitude of PTF10vqv ($-21.6$) and the faintness of the
host galaxy are also consistent with typical SLSNe-I.

{\bf SN\,2010hy} was not discovered by PTF; rather, it was recovered
in PTF survey data after it was first announced by the Lick
Observatory Supernova Search \citep[LOSS;][]{filippenko2001} with the
Katzman Automatic Imaging Telescope \citep[KAIT;][]{kodros2010}, who
noted that the target appeared to be a high-luminosity SN~Ic, although
they could not rule out a SN~Ia classification. Owing to the low
Galactic latitude ($b \approx 7^\circ$), the field was not searched
promptly as was typically the case in PTF. Nonetheless, we report
here on our spectroscopic follow-up observations of SN\,2010hy, which
also has the PTF identifier PTF10vwg. The spectra are heavily reddened 
by dust in the Galaxy, though the spectral features are consistent with
those of SLSNe-I. Our data strongly suggest the source is not a SN~Ia,
but the contrast with SNe~Ic is weaker. The first spectrum is similar
to that of SSS120810 at $t=+59$\,d. Both of our spectra favor SLSN-I
more than any other type, yet the index difference, $\Delta I_{\rm
  SLSN-I - Ic}$, is positive and only just below the cutoff
threshold. But both spectra are below the threshold, so we place this
object in the SLSN-I sample. After correcting for Galactic extinction,
the peak absolute magnitude is about $-22$\,mag, which is well above
the SLSN threshold. The host galaxy is also apparently faint and thus
similar to other SLSN-I hosts.

{\bf PTF10aagc} was flagged as a transient event on 2010 Nov. 3. From
the first spectra, obtained the following night, the target was
identified as a possible SLSN-I. The early-time spectra are similar to
those of SN\,2010gx at $t=+5$\,d. The SN is offset from a dwarf
galaxy. The peak absolute magnitude, $-20.3$, would be high for a SN~Ic
but is below the traditional dividing line for SLSNe. As we discuss
further in \S\ref{hhe}, PTF10aagc also shows hydrogen features in its
spectra, but the spectra qualitatively differ from those of 
published SLSNe-II, so we choose to keep it in the SLSN-I sample. 
PTF10aagc was also discussed by \citet{yan2015}.

{\bf SN\,2011ke} (= PTF11dij), {\bf PTF11rks}, and {\bf PTF12dam} have
all previously been published as SLSNe-I
(e.g., \citealt{inserra2013,nicholl2013, vreeswijk2017}). We confirm
that these objects have spectra more similar to those of SLSNe-I than to any
other SN type.

{\bf PTF11hrq} was originally identified as a possible variable-star
candidate owing to its compact host galaxy
(e.g., \citealt{cikota2017}). It was eventually identified as a
potential SN based on its slowly declining light curve. This
prompted spectroscopic follow-up observations about one year after the first
identification of the source that led to classification as a SLSN-I
similar to PTF10nmn above (as was anticipated from the light-curve
behavior).

{\bf PTF11mnb} was identified by the PTF survey on 2011 Sep. 19. It
was noted to have an unusually slow rise to maximum, its first
spectrum was initially suggested to show some similarities to those of
SN\,1999as (e.g., \citealt{hatano2001}), and it was internally
categorized as a likely SN~Ic. The first spectrum is noisy, but a
second spectrum taken about two weeks later is a good match to
SN\,2007gr about 1 week after maximum light. However, this spectrum is
also reasonably well matched to SLSNe-I at later light-curve phases,
such as SN\,2012il at $t=+55$\,d. Given the good matches to SNe~Ic and
the possible matches to SLSNe-I, we place this object in our possible
SLSN-I sample. PTF11mnb reached a peak absolute magnitude of about
$-18.9$, and its apparent host galaxy is a dwarf ($M_g \approx
-18$\,mag based on SDSS photometry). This object is further discussed
in a separate paper (Taddia et al., in prep.).

{\bf PTF12hni} was identified by PTF on 2012 Aug. 8, which was likely
near or after the photometric maximum. The classification of the
spectrum was initially ambiguous, with possible matches to both SNe~Ia
and SNe~Ic found, but the redshift favored by template matching
suggested that this was a relatively distant and thus luminous
source. In our analysis, the first spectrum is reasonably well matched
by the SN~Ic 2007gr at $t=-1$\,d. It may also match SN\,2003jd at
$\sim3$ weeks past photometric maximum, but to do this {\tt superfit}
requires a significantly negative $A_V$ (e.g., the templates must be
made bluer to match the data). We also find plausible matches to
SLSNe-I at even later phases, such as PTF09atu at $t=+98$\,d. For the
second spectrum, {\tt superfit} prefers matches to PTF12dam at 2--3
months after photometric maximum, but the SN~Ia-SS (a SN\,2002cx-like
event) SN\,2008A at $t=+43$\,d also provides a good match. We thus
consider PTF12hni a plausible member of the SLSN-I spectroscopic class
and place it in the possible SLSN-I sample. The target was observed to
be as bright as $-20$\,mag absolute, but, again, it was likely caught
after maximum. The SN is located on the sky in between two galaxies at
different redshifts: a large, blue galaxy at $z=0.106$ and a smaller,
redder galaxy at $z=0.185$. Both galaxies are strongly star-forming,
but only the lower-redshift and larger galaxy is at a redshift consistent
with that measured from the SN features.  Its properties are
consistent with an intermediate-mass and relatively metal-poor galaxy
undergoing rapid star formation (Taggart et al., in prep.).

{\bf PTF12gty} was first identified as a variable-star candidate on
2012 July 18, but soon thereafter it was realized that the recently
updated reference image had been constructed with images including
light from PTF12gty. The target was upgraded to a SN candidate
and a spectrum taken the following week was found to be consistent
with a SLSN-I. In particular, we find matches to PTF12dam at about 2--3
months after maximum light. The transient is well offset from a large
elliptical galaxy, but the SDSS redshift of this object, $z=0.031$,
indicates that it is an unrelated foreground galaxy. A weak
(uncataloged) source is present near the SN position in PS1 images,
indicating the true host must be very faint. The detection of
clear H~II-region lines in our late-time SN spectra indicates a
fairly high star-formation rate (Taggart et al., in prep.). The peak
absolute magnitude of PTF12gty is only $-20.0$.

{\bf PTF12mxx} was found near the end of the original PTF survey
on 2012 Dec. 15. The target was identified as a SLSN-I through
spectroscopy obtained three nights later. The spectra are well matched
by PTF12dam at $t=-21$\,d. The high peak luminosity and faint host
galaxy are further similar to typical SLSN-I hosts.

\input{ptf.speclog.tex} 

\section{SLSN-I Spectral Sequence}\label{sequence}

Traditionally, a spectrum of a given SN is compared to spectra
of other objects of the same class at a similar light-curve phase. For
homogeneous object types this works quite successfully. For a given
SN~Ia spectrum, for example, we find that the best matches in our
template library can be used to accurately predict the light-curve
phase to a precision of about $\pm2$\,d near maximum light (see
Appendix \ref{lcphasematch}). However, some SLSN-I spectra are best
matched to other SLSNe-I (or SNe~Ic) at significantly different 
light-curve phases. Here we introduce the concept of spectroscopic phase,
which can serve as an alternate indicator of the state of the SN.

We begin by assuming all SLSNe-I follow a single spectroscopic sequence
(we will test this assumption later). To build the sequence, we start
with the spectra of PTF12dam ordered by observation date. Next we add
in the spectra of a second object, one spectrum at a time, by cross
correlating the new spectrum to each of the PTF12dam spectra and
placing the new spectrum at a position along the sequence where it
best matches the PTF12dam spectra. This is repeated for each of the
spectra of the new object with the requirement that for the new object
the spectral ages increase monotonically. This may result in some
tension where one spectrum taken at a later phase than another is
actually better matched by a PTF12dam spectrum at an earlier phase. To
address this, we determine the placement of the new spectra along the
spectral sequence such that the sum of the distances between each new
spectrum and the best-matching PTF12dam spectra are minimized (subject
to monotonically increasing ages). We can then continue to add new
spectra using all other spectra assigned to the spectral sequence as
comparison nodes. In total, our spectral sequence consists of 152
spectra from the 21 objects in our spectroscopic reference set. We
include PTF10hgi in the spectral sequence because it was formally
selected as a SLSN-I by the process described in \S\ref{selection},
but as we note in \S\ref{ptfsample} and \S\ref{hhe} this object is
unique and may be better classified as a SLSN-IIb.

In practice, we actually began by arranging the spectral sequence by
hand using the cross-correlation scores as a guide. To do this, we
created PostScript files of the smoothed spectra normalized by their
continua on a logarithmic wavelength scale. Each file was identically
sized and included spectra plotted on the same scale. We then imported
these images into Keynote and positioned them by hand into a sequence,
again taking care that the age of each spectrum increased
monotonically for a given object. The transparent background of the
plots allowed us to place spectra on top of each other to visually
judge the quality of the match, and the logarithmic wavelength
scale allowed us to shift the spectra in velocity to align features as
needed.

\begin{figure*}
\begin{center}
 \includegraphics[height=\textheight]{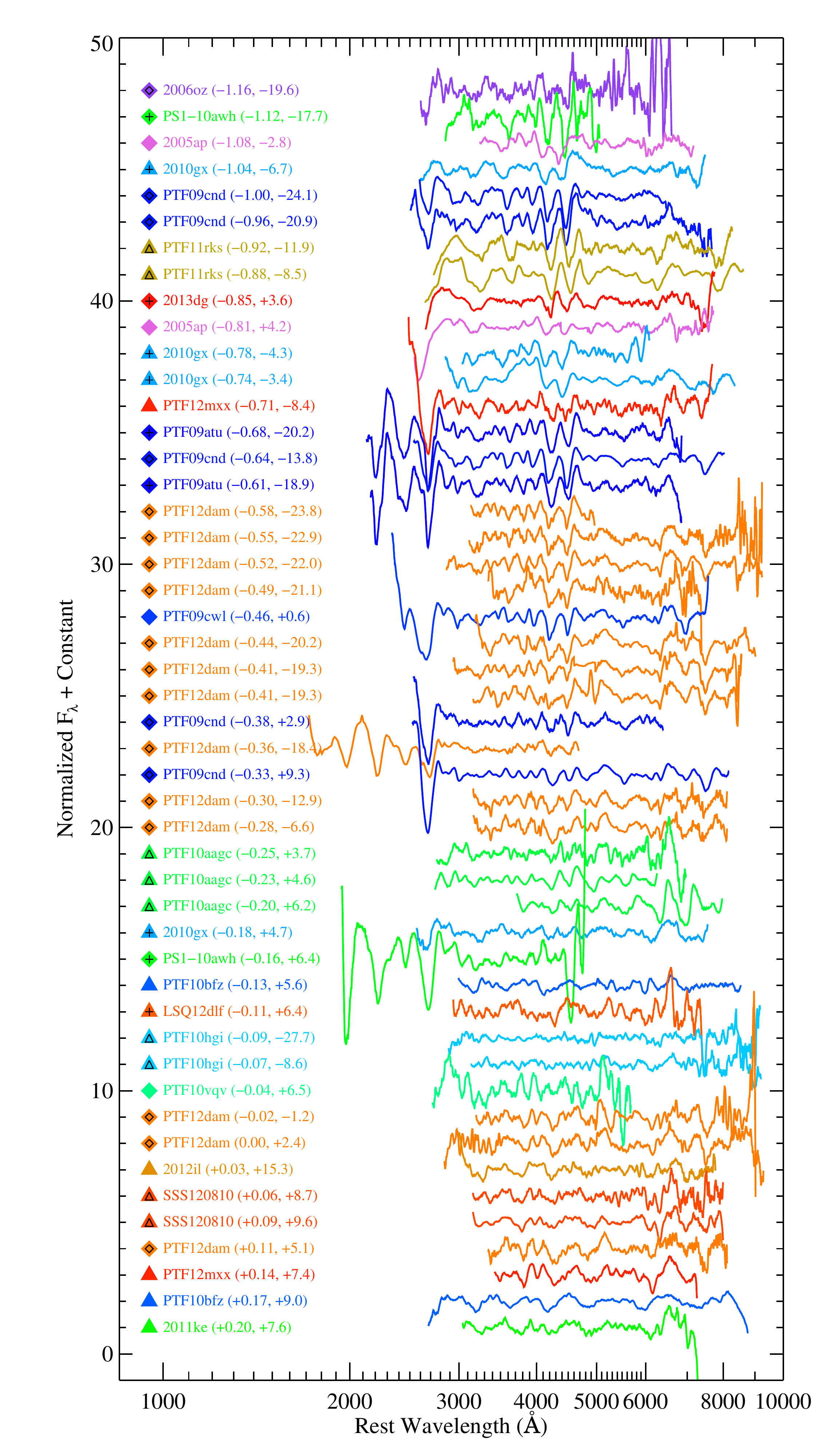}
 \caption{Spectroscopic sequence of SLSNe-I in spectral phase order
   from $\phi=-1.16$ to $+0.20$ (top to bottom). The spectra have been
   smoothed, continuum divided, and then scaled and shifted for
   clarity. The symbols next to the labels distinguish between events
   spectroscopically similar to PTF12dam (diamonds) and those more
   similar to SN\,2011ke (triangles). We have added markers inside
   some of these symbols to help distinguish between symbols of
   similar color. The numbers in parentheses next to each object name
   are the spectroscopic phase and the light-curve phase in days,
   respectively. }
   \label{fig:fid_spec1}
\end{center}
\end{figure*}

\begin{figure*}
\begin{center}
 \includegraphics[height=\textheight]{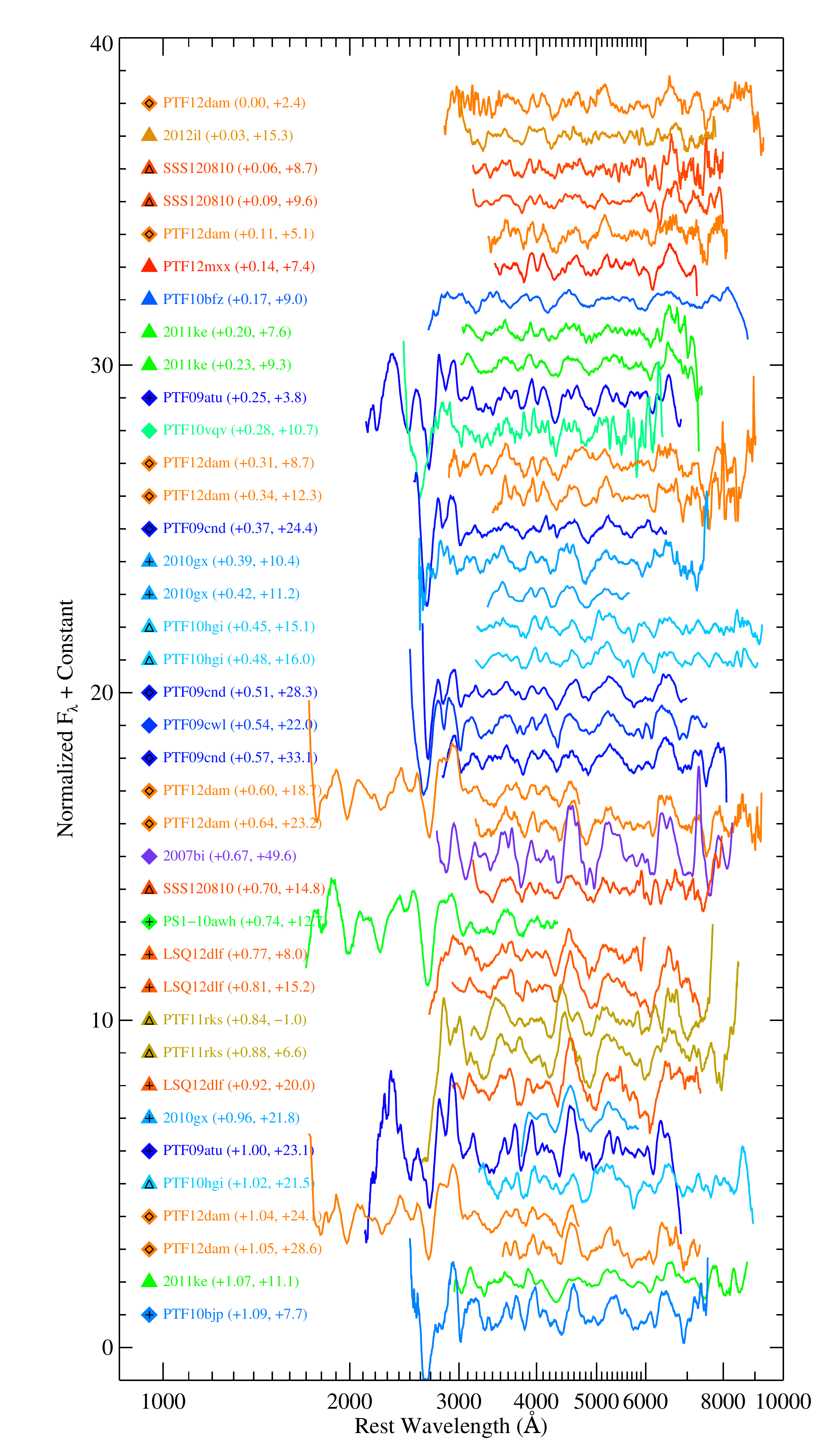}
 \caption{Similar to Figure~\ref{fig:fid_spec1} but for $\phi=0.00$ to $+1.09$.  }
   \label{fig:fid_spec2}
\end{center}
\end{figure*}

\begin{figure*}
\begin{center}
 \includegraphics[height=\textheight]{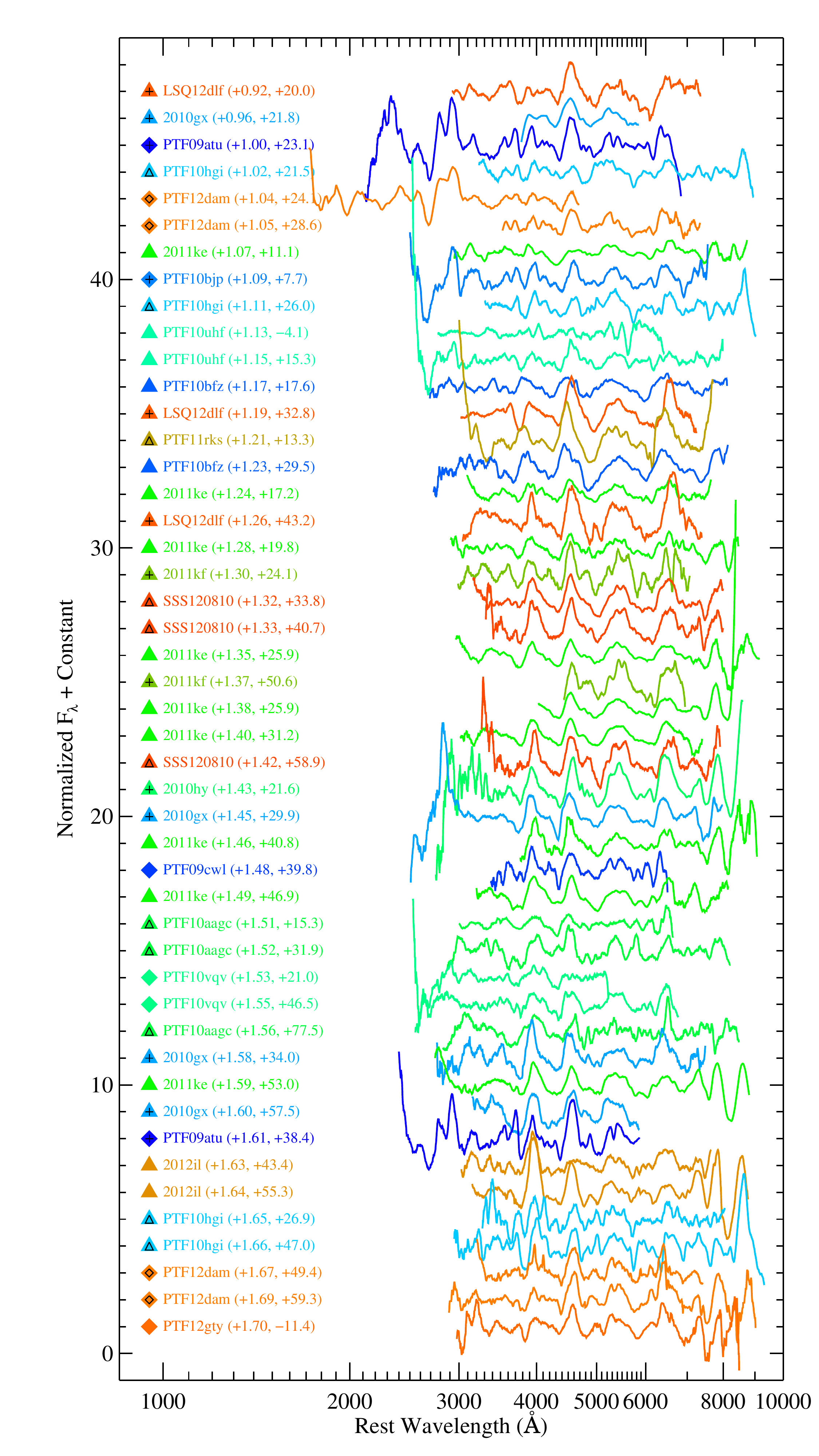}
 \caption{Similar to Figure~\ref{fig:fid_spec1} but for $\phi=+0.92$ to $+1.70$.  }
   \label{fig:fid_spec3}
\end{center}
\end{figure*}

\begin{figure*}
\begin{center}
 \includegraphics[height=\textheight]{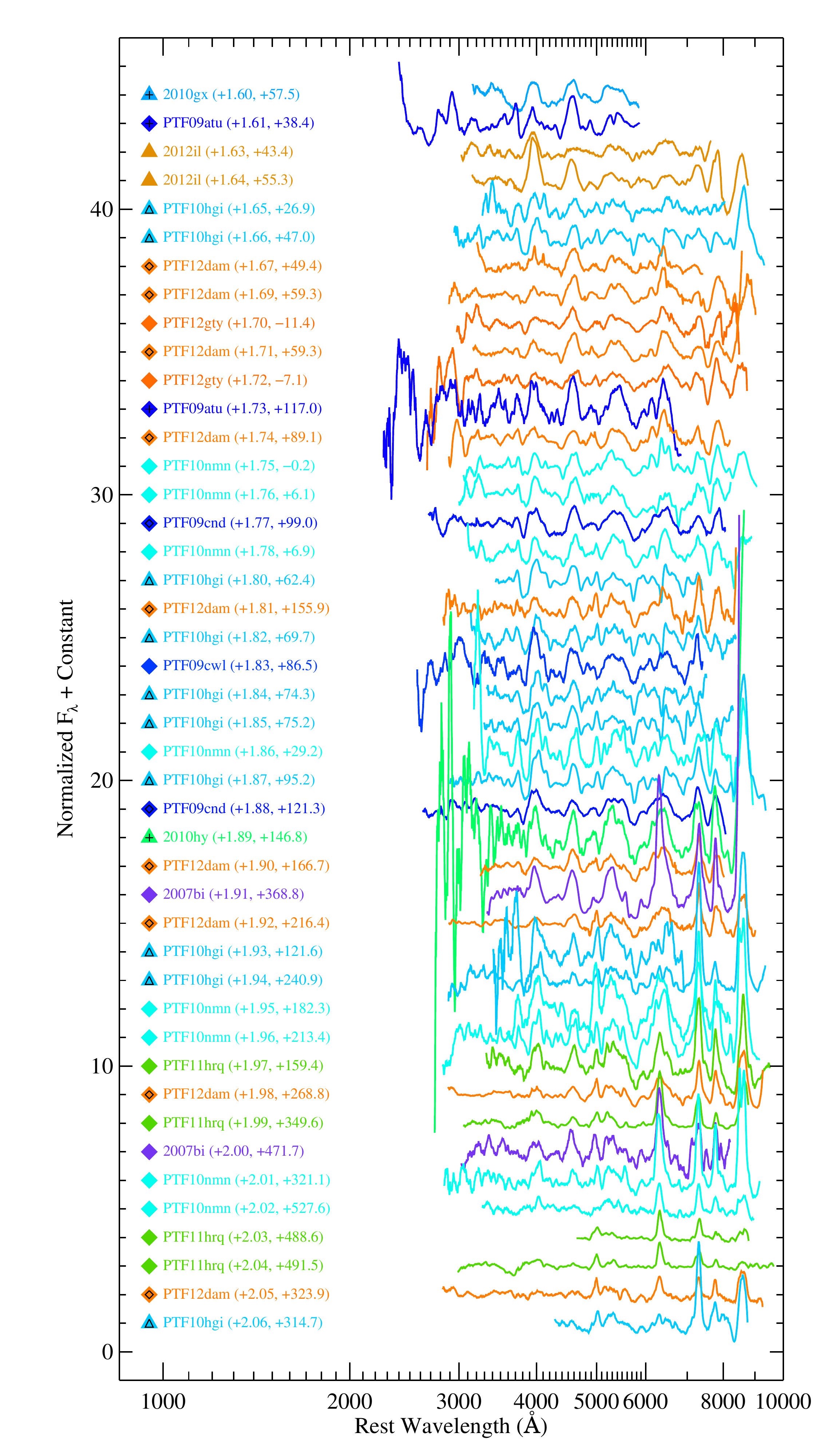}
 \caption{Similar to Figure~\ref{fig:fid_spec1} but for $\phi=+1.60$ to $+2.06$.  }
   \label{fig:fid_spec4}
\end{center}
\end{figure*}

After the initial ordering was set, we used an automated script to
reorder the spectra to minimize the total difference between the
order of each spectrum and the order of the top 5 matches found
through cross-correlation. The script first calculated the score for
the given ranking and then it randomly displaced a spectrum (maintaining
age ordering for each object) and calculated a new score. This process
was iteratively repeated until the ordering settled on a new minimum
score. Visual inspection was then used to identify any spectra
possibly stuck due to the age-ordering requirement, and the entire
process was repeated several times before settling on a final,
computer-determined ordering, which is shown in
Figures~\ref{fig:fid_spec1}--\ref{fig:fid_spec4}.

We arbitrarily assign a spectroscopic phase of $\phi=0$ to the
$t=+2.4$\,d spectrum of PTF12dam that was obtained around the time
that the \ion{O}{2} features had faded in strength, leaving a largely
featureless spectrum. Next, we assign $\phi=1$ to the spectra of
PTF09atu taken at a light-curve phase of $t=+23$\,d. These data are
found to be best matched by SN~Ic templates, such as SN\,1994I and
SN\,2004aw near photometric maximum. We assign $\phi=-1$ to the spectrum
of PTF09cnd taken $\sim24$ days before maximum light, which exhibits
strong \ion{O}{2} features. Last, we assign $\phi=2$ to the spectra of
SN\,2007bi from about 470 days after maximum light, in which nebular
features are dominant. The spectra which fall within $-1 < \phi <
1$ are then assigned fractional phases such that the change, $\Delta
\phi = \phi_2 - \phi_1$, is roughly proportional to the difference in
corresponding light-curve phases, $\Delta t = t_2 - t_1$. SLSNe-I, like
other SNe, tend to evolve more slowly as they age, and we
choose to adopt a roughly logarithmic scaling, $\Delta \phi \propto
\log \Delta t$, for $\phi > 1$. In Table~\ref{table:speclog} we give
the fiducial spectral phases for the spectra composing the spectral
sequence (Fig.~\ref{fig:fid_spec1}-\ref{fig:fid_spec4}).

Through template matching, we can now determine spectroscopic phases
for any SLSN-I (or SN~Ic) to the extent that these objects follow this
single spectroscopic sequence. We cross-correlate each spectrum with
all of the reference spectra to find the best matches and then
calculate the spectroscopic phase from the average $\phi$ of the top 5
matches. The standard deviation of these values is used as an estimate
of the uncertainty. For the cross-correlation, we focus on the
3200--7400\,\AA\ region, which is the best covered by our SLSN-I
templates (Fig.~\ref{fig:coverage}). The calculated spectroscopic
phase for each spectrum in the PTF SLSN-I sample is also given in
Table~\ref{table:speclog}.

\subsection{SLSN-I Spectral Subgroups}

During this template-matching procedure, we noticed that a number of
objects were best fit by one list of comparison objects (including
SN\,2011ke, PTF10uhf, SSS120810), while a number of other objects were
best fit by another set of comparison objects (including PTF12dam,
PTF09cnd, PTF09atu). The set of objects that drew their best
matches from both lists was surprisingly small. Motivated by this, we
attempted to classify each SLSN-I as more similar to SN\,2011ke or
PTF12dam based on how frequently their spectra appeared together with
SN\,2011ke or PTF12dam in the top 5 matches. Table~\ref{table:groups}
shows the number of times that a given object was found in the top 5
matches with SN\,2011ke and PTF12dam. To guard against spurious
results we limited the search to the 152 spectra in our spectral
sequence above. Objects that were not included in the spectral
sequence, such as PTF09q, are listed in the table but were obviously
never found in the top 5 due to this constraint. We find that there
are 81 spectra that have a match to at least one PTF12dam spectrum in
the top 5, and 56 spectra with at least one match in the top 5 to
SN\,2011ke, but there are only 10 spectra where both PTF12dam and
SN\,2011ke are found in the top 5 together.

We classify objects as spectroscopically more similar to SN\,2011ke or
PTF12dam based on how frequently a given object is in the top 5
matches of other objects with either SN\,2011ke or PTF12dam (note that
the arbitrary assignments of $\phi$ values above has no impact on the
frequency of matches). To do this, we take the fraction of cases where
the object is found in the top 5 with SN\,2011ke or PTF12dam compared
to the total number of spectra with these objects in the top 5. For
example, PTF10aagc is in the top 5 with SN\,2011ke for 4 out of 56
possible spectra ($\sim7$\% of the time), and it appears in the top 5
with PTF12dam 5 out of 81 possible spectra ($\sim6$\% of the time), so
we tentatively place it in the SN\,2011ke-like group. For objects that
are not included in the spectroscopic sequence, we assign
spectroscopic subgroups based on the frequency of matches to the
other SN\,2011ke-like and PTF12dam-like objects identified through the
procedure above.

In Figures \ref{fig:fid_spec1} to \ref{fig:fid_spec4} we mark the
SN\,2011ke-like objects with triangles and the PTF12dam-like objects
with diamonds. These figures show that the spectral sequence does not
alternate randomly between SN\,2011ke-like and PTF12dam-like
spectra. Rather, there are continuous runs of one object type or the
other. For example, the range $1.11 < \phi < 1.46$ consists of 21
SN\,2011ke-like spectra and zero PTF12dam-like spectra. These spectral
phases may thus be unique to the evolution of SN\,2011ke-like
objects. Another possibility is that PTF12dam was not well sampled
over this range (our spectra contain a 3-week gap in coverage) and it
is this lack of PTF12dam spectra over this $\phi$ range that drives
the SN\,2011ke-like versus PTF12dam-like division. We also note that
23 out of the 29 earliest phase spectra ($\phi <-0.28$) are
PTF12dam-like events, but PTF12dam itself was well observed over this
period and accounts for 10 of these spectra. Because of the lack of
early-time SN\,2011ke-like spectra and possible contamination from
host light at late times, we have only considered spectra in the $-0.3
< \phi < 1.7$ range when assigning subgroups.

\subsection{Spectral Evolution Rate of SLSN-I Subgroups}

\begin{figure*}
\begin{center}
 \includegraphics[width=\linewidth]{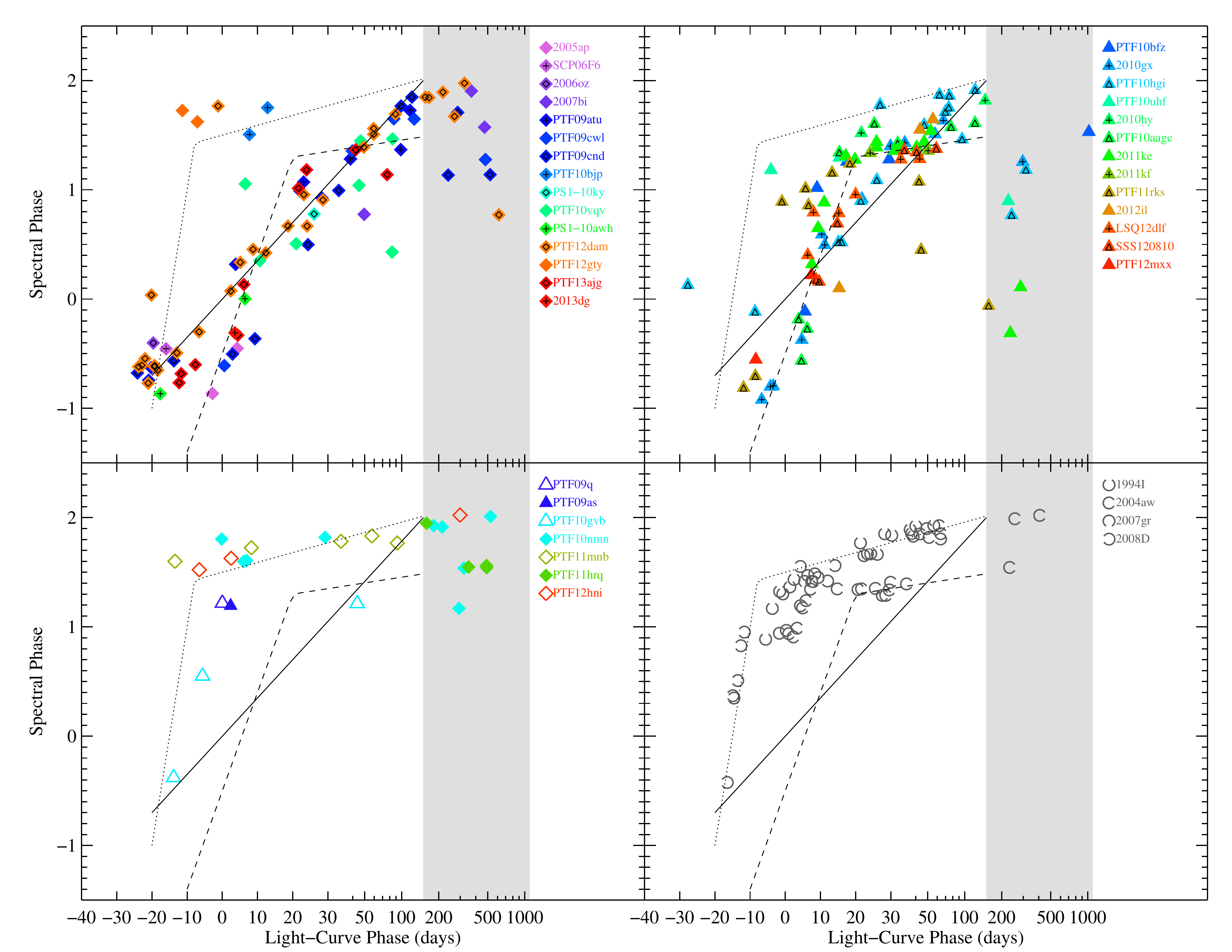}
 \caption{ Spectral phase ($\phi$) versus light-curve phase ($t$) for
   the SN samples discussed in this paper. The gray shaded
   region marks phases where the spectral matching may fail owing to
   contamination from host-galaxy light. We show separately the
   PTF12dam-like sample (upper-left panel), the SN\,2011ke-like sample
   (upper-right panel), the possible SLSNe-I and objects with poorly
   constrained light-curve phases (lower-left panel), and a few 
   normal-luminosity SNe~Ib and SNe~Ic (lower-right panel). In each panel we
   plot possible trends for the PTF12dam-like sample (solid lines),
   the SN\,2011ke-like sample (dashed lines), and the SN~Ib/c sample
   (dotted lines) for reference. See text for further details.}
   \label{fig:phase_plot}
\end{center}
\end{figure*}

We next consider how the calculated spectral phases correlate with
light-curve phase. For this comparison, we will use the standard
definition of light-curve phase for hydrogen-poor SNe, which is
the number of rest-frame days after optical maximum, often
specifically referenced to the $B$ or $g$-band maximum. This is a
convenient epoch to serve as a basis since it is quite often directly
constrained by observations; however, it is not necessarily the most
physically motivated choice. For comparison, hydrogen-rich SNe
are typically indexed by days from explosion instead. In
Figure~\ref{fig:phase_plot} we show the dependence of spectral phase
on light-curve phase for SLSNe-I and selected SNe~Ib and SNe~Ic (spectral
phases for these normal-luminosity events are determined through
cross-correlation to the SLSNe-I with assigned spectral phases as
discussed above). We separate out the SLSN-I sample into PTF12dam-like,
SN\,2011ke-like, and objects that either have poorly constrained dates
for maximum light or that were classified as possible SLSNe-I. The
abscissa in Figure~\ref{fig:phase_plot} is linear up to day 35 where
it changes to a logarithmic scale.

Starting with the upper-left panel in Figure~\ref{fig:phase_plot}, we
see that the PTF12dam-like events largely cluster along the diagonal
in the figure. As discussed above, the numerical values for spectral
phase are anchored to PTF09cnd, PTF12dam, PTF09atu, and SN\,2007bi,
which all belong to the PTF12dam-like group. It is perhaps then not a
complete surprise to find this result, although this was by no means
guaranteed. There is, however, a large amount of scatter in the
relation. Part of this is caused by a high catastrophic failure rate in
the simple cross-correlation method used to calculate spectral
phases. For example, there are two observations of PTF12dam taken
around maximum light. One of these is found through cross-correlation
to have a spectral phase of $\phi = 0.08 \pm 0.34$, which agrees quite
well with the definition of spectral phase above (PTF12dam should have
$\phi \approx 0$ near maximum light). However, the second spectrum has a
calculated value of $\phi = 1.71 \pm 0.27$, which is clearly
discrepant. Examining the fitting results, it appears that a
combination of factors including systematic error from the spectral
extraction led to this discrepant value. Visual comparison confirms
that this spectrum should have a spectral age of $\phi \approx
0$. Additionally, our cross-correlation technique does not account for
host-galaxy contamination; thus, the calculated spectral phases can be
bogus at late phases ($t>150$\,d) when the SLSNe-I have faded to or
below the background level. We retain the values of spectral phase
determined automatically through the cross-correlation analysis
described above and simply note that more advanced techniques are
required to improve the robustness of spectral phase determination.

Despite the scatter, it is clear from the upper-right plot in
Figure~\ref{fig:phase_plot} that the SN\,2011ke-like events chart a
different course through the $t$ vs. $\phi$ plane than the
PTF12dam-like events. These events navigate through the early spectral
phases ($\phi < 1$) more rapidly before they change tack and slowly
approach the PTF12dam-like sequence at about $\phi=1.4$. We note that
some of the PTF12dam-like objects may follow this trend at least at 
early times. SN\,2005ap, for example, has just two spectra available, 
but these appear to follow the SN\,2011ke-like trend better.

Moving to the lower-right plot in Figure~\ref{fig:phase_plot}, we see
further distinction between normal-luminosity SNe~Ib/c and SLSNe-I. The
lower-luminosity events tend to have significantly larger spectral
phases at earlier light-curve phases than do SLSNe-I. SN\,2008D is the
only object shown with spectral phases significantly less than $\phi
= 1$. It quickly moves through the spectral sequence, however,
reaching $\phi \approx 1.5$ after its light curve's maximum. SN\,2004aw
is also noteworthy for following a path that is not too dissimilar
from that of SN\,2011ke, especially 20--40\,d after light-curve
maximum. We note that PTF12gty is a considerable outlier from the
PTF12dam objects in the $t$ vs. $\phi$ plane, and it may better fit
with the normal-luminosity SNe~Ib/c. This is especially interesting
given the peak luminosity of PTF12gty, which is quite low compared to
the other SLSNe-I but which would be rather high for a SN~Ic. PTF10bjp
also shows advanced spectral phases at relatively early light-curve
phases, which may be more similar to the normal-luminosity SNe~Ib/c
shown.

Finally, the lower-left panel in Figure~\ref{fig:phase_plot} shows
objects from the possible SLSN-I sample and objects that have poorly
constrained dates for photometric maximum. These all tend to favor
larger spectral phases at younger light-curve phases than do most
SLSNe-I in a manner similar to the SNe~Ib/c shown. In the cases where the
date of maximum light is poorly constrained, it could simply be the
case that the true light-curve phase is significantly larger than
shown; correcting these to the true values could then shift the
observations to the PTF12dam or SN\,2011ke-like tracks.

\input{groups.table.tex} 

\begin{figure*}
\begin{center}
 \includegraphics[width=\linewidth]{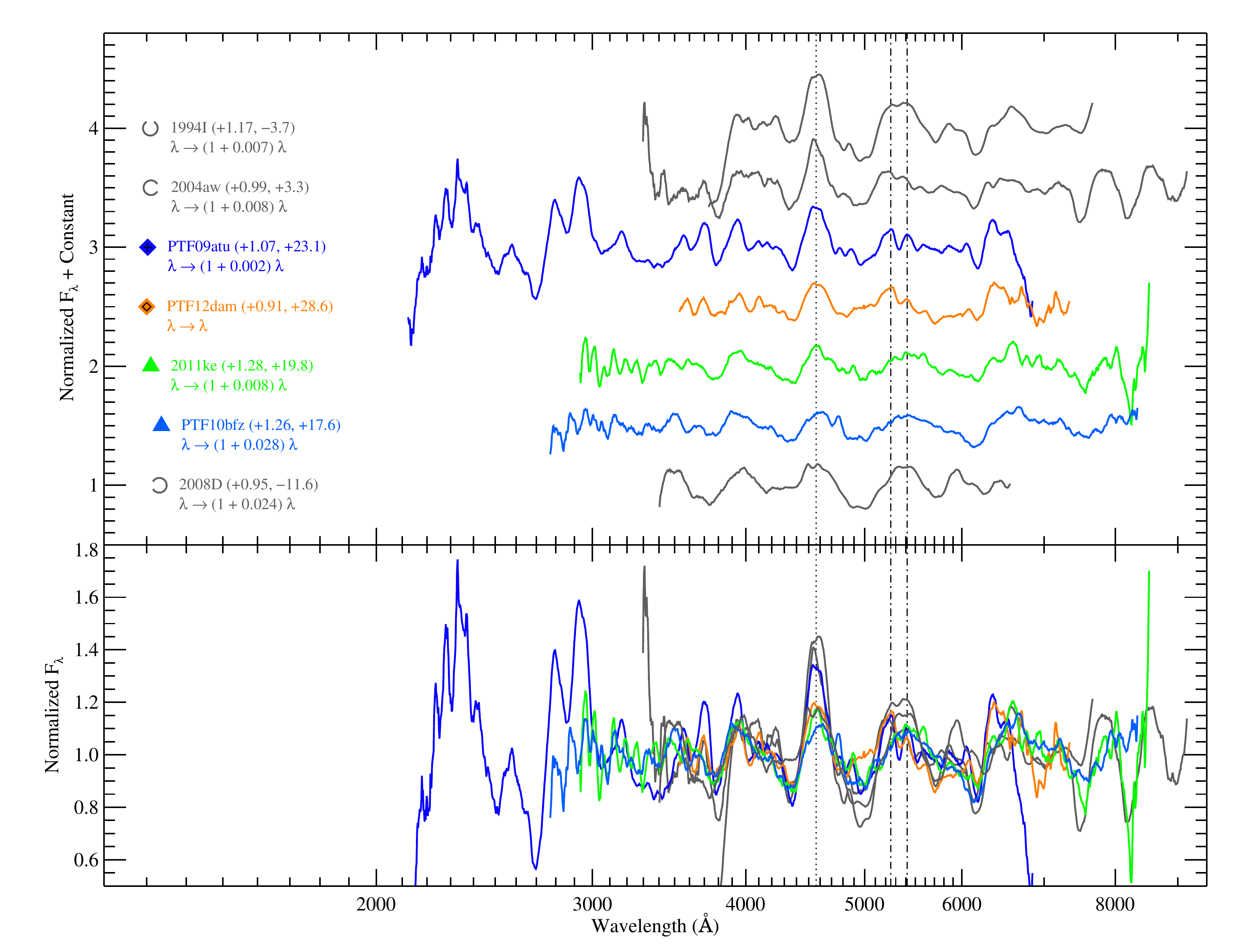}
 \caption{Spectra of SLSNe-I (colored lines) and SNe~Ib/c (gray lines) at
   spectral phases close to $\phi = 1$. Each spectrum has been shifted
   as indicated to best match the spectrum of PTF12dam in the 
   3500--5000\,\AA\ region (relative redshifts were determined through
   cross-correlation as described in the text). The vertical, dotted
   line marks a prominent peak near 4600\,\AA, and dash-dotted lines
   mark peaks seen near 5400\,\AA\ or redder in some cases. }
   \label{fig:slsn1_vs_sn1bc}
\end{center}
\end{figure*}

\section{Spectral Comparisons}\label{compare_spec}

We have shown that SLSNe-I can be spectroscopically distinguished from
normal-luminosity SNe through template comparison and that
SLSNe-I may further be divided into PTF12dam-like or SN\,2011ke-like
groups based on the relative frequency of good template matches to
these objects. In this section we look for specific spectral features
that may correlate with these divisions.

We begin by comparing in Figure~\ref{fig:slsn1_vs_sn1bc} the 
near-maximum-light spectra of the normal-luminosity SNe~Ic 1994I and 2004aw
and the premaximum-light spectra of the normal-luminosity SN~Ib 2008D
with members of the SN\,2011ke-like and PTF12dam-like SLSN-I groups,
all at a similar spectral phase ($\phi \approx 1.0$). To align spectral
features, we cross-correlate each spectrum in the 3500--5000\,\AA\ 
region to determine shifts, $\delta z$, relative to PTF12dam. In
the figure, we plot wavelength on a logarithmic scale so shifting the
spectral wavelengths by $\lambda \rightarrow (1 + \delta z)\lambda$ is just
a uniform displacement left or right as indicated by the locations of
the plotted symbols.

The spectra displayed in Figure~\ref{fig:slsn1_vs_sn1bc} show a number of
similarities, but there are noticeable differences as well. For
example, SN\,2008D has a strong absorption feature at about 5700\,\AA,
which is usually attributed to a blend of \ion{Na}{1} and
\ion{He}{1}. This feature is more weakly present in SN\,1994I but
hardly noticeable in the other spectra. As is evident in the figure,
stripped-envelope SNe comprise a heterogeneous class, and while
it is possible to identify differences between specific events
(e.g., one SN~Ic versus another SN~Ic or a SLSN-I), it is not clear from
this comparison that there is a single defining feature that
differentiates SNe~Ic from SLSNe-I. Rather, the distinction discovered in
\S\ref{selection} may be the result of a number of subtle
features in combination, such as the relative weakness of line
strengths in SLSNe-I compared to SNe~Ic or small shifts of certain
features with respect to others.

Figure~\ref{fig:slsn1_vs_sn1bc} shows a possible distinction between
SN\,2011ke-like and PTF12dam-like events: when the spectral peaks at
about 4600\,\AA\ are aligned, the broader peaks near 5400\,\AA\ extend
further to the red in the SN\,2011ke-like events shown in the
figure. It also appears that the SN\,2011ke-like events may have
broader features. The limited wavelength ranges and small number of
objects included in the comparison make it difficult to determine how
significant this difference may be, however. In order to compare
SN\,2011ke-like and PTF12dam-like events, we therefore must look at
the average spectral properties of each group over some range of
spectral phases.

\begin{figure*}
  \begin{center}

    \begin{tabular}{cc}
      
      \includegraphics[width=0.5\linewidth]{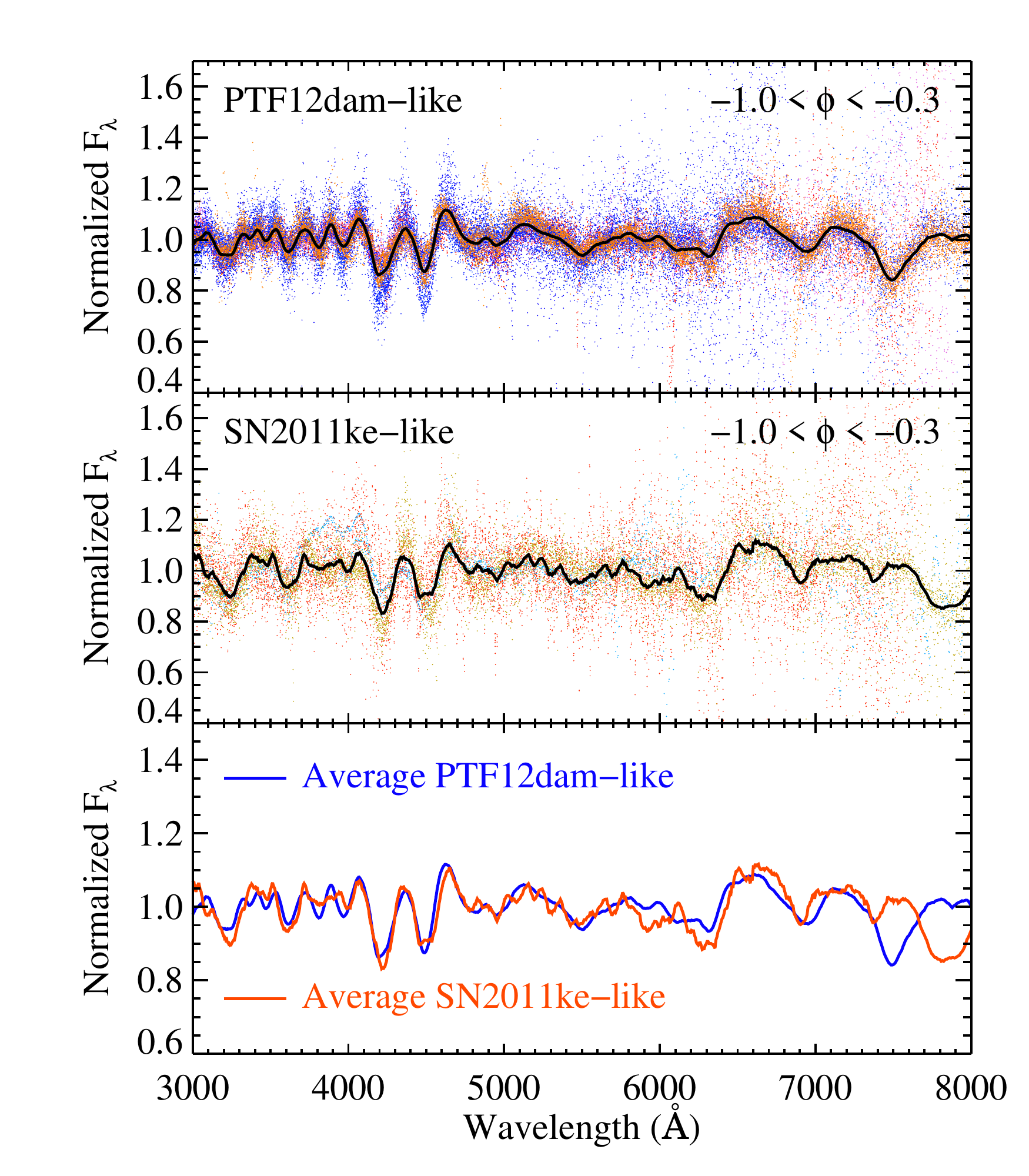} &
      \includegraphics[width=0.5\linewidth]{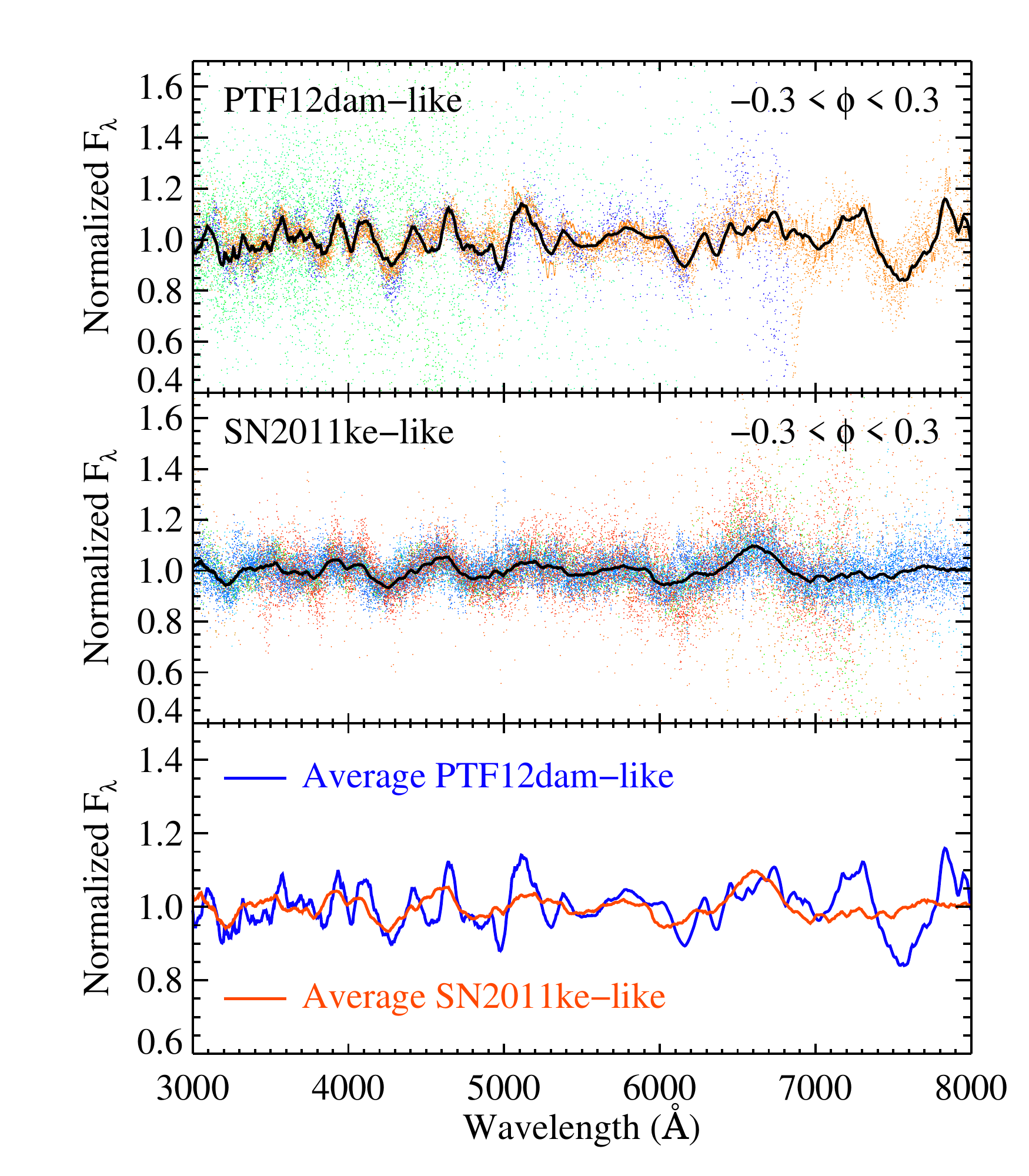}
      \\
      \includegraphics[width=0.5\linewidth]{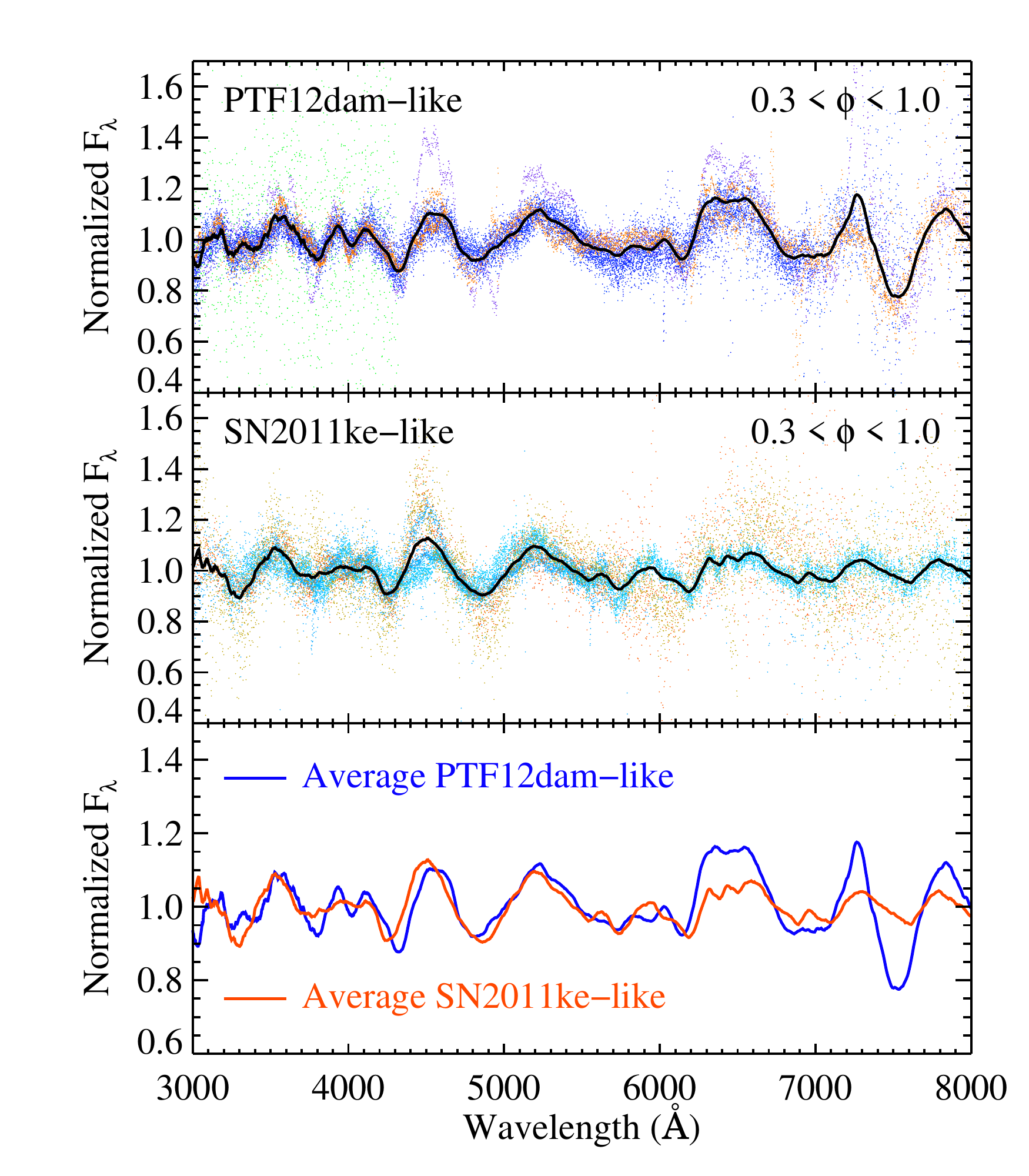} &
      \includegraphics[width=0.5\linewidth]{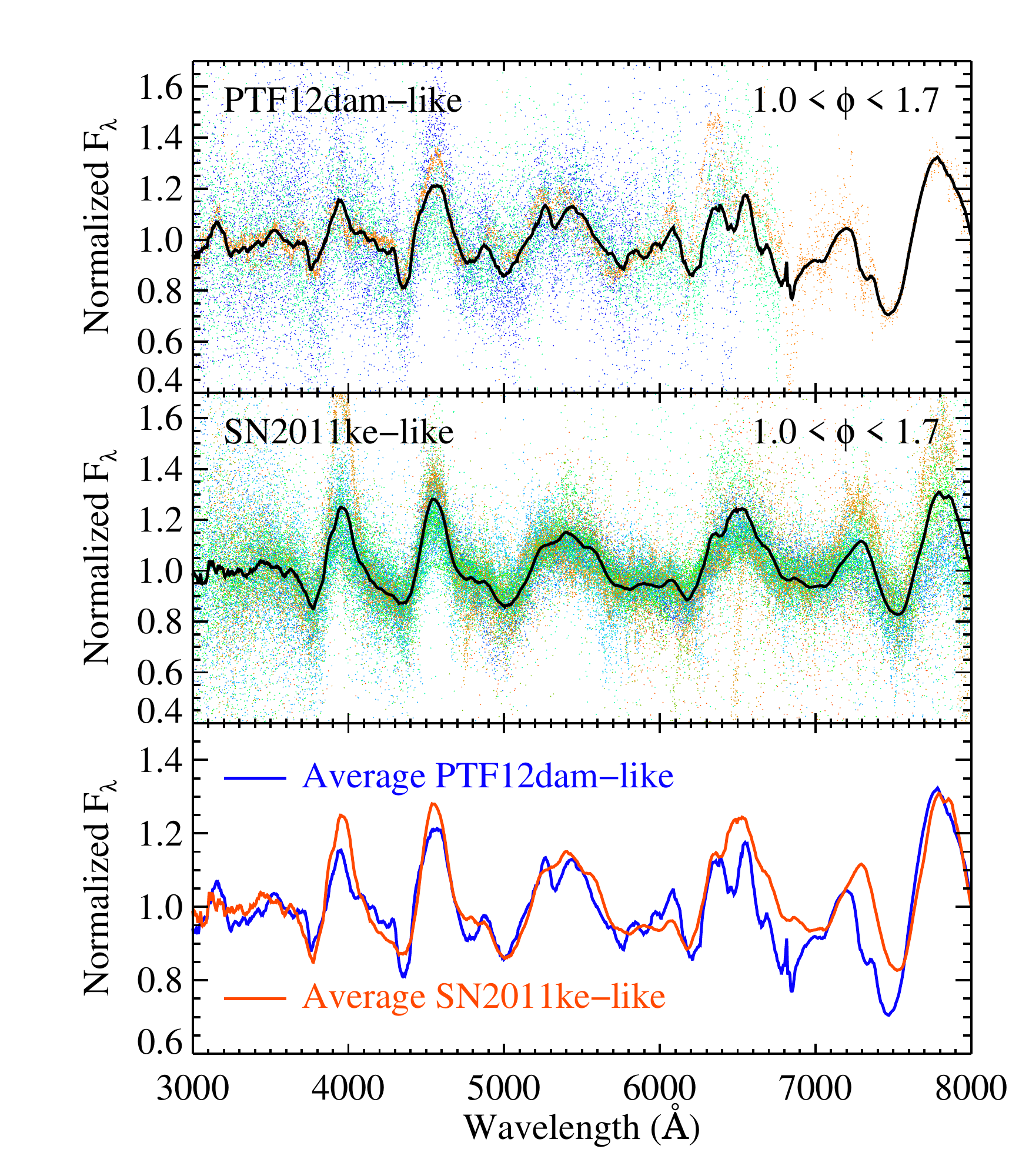}
    \end{tabular}
    \caption{ Average spectra of SLSNe-I with spectral phases in
      different ranges. Events deemed more spectroscopically similar
      to PTF12dam or SN\,2011ke are shown in the top and middle panels
      of each plot, respectively. The spectra have been continuum
      divided. Dots show measurements from individual events (color
      coded as in Figs.~\ref{fig:fid_spec1}--\ref{fig:fid_spec4}),
      and the black lines indicate these combined data smoothed with a
      generalized Savitzky-Golay filter. The lower panel in each plot
      shows only these averaged spectra for comparison.}
    \label{fig:compare_phase}
  \end{center}
\end{figure*}

In Figure \ref{fig:compare_phase} we show the spectra of the
SN\,2011ke-like and PTF12dam-like subgroups at spectral phases near
$\phi = -0.65$, 0.0, $+0.65$, and $+1.35$. The spectra are continuum
divided and color coded by object. We also compute the average spectra
for each group by using the modified Savitsky-Golay smoothing method
discussed in Appendix~\ref{sgfilter}. To do this, we first compute the
average for a given group of spectra and then determine the velocity
shift for each component spectrum relative to this initial average. We
then shift the individual spectra to align them before computing the
final average spectrum.

As shown in the upper-left plot in Figure~\ref{fig:compare_phase}, the
spectra of PTF12dam-like and SN\,2011ke-like events at early spectral
phases ($-1 < \phi < -0.3$) are broadly similar, but there are some
apparent differences. In the range 3000--5700\,\AA, the PTF12dam-like
objects exhibit a number of features usually attributed to \ion{O}{2}
(see \ref{o2lines}). In the average SN\,2011ke-like spectrum, however,
two of these features are missing and the absorption at
3600\,\AA\ appears broader. Looking back at
Figure~\ref{fig:fid_spec1}, we can see that the SN\,2011ke-like event
PTF11rks lacks the notches at about 3800 and 4000\,\AA, but these
features are clearly detected in spectra of PTF09cnd and PTF09atu at
similar spectral phases. The earliest spectrum of the SN\,2011ke-like
event SN\,2010gx also lacks these notches, although they are present in
the second spectral epoch. Additionally, these features are not
clearly present in the PTF12dam-like event SN\,2005ap. This raises the
possibility that these notches may be transient features in both
SN\,2011ke-like and PTF12dam-like events, and perhaps they are not
present at very early epochs. Another possibility is that SN\,2005ap
may be better associated with the SN\,2011ke-like group (recall also
that SN\,2005ap may evolve more similarly to SN\,2011ke-like objects
through the $\phi$ vs. $t$ diagram in Figure~\ref{fig:phase_plot}).

There also appear to be strong differences in the average
SN\,2011ke-like and PTF12dam-like spectra at 7000--8000\,\AA\ 
over the $-1 < \phi < -0.3$ range, although there is a
paucity of SN\,2011ke-like data available for the
comparison. Nevertheless, the PTF12dam-like spectra show a broad
absorption dip at $\sim7500$\,\AA\ that is absent in the
SN\,2011ke-like sample, which in turn appears to show an absorption
feature at 7800\,\AA\ that is missing in the PTF12dam-like
sample. This difference is also seen in the $-0.3 < \phi < 0.3$ range
shown in the upper right of Figure~\ref{fig:compare_phase}, although
again there are few observations contributing to this comparison.

A second difference in the $-0.3 < \phi < 0.3$ range is that the
SN\,2011ke-like spectra are noticeably smoother than the PTF12dam-like
spectra, on average. The spectra remain broadly consistent, so this
difference may suggest that the SN\,2011ke-like events have larger
expansion velocities that wash out individual features more than
PTF12dam-like events. This is suggested by the velocity shifts noted
in Figure~\ref{fig:slsn1_vs_sn1bc}, although it is not clear how well
the velocity widths of SLSN-I features correlate with expansion
velocities (see \S\ref{conclusions}).

In the spectral phase range $0.3 < \phi < 1.0$ (lower-left plot in
Fig.~\ref{fig:compare_phase}), we again find a stronger
7500\,\AA\ feature in the PTF12dam-like sample. The emission feature
at $\sim6500$\,\AA\ may also be more pronounced in the PTF12dam-like
sample. Echoing the result from Figure~\ref{fig:slsn1_vs_sn1bc}, we
see that the emission peaks at 5200\,\AA\ align, but the P-Cygni
profiles around 4400\,\AA\ are significantly offset from each
other. This may suggest that the ratio of velocities between ions in
PTF12dam-like and SN\,2011ke-like ejecta might not be constant. This
difference is not seen in the later-phase ($1.0 < \phi < 1.7$) spectra
in the lower-right plot of Figure~\ref{fig:compare_phase},
except possibly in the 7500\,\AA\ feature, which is finally present in
the SN\,2011ke-like sample. However, there are limited PTF12dam-like
observations covering this wavelength range and phase.

\section{Spectral Line Identifications}\label{lineids}

In the above sections we discussed the spectral evolution of SLSNe-I in
broad terms. In this section we attempt to identify the ions
responsible for key features in the spectra. By associating specific
features with specific ions, the composition and velocity structure of
the atmosphere are revealed. Ideally, detailed radiative transfer
calculations should be constructed and paired with a hydrodynamical
code to self-consistently model spectral features as they evolve over
time. However, such work is beyond the scope of this paper. We instead
opt for a simplistic evaluation of line features, leveraging line
identifications and numerical modeling in previous publications
\citep{quimby2011, dessart2012, howell2013, mazzali2016, yan2017}. In
some cases, we additionally use simple tools to test the validity of
these features.

Two widely used tools for the identification of features in SN
spectra are {\tt synow} \citep{jeffery_branch1990} and its more modern
incarnation, {\tt syn++}
\citep{thomas2011}\footnote{https://c3.lbl.gov/es/}. These codes
generate synthetic spectra for 1-D homologously expanding atmospheres
with a distinct (or ``sharp'') photosphere. A large number of
parameters can be adjusted to create models of observational
data. These parameters include photospheric velocity, ion species, ion
temperature, and the relative contributions of different ions. However,
the relative strengths of features arising from the same ion species
are set by assuming local thermodynamic equilibrium (LTE). 
In this manner, potential line identifications
can be tested by matching the model to a particular feature and then
checking the agreement with any additional features predicted by the
model.

One potential weakness of {\tt syn++} when dealing with multiple ion
species is that in real SNe lines from different ions may form
in different layers (and thus have different velocities). Although
{\tt syn++} has the ability to ``detach'' ions from the photosphere,
this adds further degrees of freedom to this already highly parametric
modeling process. Also, the version of {\tt syn++} available can only
model the ion distribution as an exponential function, and detached
ions are thus computed for a physically unlikely structure (Gaussian
distributions were not possible in the version available). Another
weakness of {\tt syn++} is that it is not intended for use on spectra
entering the nebular phase (where the assumption of a sharp photosphere
breaks down), although it has been used to model relatively late-phase
spectra (see \citealt{branch2008}).

We have used {\tt syn++} to generate synthetic spectra similar to the
average SLSN-I spectra at $\phi=-1$, $\phi=0$, and $\phi=+1$. These
simple models are presented in Appendices \ref{average_spec-1},
\ref{average_spec0}, and \ref{average_spec+1}, respectively. We find
that the identities of some features suggested by previous works are
problematic in that {\tt syn++} predicts additional features for these
ions that conflict with the data. Below we reconsider the
identifications of the principal SLSN-I spectral features in the
photospheric phase. We defer study of the late-phase spectra to
future work (see also \citealt{lunnan2016} and
\citealt{jerkstrand2017}).

\subsection{Study of the \ion{O}{2} Lines}\label{o2lines}

The strongest features in the early-phase ($-1.0 < \phi < -0.3$)
spectra of SLSNe-I are two broad dips typically observed around 4200
and 4500\,\AA. These were first associated with blends of \ion{C}{3},
\ion{N}{3}, and \ion{O}{3} \citep{quimby2007c}, but these are now
usually identified with a single ion, \ion{O}{2}
\citep{quimby2011}. This latter identification was first made
using {\tt synow}\footnote{The \ion{O}{2} and \ion{O}{3} entries in
  the {\tt syn++} reference line data file, ``refs.dat'', have a
  formatting error and should be edited to ``0801 7320.6640625
  3.69828e-9 3.3282802'' and ``0802 4363.2089844 4.57088e-9
  2.5170000'', respectively, to avoid spurious spectral features
  (R. Thomas, private comm.).}, and it has gained wide acceptance owing 
to the fact that this single ion can account for both of the two major
features as well as several weaker dips often observed from 3500 to
4000\,\AA. However, the simple models produced by {\tt synow} with its
default line lists also predict a relatively strong \ion{O}{2} feature
at $\sim4000$\,\AA, which does not agree well with the data
(observations do show a dip near 4000\,\AA, but it is typically
weaker and shifted to the red of the predicted
feature). \citet{mazzali2016} have produced more advanced models that
demonstrate \ion{O}{2} can account for all of the major spectral
features of early-phase SLSNe-I in the optical range, although a
similar offset in the 4000\,\AA\ feature can be seen in their Figure
5.

The \ion{O}{2} ion has not been detected in the spectra of any 
normal-luminosity SNe~Ic. Aside from SLSNe-I, only the peculiar 
SN~Ib SN\,2008D
\citep{soderberg2008,mazzali2008,modjaz2009} and the SN~Ibn OGLE-2012-SN-006
\citep{pastorello2015} have shown spectroscopic evidence for this
ion. For \ion{O}{2} to be present, oxygen must be excited to a
relatively high energy level \citep{mazzali2016}. Thus, the lack of
such observed features in normal-luminosity SNe~Ic may simply be the
product of rapid cooling through adiabatic expansion of initially
compact progenitor systems, or due to a lack of nonthermal sources of
excitation.

\begin{figure*}
\begin{center}
 \includegraphics[width=\linewidth]{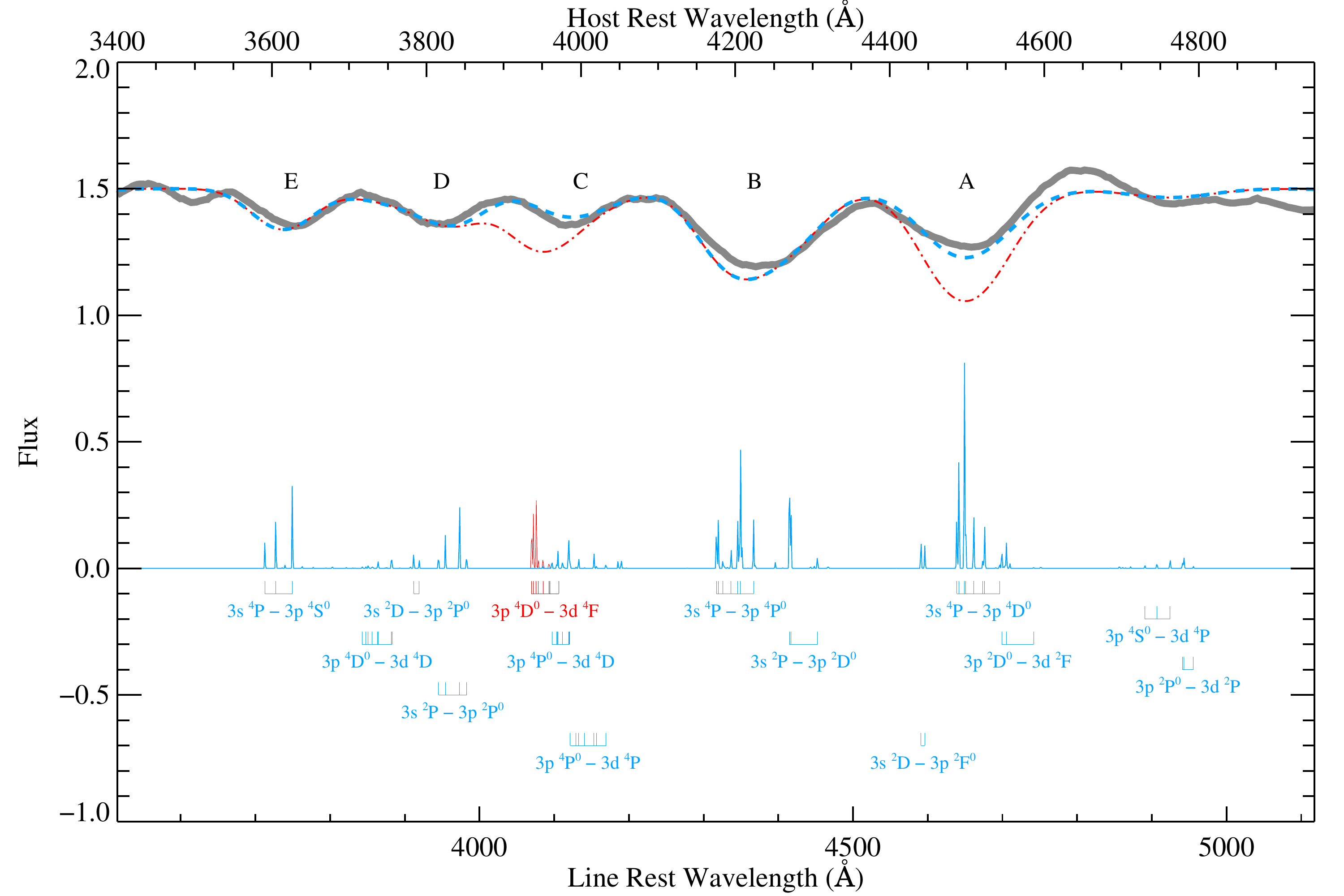}
 \caption{ Transitions of the \ion{O}{2} ion in the 3500--5000\,\AA\ 
   range (thin blue lines) compared to continuum-divided
   spectra of PTF09atu (thick gray line). Relative strengths of each
   transition from NIST are shown assuming a 15,000\,K plasma in
   LTE. Each multiplet is grouped and labeled. The lower abscissa
   indicates the laboratory wavelength for each transition while the upper
   abscissa shows wavelengths in the rest frame of PTF09atu's host
   galaxy. The latter system is blueshifted with respect to the former
   by 10,000\,km\,s$^{-1}$. The red, dot-dashed curve illustrates a
   synthetic spectrum created by Doppler broadening the multiplets (in
   absorption) by 7000\,km\,s$^{-1}$ and scaling the result to match
   the observed PTF09atu spectrum. The blue dashed curve shows a
   similar synthetic spectrum but with the
   $3p\,^{4}D^{0}$ -- $3d\,^{4}F$ multiplet (marked in red) removed and
   the $3s\,^{4}P$ -- $3p\,^{4}P^{0}$ transition weakened as described in
   the text. }
   \label{fig:O2_lines}
\end{center}
\end{figure*}

Figure \ref{fig:O2_lines} shows spectra of PTF09atu in the 
3500--5000\,\AA\ range. Five features are labeled A--E (we assign the 
letter ``A'' to the reddest feature). To investigate which \ion{O}{2}
transitions contribute to each of the features, we downloaded all of
the known \ion{O}{2} lines from
NIST\footnote{http://physics.nist.gov/PhysRefData/ASD/lines\_form.html}. Expected
relative intensities are calculated assuming the gas is in 
LTE at 15,000\,K. We can then create a simple
model spectrum where each of these lines is represented by Gaussian
absorption functions with the same width. We can Doppler shift the
lines by a uniform velocity and scale the line strengths by a uniform
factor to match the observed spectra. Although this model does not
properly account for the radiative transfer effects of an expanding
atmosphere (e.g., we model lines as pure absorption features, whereas in
reality each line should have a P-Cygni profile in accordance with the
geometry of the system), the results provide a surprisingly good fit
to the data (red dash-dotted line in Fig. \ref{fig:O2_lines}). In fact,
we recover a spectrum that is quite similar to {\tt syn++} models,
including the offset of the \ion{O}{2} ``C'' line at 4000\,\AA.

The failure of our simple absorption model to match the
4000\,\AA\ feature could mean that a second ion contributes to this
part of the spectrum, or that the relative strengths of the individual
lines in our model are inaccurate. To investigate this, we have
identified in Figure \ref{fig:O2_lines} the transitions that
contribute to each of the broad features in the spectrum. As can
readily be seen, the \ion{O}{2} features consist of a complex
blend of many individual lines (see also \citealt{mazzali2016}). In
fact, most of the five major features are blends of two or more
multiplets. For example, the ``B'' feature consists of transitions
between the $3s\,^{4}P$ and $3p\,^{4}P^{0}$ levels and also
transitions between the $3s\,^{2}P$ and $3p\,^{2}D^{0}$
levels. Looking at the ``C'' feature, we see that the absorption model
predicts that the $3p\,^{4}D^{0}$ -- $3d\,^{4}F$ transition should be
dominant, but this disagrees with the data. If we artificially remove
this transition, we find that the absorption model falls into
excellent agreement with the data (blue dashed curve). We also note
that the absorption model overpredicts the strength of the ``A''
feature. This is possibly connected to the failure of the ``C''
feature since they share a common energy level (the $3p\,^{4}D^{0}$
state). If we artificially weaken the contribution to the ``A''
feature by the same amount removed from the ``C'' feature, the overall
model is in excellent agreement with the data. (In addition to the
$3d\,^{4}F$ level, the $3p\,^{4}D^{0}$ level can also be populated
from the $3d\,^{4}D$ level, which is involved in the transition for
the C line, and the $3d\,^{2}P$ transition, which is responsible for
the weak feature to the red of the A line.) It thus seems that the
population of the $3p\,^{4}D^{0}$ state disagrees with the line
intensities reported by NIST. In the original source it is noted that
the $3p\,^{4}D$ transitions strengths were obtained through indirect
means \citep{veres_wiese1996}. Thus, it is possible that these lines
are weaker in nature than the published values suggest; given the
importance of the \ion{O}{2}, we suggest that a new laboratory
investigation of its transitions is warranted.

Artificially removing/reducing the $3p\,^{4}D^{0}$ transitions as
above, the effective wavelengths for the \ion{O}{2} A--E blends are
4650.71\,\AA, 4357.97\,\AA, 4115.17\,\AA, 3959.83\,\AA, and
3737.59\,\AA, respectively, according to our simple absorption model
assuming a temperature of 15,000\,K.

\begin{figure}
\begin{center}
 \includegraphics[width=\linewidth]{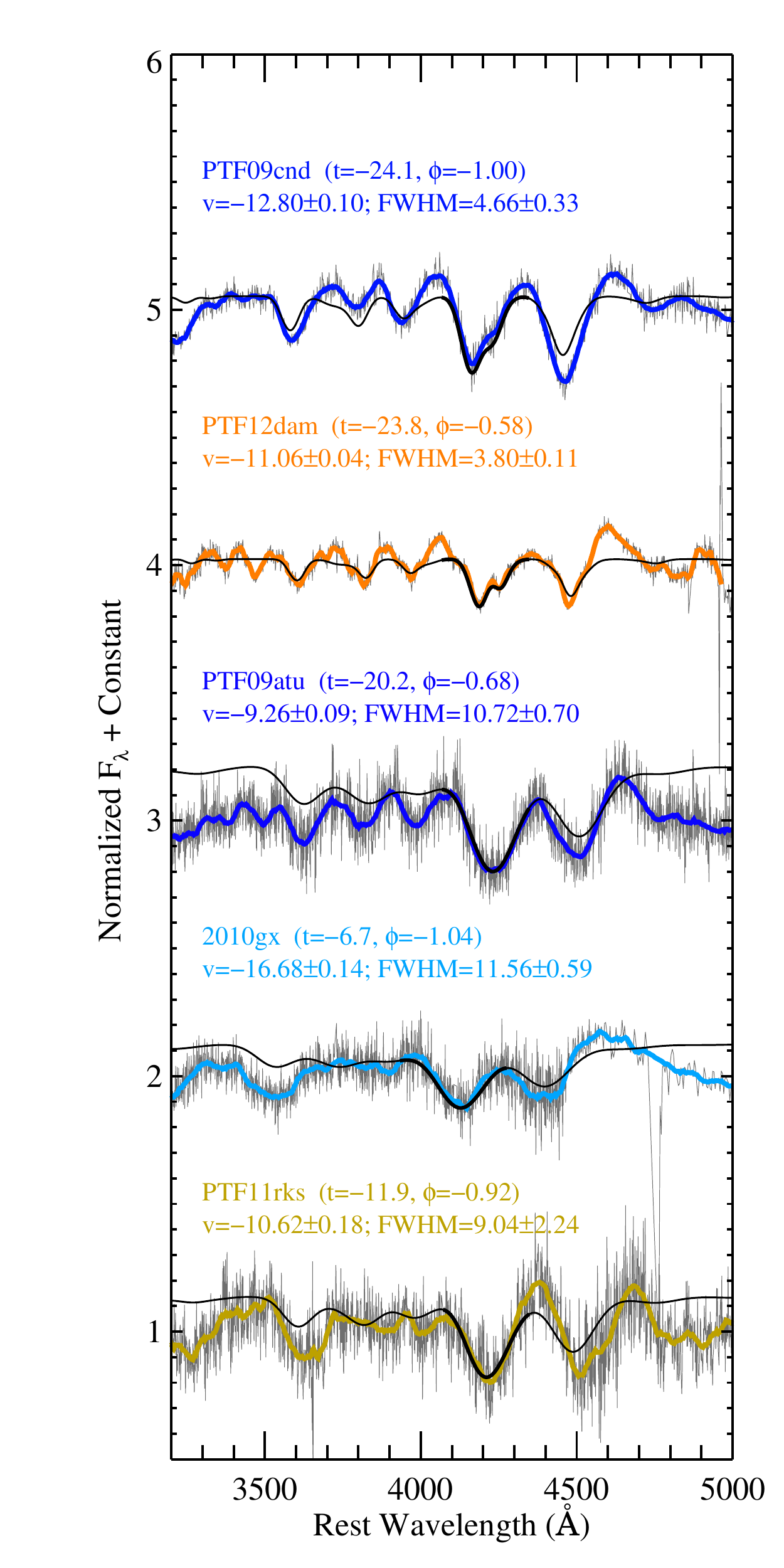}
 \caption{ Early-phase spectra of several SLSNe-I highlighting the
   \ion{O}{2} features. The spectra have been continuum divided and
   shifted for clarity. The smoothed data are shown with the thick
   colored lines. Best-fit velocities ($v$) and line widths of the
   multiplet components (FWHM) are given in 1000\,km\,s$^{-1}$. The
   fit was limited to the ``B'' feature. We show the best-fit model
   with a thick black line and an extension of this fit to other
   wavelengths with a thin black line. }
   \label{fig:compare_o2_lines}
\end{center}
\end{figure}

In Figure~\ref{fig:compare_o2_lines} we show the \ion{O}{2} features
in the early-time spectra of several SLSNe-I. The figure includes three
PTF12dam-like events (PTF09cnd, PTF09atu, and PTF12dam itself) and two
other objects (SN\,2010gx and PTF11rks) that we have tentatively
grouped with SN\,2011ke. For each of these spectra we fit our simple
\ion{O}{2} model with the $3p\,^{4}D^{0}$ transitions altered as
described above to the B feature only and then show the predicted
model spectrum at other wavelengths. We include an offset parameter in
the model fits to account for possible errors in our determination of
the continuum level.

We find that the fit to PTF12dam is excellent with the B feature
actually resolved into its two multiplet components. Resolving this
feature implies that the velocity distribution of \ion{O}{2} is
confined to a relatively narrow range. The best fit FWHM is
$3800\pm100$\,km\,s$^{-1}$. The B feature is similarly resolved for
PTF09cnd. For PTF09atu the formal fit favors a larger FWHM than we
adopted in Figure~\ref{fig:O2_lines}; the best-fit continuum level for
the model in this case is 20\% higher than our data after dividing by
the estimated continuum. If we assume our estimated continuum level is
correct then the FWHM drops to 4200\,km\,s$^{-1}$, but in this case
the B feature should be resolved. The estimated and model continuum
levels agree to $\sim2$\% for PTF12dam, and forcing the model to have a
higher continuum level quickly washes out the two local minimum in the
B feature, so the FWHM for this spectrum must be less than 
$\sim4000$\,km\,s$^{-1}$.

The FWHM of the \ion{O}{2} features of the SN\,2011ke-like objects in
Figure~\ref{fig:compare_o2_lines} is similarly dependent on the true
continuum level. Allowing the model continuum to rise $\sim15$\% above
the estimated level, SN\,2010gx and PTF11rks have FWHM of
$11,500\pm600$\,km\,s$^{-1}$ and $9000\pm2000$\,km\,s$^{-1}$,
respectively. However, if we fix the model continuum at the estimated
level, the FWHM each drop to $\sim4600$\,km\,s$^{-1}$. Although the
FWHM is strongly dependent on the estimated continuum level, the
systematic velocities are not. These velocities typically agree to
better than 500\,km\,s$^{-1}$ as the continuum level is changed (this
is also true for the PTF12dam-like objects in the figure).

The spectra of SN\,2010gx and PTF11rks shown in
Figure~\ref{fig:compare_o2_lines} are very similar despite a
systematic velocity difference of 6000\,km\,s$^{-1}$. The features are
noticeably stronger in PTF11rks, but both objects have an \ion{O}{2} E
feature that is significantly broader than in the PTF12dam-like objects
shown in the figure. Also, the C and D features in the model are not
well matched to the SN\,2011ke-like object spectra, but the two
observed spectra are similar to each other over this wavelength range.

\subsection{Spectral Features in the 5000--8000\,\AA\ Range}\label{redlines}

\begin{figure}
\begin{center}
 \includegraphics[width=\linewidth]{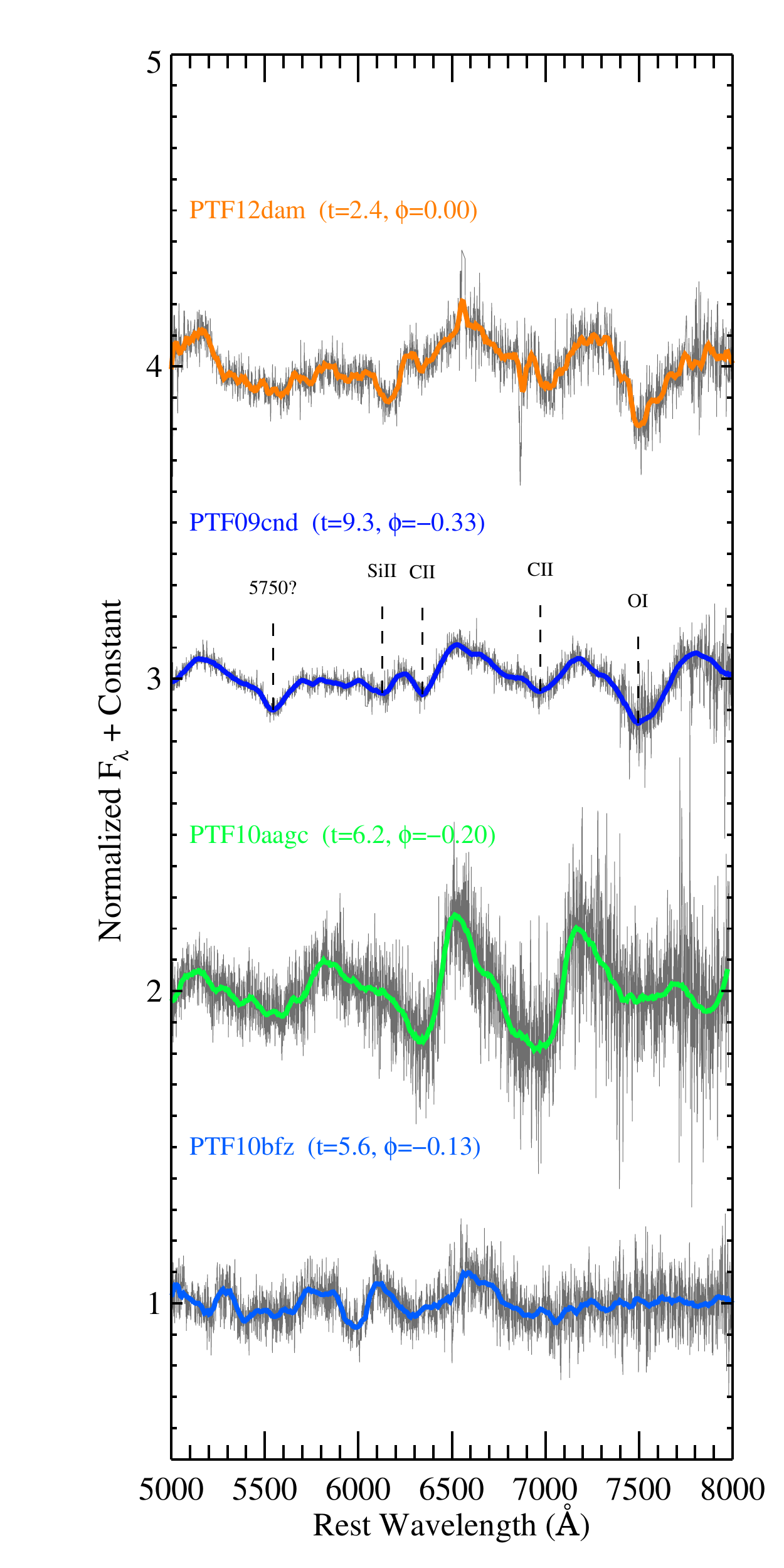}
 \caption{ Spectra of four SLSNe-I around maximum light highlighting
   weak features in the 5000--10,000\,\AA\ range. Line identifications
   are given for PTF09cnd assuming a common blueshift of
   11,000\,km\,s$^{-1}$. }
   \label{fig:o1_lines}
\end{center}
\end{figure}

In Figure~\ref{fig:o1_lines} we show spectra of PTF09cnd, PTF12dam,
PTF10aagc, and PTF10bfz around maximum light in the
5000--8000\,\AA\ range. We have classified the first two of these
objects as PTF12dam-like and the last two as SN\,2011ke-like, although
PTF10aagc is peculiar in that it exhibits hydrogen features (see
\S\ref{hhe}). In the spectra of PTF09cnd we mark the expected
locations of \ion{C}{2}, \ion{Si}{2}, and \ion{O}{1} assuming a common
velocity of 11,000\,km\,s$^{-1}$. Each of these lines is well matched
to clear local minima in the spectra, so we consider these
identifications secure (see also \citealt{nicholl2016};
\citealt{yan2017b}).

The spectra of PTF12dam displayed in Figure~\ref{fig:o1_lines} are very similar
to the PTF09cnd spectrum, but the spectra of PTF10aagc and PTF10bfz
lack clear evidence for the \ion{O}{1} $\lambda7774$
triplet. This was shown to be true for the average SN\,2011ke-like
spectrum presented in \S\ref{compare_spec}. The lack of \ion{O}{1}
$\lambda7774$ around this phase could be a defining difference
between PTF12dam-like and SN\,2011ke-like objects, although we caution
that there are relatively few spectra covering this wavelength range
at the appropriate phases.

The absorption minimum near 5550\,\AA\ in the PTF09cnd spectrum
plotted in Figure~\ref{fig:o1_lines} could not be definitively
identified. As we discuss in Appendix~\ref{average_spec0}, there are
few ions that make strong features near this wavelength without
producing stronger features that conflict with the observations. One
possibility is \ion{C}{4}, but if this is indeed the case then the
precise location of the minimum implies a velocity that is
significantly ($\sim3000$\,km\,s$^{-1}$) faster than the other
features noted in the figure.

\subsection{Spectral Features in the 1800--3500\,\AA\ Range}\label{uvlines}

We now turn to the prominent spectral features in the near-UV
band. Between 1800\,\AA\ and 2800\,\AA\ SLSNe-I typically exhibit four
relatively strong dips at roughly 2650\,\AA, 2450\,\AA, 2200\,\AA, and
1950\,\AA. We will refer to these features as UV1, UV2, UV3, and UV4,
respectively. Three of these lines were first identified through {\tt
  synapps} fits as \ion{Mg}{2} (UV1), \ion{Si}{3} (UV2), and
\ion{C}{2} (UV3) by \citet{quimby2011}. \citet{dessart2012} presented
radiation-hydrodynamical calculations that produced these same
identifications, but similar calculations by \citet{howell2013}
matched these features to blends of \ion{C}{2} and \ion{Mg}{2} (UV1);
\ion{C}{2} (UV2); \ion{C}{3} and \ion{C}{2} (UV3); and \ion{Fe}{3}
(UV4). They further find that the relative strengths of these features
depend on luminosity.  \citet{mazzali2016} constructed synthetic
spectral models for PTF13ajg and identify the four dips as \ion{C}{2}
and \ion{Mg}{2} (UV1); \ion{Ti}{3}, \ion{C}{2}, and \ion{Si}{3} (UV2);
\ion{C}{3}, \ion{C}{2}, and \ion{Ti}{3} (UV3); and \ion{Fe}{3} and
\ion{Co}{3} (UV4). For another SLSN-I, SNLS-06D4eu,
\citet{mazzali2016} favor the identifications \ion{Mg}{2} (UV1);
\ion{Si}{3} (UV2); \ion{C}{2} (UV3); and \ion{Fe}{3} (UV4).  A
consensus has yet to form on the ions responsible for these features
or if the contributions change from one event to another, although
most seem to agree that at least \ion{Mg}{2} and \ion{C}{2} are
involved.

\begin{figure}
\begin{center}
 \includegraphics[width=\linewidth]{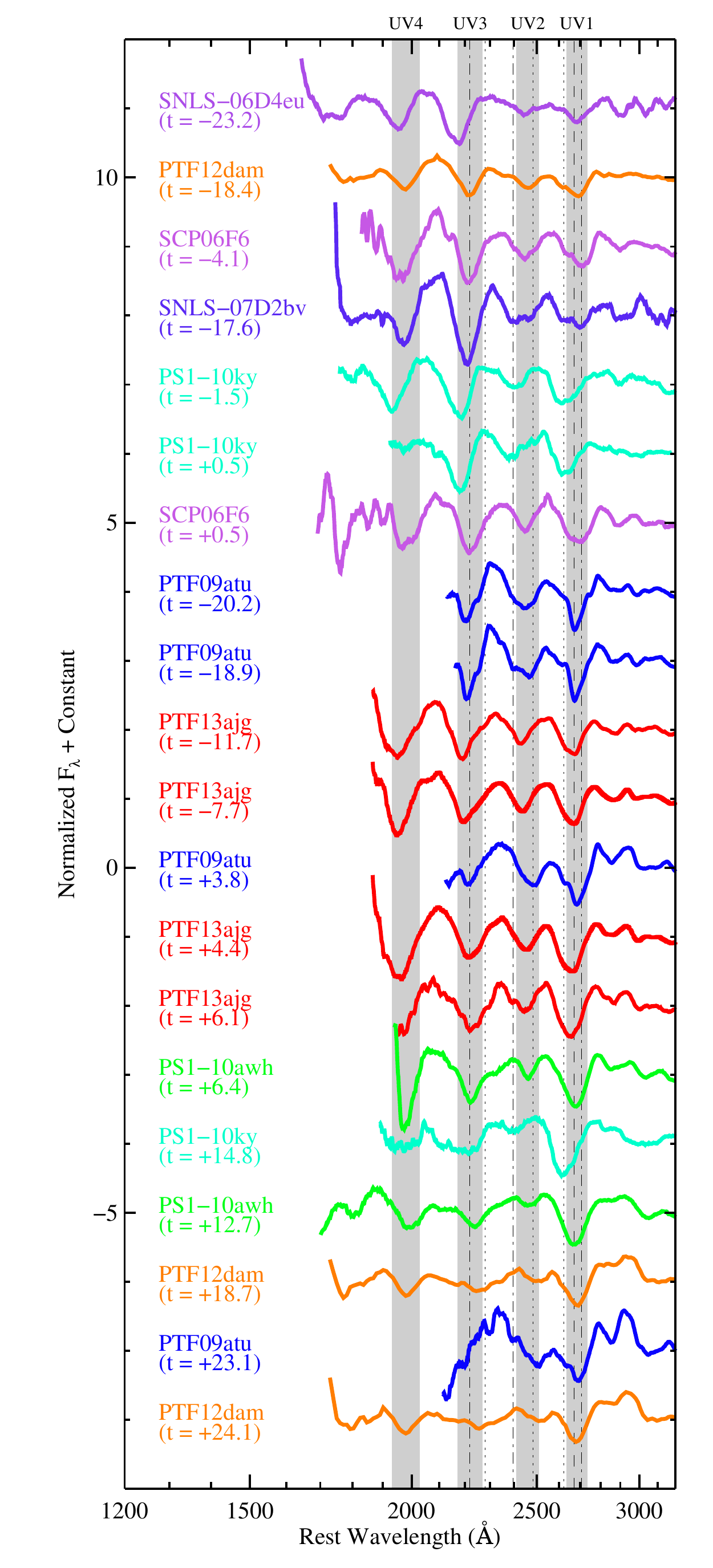}
 \caption{ Smoothed and continuum-divided spectra of SLSNe-I in the near-UV
   range. SLSN-I spectra typically show four prominent features in
   this range, which are marked by the gray shaded vertical bands. The
   expected wavelengths of several strong lines (assuming a blueshift
   of 10,000\,km\,s$^{-1}$, are marked by vertical lines: \ion{C}{2}
   with dash-dotted lines, \ion{Mg}{2} with dashes, and \ion{Fe}{2}
   with dots. Phases given for the high-redshift, non-PTF targets are based on
   rest-frame UV photometry and may be biased with respect to the optically
   derived phases of the PTF targets. }
   \label{fig:compare_nuv_spec}
\end{center}
\end{figure}

These near-UV dips are not necessarily caused by the same ions at
different phases or in different events. Some support for this is
offered by Figure~\ref{fig:compare_nuv_spec}, which shows the near-UV
spectra of several SLSNe-I at different phases. Inspection of this
figure shows that while two different spectra may exhibit dips at
similar wavelengths for one feature, other dips can be noticeably
offset. For example, the spectra of SNLS-06D4eu and the earliest near-UV
spectrum of PTF12dam both show an absorption dip near 2700\,\AA;
however, the UV3 feature in PTF12dam is centered near 2220\,\AA\ while
the closest corresponding feature in SNLS-06D4eu is centered closer to
2170\,\AA. This difference could mean that these UV3 features are the
result of different ions or that the material responsible for this
feature has a larger velocity in SNLS-06D4eu relative to other lines
(assuming a rest wavelength near 2275\,\AA\ for each spectrum, the
velocity difference would be $\sim4000$\,km\,s$^{-1}$). 
Interestingly, the later-time spectra of PTF12dam
may reveal a weak dip at a wavelength similar to the UV3 absorption in
SNLS-06D4eu. This may suggest that there are two (or more) ions
contributing to the UV3 feature and that the relative strengths of
these ions vary from event to event and with time. This especially may
be the case for the UV2 feature, which shifts considerably more 
with time in the objects observed over sufficiently
long temporal baselines (e.g., PTF09atu and PTF12dam).

To study possible line identifications for these UV features, we first
consider the spectra of PTF09atu and PTF12dam --- two SLSNe-I for which we
have already identified spectral features in the optical. We can thus
use the velocities found for the optical features to constrain the
expected wavelengths of potential lines in the near-UV. As shown
above, the optical lines of these SLSNe-I have low velocity dispersions,
helping to resolve any blends into individual components. We
present the evolution of the ion velocities in \S\ref{velocities}. 
One problem is that the velocities for ions in the
optical are not measured at all phases (e.g., \ion{O}{2} disappears
around maximum light). Thus, we must sometimes extrapolate velocity
measurements to earlier or later phases. We do this by fitting a
simple linear model to the trusted measurements and evaluating this
fit at the required epochs. This may underestimate velocities at early
and late phases as the measured velocity of normal SNe are
often observed to begin with relatively rapid declines before
flattening out at later times.

\begin{figure}
\begin{center}
 \includegraphics[width=\linewidth]{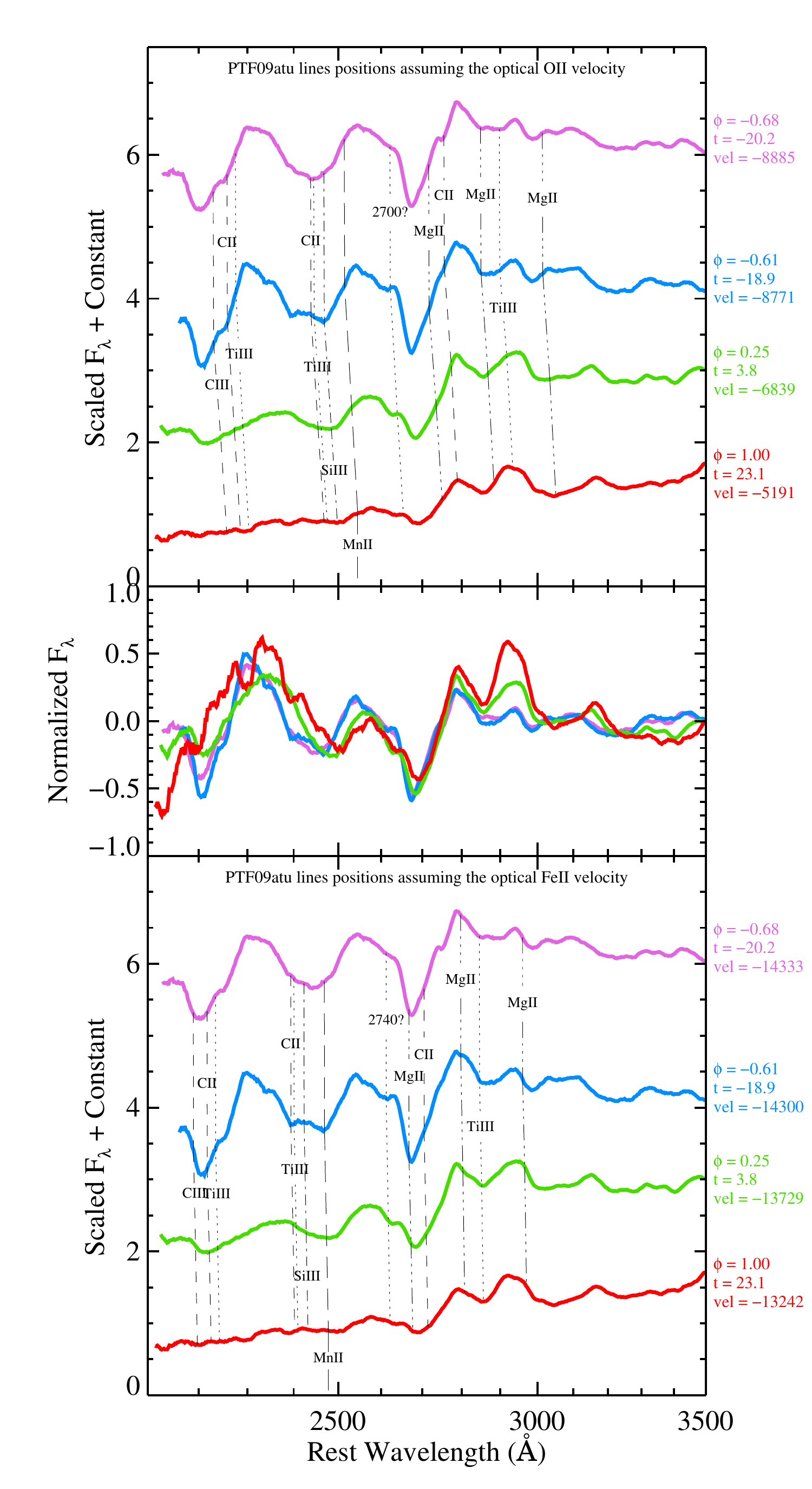}
 \caption{ Spectral evolution of PTF09atu in the near-UV region. The
   same three spectra are plotted in each panel, but the middle panel
   shows the data after division by the estimated continua. Vertical
   lines mark the expected positions of several lines based on the
   velocities derived from \ion{O}{2} (top panel) and \ion{Fe}{2}
   (bottom panel). The spectral phase ($\phi$), time since maximum
   light in days ($t$), and assumed line velocities in km\,\,s$^{-1}$
   are written to the right of the spectra.}
   \label{fig:PTF09atu_uv_lines}
\end{center}
\end{figure}

Figure~\ref{fig:PTF09atu_uv_lines} shows the spectral evolution of
PTF09atu in the near-UV. We first note that the UV1, UV2, and UV3 line
profiles are complex; it is evident from the deeper, narrower UV1
absorption compared to the shallower, broader UV2 feature and the
notches in the wings of these broad dips that these are the product of
multiple, blended lines. In the top panel we have marked the expected
positions of various line minima based on the optical \ion{O}{2}
velocities inferred from the cross-correlation method described in
\S\ref{velocities}. The weak notches in the red side of the UV1
and UV3 features in the earliest spectra are well matched to the
expected locations of \ion{C}{2}\,$\lambda\lambda$2836, 2837 and
$\lambda\lambda$2325, 2326, 2328 (respectively) at the same velocity
as \ion{O}{2}. Next we find that the location of the UV1 minimum is
significantly to the blue of the expected \ion{Mg}{2} minimum,
assuming the same velocity as \ion{O}{2}. There are two possible
interpretations of this result: (1) the UV1 feature may not be
dominated by \ion{Mg}{2}, or (2) the \ion{Mg}{2} line-forming region is
at a systematically {\em higher} velocity than \ion{O}{2}. In support
of the first possibility, we note a slight inflection in the red wing
of the UV1 feature precisely where we would expect to find \ion{Mg}{2}
if it had the same velocity distribution as \ion{O}{2}. In support of
the latter possibility, we cannot identify a better candidate to
dominate the UV1 feature (see \S\ref{uvfeatures}). Another
notable feature in Figure~\ref{fig:PTF09atu_uv_lines} is the notch
visible in all spectra on the blue wing of the UV1 feature. This notch
shifts to the red over time, suggesting it is tracing the velocity
distribution that evolves similar to \ion{O}{2}, but with a rest
wavelength near 2700\,\AA. As discussed below, we could not identify
the ion(s) responsible for this feature. Finally, we note that similar
to the ions proposed for the UV1 feature, the ions proposed for the
UV3 feature are systematically offset to the red.

In the bottom panel of Figure~\ref{fig:PTF09atu_uv_lines} we plot the
same spectra but consider a different velocity evolution --- one
following the optical \ion{Fe}{2} lines --- to predict where major
features should land. In this case, the expected position of the
\ion{Mg}{2} doublet is better aligned with the minimum of the UV1
feature, although it is now offset somewhat in the other
direction. That is, if this feature is dominated by \ion{Mg}{2}, then
it would seem that the velocity distribution of this ion is
systematically shifted to {\em lower} velocities than \ion{Fe}{2}, at
least as inferred from the line minima. The weak inflection point
observed around 2715\,\AA\ is now roughly matched by \ion{C}{2}. The
lines expected to contribute to the UV3 feature now agree more closely
with the data; however, the various lines proposed for the UV2 feature 
are offset to the blue of the observed feature.

\begin{figure}
\begin{center}
 \includegraphics[width=\linewidth]{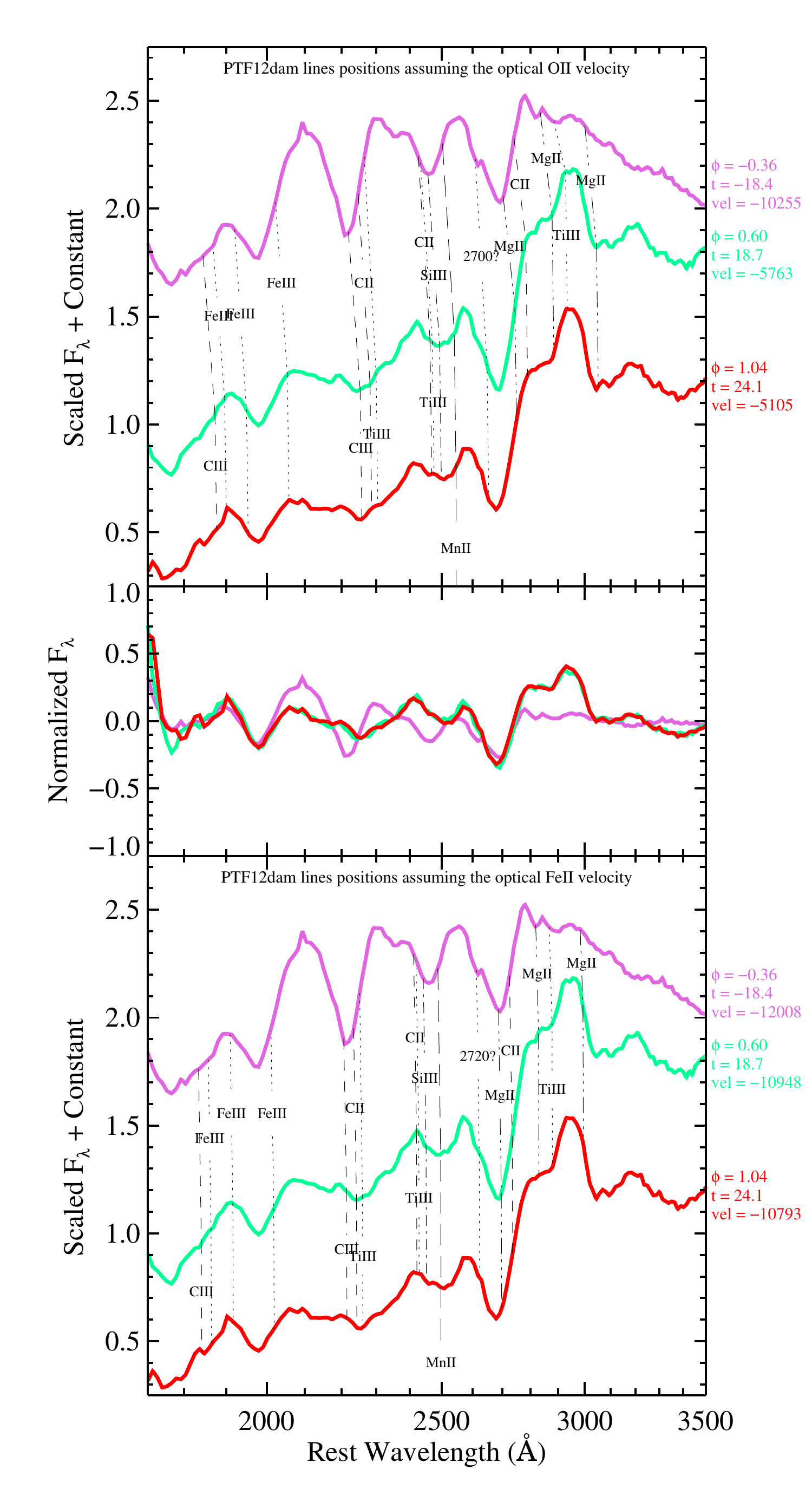}
 \caption{ Similar to Figure~\ref{fig:PTF09atu_uv_lines} but for
   PTF12dam.}
   \label{fig:PTF12dam_uv_lines}
\end{center}
\end{figure}

It seems likely from this one object that there are different velocity
distributions for different ions, with \ion{O}{2} and \ion{C}{2}
favoring lower velocities, and \ion{Fe}{2} and \ion{Mg}{2} favoring
higher velocities. To see if this is unique to PTF09atu, we now
consider the near-UV spectra of PTF12dam obtained with {\it HST}.
Figure~\ref{fig:PTF12dam_uv_lines} shows these
data; again, we mark line positions based on optical \ion{O}{2}
velocities in the top panel and optical \ion{Fe}{2} velocities in the
lower panel. The phases of observation are now different than for
PTF09atu, but there are some similar results. The unknown notch on the
blue edge of the UV1 feature is again present, although adopting the
\ion{Fe}{2} velocities, a slightly different rest wavelength is
favored (2720\,\AA\ vs. 2740\,\AA\ for PTF09atu). The minimum of the
UV1 feature is better tracked by \ion{Mg}{2} using the \ion{Fe}{2}
velocities. In addition, the proposed lines for the UV2 feature tend to be
systematically biased to the blue of the data. A key difference is
that the UV2 feature in the earliest ($\phi=-0.36$) spectrum of
PTF12dam is noticeably more narrow than the $\phi \approx -0.65$ spectra
of PTF09atu. There is a clear notch in the first PTF12dam spectra to
the blue of the UV2 feature around 2350\,\AA, which may be weakly
present in the $\phi=-0.61$ spectra of PTF09atu.

We also show in Figure~\ref{fig:PTF12dam_uv_lines} the expected
positions of strong \ion{Fe}{3} features. According to NIST there are
a number of \ion{Fe}{3} lines around 2000\,\AA\ which should blend
into two or three distinct features. However, the expected minima for
these blends are poorly matched to the UV4 feature adopting either the
\ion{O}{2} or \ion{Fe}{2} velocities. This result is in conflict with
the synthetic {\tt syn++} spectra presented in the Appendix and with
previous associations of this feature with \ion{Fe}{3}
\citep{howell2013, mazzali2016}.

\subsection{Possible Contributors to the UV Features}\label{uvfeatures}

To consider possible line identifications for the notches observed
around 2350\,\AA\ and 2600\,\AA, and to search for possibly different
ions capable of contributing to the UV1, UV2, UV3, and UV4 features,
we construct synthetic spectra using {\tt syn++}. We compare these
models to our {\it HST} UV grism spectra of PTF12dam to determine if a given
species can produce lines at wavelengths appropriate to the data
without generating strong features that are inconsistent with the data
at other wavelengths. For our model spectra, we fix the photospheric
velocity at 14,000\,km\,s$^{-1}$ because this generates model \ion{O}{2}
spectra that are roughly consistent with the data (note that the positions
of the line minima imply significantly lower velocities). We also fix
the ion temperature at 10,000\,K, but we also consider a set of
synthetic spectra generated for 15,000\,K ions as a check. The minimum
and maximum velocities for each ion were set at 10,000\,km\,s$^{-1}$ and 
40,000\,km\,s$^{-1}$, respectively, and the $e$-folding scale was set to 
1.0. We then varied the reference line opacity to find the best match 
to the data.

After manually searching through all 83 ions in the {\tt syn++}
library, we find the following possible contributors to the observed
UV features: \ion{B}{4}, \ion{C}{2}, \ion{O}{5}, \ion{Mg}{1},
\ion{Mg}{2}, \ion{Al}{2}*, \ion{Cr}{2}*, \ion{Mn}{1}, \ion{Fe}{2}, and
\ion{Fe}{4}* (UV1); \ion{B}{1}, \ion{C}{1}, \ion{C}{2}, \ion{Mg}{1},
\ion{Si}{3}, \ion{P}{1}, \ion{Ti}{3}, \ion{V}{3}*, \ion{Mn}{2},
\ion{Fe}{1}, \ion{Fe}{2}, and \ion{Co}{2}* (UV2); \ion{C}{2},
\ion{C}{3}, \ion{N}{4}, \ion{Co}{2}*, \ion{Ni}{1}, and \ion{Ba}{2}
(UV3); and \ion{B}{3}, \ion{N}{2}, \ion{Mg}{1}, \ion{S}{2},
\ion{Cr}{2}*, \ion{Mn}{3}*, \ion{Fe}{3} (UV4).  Asterisks mark cases where
additional, strong features are predicted that at least partially
disagree with the PTF12dam spectra. Detailed models are required to
determine if any of these species are truly compatible with the
observations. Many of these ions produce strong features at shorter
wavelengths. Further observations constraining the far-UV spectra of
SLSNe-I may prove useful in determining the true identifications of the
near-UV spectral features (e.g., \citealt{yan2015}).

\subsection{Identification of \ion{Mn}{2}}

Again, the UV features are likely blends of multiple ions, and the
relative strengths of these ions may ebb and flow as the spectra
evolve. The most likely candidate for such a change in the dominant
ion is the UV2 feature, which shifts significantly more to the red
than do the other features. Most of the possible UV2 contributors
listed above cannot explain the location of the UV2 minimum at later
phases (e.g., the $\phi=0.60$ spectra of PTF12dam) without invoking
impractically low velocities. We find that the best identification for
the UV2 feature at these later phases is likely \ion{Mn}{2}. This ion
has a resonant transitions at 2576, 2593, and 2605\,\AA\ that agree
well with the later-phase position of the UV2 line (assuming a
velocity common to other lines; see Figures
\ref{fig:PTF09atu_uv_lines} and \ref{fig:PTF12dam_uv_lines}), and
there are no stronger lines expected over the available data
range. The spectra may also show \ion{Mn}{2} $\lambda\lambda$2939, 2949
and a blend of \ion{Mn}{2} $\lambda\lambda$3460, 3474, 3482, but these
features are weaker and possibly blended with other ions.  No strong
lines are expected in the optical, but \ion{Mn}{2} has possibly been
detected in the near-infrared spectra of other SNe
(e.g., \citealt{marion2009}). The Mn may have been synthesized in the
explosion or in the final burning stages of the progenitor star, or it
may reflect the metallicity of the progenitor. This feature may thus
yield constraints on the progenitor or burning during the explosion.

\subsection{Evidence for Hydrogen and Helium}\label{hhe}

As previously noted, the spectra of SLSNe-I most closely resemble those
of SNe~Ic,
which exhibit little to no hydrogen and helium in their spectra. We
now examine our spectroscopic SLSN-I sample to search for any signs of
these elements.

Previous works suggest possible links between SLSNe-I and hydrogen-rich 
or helium-rich events (\citealt{benetti2014, inserra2016}; see, however,
possible host differences noted by \citealt{leloudas2015} and 
\citealt{perley2016}). These include the high-luminosity SN\,2008es, which
exhibited broad hydrogen features in its spectra but without the
narrow emission lines characteristic of SNe~IIn and other SLSNe-II
\citep{miller2009, gezari2009}. It has been suggested that SN\,2008es
may actually be a relative of SLSNe-I that retained some of its
hydrogen envelope at the time of its explosion. There have also been
SLSNe-I reported that appeared hydrogen poor at maximum light but which
developed hydrogen emission lines in their late-phase spectra
\citep{yan2015, yan2017b}. For these objects, it is possible that the
hydrogen detached from the progenitor just prior to the SN
explosion. In this scenario helium should also be expected to be
present, and it is perhaps present in other SLSN-I atmospheres as
well.  There have been reports of helium in the spectra of the SLSN-I
SN\,2012il \citep{inserra2013}, but analysis of the spectra shows that
the feature proposed to be \ion{He}{1} $\lambda10830$ is actually
significantly offset to the red (see Appendix~\ref{2012il}); thus, this
feature is unlikely to be \ion{He}{1}.

\begin{figure}
\begin{center}
 \includegraphics[width=\linewidth]{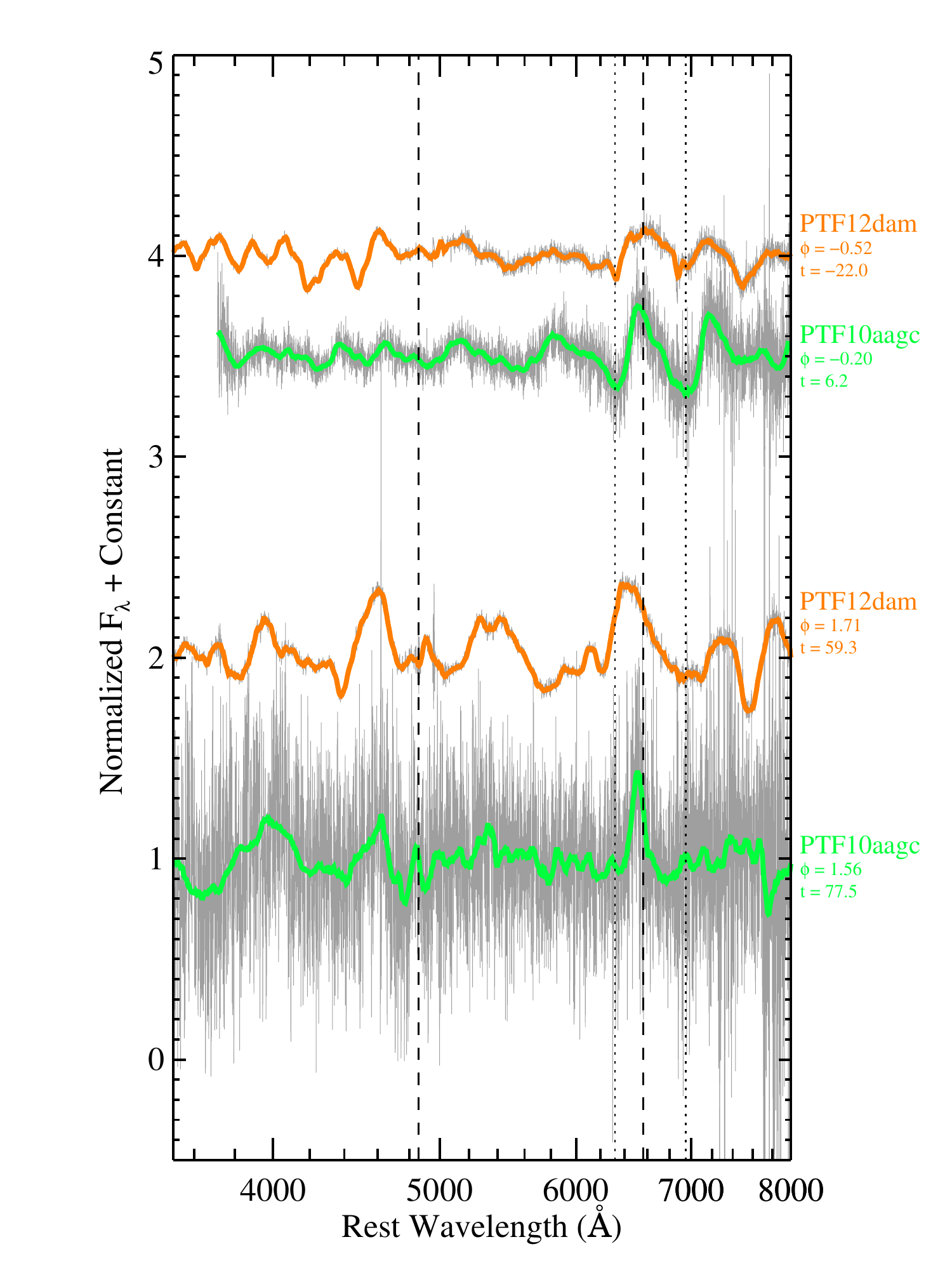}
 \caption{ Continuum-divided spectra of PTF12dam and PTF10aagc with
   the rest wavelengths of H$\alpha$ and H$\beta$ marked with vertical
   dashed lines and the expected positions of 12,000\,km\,s$^{-1}$
   blueshifted \ion{C}{2} lines marked with vertical dotted
   lines. Smoothed versions of the spectra are plotted with the thick
   colored lines. }
   \label{fig:hlines}
\end{center}
\end{figure}

We visually inspected the spectral time series for each of our SLSNe-I
to look for signs of hydrogen. This is complicated by the potential
presence of other ions that may produce features at similar
wavelengths. In particular, \ion{C}{2} $\lambda\lambda$6578.05, 
6582.88 and $\lambda\lambda$7231.33, 7236.42 bracket the
expected location of H$\alpha$ emission. Together, the P-Cygni profiles
from these lines can potentially mimic the presence of H$\alpha$ (the
6578--82 doublet in particular could form a P-Cygni feature very close
to what might be expected for H$\alpha$). In Figure~\ref{fig:hlines}
we show the spectra of PTF12dam at $t=-22$\,d ($\phi=-0.52$) with
the possible locations of these \ion{C}{2} features marked assuming a
blueshift of 12,000\,km\,s$^{-1}$ (see also
Fig.~\ref{fig:o1_lines}). Vertical dashed lines mark the rest
wavelengths of H$\alpha$ and H$\beta$. While there is a possible broad
emission feature roughly centered where H$\alpha$ would be, the
absorption dips to either side of this are well matched by \ion{C}{2}
at a velocity similar to the \ion{O}{2} lines discussed in
\S\ref{o2lines} and the \ion{C}{2} features discussed in
\S\ref{uvlines}. Thus, it is not clear if hydrogen is required to
explain the spectra of PTF12dam.

Figure~\ref{fig:hlines} also shows spectra of PTF10aagc at an early
($\phi=-0.20$) and later ($\phi=+1.56$) phase. In this case the
emission peak near H$\alpha$ is much stronger in the early-phase
data. In the later-phase spectrum, there is an emission feature
slightly blueshifted from rest H$\alpha$ similar to what has been
observed for other SLSNe-I with late-time hydrogen emission
\citep{yan2015,yan2017b}. This offset is seen for both H$\alpha$ and
the weaker H$\beta$ line, which is apparent in the smoothed data as
well. We thus conclude that at least some hydrogen is present in the
envelope of PTF10aagc at the time of explosion. The later-phase
spectra of PTF12dam do show broad emission to the blue of rest
H$\alpha$; however, the line center implies a blueshift of
6000\,km\,s$^{-1}$, which tends to exclude this possibility. An
alternative identification is [\ion{O}{1}] $\lambda\lambda$6300,
6364, although the observed feature is shifted to the red of this. We
also note that the later spectra of PTF12dam show a broad absorption
minimum near 7550\,\AA, which we identified as the \ion{O}{1}
$\lambda$7774 triplet. This feature appears to be absent in the
PTF10aagc spectra, perhaps consistent with more envelope
stripping in the case of PTF12dam (see \citealt{sun_galyam2017} and
references therein). Spectra of other SLSNe-I do not provide clear
evidence for hydrogen. The late-phase spectra of SN\,2011ke do appear
to show a broad feature centered at the wavelength of H$\alpha$ (see
Fig.~\ref{fig:spec07} in Appendix~\ref{specplots}), but these spectra
are dominated by host-galaxy light and it is possible that this
feature is simply made from the wings of the H$\alpha$ profile from
the host.

\begin{figure}
\begin{center}
 \includegraphics[width=\linewidth]{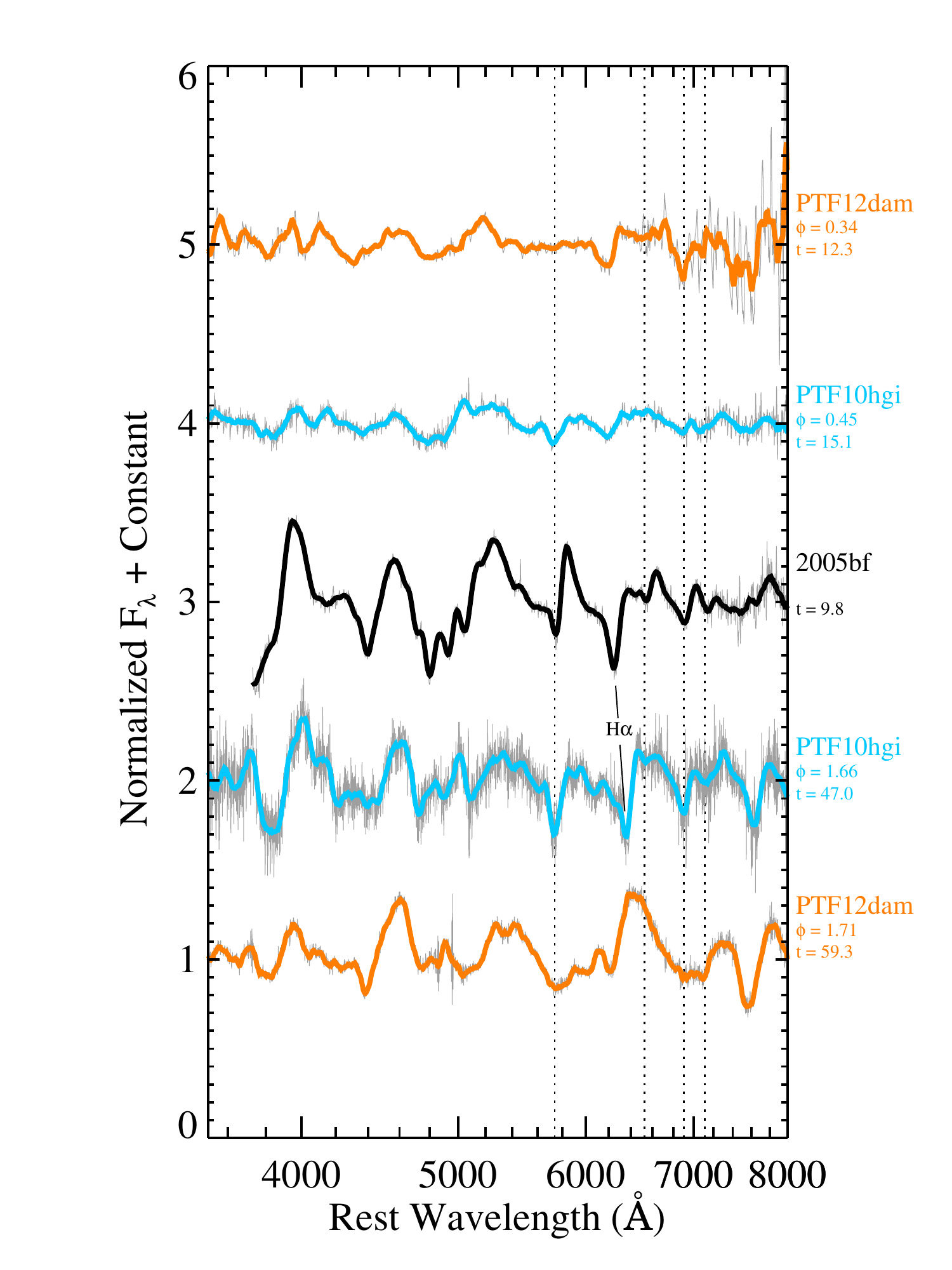}
 \caption{ Spectra of the SN~Ib 2005bf compared to the
   SLSNe-I PTF10hgi and PTF12dam. The spectra have been continuum 
   divided, and the thick lines plot smooth versions of the data. The
   vertical dotted lines mark the expected position of prominent
   \ion{He}{1} lines assuming a blueshift of 7000\,km\,s$^{-1}$. All
   four of these lines are clearly detected in the SN\,2005bf spectra
   and they can also be seen in PTF10hgi. However, the presence of
   \ion{He}{1} lines in the PTF12dam spectra is less obvious.}
   \label{fig:helines}
\end{center}
\end{figure}

We examined the temporal series of spectra of our SLSN-I sample and find
one case where helium is clearly detected. PTF10hgi not only shows a
much stronger absorption feature near 5750\,\AA\ (which is sometimes
identified as blueshifted \ion{He}{1} $\lambda5875$ but can be
blended or dominated by \ion{Na}{1} D) compared to other SLSNe-I, but
it also exhibits absorption dips at a consistent velocity for the 6678,
7065, and 7281\,\AA\ lines of \ion{He}{1}. In Figure~\ref{fig:helines}
we compare the spectra of PTF10hgi to the peculiar SN~Ib
2005bf (e.g., \citealt{folatelli2006}). \ion{He}{1} lines show up
clearly in the spectra of PTF10hgi by $t=+15$\,d ($\phi=0.45$) and
they persist until at least $t=+95$\,d ($\phi=1.87$). Some other
SLSNe-I may show weak evidence for one or two of these lines, but the
presence of helium in these other cases is far less definitive. For
example, the spectra of SSS120810 show possible \ion{He}{1} $\lambda$7065 
and \ion{He}{1} $\lambda$7281 lines blueshifted by about 
12,000\,km\,s$^{-1}$, but if the 5875\,\AA\ feature is 
present it is rather weak. PTF10aagc also shows
a relatively strong feature around 5550\,\AA\ in its $\phi=-0.20$
spectrum that could be \ion{He}{1} $\lambda$5875 blueshifted by about
15,000\,km\,s$^{-1}$. However, this feature could be from another line
(e.g., \ion{Na}{1} D), and it is unclear if the other \ion{He}{1} lines
are detected or if the features at roughly the appropriate wavelengths
are the result of other ions such as \ion{C}{2} (see also
\citealt{yan2017b}).

Strong helium lines appear to be a unique feature of PTF10hgi. For
comparison, we show spectra of PTF12dam in Figure~\ref{fig:helines} at
similar phases to the PTF10hgi spectra. Similar to SSS120810, in the
$t=+57.8$\,d ($\phi=1.71$) spectrum of PTF12dam there are
dips around 6900 and 7100\,\AA\ that might suggest the presence of
helium, but the expected \ion{He}{1} $\lambda 5875$ feature is
weak or absent. There are also other possibilities for the potential 
\ion{He}{1} features including contributions from \ion{C}{2}, \ion{O}{1}, 
or a false absorption feature created by the emerging nebular lines of
\ion{O}{1} and \ion{Ca}{2}.

\citet{folatelli2006} have argued that the 6260\,\AA\ dip in the
spectra of SN\,2005bf shown in Figure~\ref{fig:helines} is
high-velocity H$\alpha$. Spectra of PTF10hgi contain a similar 
feature that strengthens up to the $t=+95$\,d spectrum before
subsiding. However, for PTF10hgi the minimum of the dip is closer to
6350\,\AA. It is possible that this, too, is H$\alpha$ with an
absorption minimum velocity merely 3000\,km\,s$^{-1}$ faster than the
\ion{He}{1} lines (instead of $\sim7000$\,km\,s$^{-1}$ faster for
SN\,2005bf). This identification is supported by the detection of
possible H$\beta$, H$\gamma$, and H$\delta$ lines at a similar velocity. 
The spectra of PTF10hgi bear a striking resemblance to those of the
Type~IIb SN\,1993J, which also shows hydrogen lines offset to higher
velocities than its helium lines \citep{barbon1995}. We consider the
identification of hydrogen and helium in the spectra of PTF10hgi
secure, and these features remain clearly visible in the spectra for
months. Thus, PTF10hgi may be the first example of a SLSN-IIb.

Our automated classification system may merely have flagged PTF10hgi as
a SLSN-I owing to a lack of SN~IIb comparison templates. But the
PTF10hgi spectra do show some resemblance to those of SLSNe-I and SNe~Ic.  
Nine of the eleven PTF10hgi spectra have $\Delta I_{\rm Ic - Ib}$ scores
that favor classification as a SN~Ic over a SN~Ib. But again, based on
the traditional classification scheme, the clear presence of hydrogen
and \ion{He}{1} lines should result in a SN~IIb classification.

\section{Velocity Measures}\label{velocities}

In principle, the velocity structure of the ejecta can be determined
through multi-epoch spectroscopic measurements. Since SNe
produce broad P-Cygni profiles, it is common to measure the wavelength
of the absorption trough (the minimum of the line profile) and to
quote an ion's absorption velocity using the known rest wavelength of
the line and the (relativistic) Doppler formula. For SLSNe-I this
procedure is complicated by two facts. First, the line features are
very weak in comparison to those of normal-luminosity SNe. Second, there
are multiple, blended lines present and unknown velocities that make
association of a rest wavelength with any particular feature
troublesome. Thus, in practice it is more difficult to assign
velocities to particular ions from the spectra of SLSNe-I than for
SNe~Ic (see for example \citealt{liu2016}).

For normal-luminosity SNe it is common to quote \ion{Fe}{2}
velocities based on the absorption minimum of the
5169\,\AA\ feature. This line is clearly detected in normal
SN spectra and can be readily identified by the presence of two
other \ion{Fe}{2} lines at 4923\,\AA\ and 5018\,\AA. Usually the
5169\,\AA\ line is the strongest of these three, and it has been
determined to be a good tracer of the photospheric velocity
\citep{dessart_hillier2005}. However, this is not always the case. In
SNe~Ic the 5169\,\AA\ line is sometimes weaker than the two bluer lines, or
the three may be blended together into a single, broad dip
\citep{modjaz2014}. It is possible for these lines to form above the
photosphere and thus give a biased measure of the true photospheric
velocity.

We have attempted to measure \ion{Fe}{2} absorption velocities from
the spectra in our sample, but clear measurements are not always
possible given the difficulties mentioned above. We therefore attempt
to measure velocities using several techniques and present the values
obtained from each of these to give a sense of the systematic errors
in these measurements.

We focus our discussion of velocity measurements on PTF12dam, which
has excellent spectral coverage. The first technique we attempt is to
identify the \ion{Fe}{2} $\lambda\lambda$4923, 5018, and 5169 
triplet by eye and
then fit second-order polynomials to the data near each minimum. In
the spectra taken near maximum light (spectroscopic phases $-0.3 <
\phi < +0.11$), all three of these lines can be clearly resolved. The
formal fits result in relative velocity offsets of
$\sim500$\,km\,s$^{-1}$. This is consistent enough that we feel
confident identifying these features as the \ion{Fe}{2} triplet. Some
of the earlier-phase data also show one or two absorption features that
appear to correspond to \ion{Fe}{2} triplet components, but we note
that the data are possibly contaminated by other features. At later
phases the lines blend together. This may indicate an increase in the
velocity width of the features. Also, the 5169\,\AA\ line appears
to weaken relative to the two bluer components. In general, the
5169\,\AA\ feature does not appear to dominate this region of the
spectra, implying that previous measurements that assumed the
minimum of the broad feature corresponds to the 5169\,\AA\ line may
be systematically in error \citep{modjaz2014}. The 5018\,\AA\ feature
may be the best tracer of the \ion{Fe}{2} ion velocity in this case.

In Table \ref{table:PTF12dam_fe2} we present the velocities for
PTF12dam as measured individually from each of the \ion{Fe}{2} triplet
components. There is considerable scatter for individual lines (around
500\,km\,s$^{-1}$), but the data favor \ion{Fe}{2} velocities of about
$-11,500$\,km\,s$^{-1}$ at the earliest epochs and a possible decline
to lower blueshifts over time. The last epoch for which we find a
reasonably well-resolved \ion{Fe}{2} line minimum ($t \approx +59$\,days)
suggests a sudden drop in \ion{Fe}{2} ion velocity to about
$-10,500$\,km\,s$^{-1}$.

\input{PTF12dam.fe2.vels.tex} 

We next attempt to measure \ion{Fe}{2} ion velocities from the
PTF12dam spectra by fitting a simple model to the spectra. The model
is created using the wavelengths and relative intensities of
\ion{Fe}{2} lines from NIST assuming a 10,000\,K plasma in LTE.
Each line is represented by a
Gaussian. Similar to the analysis in \S\ref{o2lines}, the
wavelengths of the lines are shifted using the relativistic Doppler
formula, and the line widths and absolute intensities are fitted
parameters (applied equally to all lines). The models are fit to the
flattened spectra in the 4600--5200\,\AA\ range. Visual inspection
suggests that the near-maximum-light ($t = +2.4$\,day; spectroscopic
phase $\phi=0$) spectra show three features with minimum reasonably
matched to our simple model with a blueshift of
11,760\,km\,s$^{-1}$. Although the overall fit is imperfect, the
location of the minimum is relatively robust. We can compare this
blueshift to the velocities found by fitting the 5018 and
5196\,\AA\ lines individually, which are 11,640 and
11,780\,km\,s$^{-1}$, respectively. This supports an \ion{Fe}{2}
absorption velocity of about 11,700\,km\,s$^{-1}$ at maximum light for
PTF12dam. The FWHM of the fit is dependent on the local continuum (see
\S\ref{o2lines}) and the 5018, and 5169\,\AA\ lines appear to be
blended with other features, but the 4923\,\AA\ line suggests a
FWHM of less than 5000\,km\,s$^{-1}$. Some of the additional PTF12dam
spectra can be adequately fit using this simple model, but the model
fitting breaks down when the three primary features are not all
clearly detected. The best-fit values are presented in
Table~\ref{table:PTF12dam_fe2}.

\begin{figure}
\begin{center}
 \includegraphics[width=\linewidth]{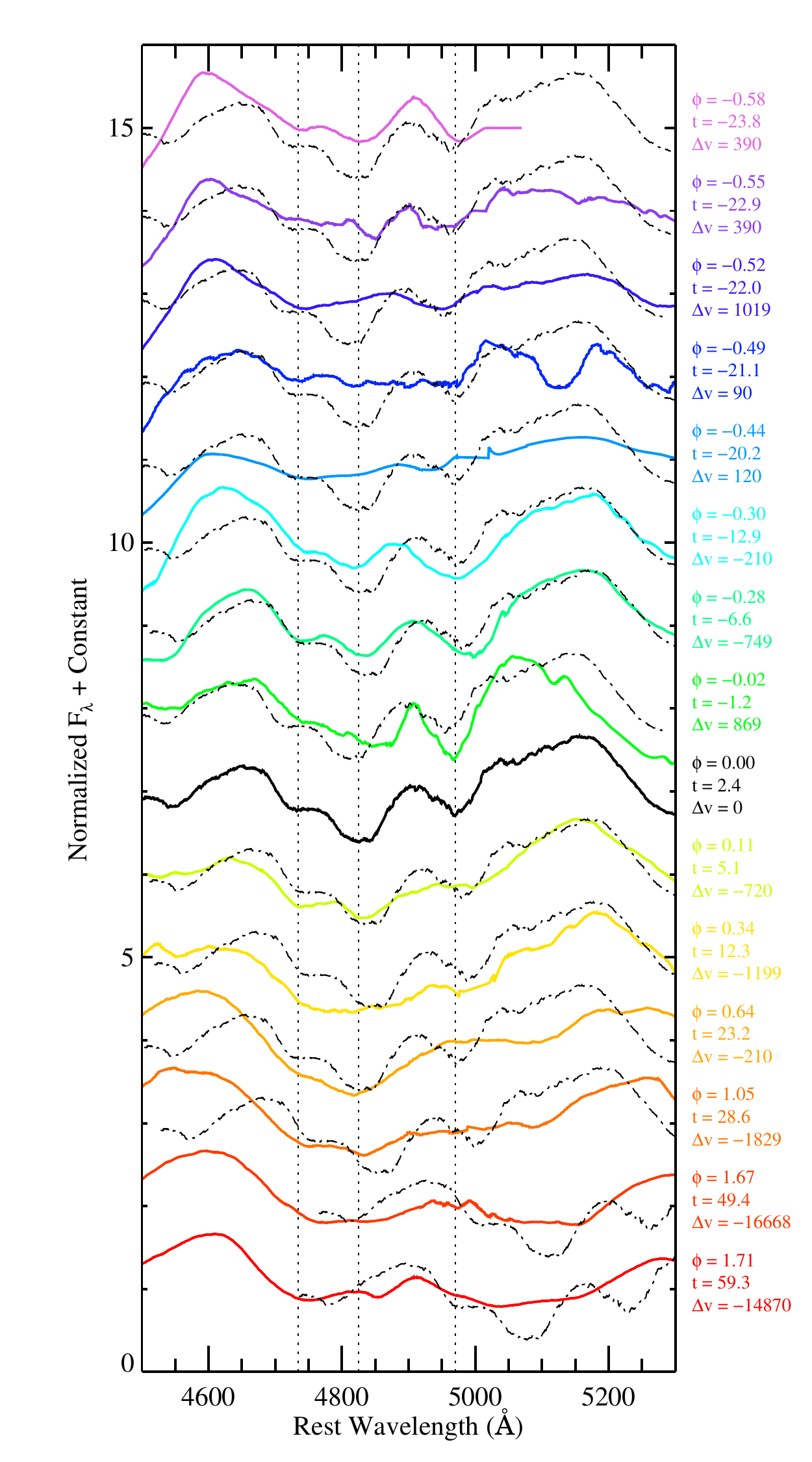}
 \caption{ Spectra of PTF12dam highlighting the region of \ion{Fe}{2}
   $\lambda\lambda$4923, 5018, 5196. Thick curves are the smoothed and
   flattened data. The right margin shows the spectroscopic phase
   ($\phi$), light-curve phase ($t$), and velocity shift ($\Delta v$)
   in km\,s$^{-1}$ relative to the maximum-light spectrum in black (positive
   values mean greater blueshifts). The dash-dotted lines show the
   maximum-light spectra shifted by the relative velocity found
   through cross-correlation. The vertical dotted lines mark the
   positions of the \ion{Fe}{2} lines in the maximum-light spectrum for
   reference ($v_{\rm abs} = 11,760$\,km\,s$^{-1}$).}
   \label{fig:PTF12dam_FeII_spectra}
\end{center}
\end{figure}

An alternative technique is to measure the velocity difference between
two spectra directly using cross-correlation
(e.g., \citealt{modjaz2014}). This gives the relative velocity shift
between the spectra, but if one of the input spectra has a known
velocity measured from the techniques above, this can be added to the
relative shift to determine the absolute velocity. In Figure
\ref{fig:PTF12dam_FeII_spectra} we show spectra of PTF12dam in the
4500--5300\,\AA\ range to highlight the \ion{Fe}{2} features. We select the
spectrum taken near maximum light and cross-correlate this against the
other spectra shown in the figure to determine the relative velocity
shifts. The dashed-dotted lines show the maximum-light spectrum
shifted according to the cross-correlation results for
comparison. There is clear evolution in the line features over the
time interval considered. At early times the emission peak from the
\ion{O}{2} P-Cygni profiles likely affects features at shorter
wavelengths, and at later times the emerging \ion{Mg}{2} emission likely
overwhelms the \ion{Fe}{2} 5169\,\AA\ absorption. Nonetheless, the 
cross-correlation appears satisfactory until $t = +29$\,days. Assuming the
reference spectrum has a velocity of 11,760\,km\,s$^{-1}$, we give
the cross-correlation velocity evolution in
Table~\ref{table:PTF12dam_fe2}.

\begin{figure}
\begin{center}
 \includegraphics[width=\linewidth]{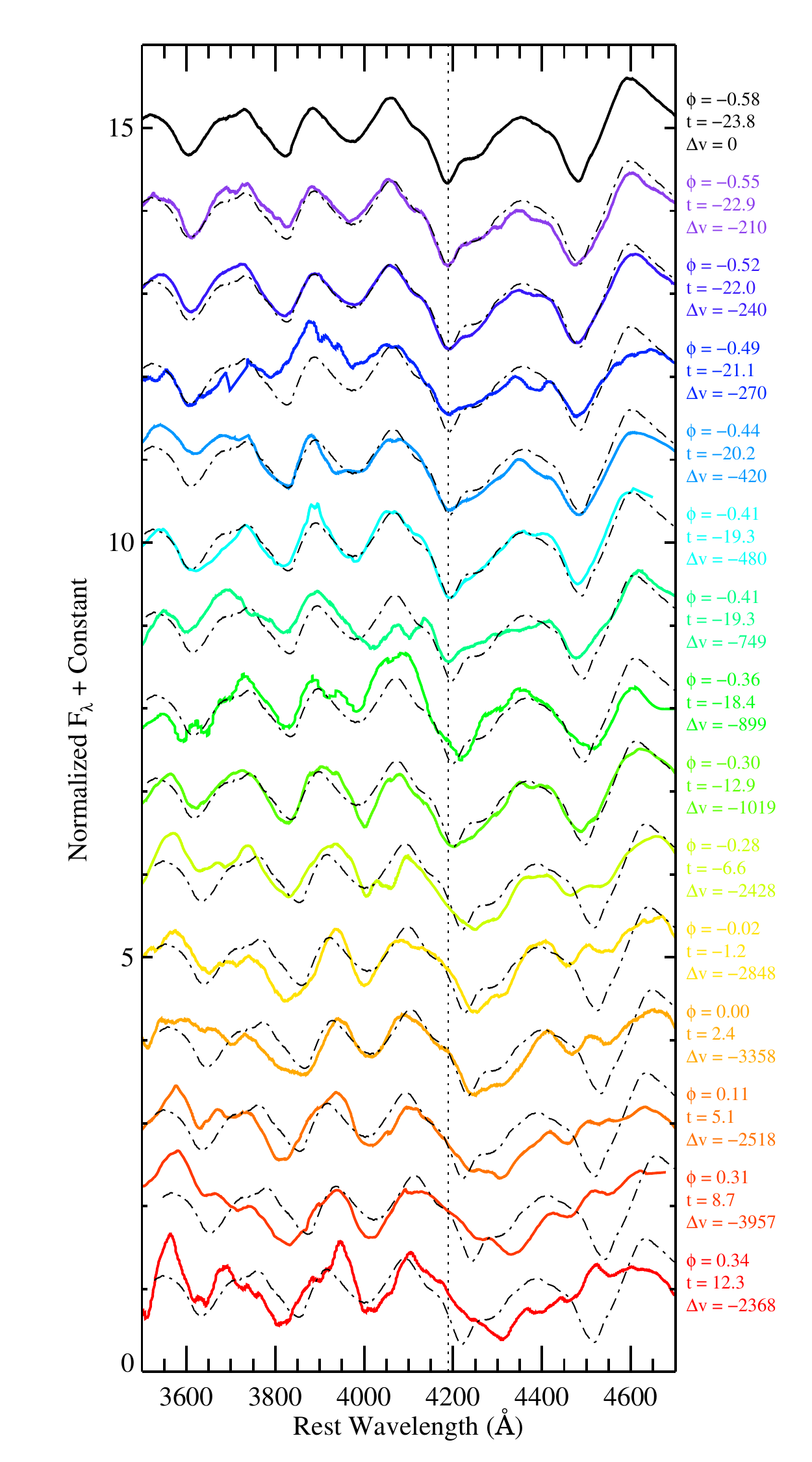}
 \caption{ Similar to Figure \ref{fig:PTF12dam_FeII_spectra} but
   highlighting the \ion{O}{2} multiplets in the time-series spectra
   of PTF12dam. The vertical dotted line marks the
   position of the \ion{O}{2} ``B'' line in the first spectrum for
   reference ($v_{\rm abs} = 11,100$\,km\,s$^{-1}$).}
 \label{fig:PTF12dam_OII_spectra}
\end{center}
\end{figure}

We can use the same techniques to measure the velocity evolution
of \ion{O}{2}. The measurements for PTF12dam are given in Table~\ref{table:PTF12dam_o2}
and the cross-correlated spectra are shown in
Figure~\ref{fig:PTF12dam_OII_spectra}.

\input{PTF12dam.o2.vels.tex} 

\begin{figure}
\begin{center}
 \includegraphics[width=\linewidth]{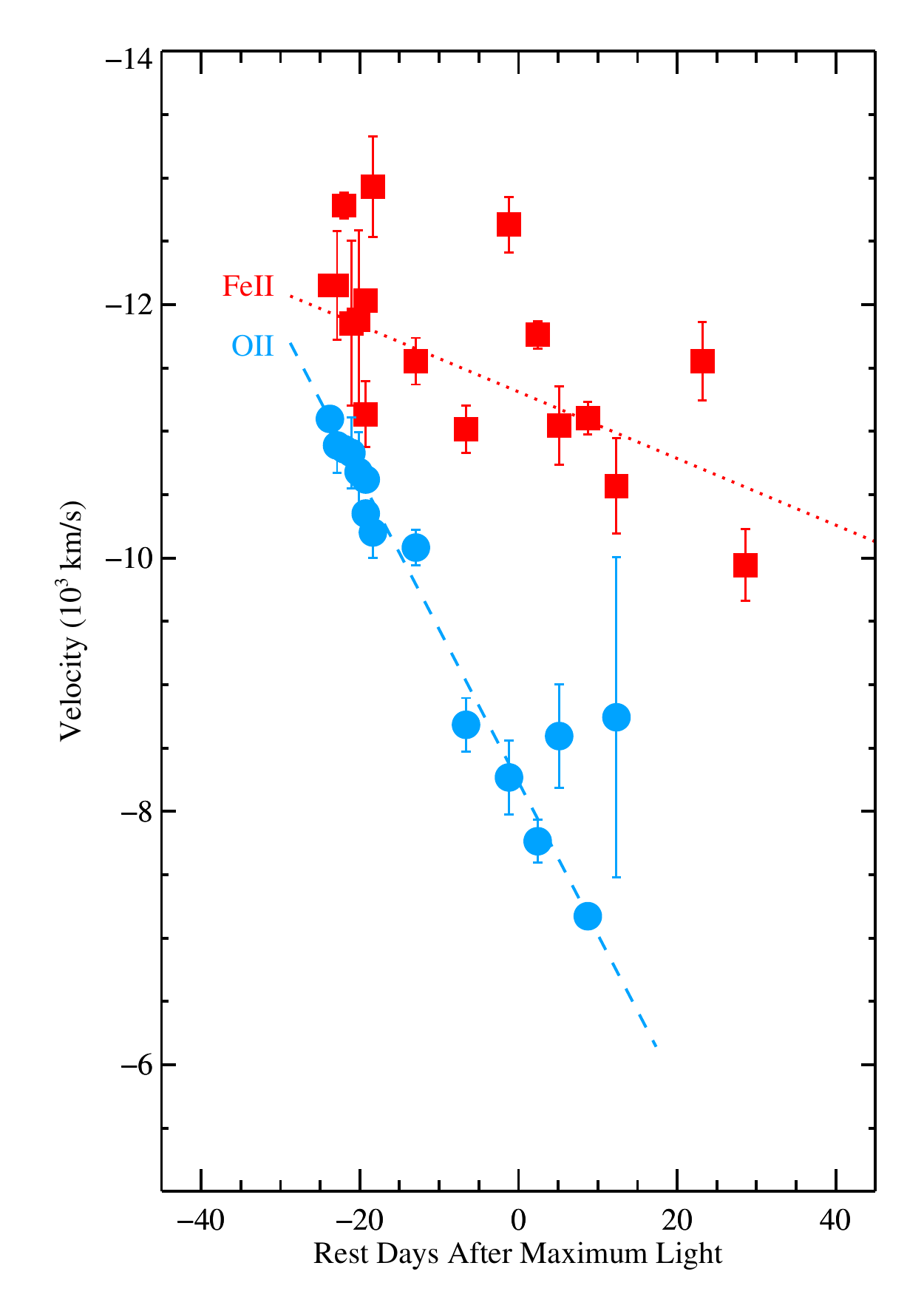}
 \caption{ PTF12dam \ion{Fe}{2} and \ion{O}{2} line velocities as
   determined from cross-correlation against a spectrum with a
   reference velocity measured by fitting an absorption-line model to
   the data. Red squares show velocities from \ion{Fe}{2} and blue
   circles indicate velocities from \ion{O}{2}. The dotted and dashed
   lines show linear fits to the \ion{Fe}{2} and \ion{O}{2} line
   velocities, respectively. }
   \label{fig:linevel}
\end{center}
\end{figure}

In Figure~\ref{fig:linevel} we compare the velocity evolution of the
\ion{Fe}{2} and \ion{O}{2} ions in the spectra of PTF12dam. The
\ion{O}{2} velocities of PTF12dam are significantly
($\sim1000$\,km\,s$^{-1}$) slower than the \ion{Fe}{2} velocities at
the earliest phases, and they decline much more rapidly. The low
velocities for \ion{O}{2} imply that these lines form in lower
levels of the ejecta, presumably closer to the photosphere than
\ion{Fe}{2}. Relatively high energies are required to coax oxygen into
an ionized state where it may produce \ion{O}{2} absorption
\citep{mazzali2016}, so these features may form closer to the main
source of power than do the \ion{Fe}{2} lines. Thus, the \ion{O}{2} ion
may better trace the photospheric velocities when they are the
dominant spectral features.

\section{Conclusions}\label{conclusions}

We have shown that SLSNe-I are spectroscopically distinct from other
types of SNe. Photospheric-phase spectra of SLSNe-I lack the
strong hydrogen lines of SNe~II, the strong \ion{Si}{2} features
characteristic of SNe~Ia, and (usually) the strong \ion{He}{1} features
of SNe~Ib. Under the classic classification scheme SLSNe-I could thus be
considered SNe~Ic, but this default category may encompass a wide
variety of potentially physically distinct events. We do find,
however, that there are spectroscopic differences between SLSNe-I and
normal-luminosity SNe~Ic that can be used to delineate these
groups. SLSNe-I tend to have much bluer continua. To match SN~Ic
spectral templates to those of SLSNe-I, the templates typically need to be
dereddened by $A_V \approx -2$\,mag. Because normal-luminosity SNe~Ic
include events with a wide variety of spectral behaviors, we do not
find a single feature that is always present in all SLSNe-I but never
in normal-luminosity SNe~Ic or vice versa. Rather, the spectroscopic
distinction between SLSNe-I and SNe~Ic at matching spectral phases, aside
from continuum differences, is the combination of multiple minor
differences. SLSNe-I tend to have weaker spectral features, and when a
SLSN-I and a SN~Ic are aligned in velocity space to one common feature,
other features sometimes remain noticeably offset.

One obvious difference between SLSNe-I and SNe~Ic is that \ion{O}{2}
features have never been observed in the spectra of the latter but are
common in the early-time spectra of the former. Among all normal-luminosity
SNe, only two peculiar events, the SN~Ib SN\,2008D
\citep{soderberg2008,mazzali2008,modjaz2009} and the SN~Ibn OGLE-2012-SN-006
\citep{pastorello2015}, have shown these features. The long-lived
presence of \ion{O}{2} lines could be another defining feature of
SLSNe-I, although there are several SLSNe-I that either did not show
them or that were simply not observed at sufficiently early
phases for these features to be seen
(e.g., SN\,2011ke). However, these SLSNe-I can still be separated from
normal-luminosity SNe~Ic by their later-phase spectra.

Aside from subtle differences, the spectral features of SLSNe-I around
a month after maximum light are similar to those seen in SN~Ic spectra
near maximum light \citep{pastorello2010}. Instead of comparing
spectra of different SLSNe-I and SNe~Ic to each other at the same 
light-curve phases, we study the differences in spectra at matched spectral
phases. We define a spectral sequence using SLSNe-I which ranges from
$\phi=-1$ for early-phase spectra with strong \ion{O}{2} lines, to
$\phi=0$ near the phase when these lines disappear, to $\phi=+1$ when
the spectra resemble SNe~Ic near maximum light, and finally to $\phi=+2$
when nebular emission lines begin to dominate. A given SN~Ic or SLSN-I
spectrum can be assigned a spectral phase by comparing the spectra to
the ordered sequence shown in Figures 
\ref{fig:fid_spec1}--\ref{fig:fid_spec4} and identifying the closest match.

When comparing SLSN-I spectra to each other, we found that some
objects had spectra similar to PTF12dam while other objects more
closely resembled SN\,2011ke. We compared the average spectra of these
two groups and find that while the spectra are similar overall, there
may be some systematic differences. SN\,2011ke-like objects may
exhibit smoother spectral features around $-0.3 < \phi < +0.3$ than do
PTF12dam-like objects. One interpretation of this is that the
SN\,2011ke-like objects have broader velocity distributions that tend
to blend and smooth out individual ion features. This is supported by
cross-correlation of SN\,2011ke-like objects against the spectra of
PTF12dam, which suggests higher expansion velocities for
SN\,2011ke-like objects. However, the systematic velocity shifts may
only weakly correlate with feature width. For example, the
SN\,2011ke-like object PTF10bfz around $t=+18$\,d is very similar to the
spectra of SN\,2011ke at $t\sim+25$\,d, but the spectra of PTF10bfz have
a systematic blueshift that is about 12,000\,km\,s$^{-1}$ larger than
that of SN\,2011ke. Thus, the systematic velocity of PTF10bfz is much higher
than for SN\,2011ke, but the two objects appear to have similar
velocity widths. We find a similar result for SN\,2010gx, which is
spectrally similar to PTF11rks at $\phi=-1$ but with a
6000\,km\,s$^{-1}$ faster systematic blueshift.

Spectroscopic phases were defined under the assumption that the
spectra of all SLSNe-I evolve in a consistent manner, but this may be
undermined by the preference of some objects to better match either
SN\,2011ke or PTF12dam. In the full spectral sequence shown in Figures
\ref{fig:fid_spec1}--\ref{fig:fid_spec4}, it is evident that certain
phases consist of spectra from one of these groups or the other. For
example, in our spectral sequence the $1.11 < \phi < 1.46$ range has
21 SN\,2011ke-like spectra but no PTF12dam-like spectra, and most of
the spectra at $\phi<0$ (with strong \ion{O}{2} lines) belong to the
PTF12dam-like group. Perhaps we simply lack spectra of objects from
the other groups at these phases (there are significant gaps in the
temporal coverage of most SLSNe-I), but alternatively these groups may go
through certain spectral phases that are not experienced by the other
group (or only fleetingly experienced by the other group).

In addition to the \ion{O}{2} lines, we find good evidence for
\ion{C}{2}, \ion{O}{1}, \ion{Mg}{2}, \ion{Si}{2}, \ion{Ca}{2},
\ion{Ti}{2}, and \ion{Mn}{2} in photospheric-phase spectra of
SLSNe-I. Other ions likely contribute to the spectra as well, but these
are difficult to securely identify using relatively primitive tools
such as {\tt syn++}. In particular, there is almost certainly
\ion{Fe}{2} $\lambda\lambda$4923, 5018, 5169 (other ions may contribute at these
wavelengths as well), but {\tt syn++} predicts stronger \ion{Fe}{2}
features in the UV that conflict with the data. Non-LTE effects may
need to be taken into account to properly model these features. The
identification of the ions responsible for the four broad dips in the
near-UV bands is similarly difficult to confirm with {\tt
  syn++}. Previously suggested contributions from \ion{C}{3},
\ion{Si}{3}, \ion{Ti}{3}, and \ion{Fe}{3} remain plausible, but a
simple comparison of the observed line minima against the predicted
locations of these ions for a given blueshift leaves the true
identities of these features in doubt. A complication to this picture
is that these features likely evolve over time. This is especially
apparent for the UV2 feature, which we find is most probably dominated
by \ion{Mn}{2} at late times but other ions at earlier phases. Far-UV
spectra may hold the key to resoling these line identifications.

Even at similar spectral phases, the strengths of various features
vary considerably among SLSNe-I. For example, PTF09cnd shows
significantly stronger \ion{O}{2} than PTF12dam around
$\phi=-0.6$. Another example, LSQ12dlf, exhibits much larger peak-to-trough 
variations around $\phi=+1.25$ than does SN\,2011ke. Our simple
absorption-line modeling of the \ion{O}{2} features between
3500\,\AA\ and 4800\,\AA\ (which blend into 5 dips that we label A--E
from longest to shortest wavelength) provides a reasonably good fit to 
the spectra of PTF12dam and PTF09cnd considering it does not
account for P-Cygni emission, but the models do not fare as well for
SN\,2010gx and PTF11rks. These objects have very weak \ion{O}{2} C and
D features and relatively strong E features around $\phi=-1$. When the
models are fit to the \ion{O}{2} B feature of PTF11rks, the predicted 
minima do not agree well with the data for the A, C, D, or E features. 
The culprit may be contamination from other lines as suggested by the
slightly resolved A feature in PTF11rks.

As we have shown, it is difficult to measure accurate velocity curves
in the spectra of SLSNe-I because the features are weak
blends of more than one ion, and most features can only be detected
over a limited range of phases. However, we have presented a method
for measuring the velocity of the \ion{O}{2} features, which are
prominent in SLSN-I spectra up to around maximum light, by modeling
them as simple Gaussian blends of multiplet lines. In doing so, we can
determine a velocity matched to the absorption minima and a FWHM for
the velocity distribution. The blueshift of the \ion{O}{2} absorption
minima in PTF12dam is measured through this technique to be about
11,000\,km\,s$^{-1}$ in the earliest spectra and it has a FWHM of only
3000--4000\,km\,s$^{-1}$. This relatively narrow velocity distribution
supports the idea that \ion{O}{2} is confined to a thin shell. The
derived FWHM is sensitive to the assumed continuum level, but for
PTF12dam the resolution of the \ion{O}{2} B feature excludes velocity
widths greater than $\sim4000$\,km\,s$^{-1}$. We also find that the
\ion{O}{2} absorption minima velocity of PTF12dam declines slowly by
about 120\,km\,s$^{-1}$\,d$^{-1}$ (assuming a linear decay) by
cross-correlating its spectra with a reference spectrum and noting the
velocity shift. The \ion{O}{2} velocities appear to be consistent with
weak absorption features from \ion{C}{2}, which we identify at both UV
and optical wavelengths.

The subset of SLSNe-I with high systematic blueshifts (e.g., PTF10bfz)
may result from particular viewing angles of highly asymmetric events,
such as looking directly at jets of fast-moving material. Spectral
polarimetry studies have begun to look for such geometries, and the
implied asymmetries have thus far been low or similar to those of GRB-associated
SNe~Ic \citep{leloudas2015c,inserra2016b}. For some SLSNe-I, the
velocities are not only lower but the velocity widths of the features
are also very low. This is most readily evident in PTF12dam, for which
the \ion{O}{2} B is clearly resolved into two components at early
phases, as discussed above. Thus, in contrast to \citet{liu2016}, we
find at least some SLSNe-I have ion distributions that are rather
narrow as compared to the broad spectral features that define SNe~Ic-bl
(but note that \citealt{liu2016} also find that the \ion{Fe}{2} width
of SN\,2007bi is not only less than for SNe~Ic-bl, it is less than the
average SN~Ic as well). We speculate that events like PTF10bfz may
reflect the high-luminosity tail of the SN~Ic-bl distribution, but
events like PTF12dam that are far from broad lined may have an
important physical difference. Late-time spectra of these events may
help to determine if such SN\,2011ke-like and PTF12dam-like events are
physically distinct or if viewing-angle geometries are the deciding
factor.

We show that there are other ions including \ion{Mg}{2} and
\ion{Fe}{2} that appear to favor velocities higher than those of \ion{C}{2} and
\ion{O}{2}. The velocities we measure for \ion{Fe}{2} in the spectra
of PTF12dam begin about 1000\,km\,s$^{-1}$ faster and decline more
slowly (only about 30\,km\,s$^{-1}$\,d$^{-1}$), although there is
considerable scatter in the measurements. \ion{O}{2} and \ion{C}{2}
may be better tracers of the photospheric velocity. If we assume spherical symmetry, then the
presence of \ion{Mg}{2} and \ion{Fe}{2} at significantly higher
velocities than the photosphere may disfavor the interaction model.
This is because the photosphere would likely be located close to the
fastest-moving ejecta, which are interacting with the slower-moving
CSM, and this would leave little room for lines to form at higher
velocities. Alternatively, we could be observing asymmetric explosion with jets of fast-moving
\ion{Mg}{2} and \ion{Fe}{2} pointed at us. In this case we would predict that other events may
have jets pointed away from the observer and thus lower apparent  \ion{Mg}{2} and \ion{Fe}{2}
velocities.

In the magnetar model the material is energized from within, in which
case one might expect to see more highly ionized ions at later
times. \citet{metzger2014} suggest that some SLSNe-I may have an
ionization front that moves out through the ejecta, eventually causing
the ejecta to become transparent to X-rays. But this is not evident
from the spectra in our SLSN-I sample. In particular, \ion{O}{2} is
only observed at early phases and \ion{O}{1} strengthens at later
phases. If these events are powered by magnetars, it would seem that
the energy released by these central engines decreases substantially
over time so that even as the ejecta expand and become less dense, the
ionization front continues to recede with time.

Lacking adequate SLSN-II templates for our automated spectral
classification, we have simply removed objects with obvious hydrogen
emission consistent with SNe~IIn from consideration as SLSNe-I. However,
the relatively weak signs of hydrogen in the photospheric-phase
spectra of PTF10aagc and otherwise plausible spectral similarities to
SLSNe-I secured PTF10aagc a SLSN-I designation. This object is clearly
distinct from others in the sample at least for its strong, broad
H$\alpha$ at late phases. Similarly, our automated classification of
PTF10hgi places it in the SLSN-I category as well (and this object has
previously been published as a SLSN-I), but it is clearly
distinct from other SLSNe-I in that it has clear signs of \ion{He}{1}
as well as hydrogen, suggesting a SLSN-IIb classification may be more
appropriate. These outlying objects might favor a continuum between
SLSNe-I and their hydrogen-rich SLSN-II cousins, but another
possibility is that these are simply relatively high-luminosity
examples of other, unrelated phenomena and we lack sufficient
examples of each type to identify this division with our current
tools.

Because the previously published SLSNe-I were based simply on peak
luminosity and lack of obvious spectroscopic evidence for hydrogen,
there may very well be objects in this group that meet these criteria
but have physically different origins. This is one possible
explanation for the preference of certain objects to have spectra
more similar to those of either PTF12dam or SN\,2011ke. We speculate that the
PTF12dam-like objects with long-lived \ion{O}{2} may form through one
progenitor channel while more spectrally diffuse objects like PTF10bfz
may represent the tail of the SN~Ic-bl distribution. It may further be
worth considering if PTF10hgi is a higher-luminosity relative of a
SN\,2005bf-like explosion, perhaps in the same family as the classic
SN~IIb SN\,1993J.
  
Alternatively, some peculiar, lower-luminosity SNe may represent
the low-luminosity tail of the SLSN-I/II distribution, and the
longevity of \ion{O}{2} features may be key to this connection. There
have only been two lower-luminosity events, the SN~Ib SN\,2008D
\citep{soderberg2008,mazzali2008,modjaz2009} and the SN~Ibn OGLE-2012-SN-006
\citep{pastorello2015}, that have shown these features. In both cases
it has been argued that ejecta/CSM interactions are at play, and these
may be important factors in maintaining the high temperatures (or at
least the high levels of energy) required to ionize
oxygen. \citet{pastorello2015} favor CSM interaction as the dominant
power source for OGLE-2012-SN-006, in which case oxygen must be either
thermally excited or excited by hard radiation from the shocked
material in order to explain the presence of \ion{O}{2}.
\citet{mazzali2016} have argued that oxygen cannot be sufficiently
excited thermally to produce the \ion{O}{2} features in SLSNe-I and
take these features as evidence of nonthermal excitation by a central
magnetar. However, some studies of the bolometric light-curves of
SLSNe-I favor power from CSM interaction (e.g.,
\citealt{wheeler2017, tolstov2017b}). Low-luminosity objects like OGLE-2012-SN-006
that have both clear signs of CSM interaction and strong \ion{O}{2}
spectroscopic features may make it worth considering what role such a
process may play in the ionization state of SLSN-I envelopes.

The data presented here can further be used to help identify the
underlying source of SLSN-I power and they may aid future studies of
the possible connection between SNe~Ic and SLSNe-I. Spectral properties
such as the equivalent widths or velocities of particular features can
be correlated with light-curve parameters such as peak luminosity,
rise time, and fading rates. For example, we find that the time
required for PTF12dam-like events to fade by 1\,mag from maximum
may be longer than for SN\,2011ke-like events using the results of
\citet{decia2017}. The average fall time for seven PTF12dam-like events
(PTF09atu, PTF09cnd, PTF10nmn, PTF10vqv, PTF11hrq, PTF12dam, and
PTF12gty) is $54\pm12$\,d, whereas the average fall time for eight
SN\,2011ke-like events (PTF09as, PTF10aagc, PTF10bfz, 2010gx,
PTF10hgi, PTF10uhf, 2011ke, and PTF11rks) is $24\pm8$\,d, with most of
these later events clustered around fall times of 29\,d. The $p$-value
from a formal KS test is $3\times10^{-4}$, which strongly rejects the
null hypothesis that these two populations, divided by spectral
properties, are drawn from the same light-curve distribution.

Our high-quality time-series spectra can be used to check predictions
made by theoretical models regarding the velocity distribution and
evolution of ejecta components. If a given model can naturally explain
these observables for multiple SLSNe-I and for SNe~Ic as well, then this
might finally resolve the physical nature of these events. The
limiting factor in this comparison may actually be a relative lack of
SN~Ic observations. There appears to be considerable diversity among
so-called stripped-envelope SNe, and few have high-quality spectra
extending into the rest-frame UV that can be compared against the
higher-redshift SLSN-I sample. There could be multiple
progenitor channels that lead to stellar explosions with relatively
weak hydrogen, silicon, and helium in their spectra, so we may need to
assemble a large sample of these objects with spectroscopic
observations taken over a range of similar spectral phases and 
rest-wavelength ranges to properly compare and discriminate among
them. Upcoming surveys, such as the Zwicky Transient Facility, will
easily generate the requisite candidate pool for such work; the
challenge will be to efficiently monitor these detections to
generate new discoveries.

\acknowledgments 

The data presented herein were obtained in part at
the W.M. Keck Observatory, which is operated as a scientific
partnership among the California Institute of Technology, the
University of California, and the National Aeronautics and Space
Administration (NASA). The Observatory was made possible by the generous
financial support of the W.M. Keck Foundation. 
We thank the following for assistance with some of the Lick and Keck
observations and reductions: Kelsey Clubb, Ryan Foley, Christopher
Griffith, Pat Kelly, Michael Kandrashoff, and
Isaac Shivvers. The William Herschel
Telescope is operated on the island of La Palma by the Isaac Newton
Group of Telescopes in the Spanish Observatorio del Roque de los
Muchachos of the Instituto de Astrof{\'i}sica de Canarias.
Research at Lick Observatory is supported in part by a generous
gift from Google.  
Support for {\it HST} programs GO-12223 and GO-12524 was provided by NASA
through a grant from STScI, which is
operated by the Association of Universities for Research in Astronomy,
Inc., under NASA contract NAS 5-26555

M.S. acknowledges support from EU/FP7-ERC grant no [615929]. A.D.C.
acknowledges support by the Weizmann Institute of Science Koshland
Center for Basic Research.
A.V.F.'s supernova research group at U.C. Berkeley is grateful for            
generous financial assistance from the Christopher R. Redlich Fund,   
the TABASGO Foundation, NSF grant AST-1211916, and the Miller 
Institute for Basic Research in Science (U.C. Berkeley).
M.M.K. acknowledges support by the GROWTH project funded by
NSF grant AST-1545949.
J.C. acknowledges support from the Australian Research Council Future Fellowship, grant FT130101219.
A.G.-Y. is supported by the EU via ERC grant No. 725161,
the Quantum Universe I-Core program, the ISF, the BSF Transformative program and by a Kimmel award.
We thank Melissa Graham for providing
spectra of PTF12dam, PTF12gty, and PTF12hni.

\software{
  superfit \citep{howell2006},
  SNID \citep{blondin2007},
  synow \citep{jeffery_branch1990},
  syn++ \citep{thomas2011},
  IRAF \citep{tody1986, tody1993},
  bspline\_iterfit.pro,
  aXe \citep{kummel2009}}

\bibliographystyle{apj}
\bibliography{paper}

\appendix

\section{Generalized Savitzky-Golay Smoothing of Spectra}\label{sgfilter}

SN spectra are often noisy and it is necessary to smooth them
for use as comparison templates or simply for display purposes. The
act of smoothing removes information from the data and should thus not
be performed for some analyses. However, if most of the information
lost is connected to noise, smoothing spectra can simplify the
analysis. It is important, however, to maintain the integrity of the
true signal. Because the spectral features of SNe are broadened
by thousands of km\,s$^{-1}$, whereas most instruments sample the spectra in
bins only tens of km\,s$^{-1}$ wide, it is relatively easy in most cases to
isolate the signal from the bulk of the noise.

Standard techniques for smoothing spectra include the boxcar method,
fast Fourier transform (FFT) filtering, 
and local polynomial fitting (e.g., Savitzky-Golay). In
the boxcar method, the value in the $i^{\rm th}$ bin is replaced with
the average of the values in bins $i-n$ to $i+n$. Boxcar smoothing is
widely used because of its simplicity to implement, but it has the
well-documented problem of washing out the true peaks and valleys of
the signal. For FFT filtering, the spectrum is transformed to wave-number 
space, bins with wave numbers higher than expected for the
broad SN features are muted, and the smoothed spectrum is
recovered with application of a reverse FFT transform. FFT filtering
often performs better than boxcar smoothing, but it requires the input
spectra, which are usually sampled into bins with a constant
wavelength width, $\delta\lambda$, to be resampled into bins of constant
velocity width, $\delta v$. Any missing wavelength bins produced by gaps in
coverage or rejection due to artifacts must first also be interpolated
over. Thus, a nontrivial amount of preprocessing is required before
the FFT filter can be applied. It is also not straightforward to
account for measurement errors when FFT filtering a spectrum.

The standard Savitzky-Golay filter \citep{savitky_golay1964} is
similar to the boxcar method in that the value in the $i^{\rm th}$ bin
is replaced based on the values in bins $i-n$ to $i+n$, but instead of
taking an average, a polynomial is fit to the chosen bins and the
value of this polynomial in the $i^{\rm th}$ bin is reserved. The
order of the polynomial fit can be varied to better fit faster
variations in the input spectrum (if set to zero, boxcar smoothing is
recovered). Thus, like the FFT method, Savitzky-Golay filtering can do
a better job of preserving the signal. This filtering technique has
many applications, including the smoothing of SN spectra
(see, e.g., \citealt{silverman2012b}).

When applied to SN spectra, we can generalize the
Savitzky-Golay filtering technique to take advantage of the extra
information available. Each spectral bin has not just one number but
two or three: the wavelength, flux, and (often) the flux error
values. Applying this additional information, we can naturally
interpolate over missing data from bad pixels (e.g., cosmic-ray hits)
and, to some extent, larger gaps caused by, for example, nonoverlapping
two-channel observations. When the error spectrum is available, we can
also use this to weight the fit. This can significantly improve the
performance of the smoothing in regions subject to strong night-sky 
lines (with correspondingly large flux errors in limited bins).

As further generalization of the Savitzky-Golay filter, we can allow
the number of bins included in the polynomial fitting to vary. This is
desirable because most spectra are dispersed by gratings and thus have
more bins per resolution element in the red as compared to the blue.
Because the spectra are sampled in uniformly sized wavelength bins,
SN features broadened by, say, 10,000\,km\,s$^{-1}$ cover twice as many
bins at 10,000~\AA\ as compared to 5000\,\AA. We thus let the number of
bins vary with wavelength to maintain constant sampling in velocity
space.

\begin{figure}
\begin{center}
 \includegraphics[width=\linewidth]{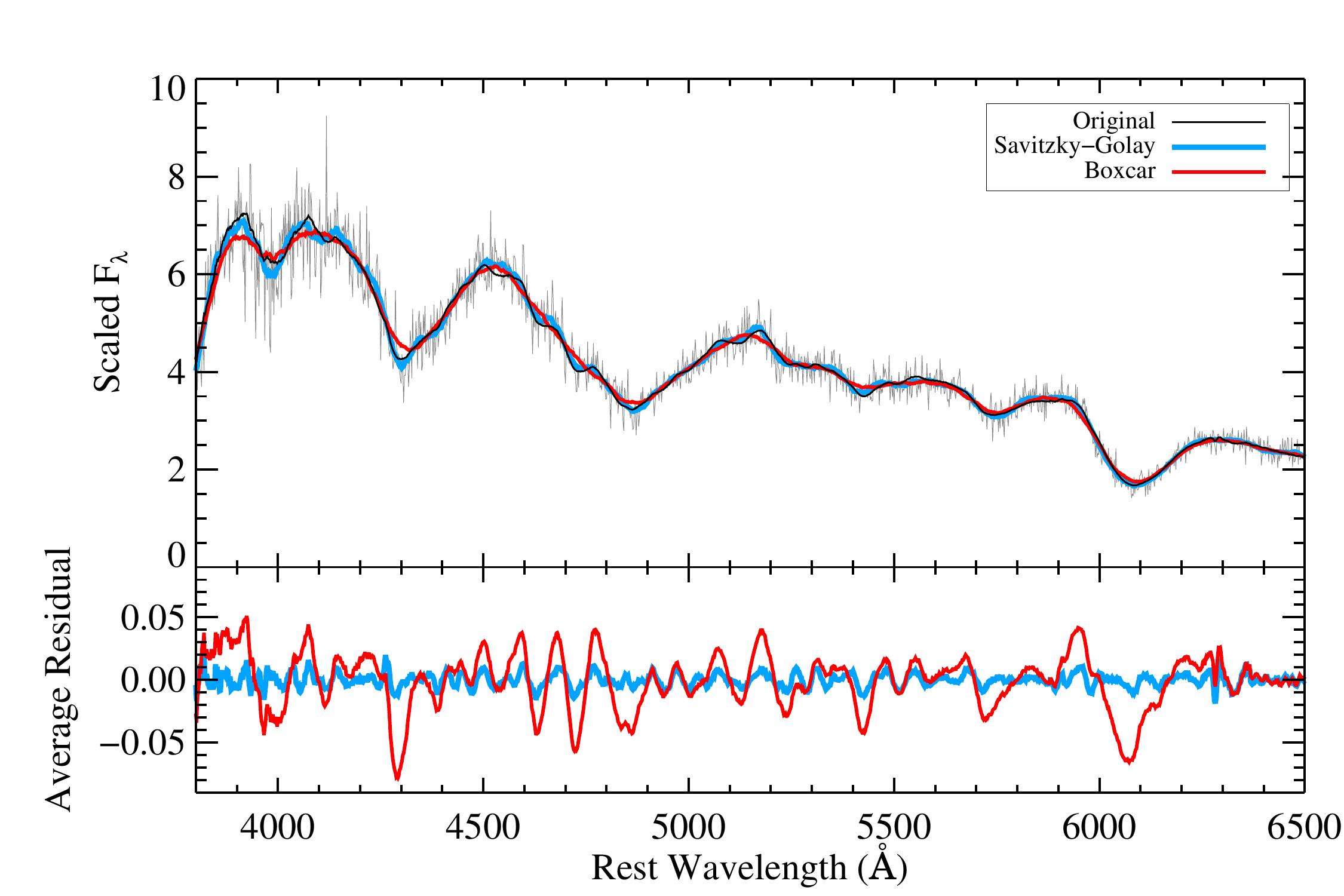}
 \caption{An example of generalized Savitzky-Golay and boxcar
   smoothing applied to spectra of a SN~Ia. In the upper
   panel, the thick black line is a high-S/N ratio
   spectrum of SN\,2011fe \citep{maguire2014}. The thin gray line gives
   a realization of this spectrum with noise added. The thick red
   line shows the result of boxcar smoothing this latter, noisy
   spectrum, and the thicker blue line shows the result from
   generalized Savitzky-Golay filtering. In the lower panel we show
   the fractional difference between the original spectrum and the
   smoothed spectra averaged over 1000 realizations. It is apparent
   that the boxcar method performs poorly near the peaks and troughs
   of the spectral lines while the generalized Savitzky-Golay method
   handles these better.} \label{fig:check_sg}
\end{center}
\end{figure}

Figure~\ref{fig:check_sg} shows the effects of smoothing on real data
(with simulated noise). For each wavelength, $\lambda$, a second-order
polynomial fit is performed to all data in the range $\lambda -
\lambda / 100 < \lambda < \lambda + \lambda / 100$. In order to
facilitate comparison of the smoothed data to the original data, the
smoothed data are output in identical wavelength bins; however, note
that another advantage of this generalized Savitzky-Golay filtering is
that the output wavelength scale need not match the input one
exactly. We can thus in one step smooth and register spectra onto a
desired wavelength scale. Similarly, the procedure can conveniently be
used to stack a set of spectra even if they are each on shifted
wavelength scales.

\clearpage
\section{Spectral Template Matches to SNe~Ic at $+30$d}\label{latematch}

\begin{figure}
\begin{center}
 \includegraphics[width=0.5\textheight]{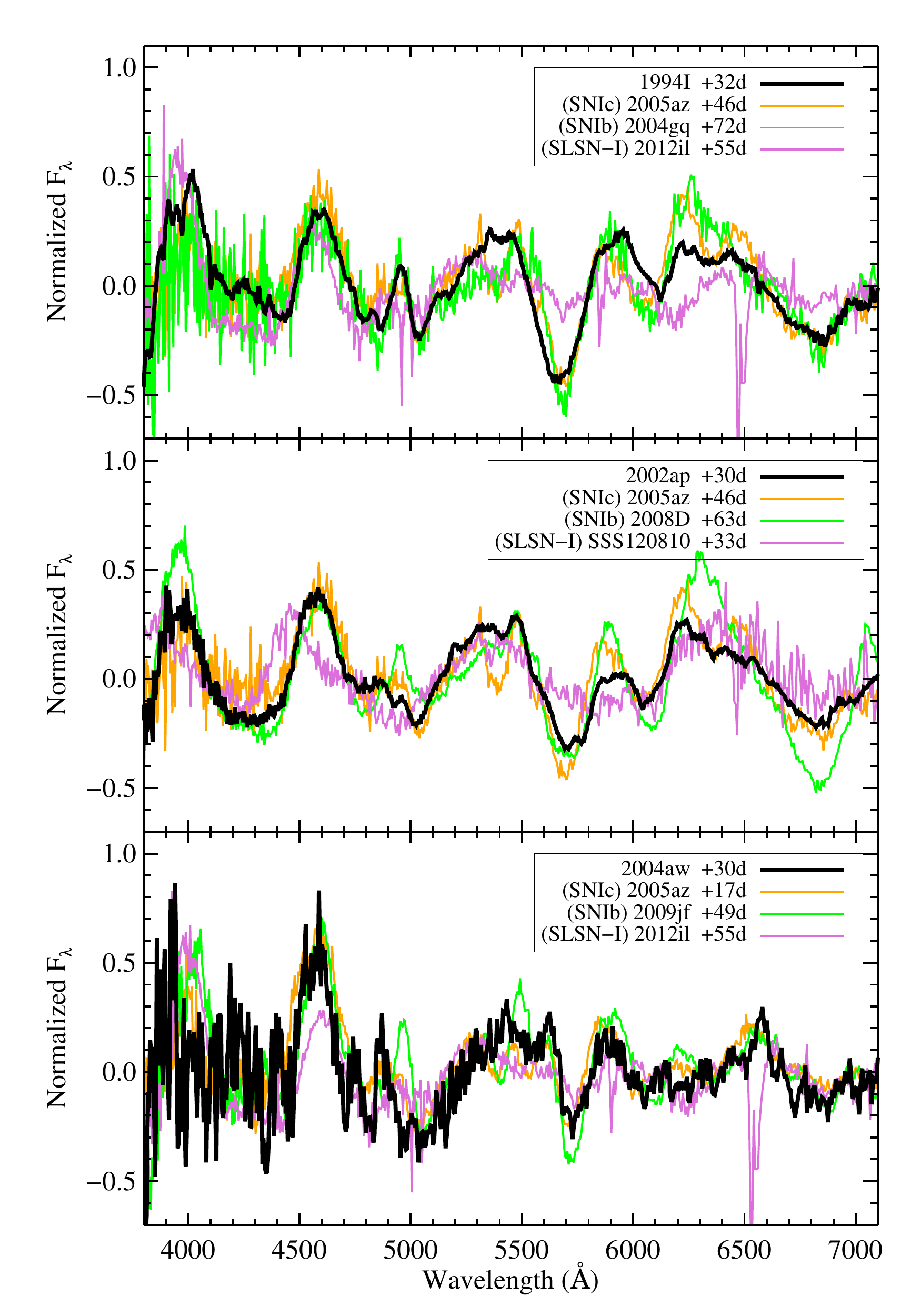}
 \caption{Best-fitting SN~Ic, SN~Ib, and SLSN-I spectral templates to
   the Type~Ic supernovae SN~1994I, SN~2002ap, and SN~2004aw. The
   spectra and templates have been continuum divided and the templates
   have been shifted in velocity. } \label{fig:latematch}
\end{center}
\end{figure}

Following the procedure detailed in \S\ref{selection}, we have
matched the spectra of three SNe~Ic taken about one month after
maximum light to our spectral template libraries. In
Figure~\ref{fig:latematch} we plot the best matching SN~Ic, SN~Ib, and
SLSN-I to each of these spectra for comparison. Some later phase
SLSN-I show many of the same spectral features as ordinary luminosity
SNe~Ic at this phase, but the SLSNe-I tend to have weaker features as
perhaps best shown by the absorption dip near 5700\,\AA. SNe~Ic and
SNe~Ib templates typically provide better matches during this phase
according to the $\Delta I_{\rm Ic - X}$ scores.

\clearpage
\section{Light-Curve Phase Estimates from Spectral Matching}\label{lcphasematch}

Using our library of spectral templates taken at known light-curve
phases we can test the precision and accuracy to which the light-curve
phases of test spectra can be recovered. We use {\tt superfit} to
match the spectral templates to the input spectrum as described in
\S\ref{selection} and then compare the actual light-curve phase of the
input spectrum to the average light-curve phase of the top three matches
of the correct type.

\begin{figure}
\begin{center}
 \includegraphics[width=\linewidth]{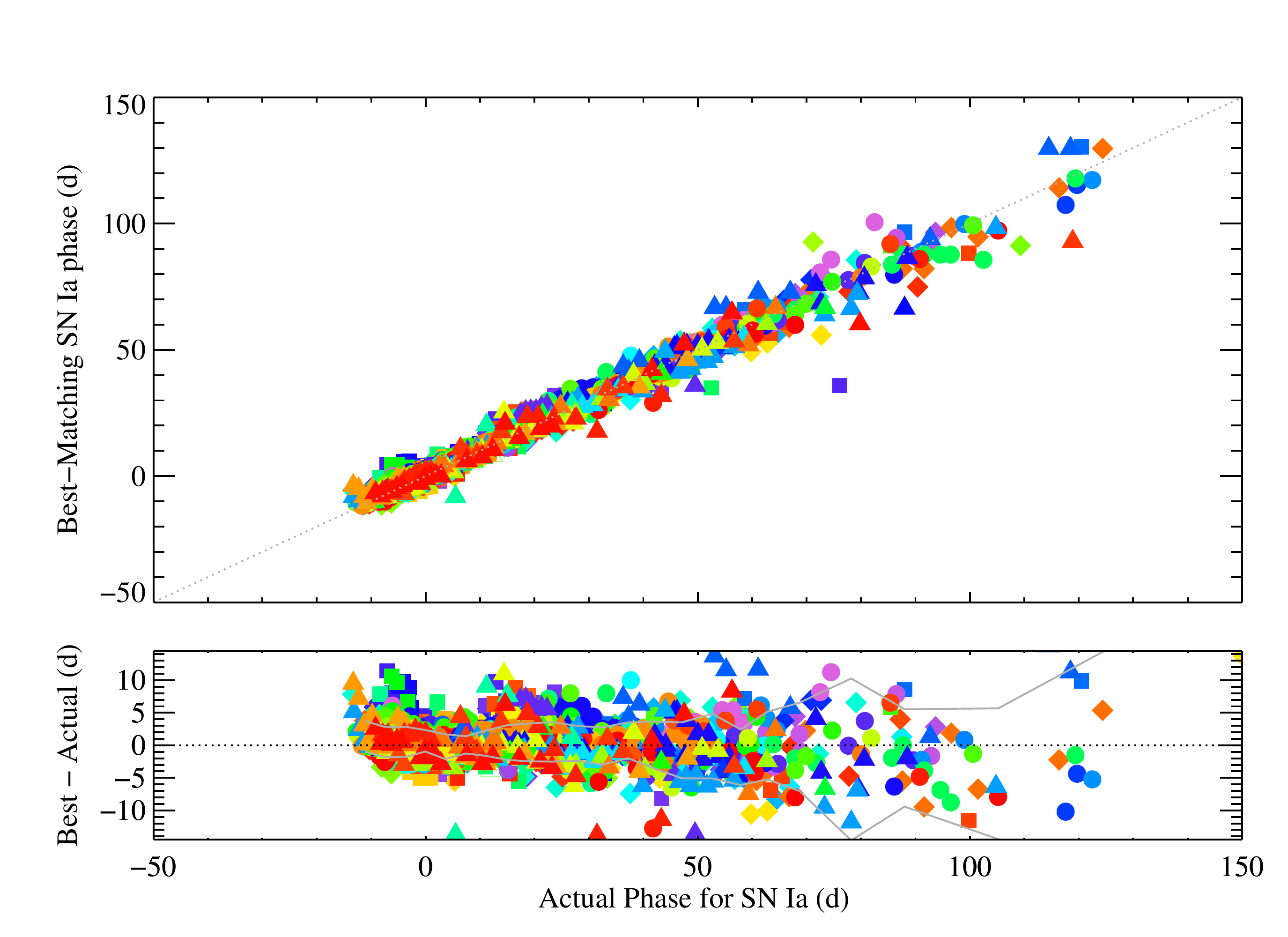}
 \caption{ Precision and accuracy of SN~Ia light-curve phases derived
   from spectral template matching. The top panel plots the real phase
   against the derived phase and the lower panel shows the residual
   from the true values. The gray lines show the $1\sigma$ range of
   the residuals evaluated in several phase bins. All spectra for a
   given object are plotted with the same color, and a different color
   is used for each object. Marker shapes reflect the subtype of the
   SN~Ia: triangles for SN~Ia-SS, squares for SN~Ia-CL, diamonds for
   SN~Ia-BL, and circles for SN~Ia-CN. }
   \label{fig:SNIa_age}
\end{center}
\end{figure}

Figure~\ref{fig:SNIa_age} shows the actual phase of the SN~Ia spectra
tested vs. the average phase derived from the best {\tt superfit}
matches. We find that the averaged {\tt superfit} phases are typically
biased by less than 0.5\,d (with {\tt superfit} preferring slightly
later phases) between about 1 week before maximum light to about 1
month after. Over this period the standard deviation is around 2.3\,d
with measurements closer to maximum light faring better. This result
is similar to that of \citet{riess1997}, who find a precision of 1.4\,d over 
a comparable range of phases, and \citet{blondin2012}, who find a 2.9\,d
dispersion in spectral ages for SNe~Ia within about 10 days of
maximum light. We find that the accuracy remains good to at least two months
after maximum light but the standard deviation degrades to about 5\,d.

We also tested the precision and accuracy of SN~Ic and SLSN-I phases
derived from spectral template matching. The median absolute bias is
2.2\,d for the SN~Ic sample and 8.0\,d for the SLSNe-I. The standard
deviations of the difference between the actual light-curve phases and
the averaged values of the best {\tt superfit} matches are also
significantly higher for the SNe~Ic and SLSNe-I than for the SN~Ia
sample. For SNe~Ic the standard deviation rises from about 6\,d one week
before maximum light to about 17\,d one month after maximum. The
standard deviations of the SLSN-I sample are about 15\,d over a
similar time period. These larger values could reflect the smaller
sample sizes or greater intrinsic differences in the spectra and how
these vary with light-curve phase.

\clearpage
\section{On the \ion{He}{1} $\lambda10830$ Line in SN\,2012il}\label{2012il}

\begin{figure}
\begin{center}
 \includegraphics[width=0.5\linewidth]{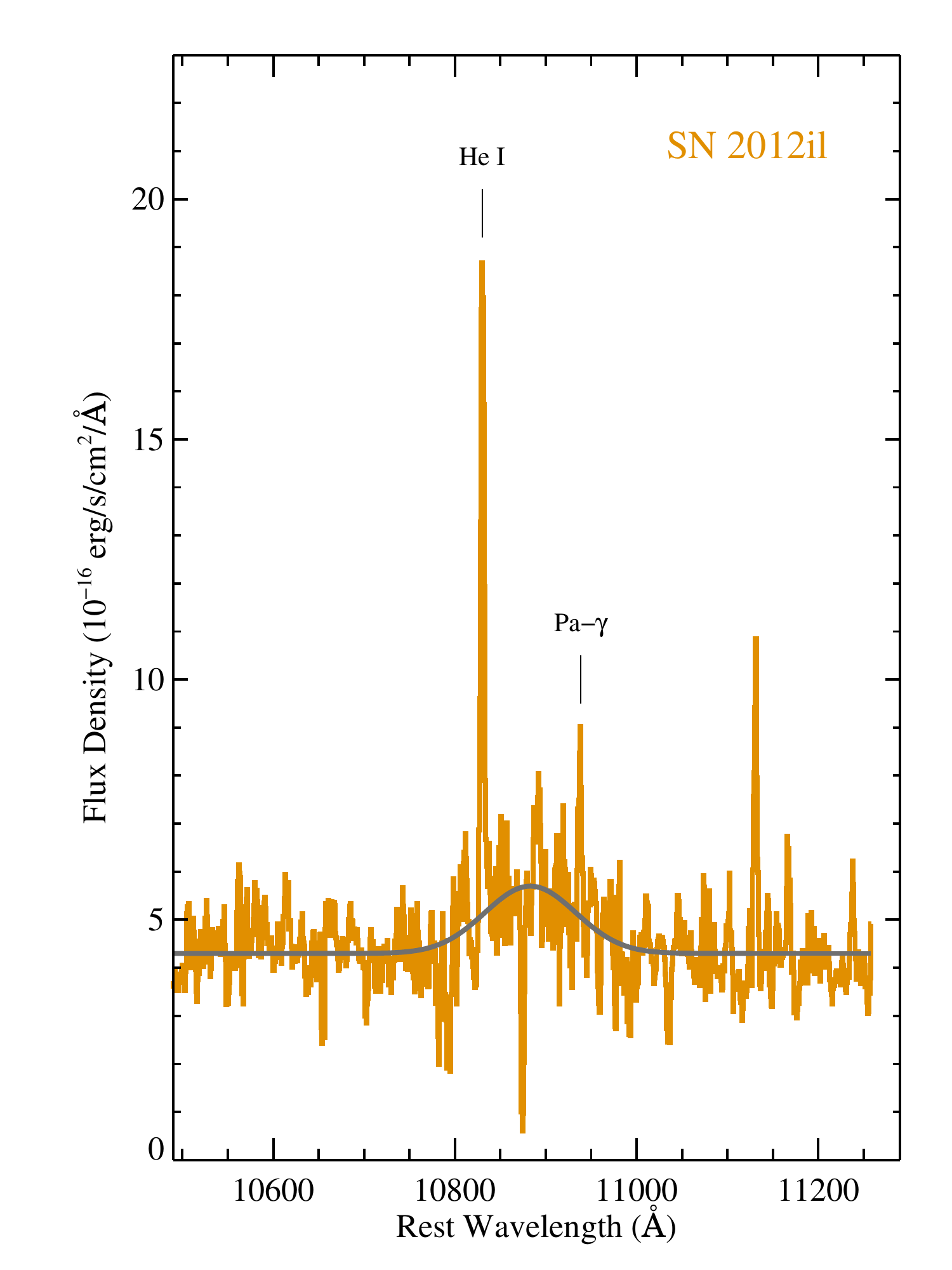}
 \caption{ Near-infrared spectra of SN\,2012il highlighting the \ion{He}{1}
   $\lambda10830$ region. Data are from \citet{inserra2013}. The
   gray curve shows a Gaussian function fit to the data. The Gaussian
   is centered at 10,883\,\AA, which is redshifted by about
   1500\,km\,s$^{-1}$ from the expected position of the \ion{He}{1}
   line. Narrow emission lines matching the wavelengths of \ion{He}{1}
   $\lambda10830$ and hydrogen Pa-$\gamma$ in the host-galaxy frame
   are labeled.}
   \label{fig:2012il}
\end{center}
\end{figure}

To investigate the reported discovery of helium in the spectra of
SN\,2012il by \citet{inserra2013}, we downloaded the 2012 March 17
VLT+XSHOOTER spectra of this object. In studying the spectra we found
that the data agree well with what is plotted in the main part of
Figure 9 from \citet{inserra2013}, but we do not reproduce the results
from the right panel of this figure, which is intended to show a
smoothed version of the reported \ion{He}{1} $\lambda10830$ line. 
Instead, we find that there is a broad emission bump in this
region, but it is significantly offset to the red of 10,830\,\AA\ by about
1500\,km\,s$^{-1}$. In Figure~\ref{fig:2012il} we show the data in
this region without applying any smoothing. As can be seen, there are
narrow emission lines at 10,830\,\AA\ and 10,938\,\AA, which are
likely \ion{He}{1} and hydrogen Pa-$\gamma$ from the host galaxy. The
broad emission bump is centered in between these. The detection of the
narrow host lines shows that the wavelength solution is accurate and
that the broad bump is thus not centered at 10,830\,\AA\ as was
reported. Given the offset, we find it unlikely that this feature is
associated with helium. An association with hydrogen may be possible
(nebular hydrogen lines are sometimes blueshifted owing to geometric
effects; e.g., \citealt{yan2015}), but in this case the lack of
hydrogen Balmer lines in the spectra is puzzling.

\clearpage
\section{Spectra of PTF SLSN-I}\label{specplots}

\begin{figure}
\begin{center}
 \includegraphics[height=\textheight]{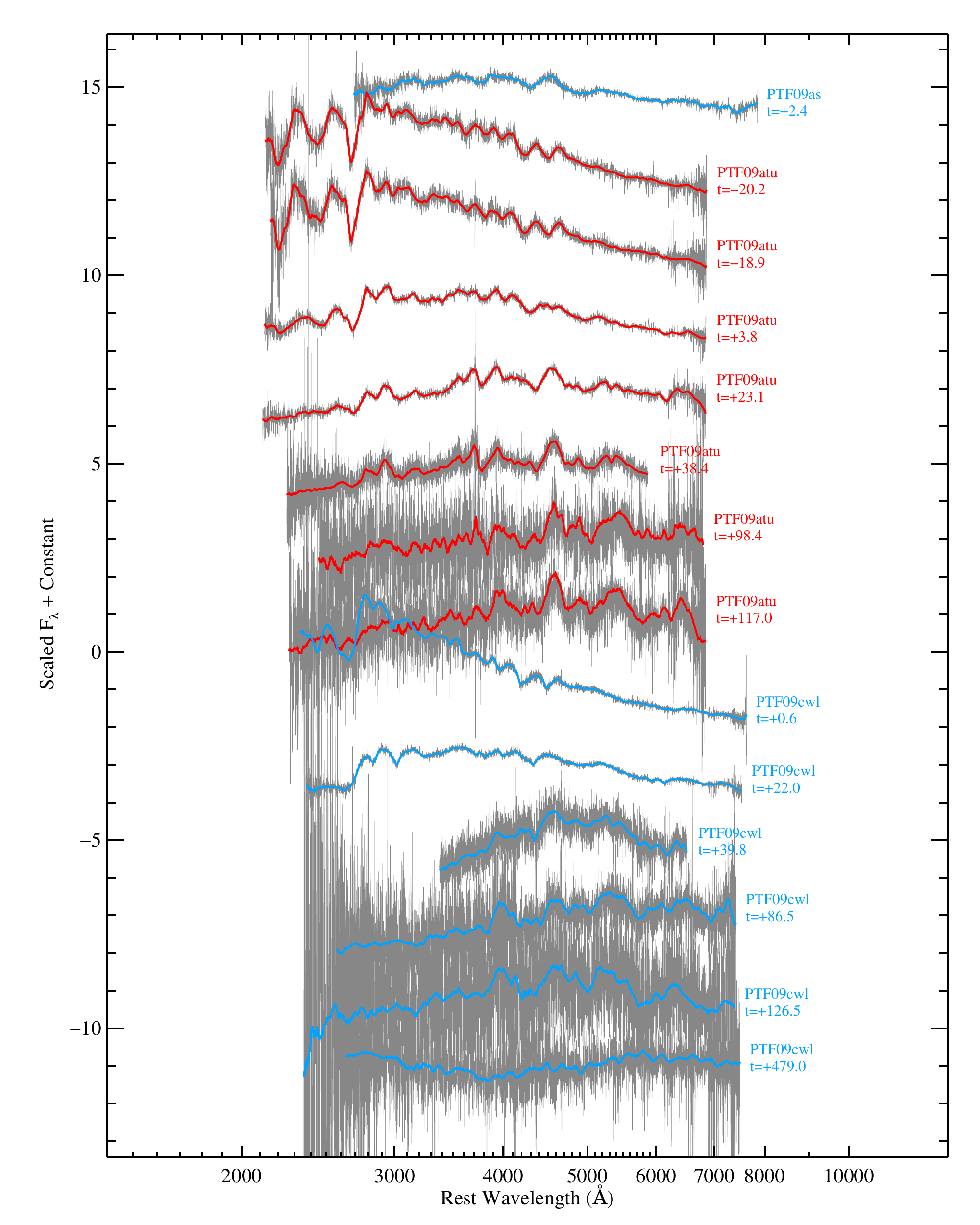}
 \caption{ Spectra of PTF09as, PTF09atu, and PTF09cwl. }
   \label{fig:spec01}
\end{center}
\end{figure}

\begin{figure}
\begin{center}
 \includegraphics[height=\textheight]{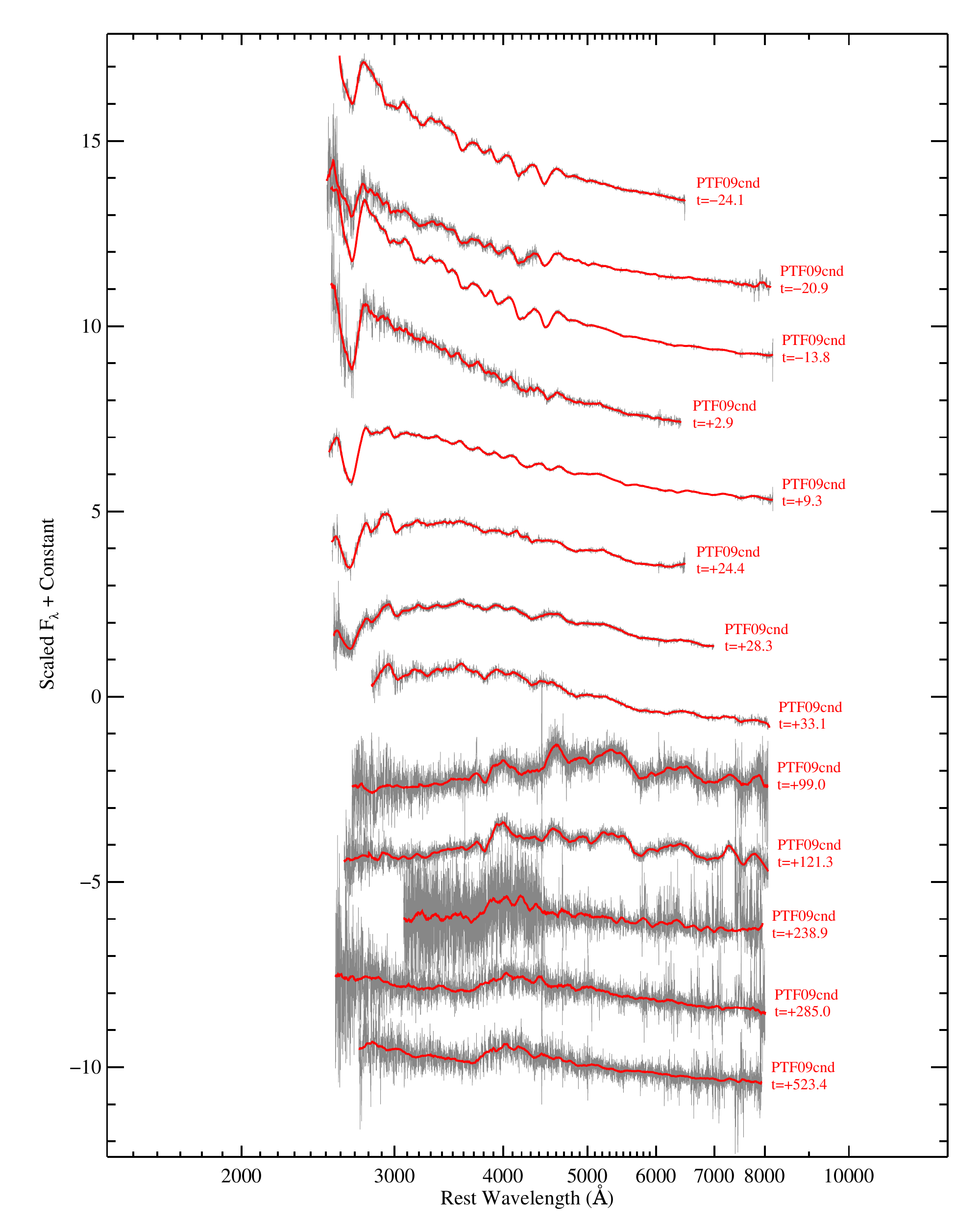}
 \caption{ Spectra of PTF09cnd. }
   \label{fig:spec02}
\end{center}
\end{figure}

\begin{figure}
\begin{center}
  \includegraphics[height=\textheight]{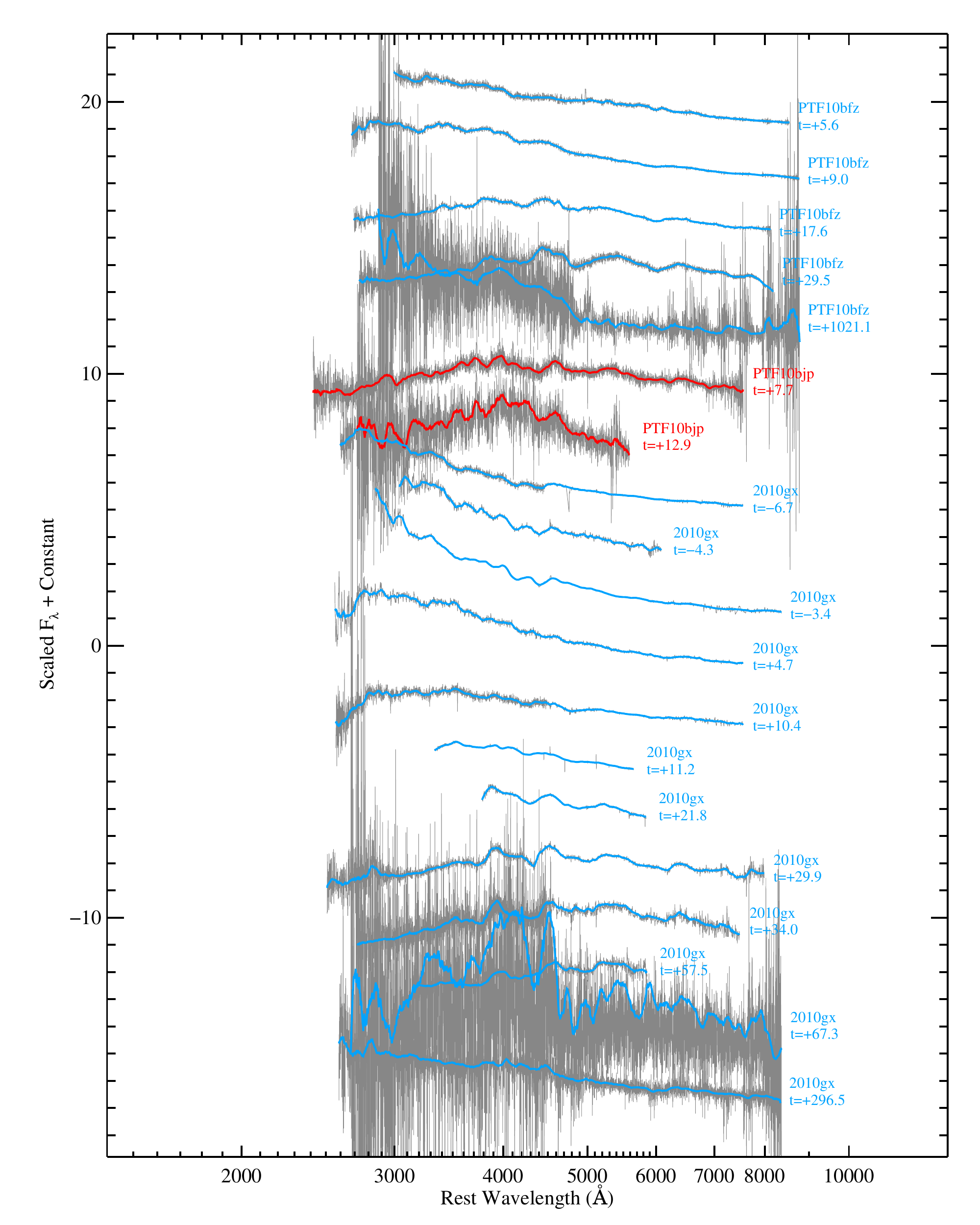}
  \caption{ Spectra of PTF10bfz, PTF10bjp, and SN\,2010gx. }
   \label{fig:spec03}
\end{center}
\end{figure}

\begin{figure}
\begin{center}
  \includegraphics[height=\textheight]{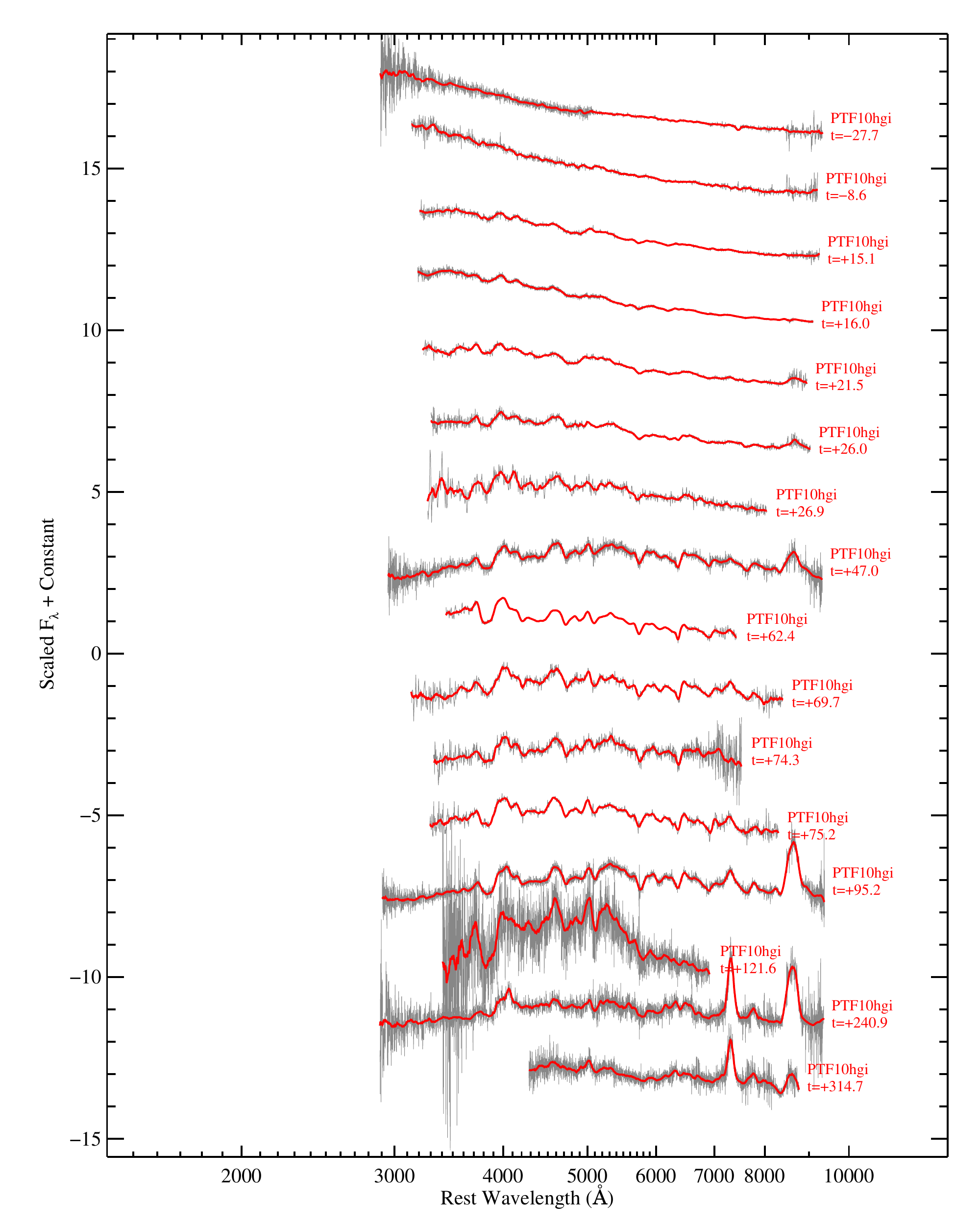}
   \caption{ Spectra of PTF10hgi. }
   \label{fig:spec04}
\end{center}
\end{figure}

\begin{figure}
\begin{center}
  \includegraphics[height=\textheight]{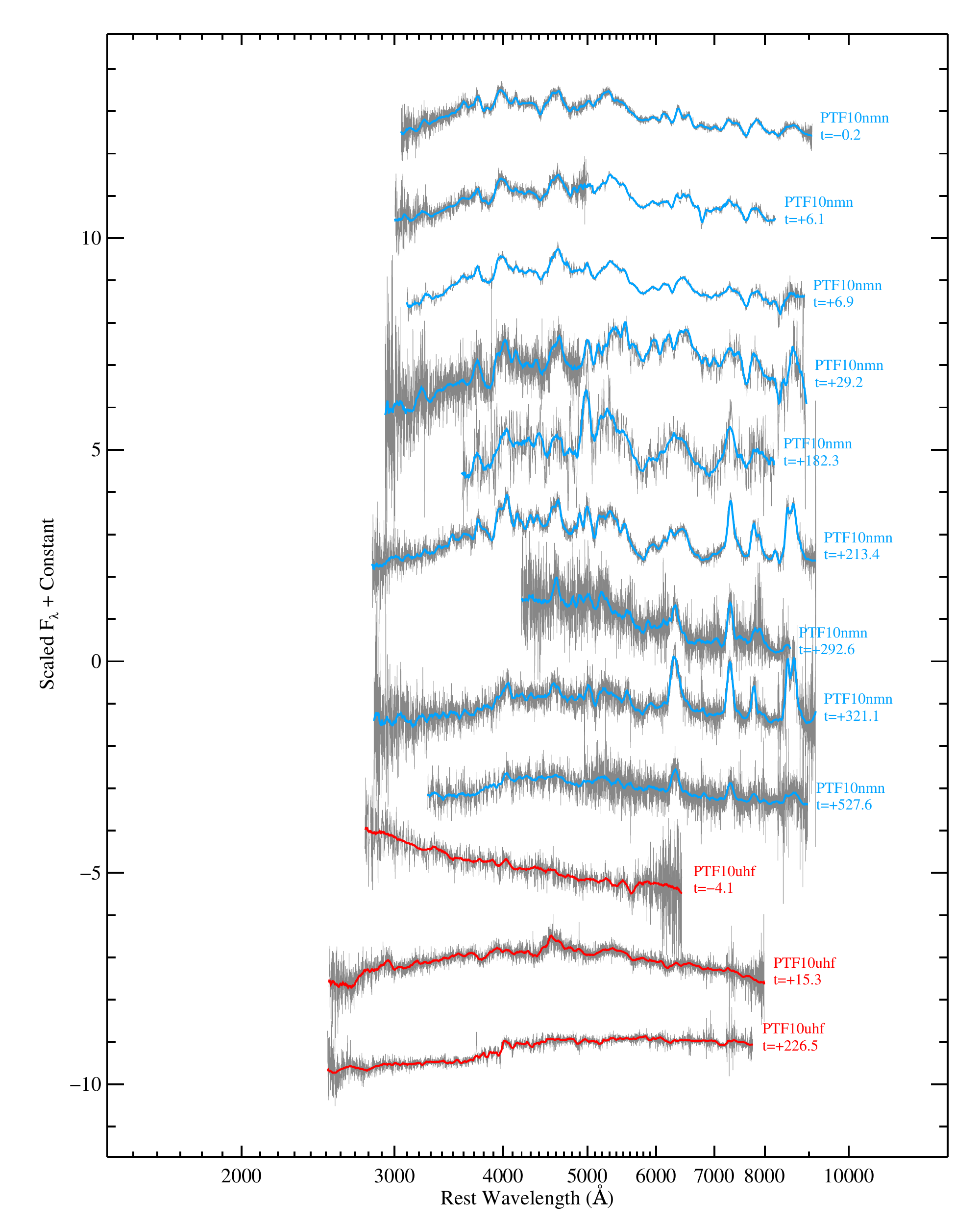}
   \caption{ Spectra of PTF10nmn and PTF10uhf. }
   \label{fig:spec05}
\end{center}
\end{figure}

\begin{figure}
\begin{center}
  \includegraphics[height=\textheight]{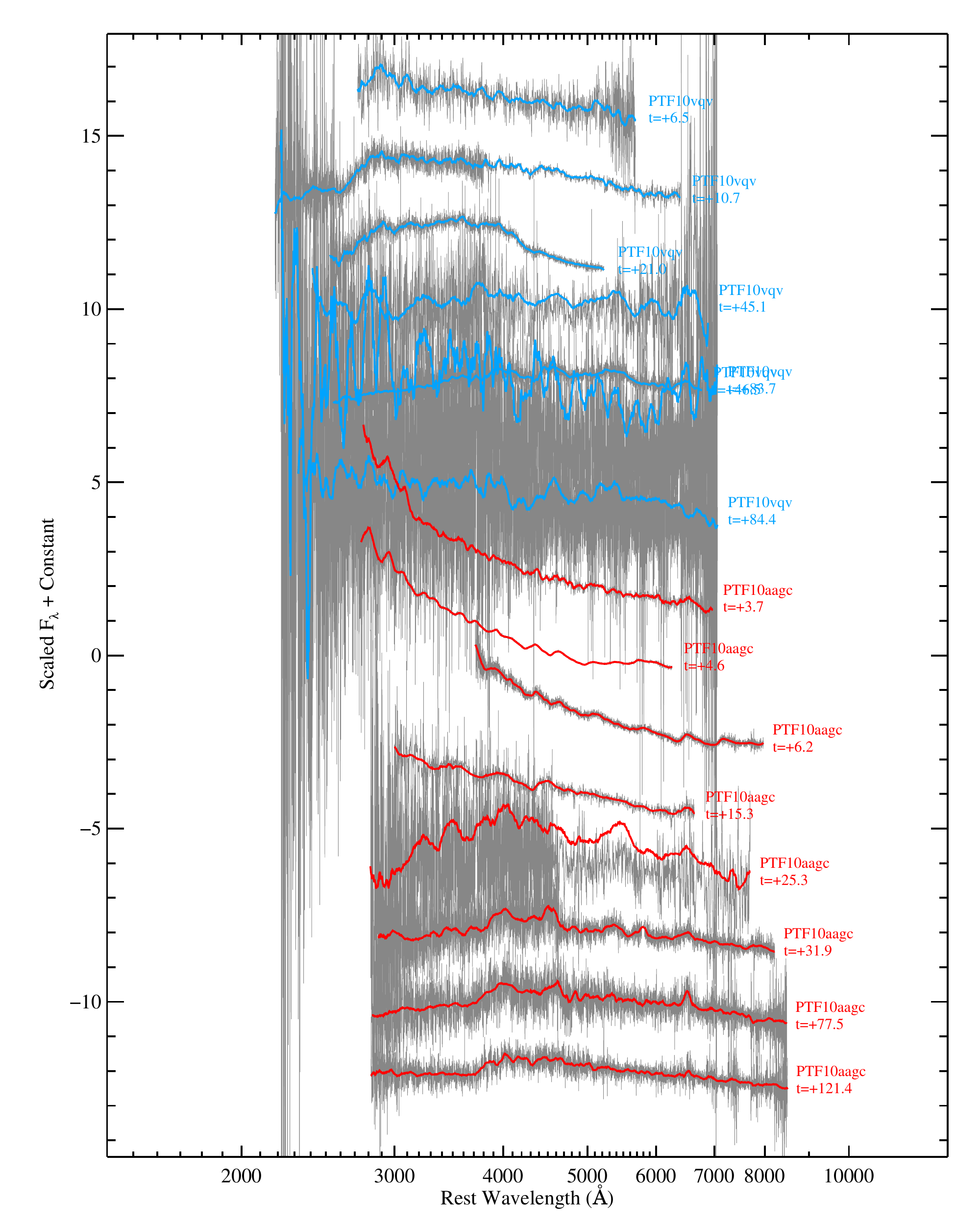}
   \caption{ Spectra of PTF10vqv and PTF10aagc. }
   \label{fig:spec06}
\end{center}
\end{figure}

\begin{figure}
\begin{center}
  \includegraphics[height=\textheight]{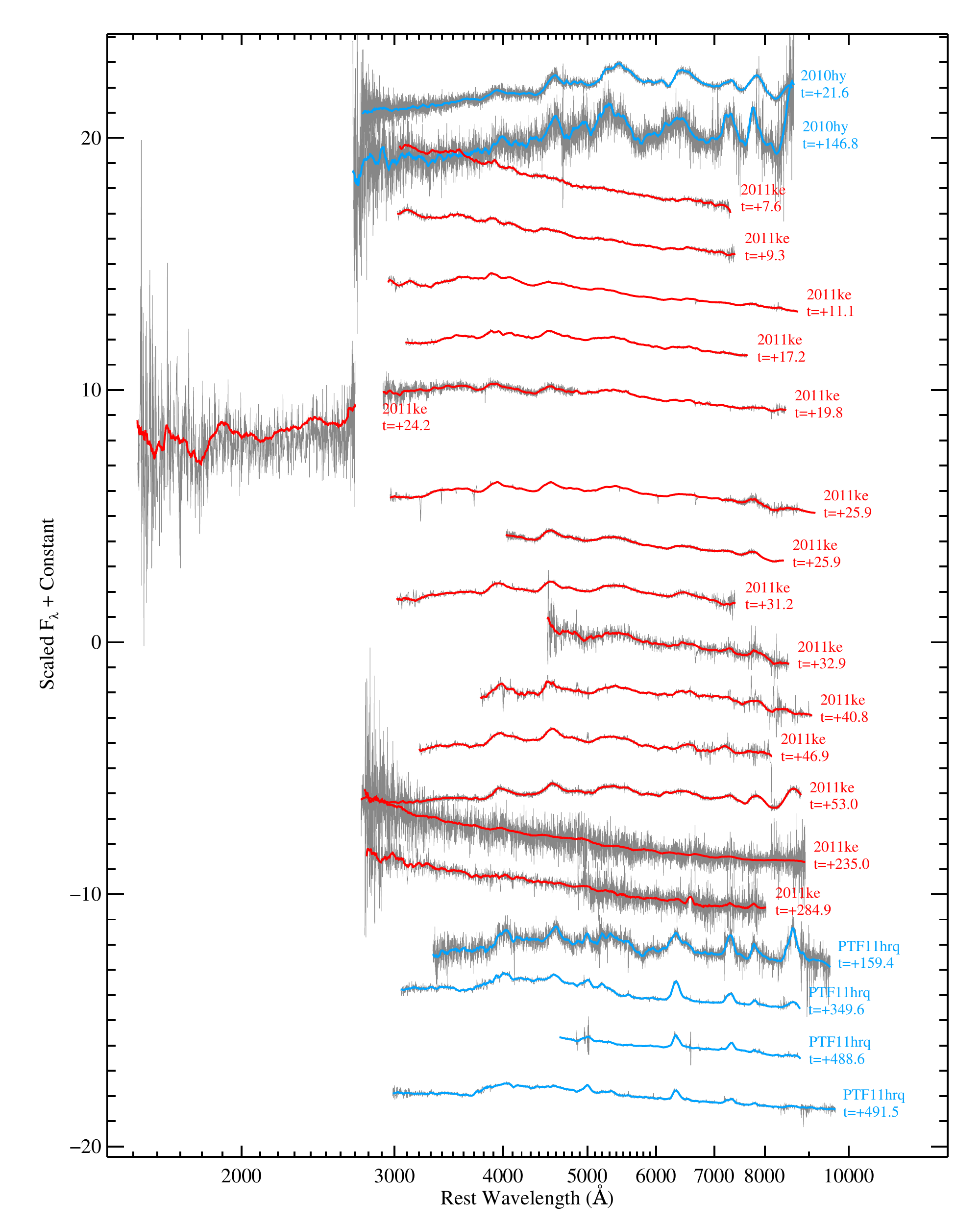}
   \caption{ Spectra of SN\,2010hy, SN\,2011ke, and PTF11hrq. }
   \label{fig:spec07}
\end{center}
\end{figure}

\begin{figure}
\begin{center}
  \includegraphics[height=\textheight]{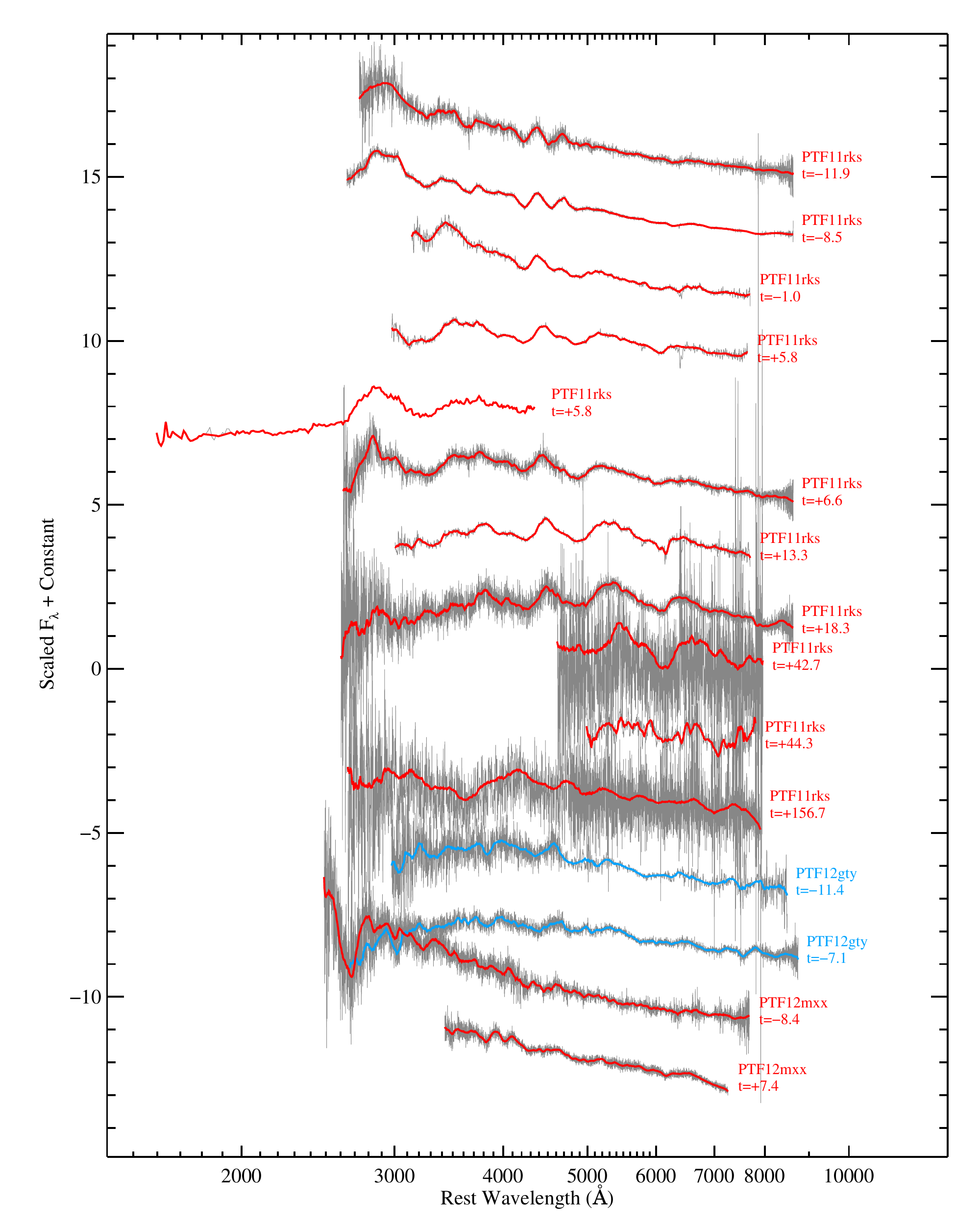}
   \caption{ Spectra of PTF11rks, PTF12gty, and PTF12mxx. }
   \label{fig:spec08}
\end{center}
\end{figure}

\begin{figure}
\begin{center}
  \includegraphics[height=\textheight]{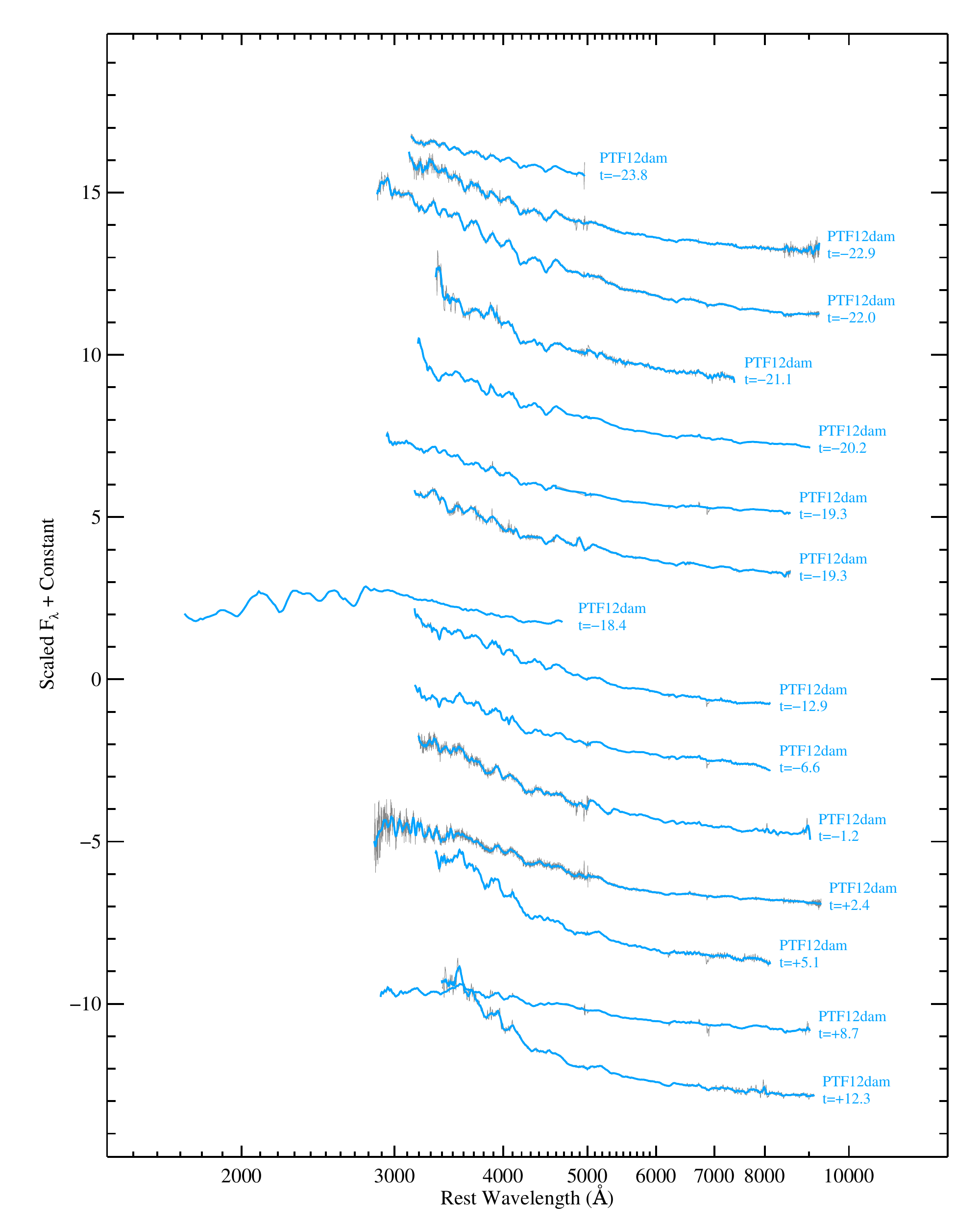}
   \caption{ Spectra of PTF12dam from $t=-25.3$ through $t=+10.8$\,d. }
   \label{fig:spec09}
\end{center}
\end{figure}

\begin{figure}
\begin{center}
  \includegraphics[height=\textheight]{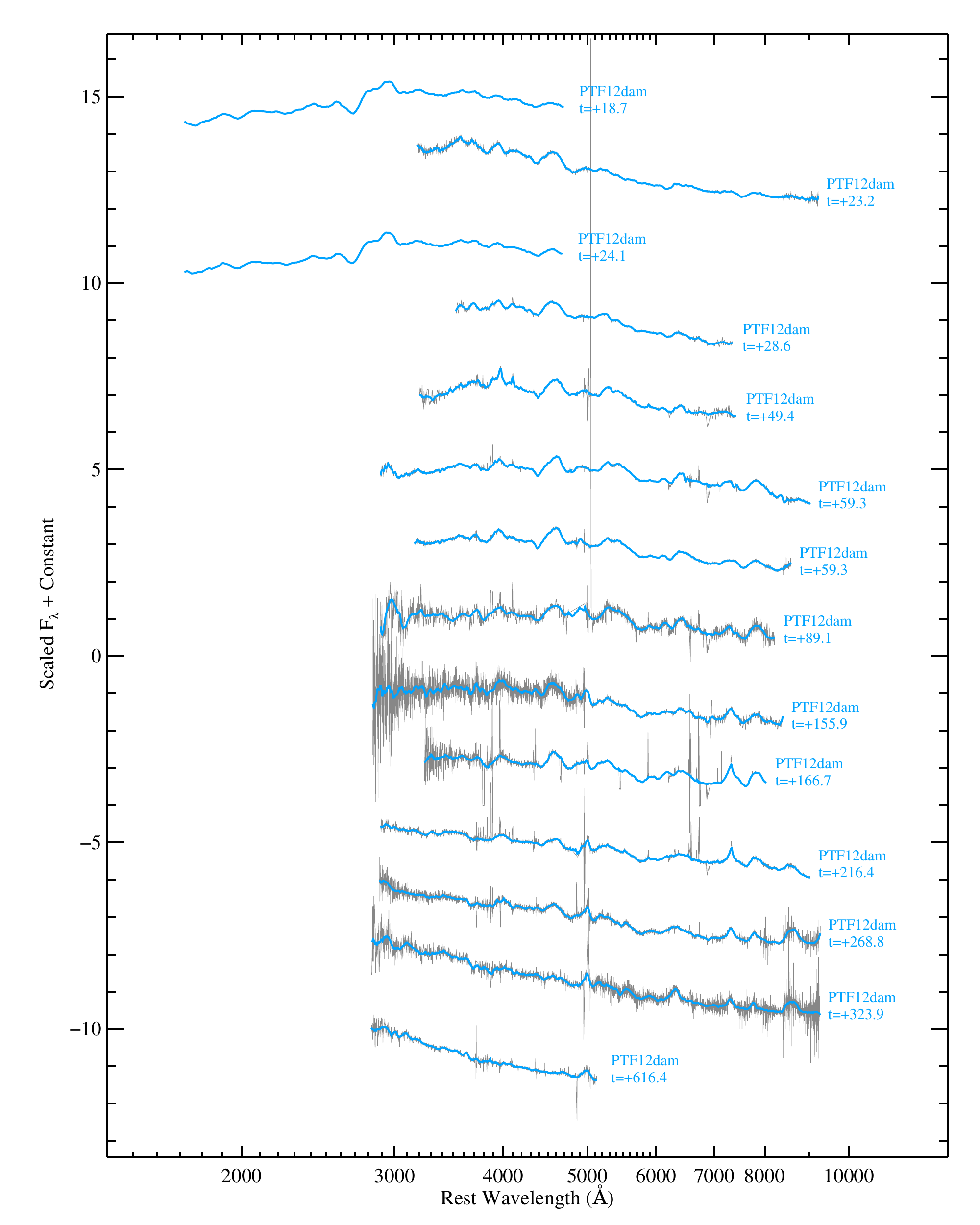}
   \caption{ Spectra of PTF12dam from $t=+17.2$ through $t=+614.9$\,d. }
   \label{fig:spec10}
\end{center}
\end{figure}

\begin{figure}
\begin{center}
  \includegraphics[height=\textheight]{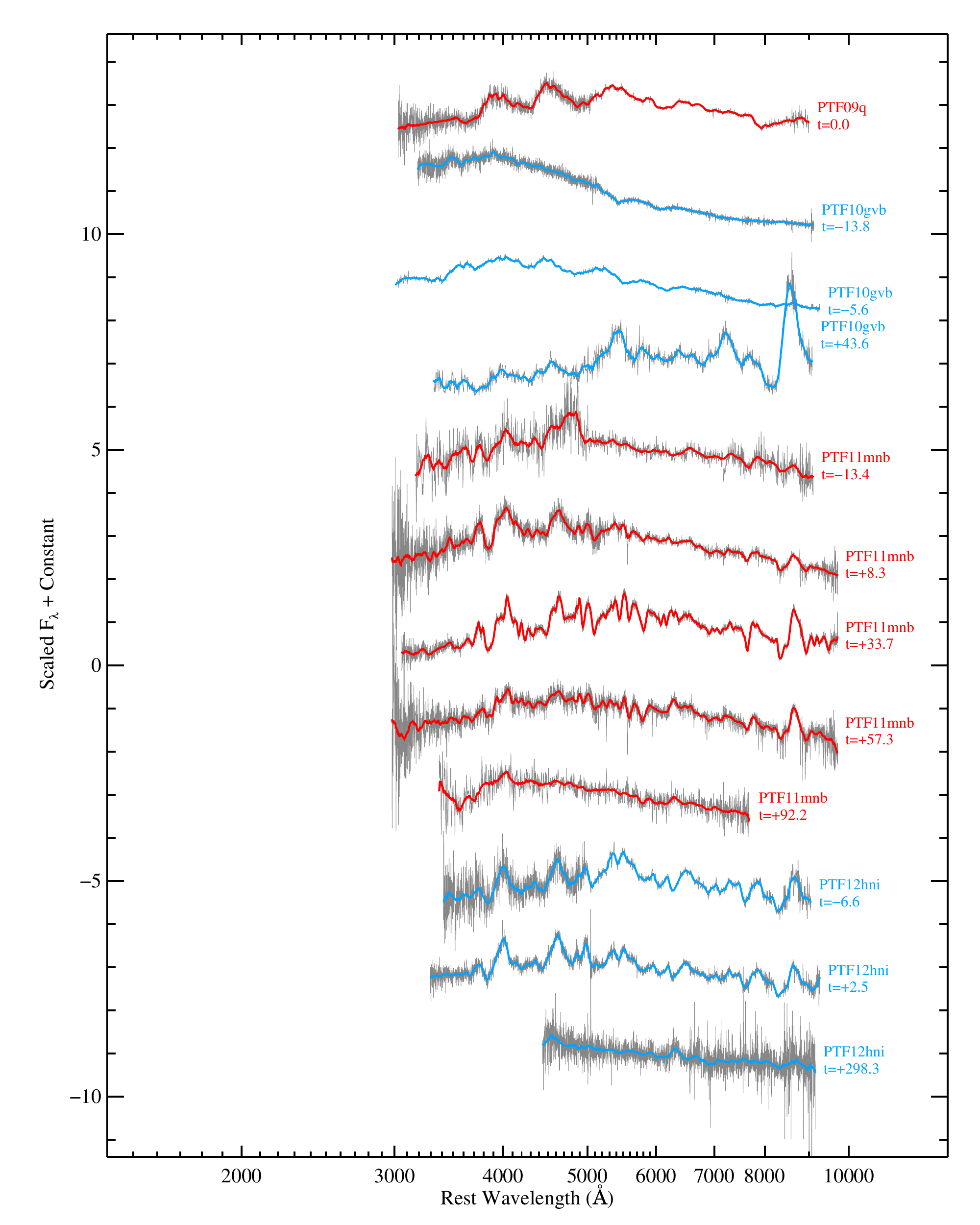}
   \caption{ Spectra of the possible SLSNe-I PTF09q, PTF10gvb, PTF11mnb, and PTF12hni. }
   \label{fig:spec11}
\end{center}
\end{figure}

In Figures~\ref{fig:spec01}--\ref{fig:spec11} we plot the 166
spectra of likely and possible SLSNe-I recorded by the PTF
survey. Included are 34 spectra previously published for SN\,2010gx,
SN\,2011ke, PTF10hgi, PTF11rks, and PTF12dam
\citep{pastorello2010,quimby2011,nicholl2013,inserra2013}. Each
extracted spectrum has been processed to remove narrow emission lines
from the host environment. Similar to the discussion in
\S\ref{selection}, lines are removed by simultaneously fitting
Gaussians at the expected wavelengths of host-galaxy lines, with all lines in
a given fit required to have the same width. The fitting is done
section by section (e.g., lines near H$\alpha$, then H$\beta$, etc.). We
shift the spectra to the host-galaxy rest frame and vertically for
clarity, and we overplot a smoothed version of the spectra (see
\S\ref{sgfilter}). Each spectrum is labeled with the object name and
light-curve phase. Other than the narrow emission lines, host-galaxy light
has not been removed from these spectra, and this may dominate the
signal at later phases in some cases.

\clearpage
\section{Average Spectral Features at $\phi \approx -1.0$}\label{average_spec-1}

\begin{figure*}
\begin{center}
 \includegraphics[width=\linewidth]{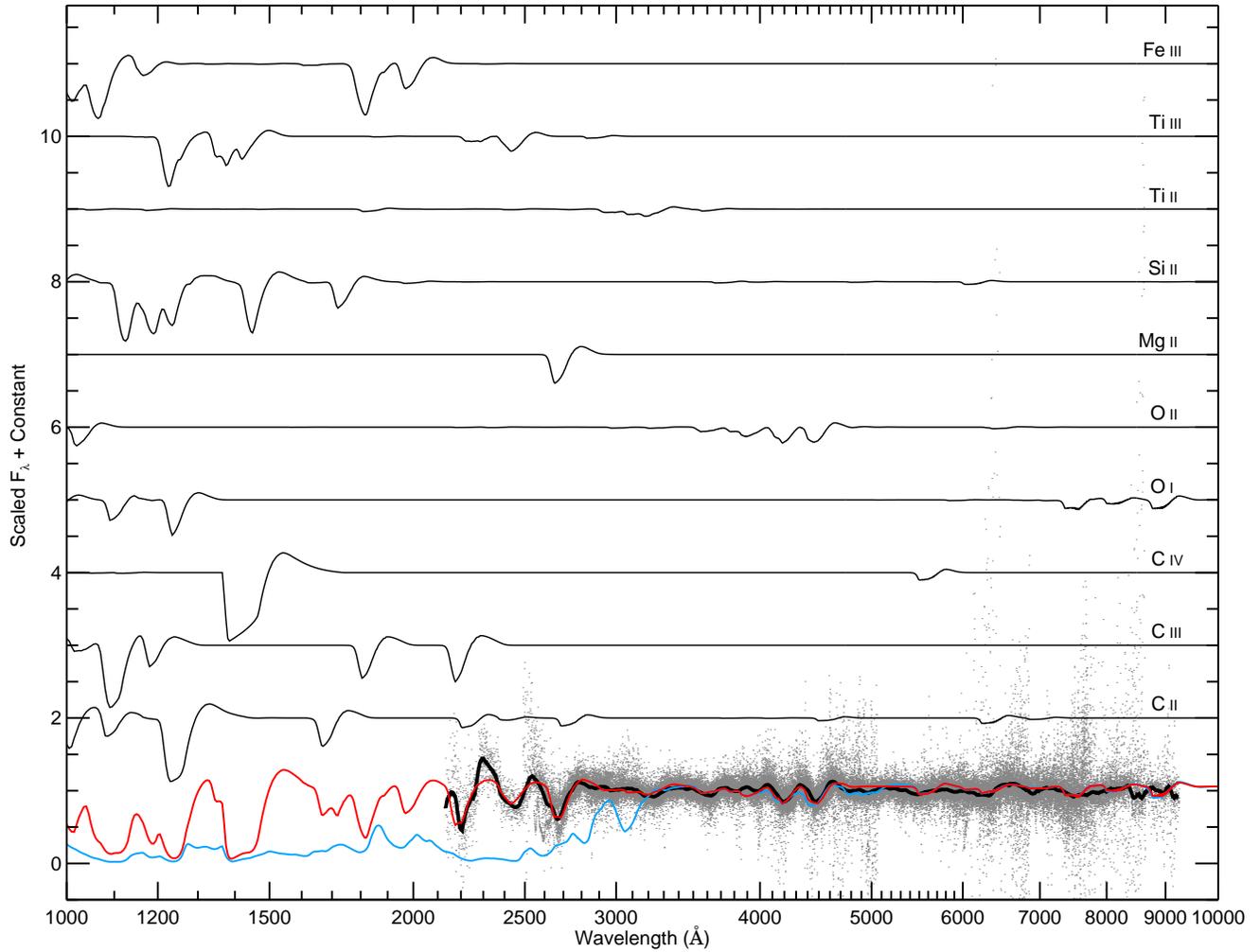}
 \caption{ Synthetic spectral model (red) from {\tt syn++} compared to
   the average of SLSN-I spectra in the range $-1.5 < \phi < -0.5$
   (thick black line). The spectra have been continuum divided to
   emphasize the features. Gray points show contributions to the
   average spectrum from individual object spectra. Contributions to
   the {\tt syn++} fit from each individual ion are plotted with thin
   black lines and shifted vertically for clarity. The blue curve
   shows the effect of adding \ion{Fe}{2} to the model. The absorption
   near 4900\,\AA, can be well fit with this change, but in doing so
   the {\tt syn++} models predict a strong feature at about
   3050\,\AA\ and severe line blanketing below 3000\,\AA, which is not
   observed. }
   \label{fig:syn++_-1}
\end{center}
\end{figure*}

In Figure \ref{fig:syn++_-1} we show the average spectrum of SLSNe-I at
early phases ($-1.5 < \phi < -0.5$) constructed as described in
\S\ref{compare_spec}. A {\tt syn++} model was constructed using
ions previously associated with SNe, especially SLSNe-I. For all
ion species we adopt the same velocity distributions to eliminate some
degree of freedom. We essentially fit only for the strength of the
ion and retain only those ions which appear to contribute
significantly to the average observed spectrum while not introducing
superfluous features into the model. We adopt an ion temperature of
10,000\,K for the models, except for the oxygen lines for which
15,000\,K seems to produce better results.

The strongest features in the average spectrum ----three broad dips
in the range 2100--2800\,\AA\ --- can be reasonably well fit by blends
of \ion{C}{2}, \ion{C}{3}, \ion{Si}{3}, \ion{Ti}{3}, and \ion{Mg}{2}
\citep{dessart2012, howell2013, mazzali2016}. A weak, broad dip
at 2900--3300\,\AA\ can be explained by \ion{Ti}{2} with
a minor contribution from \ion{O}{2} (a check using {\tt syn++} found
no other obvious ion species that could explain this dip without
introducing other, stronger features that conflict with the data). The
strongest features in the optical range --- two dips in the range
4000--4700\,\AA\ ---a re well fit by \ion{O}{2}
\citep{quimby2011}. As previously found, \ion{O}{2} is also the main
contributor to other, weaker dips to the blue of this ``W''
feature. However, the {\tt syn++} model significantly overpredicts the
absorption around 3900\,\AA. This discrepancy is investigated further
in \S\ref{o2lines}. Note we do not include \ion{Ca}{2} at this phase,
as it does not improve the quality of the fit. However, at later
phases this ion may become dominate around 3800\,\AA. The \ion{O}{2}
features may also be blended with \ion{C}{2} and possibly
\ion{Si}{2}. The latter ion is typically invoked to explain a common
feature at 6150\,\AA\ in SN spectra. For SLSNe-I, this feature
may be blended with \ion{C}{2}. If \ion{Si}{2} is present, then {\tt
  syn++} predicts there should be stronger lines in the UV around
1450\,\AA\ and 1720\,\AA\ as well. In addition to the \ion{C}{2}
lines near 6300\,\AA\ and 6940\,\AA\ \citep{quimby2011,yan2017},
there are likely features from \ion{O}{1} including the triplet around
7774\,\AA\ \citep{nicholl2016}.

\clearpage
\section{Average Spectral Features at $\phi \approx 0$}\label{average_spec0}

\begin{figure*}
\begin{center}
 \includegraphics[width=\linewidth]{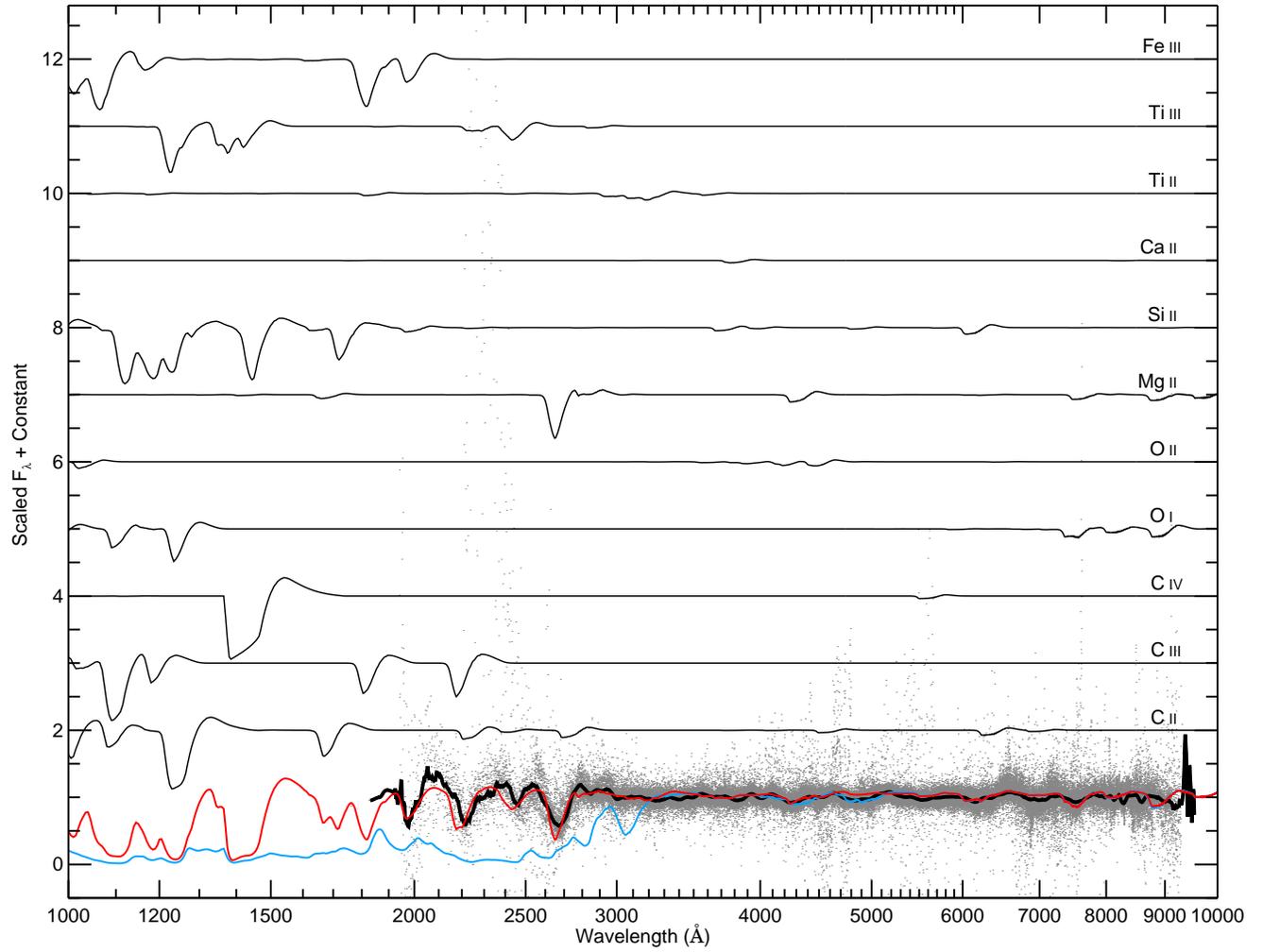}
 \caption{ Similar to Figure \ref{fig:syn++_-1} but for the average
   SLSN-I spectrum in the $-0.5 < \phi < +0.5$ range. }
   \label{fig:syn++_+0}
\end{center}
\end{figure*}

In Figure \ref{fig:syn++_+0} we show the average SLSN-I spectrum in
the $-0.5 < \phi < +0.5$ range. The {\tt syn++} fit includes many of
the same ions as in the earlier-phase spectra, but some of these have
been adjusted in strength. In particular, the \ion{O}{2} features may
persist to this phase, but they appear weaker and are more noticeably
blended with other ions. \ion{Ca}{2} may be weakly detected at this
phase, and \ion{Mg}{2} may start to dominate near 4300\,\AA. A
predicted \ion{Mg}{2} feature near 8900\,\AA, possibly blended with
\ion{O}{2}, is not well matched to the data. There is a weak, broad
dip near 4850\,\AA\ that can be well fit by \ion{Fe}{2}; however, to
match this feature's strength {\tt syn++} predicts a strong feature at
3050\,\AA\ and severe line blanketing at shorter wavelengths, which is
not observed. Following numerical modeling of other SNe, most
likely the 4850\,\AA\ feature is a blend of \ion{Fe}{2}. The poor {\tt
  syn++} match to the UV portion of the spectrum could presumably be
due to non-LTE effects, but investigation of this discrepancy is left
to future works. It is also notable that while the strong features in
the UV near 1950\,\AA\ and 2650\,\AA\ are well matched by the {\tt
  syn++} model, the positions of the minima for the 2200\,\AA\ and
2400\,\AA\ features are systematically offset. This could indicate
incorrect line identifications, which is investigated further in
\S\ref{lineids}.

We also note that the observations suggest a weak absorption feature
near 5500\,\AA. Using {\tt syn++} to check possible ion species we
find few options for its production without the
addition of stronger features as well in the wavelength range
constrained by our spectral sample. One possibility shown in Figure
\ref{fig:syn++_+0} is that this absorption is caused by \ion{C}{4}. If
this identification is correct, then {\tt syn++} predicts that a much
stronger line should also form at $\sim1400$\,\AA. Recent, rest-frame
far-UV observations of the SLSN-I Gaia16apd have revealed a strong
feature near this wavelength \citep{yan2017}, which may match the
one predicted by {\tt syn++}. Detailed modeling is required to
test the validity of this tentative assignment.

\clearpage
\section{Average Spectral Features at $\phi \approx +1.0$}\label{average_spec+1}

\begin{figure*}
\begin{center}
 \includegraphics[width=\linewidth]{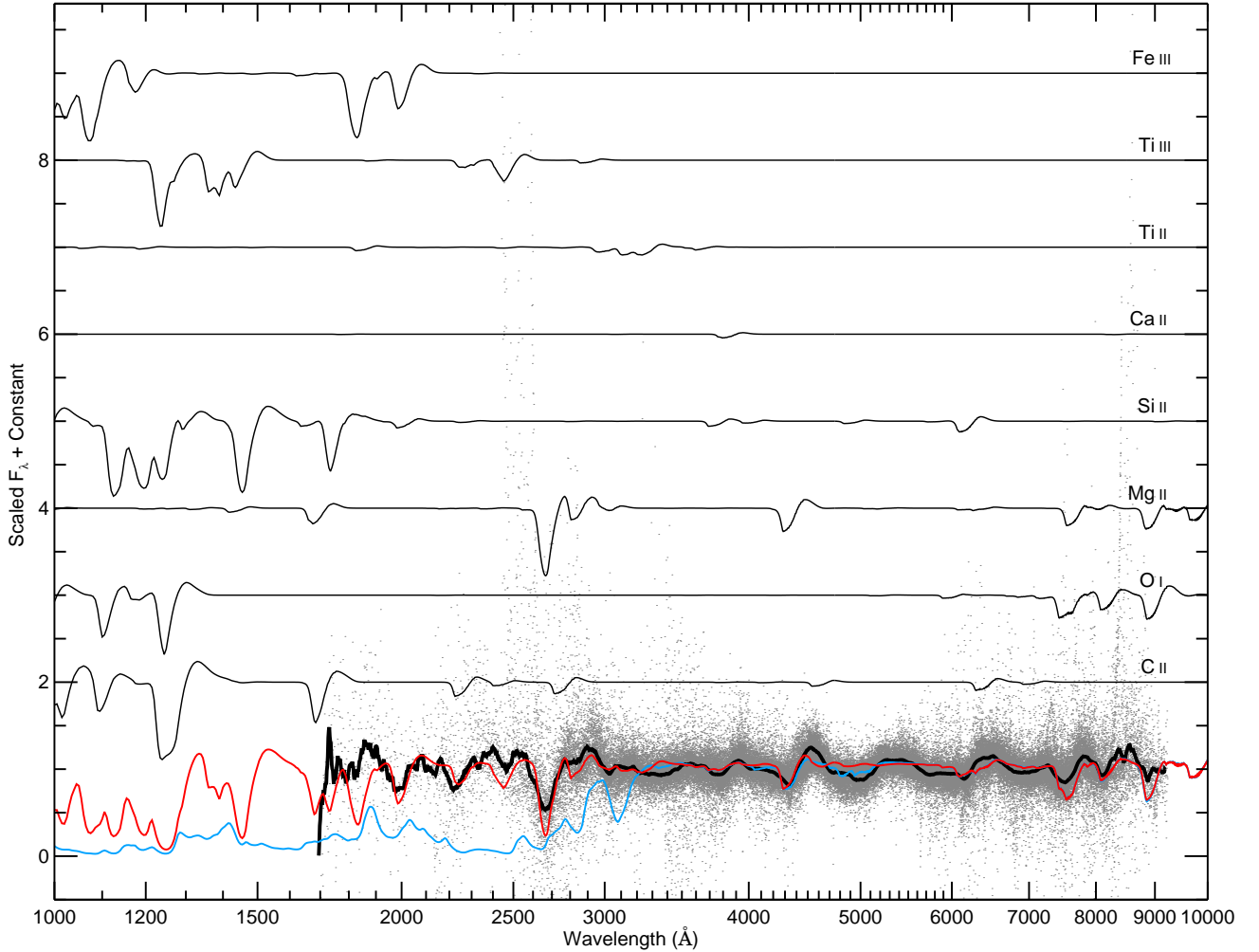}
 \caption{ Similar to Figures \ref{fig:syn++_-1} and
   \ref{fig:syn++_+0} but for the average SLSN-I spectrum in the $+0.5
   < \phi < +1.5$ range. }
   \label{fig:syn++_+1}
\end{center}
\end{figure*}

In Figure \ref{fig:syn++_+1} we show the average SLSN-I spectrum in
the $+0.5 < \phi < +1.5$ range with a {\tt syn++} model. To fit the
2000\,\AA\ absorption feature with \ion{Fe}{3} as has previously been
suggested \citep{howell2013} {\tt syn++} predicts a stronger feature
at 1850\,\AA\ that conflicts with the data. Again, the inclusion of
\ion{Fe}{2} in the fit greatly improves the agreement in the
optical portion of the spectrum near 4800\,\AA\ and also near
4100\,\AA; however, the model differs significantly from the average
observed spectrum in the UV. This includes a strong predicted feature
around 3050\,\AA\ and near total absorption around 2400\,\AA. We do
note, however, some rough agreement between peaks in the model flux
(including \ion{Fe}{2}) in the range 1700--2200\,\AA. Thus, it is
possible that \ion{Fe}{2} does contribute significantly to the UV
portion of the spectrum at this phase. Some of the discrepancy between
the data and the model including \ion{Fe}{2} in Figure
\ref{fig:syn++_+1} may be due to the way the observational data were
continuum divided, but this would not explain the 3050\,\AA\ feature
nor the relatively high luminosity observed in the UV bands.

The model fit indicates the \ion{O}{2} lines have subsided and that
the dominant features in the optical range are now from \ion{Mg}{2},
\ion{Fe}{2}, and \ion{Ca}{2}. Moving to the near-infrared, there is 
\ion{O}{1}, which produces three dips at 7500\,\AA, 8100\,\AA, and
8900\,\AA. The \ion{Mg}{2} also has relatively strong features in this
range which may contribute to these dips. A further contributor may be
\ion{Fe}{2}, although the strengths of its lines are uncertain from the
non-LTE {\tt syn++} models as noted above.

\end{document}

%% file: library_refs.tex
\begin{table*}
\centering
\caption{Spectral Template Libraries}
\label{table:libraries}
\begin{tabular}{| p{0.15\linewidth} | p{0.15\linewidth} | p{0.5\linewidth} |}
  \hline
  Group & Number of Spectra & References \\
\hline
\hline
SNIa & 1924 & \citet{blondin2012}
\\

\hline
SNII/SNIIn & 373 & \citet{pun1995,leonard2000,fassia2001,leonard2002,pastorello2004,fransson2005,vinko2006,pastorello2006,sahu2006,quimby2007b,bufano2009,pastorello2009,stritzinger2012,taddia2013,bayless2013,dallora2014,zhang2014,takats2014,spiro2014,jerkstrand2014}
\\
\hline
SNIb, SNIc, SNIc-bl, SNIbn, SNIIb
& 587 & \citet{barbon1995,filippenko1995,matheson2001,patat2001,branch2002,foley2003,folatelli2006,taubenberger2006,pastorello2007,harutyunyan2008,valenti2008,pastroello2008,malesani2009,modjaz2009,taubenberger2009,taubenberger2011,modjaz2014}
\\
\hline
SLSN-I & 118 & {\bf this work}; PESSTO-DR1; \citet{quimby2007c,barbary2009,galyam2009,pastorello2010,chomiuk2011,inserra2013,howell2013}
\\
\hline
SLSN-II & 22 &
\citet{smith2007,smith2010,chatzopoulos2011}
\\

\hline
\end{tabular}
\end{table*}

%% file: ref.SLSN-I.table.tex
\begin{table*}
\centering
\caption{SLSN-I Spectroscopic Reference Set}
\label{table:published}
\begin{tabular}{llrl}
\hline
Name & $z$ & $N_{\rm spec}$ & Reference \\

\hline
         SN\,2005ap   &  0.283 &    2 & \citet{quimby2007c} \\ 
        SCP06F6   &  1.189 &    3 & \citet{barbary2009} \\ 
    SNLS-06D4eu   &  1.588 &    1 & \citet{howell2013} \\ 
         SN\,2006oz   &  0.396 &    1 & \citet{leloudas2012} \\ 
         SN\,2007bi   &  0.128 &    3 & \citet{galyam2009} \\ 
    SNLS-07D2bv   &  1.500 &    1 & \citet{howell2013} \\ 
       PTF09atu   &  0.501 &    7 &  {\bf this work}, \citet{quimby2011} \\ 
       PTF09cwl   &  0.350 &    6 &  {\bf this work}, \citet{quimby2011} \\ 
       PTF09cnd   &  0.259 &   13 &  {\bf this work}, \citet{quimby2011} \\ 
SN\,2010gx (PTF10cwr) &  0.230 &   12 &  {\bf this work}, \citet{pastorello2010}, \citet{quimby2011} \\ 
       PTF10hgi   &  0.098 &   16 &  {\bf this work}, \citet{inserra2013} \\ 
       PS1-10ky   &  0.956 &    4 & \citet{chomiuk2011} \\ 
      PS1-10awh   &  0.908 &    3 & \citet{chomiuk2011} \\ 
SN\,2011ke (PTF11dij) &  0.143 &   15 &  {\bf this work}, \citet{inserra2013} \\ 
         SN\,2011kf   &  0.245 &    2 & \citet{inserra2013} \\ 
       PTF11rks   &  0.192 &   10 &  {\bf this work}, \citet{inserra2013} \\ 
         SN\,2012il   &  0.175 &    3 & \citet{inserra2013} \\ 
       PTF12dam   &  0.108 &   28 &  {\bf this work}, \citet{nicholl2013} \\ 
       LSQ12dlf   &  0.250 &    6 & \citet{nicholl2014} \\ 
      SSS120810   &  0.156 &    6 & \citet{nicholl2014} \\ 
         SN\,2013dg   &  0.265 &    1 & \citet{nicholl2014} \\ 
\hline
\end{tabular}


\end{table*}

%% file: ptf.SLSN-I.table.tex
\begin{table*}
\centering
\caption{PTF Spectroscopic SLSN-I Sample}
\label{table:sample}
\begin{tabular}{lllllrrll}
\hline
Name & $\alpha$(J2000) & $\delta$(J2000) & $z$ & MJD$_{\rm peak}$ & $M_{\rm peak}$ & $N_{\rm spec}$ & Spec. Class & \\

\hline
        PTF09as   &  12:59:15.78 &  $+$27:16:38.5 & 0.1864 & 54918.2 & \nodata  &  1 &   SLSN-I &  \\ 
       PTF09atu   &  16:30:24.55 &  $+$23:38:25.0 & 0.5014 & 55062.3 & $-21.75$ &  7 &   SLSN-I &  \\ 
       PTF09cnd   &  16:12:08.94 &  $+$51:29:16.2 & 0.2585 & 55085.3 & $-22.18$ & 13 &   SLSN-I &  \\ 
       PTF09cwl   &  14:49:10.08 &  $+$29:25:11.4 & 0.3502 & 55067.2 & $-21.92$ &  6 &   SLSN-I &  \\ 
       PTF10bfz   &  12:54:41.27 &  $+$15:24:17.0 & 0.1699 & 55227.5 & $-21.06$ &  5 &   SLSN-I &  \\ 
       PTF10bjp   &  10:06:34.30 &  $+$67:59:19.0 & 0.3585 & 55251.5 & $-20.54$ &  2 &   SLSN-I &  \\ 
SN\,2010gx (PTF10cwr) &  11:25:46.67 &  $-$08:49:41.2 & 0.2301 & 55281.2 & $-21.62$ &  5 &   SLSN-I &  \\ 
       PTF10hgi   &  16:37:47.04 &  $+$06:12:32.3 & 0.0982 & 55367.4 & $-20.35$ & 13 &   SLSN-I$^a$ &  \\ 
       PTF10nmn   &  15:50:02.79 &  $-$07:24:42.1 & 0.1236 & 55384.2 & $-20.53$ &  9 &   SLSN-I &  \\ 
       PTF10uhf   &  16:52:46.68 &  $+$47:36:22.0 & 0.2879 & 55452.3 & $-20.68$ &  3 &   SLSN-I &  \\ 
       PTF10vqv   &  03:03:06.84 &  $-$01:32:34.9 & 0.4520 & 55470.5 & $-21.55$ &  7 &   SLSN-I &  \\ 
SN\,2010hy (PTF10vwg) &  18:59:32.86 &  $+$19:24:25.7 & 0.19   & 55457.3 & $-22.09$ &  2 &   SLSN-I &  \\ 
      PTF10aagc   &  09:39:56.93 &  $+$21:43:16.9 & 0.2067 & 55499.5 & $-20.25$ &  8 &   SLSN-I &  \\ 
SN\,2011ke (PTF11dij) &  13:50:57.77 &  $+$26:16:42.8 & 0.1429 & 55683.4 & $-21.39$ &  9 &   SLSN-I &  \\ 
       PTF11hrq   &  00:51:47.22 &  $-$26:25:10.0 & 0.0571 & 55753.5 & $-20.05$ &  4 &   SLSN-I &  \\ 
       PTF11rks   &  01:39:45.51 &  $+$29:55:27.0 & 0.1924 & 55936.1 & $-20.92$ &  7 &   SLSN-I &  \\ 
       PTF12dam   &  14:24:46.20 &  $+$46:13:48.3 & 0.1075 & 56093.3 & $-21.69$ & 15 &   SLSN-I &  \\ 
       PTF12gty   &  16:01:15.23 &  $+$21:23:17.4 & 0.1768 & 56143.4 & $-20.04$ &  2 &   SLSN-I &  \\ 
       PTF12mxx   &  22:30:16.68 &  $+$27:58:21.9 & 0.3274 & 56290.1 & $-21.61$ &  2 &   SLSN-I &  \\ 
\hline                                                              
         PTF09q   &  12:24:50.11 &  $+$08:25:58.8 & 0.09   & 54910.0 &  \nodata &  1 & possible SLSN-I &  \\
       PTF10gvb   &  12:15:32.28 &  $+$40:18:09.5 & 0.098  & 55337.2 & $-19.58$ &  3 & possible SLSN-I &  \\
       PTF11mnb   &  00:34:13.25 &  $+$02:48:31.4 & 0.0603 & 55855.3 & $-18.91$ &  5 & possible SLSN-I &  \\ 
       PTF12hni   &  22:31:55.86 &  $-$06:47:49.0 & 0.1056 & 56155.3 & $-19.96$ &  3 & possible SLSN-I &  \\ 
\hline
\end{tabular}

$^a$Spectra of PTF10hgi show clear evidence for hydrogen and
helium. Based on this it may better be classified as a SLSN-IIb.

\end{table*}

%% file: ptf.speclog.tex
\begin{longtable*}{llrlcrrl}
\caption{Log of Spectroscopic Observations}
\label{table:speclog}
\\
\hline
\multicolumn{1}{c}{Name}
& \multicolumn{1}{c}{Date}
& \multicolumn{1}{c}{LC Phase}
& \multicolumn{1}{c}{Inst.}
& \multicolumn{1}{c}{Range}
& \multicolumn{1}{c}{Spec. Phase}
& \multicolumn{1}{c}{Spec. Phase}
\\
\multicolumn{1}{c}{}
& \multicolumn{1}{c}{(UT)}
& \multicolumn{1}{c}{(days)}
& \multicolumn{1}{c}{}
& \multicolumn{1}{c}{(\AA)}
& \multicolumn{1}{c}{(fiducial)}
& \multicolumn{1}{c}{(fit)}
\\
\hline
\endfirsthead
\hline
\multicolumn{7}{c}%
{{\bfseries \tablename\ \thetable{} -- continued from previous page}} \\
\multicolumn{1}{c}{Name}
& \multicolumn{1}{c}{Date}
& \multicolumn{1}{c}{LC Phase}
& \multicolumn{1}{c}{Inst.}
& \multicolumn{1}{c}{Range}
& \multicolumn{1}{c}{Spec. Phase}
& \multicolumn{1}{c}{Spec. Phase}
\\
\multicolumn{1}{c}{}
& \multicolumn{1}{c}{(UT)}
& \multicolumn{1}{c}{(days)}
& \multicolumn{1}{c}{}
& \multicolumn{1}{c}{(\AA)}
& \multicolumn{1}{c}{(fiducial)}
& \multicolumn{1}{c}{(fit)}
\\
\hline
\endhead
\hline \multicolumn{7}{|r|}{{Continued on next page}} \\ \hline
\endfoot
\hline \hline
\endlastfoot
         PTF09q & 2009-03-20 & $         -0.0$ & Palomar-5m/DBSP & $ 3016-10204$ &  \nodata  & $  1.27 \pm   0.22$ & \\ 
        PTF09as & 2009-03-31 & $         +2.4$ &     Keck-I/LRIS & $ 3026- 9316$ &  \nodata  & $  1.35 \pm   0.09$ & \\ 
       PTF09atu & 2009-07-20 & $        -20.2$ &     Keck-I/LRIS & $ 3001-10297$ & $ -0.68$  & $ -0.58 \pm   0.25$ & \\ 
       PTF09atu & 2009-07-22 & $        -18.9$ &     Keck-I/LRIS & $ 3001-10285$ & $ -0.61$  & $ -0.58 \pm   0.25$ & \\ 
       PTF09atu & 2009-08-25 & $         +3.8$ &     Keck-I/LRIS & $ 3005-10296$ & $  0.25$  & $  0.20 \pm   0.09$ & \\ 
       PTF09atu & 2009-09-23 & $        +23.1$ &     Keck-I/LRIS & $ 3001-10296$ & $  1.00$  & $  1.21 \pm   0.38$ & \\ 
       PTF09atu & 2009-10-16 & $        +38.4$ &     Keck-I/LRIS & $ 3106- 8811$ & $  1.61$  & $  1.43 \pm   0.26$ & \\ 
       PTF09atu & 2010-01-14 & $        +98.4$ &     Keck-I/LRIS & $ 3001-10296$ &  \nodata  & $  1.40 \pm   0.16$ & \\ 
       PTF09atu & 2010-02-11 & $       +117.0$ &     Keck-I/LRIS & $ 3001-10264$ & $  1.73$  & $  1.77 \pm   0.08$ & \\ 
       PTF09cwl & 2009-08-25 & $         +0.6$ &     Keck-I/LRIS & $ 3005-10296$ & $ -0.46$  & $ -0.49 \pm   0.13$ & \\ 
       PTF09cwl & 2009-09-23 & $        +22.0$ &     Keck-I/LRIS & $ 3001-10296$ & $  0.54$  & $  0.78 \pm   0.27$ & \\ 
       PTF09cwl & 2009-10-17 & $        +39.8$ &     Keck-I/LRIS & $ 3105- 8796$ & $  1.48$  & $  1.35 \pm   0.23$ & \\ 
       PTF09cwl & 2009-12-19 & $        +86.5$ &     Keck-I/LRIS & $ 3001-10296$ & $  1.83$  & $  1.65 \pm   0.32$ & \\ 
       PTF09cwl & 2010-02-11 & $       +126.5$ &     Keck-I/LRIS & $ 3001-10265$ &  \nodata  & $  1.72 \pm   0.36$ & \\ 
       PTF09cwl & 2011-06-02 & $       +479.0$ &     Keck-I/LRIS & $ 3001-10297$ &  \nodata  & $  1.37 \pm   0.14$ & \\ 
       PTF09cnd & 2009-08-12 & $        -24.1$ &        WHT/ISIS & $ 3261- 8159$ & $ -1.00$  & $ -0.66 \pm   0.25$ & \\ 
       PTF09cnd & 2009-08-16 & $        -20.9$ & Palomar-5m/DBSP & $ 3001-10237$ & $ -0.96$  & $ -0.66 \pm   0.23$ & \\ 
       PTF09cnd & 2009-08-25 & $        -13.8$ &     Keck-I/LRIS & $ 3005-10297$ & $ -0.64$  & $ -0.56 \pm   0.08$ & \\ 
       PTF09cnd & 2009-09-15 & $         +2.9$ &        WHT/ISIS & $ 3151- 8179$ & $ -0.38$  & $ -0.50 \pm   0.14$ & \\ 
       PTF09cnd & 2009-09-23 & $         +9.3$ &     Keck-I/LRIS & $ 3001-10296$ & $ -0.33$  & $ -0.36 \pm   0.23$ & \\ 
       PTF09cnd & 2009-10-12 & $        +24.4$ &        WHT/ISIS & $ 3151- 8163$ & $  0.37$  & $  0.50 \pm   0.15$ & \\ 
       PTF09cnd & 2009-10-17 & $        +28.3$ &     Keck-I/LRIS & $ 3105- 8809$ & $  0.51$  & $  0.92 \pm   0.49$ & \\ 
       PTF09cnd & 2009-10-23 & $        +33.1$ &     Keck-I/LRIS & $ 3481-10195$ & $  0.57$  & $  0.99 \pm   0.46$ & \\ 
       PTF09cnd & 2010-01-14 & $        +99.0$ &     Keck-I/LRIS & $ 3001-10297$ & $  1.77$  & $  1.64 \pm   0.31$ & \\ 
       PTF09cnd & 2010-02-11 & $       +121.3$ &     Keck-I/LRIS & $ 3001-10263$ & $  1.88$  & $  1.85 \pm   0.07$ & \\ 
       PTF09cnd & 2010-07-09 & $       +238.9$ &     Keck-I/LRIS & $ 3117-10230$ &  \nodata  & $  1.97 \pm   0.08$ & \\ 
       PTF09cnd & 2010-09-05 & $       +285.0$ &     Keck-I/LRIS & $ 3003-10228$ &  \nodata  & $  1.57 \pm   0.68$ & \\ 
       PTF09cnd & 2011-07-02 & $       +523.4$ &     Keck-I/LRIS & $ 3001-10074$ &  \nodata  & $  1.12 \pm   1.05$ & \\ 
       PTF10bfz & 2010-02-07 & $         +5.6$ &     Keck-I/LRIS & $ 3499- 9998$ & $ -0.13$  & $  0.38 \pm   0.72$ & \\ 
       PTF10bfz & 2010-02-11 & $         +9.0$ &     Keck-I/LRIS & $ 3001-10262$ & $  0.17$  & $  0.86 \pm   0.62$ & \\ 
       PTF10bfz & 2010-02-21 & $        +17.6$ &        WHT/ISIS & $ 3149- 9499$ & $  1.17$  & $  1.33 \pm   0.07$ & \\ 
       PTF10bfz & 2010-03-07 & $        +29.5$ &     Keck-I/LRIS & $ 3001- 9582$ & $  1.23$  & $  1.33 \pm   0.08$ & \\ 
       PTF10bfz & 2013-05-10 & $      +1021.1$ &     Keck-I/LRIS & $ 3149-10274$ &  \nodata  & $  1.21 \pm   0.46$ & \\ 
       PTF10bjp & 2010-03-07 & $         +7.7$ &     Keck-I/LRIS & $ 3001-10272$ & $  1.09$  & $  1.51 \pm   0.54$ & \\ 
       PTF10bjp & 2010-03-14 & $        +12.9$ &     Keck-I/LRIS & $ 3253- 7602$ &  \nodata  & $  1.28 \pm   0.62$ & \\ 
         SN\,2010gx & 2010-03-18 & $         -6.7$ & Palomar-5m/DBSP & $ 3007- 9298$ & $ -1.04$  & $ -0.79 \pm   0.21$ & \\ 
         SN\,2010gx & 2010-04-08 & $        +10.4$ & Palomar-5m/DBSP & $ 3003- 9299$ & $  0.39$  & $  0.79 \pm   0.35$ & \\ 
         SN\,2010gx & 2010-05-07 & $        +34.0$ & Palomar-5m/DBSP & $ 3029- 9299$ & $  1.58$  & $  1.44 \pm   0.10$ & \\ 
         SN\,2010gx & 2010-06-17 & $        +67.3$ &     Keck-I/LRIS & $ 3053-10296$ &  \nodata  & $  1.64 \pm   0.20$ & \\ 
         SN\,2010gx & 2011-03-26 & $       +296.5$ &     Keck-I/LRIS & $ 3012-10260$ &  \nodata  & $  1.03 \pm   0.50$ & \\ 
       PTF10hgi & 2010-05-21 & $        -27.7$ & Palomar-5m/DBSP & $ 3001-10288$ & $ -0.09$  & $  0.26 \pm   0.72$ & \\ 
       PTF10hgi & 2010-06-11 & $         -8.6$ &    Lick-3m/Kast & $ 3443-10157$ & $ -0.07$  & $ -0.12 \pm   0.32$ & \\ 
       PTF10hgi & 2010-07-07 & $        +15.1$ &    Lick-3m/Kast & $ 3429-10163$ & $  0.45$  & $  0.62 \pm   0.27$ & \\ 
       PTF10hgi & 2010-07-08 & $        +16.0$ &     Keck-I/LRIS & $ 3499- 9999$ & $  0.48$  & $  0.62 \pm   0.27$ & \\ 
       PTF10hgi & 2010-07-14 & $        +21.5$ & Palomar-5m/DBSP & $ 3549- 9899$ & $  1.02$  & $  0.90 \pm   0.51$ & \\ 
       PTF10hgi & 2010-07-19 & $        +26.0$ &    Lick-3m/Kast & $ 3477- 9929$ & $  1.11$  & $  1.09 \pm   0.55$ & \\ 
       PTF10hgi & 2010-08-11 & $        +47.0$ &     Keck-I/LRIS & $ 3004-10236$ & $  1.66$  & $  1.59 \pm   0.15$ & \\ 
       PTF10hgi & 2010-09-05 & $        +69.7$ & Palomar-5m/DBSP & $ 3439- 9849$ & $  1.82$  & $  1.71 \pm   0.14$ & \\ 
       PTF10hgi & 2010-09-10 & $        +74.3$ &  KPNO-4m/RCspec & $ 3319- 8397$ & $  1.84$  & $  1.75 \pm   0.14$ & \\ 
       PTF10hgi & 2010-10-03 & $        +95.2$ &     Keck-I/LRIS & $ 3001-10297$ & $  1.87$  & $  1.76 \pm   0.13$ & \\ 
       PTF10hgi & 2010-11-01 & $       +121.6$ &     Keck-I/LRIS & $ 3218- 7591$ & $  1.93$  & $  1.92 \pm   0.07$ & \\ 
       PTF10hgi & 2011-03-12 & $       +240.9$ &     Keck-I/LRIS & $ 3001-10297$ & $  1.94$  & $  1.93 \pm   0.04$ & \\ 
       PTF10hgi & 2011-06-01 & $       +314.7$ &  Keck-II/DEIMOS & $ 4652- 9625$ & $  2.06$  & $  1.97 \pm   0.05$ & \\ 
       PTF10gvb & 2010-05-06 & $        -13.8$ &     Keck-I/LRIS & $ 3500-10000$ &  \nodata  & $ -0.17 \pm   0.13$ & \\ 
       PTF10gvb & 2010-05-15 & $         -5.6$ &     Keck-I/LRIS & $ 3300-10180$ &  \nodata  & $  1.12 \pm   0.52$ & \\ 
       PTF10gvb & 2010-07-08 & $        +43.6$ &     Keck-I/LRIS & $ 3120-10040$ &  \nodata  & $  1.39 \pm   0.59$ & \\ 
       PTF10nmn & 2010-07-07 & $         -0.2$ &     Keck-I/LRIS & $ 3299-10200$ & $  1.75$  & $  1.80 \pm   0.09$ & \\ 
       PTF10nmn & 2010-07-14 & $         +6.1$ & Palomar-5m/DBSP & $ 3200- 9461$ & $  1.76$  & $  1.33 \pm   0.24$ & \\ 
       PTF10nmn & 2010-07-15 & $         +6.9$ & Palomar-5m/DBSP & $ 3480-10000$ & $  1.78$  & $  1.78 \pm   0.07$ & \\ 
       PTF10nmn & 2010-08-09 & $        +29.2$ & Palomar-5m/DBSP & $ 3001-10044$ & $  1.86$  & $  1.83 \pm   0.06$ & \\ 
       PTF10nmn & 2011-01-28 & $       +182.3$ & Palomar-5m/DBSP & $ 3490- 9750$ & $  1.95$  & $  1.94 \pm   0.04$ & \\ 
       PTF10nmn & 2011-03-04 & $       +213.4$ &     Keck-I/LRIS & $ 3011-10297$ & $  1.96$  & $  1.90 \pm   0.06$ & \\ 
       PTF10nmn & 2011-06-01 & $       +292.6$ &  Keck-II/DEIMOS & $ 4652- 9626$ &  \nodata  & $  1.38 \pm   1.42$ & \\ 
       PTF10nmn & 2011-07-03 & $       +321.1$ &     Keck-I/LRIS & $ 3001-10297$ & $  2.01$  & $  2.00 \pm   0.06$ & \\ 
       PTF10nmn & 2012-02-20 & $       +527.6$ &     Keck-I/LRIS & $ 3081-10271$ & $  2.02$  & $  1.57 \pm   1.01$ & \\ 
       PTF10uhf & 2010-09-08 & $         -4.1$ &  KPNO-4m/RCspec & $ 3499- 8435$ & $  1.13$  & $  0.98 \pm   0.52$ & \\ 
       PTF10uhf & 2010-10-03 & $        +15.3$ &     Keck-I/LRIS & $ 3001-10292$ & $  1.15$  & $  1.36 \pm   0.08$ & \\ 
       PTF10uhf & 2011-07-02 & $       +226.5$ &     Keck-I/LRIS & $ 3001-10076$ &  \nodata  & $  1.38 \pm   0.03$ & \\ 
         SN\,2010hy & 2010-10-14 & $        +21.6$ &     Keck-I/LRIS & $ 3105-10239$ & $  1.43$  & $  1.44 \pm   0.16$ & \\ 
         SN\,2010hy & 2011-03-12 & $       +146.8$ &     Keck-I/LRIS & $ 3001-10296$ & $  1.89$  & $  1.84 \pm   0.08$ & \\ 
       PTF10vqv & 2010-10-11 & $         +6.5$ &  KPNO-4m/RCspec & $ 3840- 8400$ & $ -0.04$  & $  0.88 \pm   0.88$ & \\ 
       PTF10vqv & 2010-10-17 & $        +10.7$ & Palomar-5m/DBSP & $ 3001- 9966$ & $  0.28$  & $  0.47 \pm   0.10$ & \\ 
       PTF10vqv & 2010-11-01 & $        +21.0$ &     Keck-I/LRIS & $ 3219- 7591$ & $  1.53$  & $  1.33 \pm   0.76$ & \\ 
       PTF10vqv & 2010-12-06 & $        +45.1$ & Palomar-5m/DBSP & $ 3500-10000$ &  \nodata  & $  1.04 \pm   0.72$ & \\ 
       PTF10vqv & 2010-12-08 & $        +46.5$ &     Keck-I/LRIS & $ 3299-10000$ & $  1.55$  & $  1.40 \pm   0.19$ & \\ 
       PTF10vqv & 2011-01-31 & $        +83.7$ &     Keck-I/LRIS & $ 3001-10252$ &  \nodata  & $  0.60 \pm   0.27$ & \\ 
       PTF10vqv & 2011-02-01 & $        +84.4$ &     Keck-I/LRIS & $ 3004-10248$ &  \nodata  & $  1.58 \pm   0.28$ & \\ 
      PTF10aagc & 2010-11-04 & $         +3.7$ &  KPNO-4m/RCspec & $ 3329- 8459$ & $ -0.25$  & $ -0.17 \pm   0.15$ & \\ 
      PTF10aagc & 2010-11-05 & $         +4.6$ &     Keck-I/LRIS & $ 3299- 7633$ & $ -0.23$  & $ -0.56 \pm   0.22$ & \\ 
      PTF10aagc & 2010-11-07 & $         +6.2$ &  Keck-II/DEIMOS & $ 4480- 9631$ & $ -0.20$  & $  0.04 \pm   0.64$ & \\ 
      PTF10aagc & 2010-11-18 & $        +15.3$ &     TNG/DOLORES & $ 3303- 8097$ & $  1.51$  & $  1.41 \pm   0.06$ & \\ 
      PTF10aagc & 2010-11-30 & $        +25.3$ & Palomar-5m/DBSP & $ 3001- 9299$ &  \nodata  & $  1.60 \pm   0.23$ & \\ 
      PTF10aagc & 2010-12-08 & $        +31.9$ &     Keck-I/LRIS & $ 3299-10000$ & $  1.52$  & $  0.99 \pm   0.54$ & \\ 
      PTF10aagc & 2011-02-01 & $        +77.5$ &     Keck-I/LRIS & $ 3004-10244$ & $  1.56$  & $  1.55 \pm   0.07$ & \\ 
      PTF10aagc & 2011-03-26 & $       +121.4$ &     Keck-I/LRIS & $ 3013-10256$ &  \nodata  & $  1.45 \pm   0.13$ & \\ 
         SN\,2011ke & 2011-05-11 & $         +7.6$ &  KPNO-4m/RCspec & $ 3473- 8357$ & $  0.20$  & $  0.71 \pm   0.54$ & \\ 
         SN\,2011ke & 2011-05-13 & $         +9.3$ &  KPNO-4m/RCspec & $ 3449- 8450$ & $  0.23$  & $  0.71 \pm   0.53$ & \\ 
         SN\,2011ke & 2011-05-25 & $        +19.8$ & Palomar-5m/DBSP & $ 3001-10256$ & $  1.28$  & $  1.34 \pm   0.11$ & \\ 
         SN\,2011ke & 2011-05-30 & $        +24.2$ &        HST/STIS & $ 1580- 3150$ &  \nodata  & $ -0.09 \pm   1.34$ & \\ 
         SN\,2011ke & 2011-06-01 & $        +25.9$ &  Keck-II/DEIMOS & $ 4530- 9625$ & $  1.38$  & $  1.39 \pm   0.06$ & \\ 
         SN\,2011ke & 2011-06-07 & $        +31.2$ &  KPNO-4m/RCspec & $ 3390- 8459$ & $  1.40$  & $  1.39 \pm   0.06$ & \\ 
         SN\,2011ke & 2011-07-02 & $        +53.0$ &     Keck-I/LRIS & $ 3001-10077$ & $  1.59$  & $  1.49 \pm   0.19$ & \\ 
         SN\,2011ke & 2012-01-26 & $       +235.0$ &     Keck-I/LRIS & $ 3001-10293$ &  \nodata  & $ -0.56 \pm   0.48$ & \\ 
         SN\,2011ke & 2012-03-23 & $       +284.9$ &     Keck-I/LRIS & $ 3181- 9325$ &  \nodata  & $  0.61 \pm   0.69$ & \\ 
       PTF11mnb & 2011-10-07 & $        -13.4$ &      UH88/SNIFS & $ 3301- 9701$ &  \nodata  & $  1.62 \pm   0.11$ & \\ 
       PTF11mnb & 2011-10-30 & $         +8.3$ & Palomar-5m/DBSP & $ 3001-10293$ &  \nodata  & $  1.72 \pm   0.13$ & \\ 
       PTF11mnb & 2011-11-26 & $        +33.7$ &     Keck-I/LRIS & $ 3001-10296$ &  \nodata  & $  1.78 \pm   0.11$ & \\ 
       PTF11mnb & 2011-12-21 & $        +57.3$ & Palomar-5m/DBSP & $ 3001-10296$ &  \nodata  & $  1.88 \pm   0.08$ & \\ 
       PTF11mnb & 2012-01-27 & $        +92.2$ &  KPNO-4m/RCspec & $ 3574- 8137$ &  \nodata  & $  1.84 \pm   0.14$ & \\ 
       PTF11rks & 2011-12-27 & $        -11.9$ & Palomar-5m/DBSP & $ 3175-10296$ & $ -0.92$  & $ -0.73 \pm   0.32$ & \\ 
       PTF11rks & 2011-12-31 & $         -8.5$ &     Keck-I/LRIS & $ 3001-10297$ & $ -0.88$  & $ -0.60 \pm   0.16$ & \\ 
       PTF11rks & 2012-01-17 & $         +5.8$ &        HST/WFC3 & $ 1903- 5185$ &  \nodata  & $ -0.09 \pm   1.34$ & \\ 
       PTF11rks & 2012-01-18 & $         +6.6$ & Palomar-5m/DBSP & $ 3022-10296$ & $  0.88$  & $  1.12 \pm   0.25$ & \\ 
       PTF11rks & 2012-02-01 & $        +18.3$ & Palomar-5m/DBSP & $ 3001-10295$ &  \nodata  & $  1.31 \pm   0.10$ & \\ 
       PTF11rks & 2012-03-01 & $        +42.7$ &        WHT/ISIS & $ 5501- 9499$ &  \nodata  & $  1.32 \pm   0.09$ & \\ 
       PTF11rks & 2012-07-15 & $       +156.7$ &     Keck-I/LRIS & $ 3002- 9443$ &  \nodata  & $  0.88 \pm   1.05$ & \\ 
       PTF11hrq & 2011-12-27 & $       +159.4$ & Palomar-5m/DBSP & $ 3178-10295$ & $  1.97$  & $  1.93 \pm   0.05$ & \\ 
       PTF11hrq & 2012-07-15 & $       +349.6$ &     Keck-I/LRIS & $ 3003- 9440$ & $  1.99$  & $  1.55 \pm   0.99$ & \\ 
       PTF11hrq & 2012-12-09 & $       +488.6$ &  Keck-II/DEIMOS & $ 4907- 9300$ & $  2.03$  & $  1.45 \pm   0.73$ & \\ 
       PTF11hrq & 2012-12-12 & $       +491.5$ &     Keck-I/LRIS & $ 3002-10264$ & $  2.04$  & $  1.54 \pm   0.99$ & \\ 
       PTF12dam & 2012-05-20 & $        -23.8$ &    Lick-3m/Kast & $ 3465- 5558$ & $ -0.58$  & $ -0.62 \pm   0.22$ & \\ 
       PTF12dam & 2012-05-21 & $        -22.9$ &    Lick-3m/Kast & $ 3447-10239$ & $ -0.55$  & $ -0.47 \pm   0.22$ & \\ 
       PTF12dam & 2012-05-22 & $        -22.0$ &     Keck-I/LRIS & $ 3003-10247$ & $ -0.52$  & $ -0.54 \pm   0.15$ & \\ 
       PTF12dam & 2012-05-25 & $        -19.3$ &        WHT/ISIS & $ 3500- 9498$ & $ -0.41$  & $ -0.66 \pm   0.27$ & \\ 
       PTF12dam & 2012-05-26 & $        -18.4$ &        HST/WFC3 & $ 1903- 5185$ & $ -0.36$  & $ -0.74 \pm   0.20$ & \\ 
       PTF12dam & 2012-06-14 & $         -1.2$ &    Lick-3m/Kast & $ 3499- 9999$ & $ -0.02$  & $  1.71 \pm   0.27$ & \\ 
       PTF12dam & 2012-06-18 & $         +2.4$ & Palomar-5m/DBSP & $ 3001-10295$ & $  0.00$  & $  0.08 \pm   0.34$ & \\ 
       PTF12dam & 2012-07-06 & $        +18.7$ &        HST/WFC3 & $ 1903- 5198$ & $  0.60$  & $  0.62 \pm   0.41$ & \\ 
       PTF12dam & 2012-07-11 & $        +23.2$ &    Lick-3m/Kast & $ 3529-10221$ & $  0.64$  & $  0.80 \pm   0.52$ & \\ 
       PTF12dam & 2012-07-12 & $        +24.1$ &        HST/WFC3 & $ 1903- 5185$ & $  1.04$  & $  0.73 \pm   0.45$ & \\ 
       PTF12dam & 2012-08-20 & $        +59.3$ &        WHT/ISIS & $ 3500- 9499$ & $  1.71$  & $  1.55 \pm   0.29$ & \\ 
       PTF12dam & 2012-12-05 & $       +155.9$ & Palomar-5m/DBSP & $ 3001- 9298$ & $  1.81$  & $  1.86 \pm   0.08$ & \\ 
       PTF12dam & 2013-04-09 & $       +268.8$ &     Keck-I/LRIS & $ 3060-10288$ & $  1.98$  & $  1.97 \pm   0.04$ & \\ 
       PTF12dam & 2013-06-09 & $       +323.9$ &     Keck-I/LRIS & $ 3002-10266$ & $  2.05$  & $  1.98 \pm   0.04$ & \\ 
       PTF12dam & 2014-04-29 & $       +616.4$ &     Keck-I/LRIS & $ 3068- 5677$ &  \nodata  & $  1.16 \pm   1.15$ & \\ 
       PTF12gty & 2012-07-22 & $        -11.4$ &    Lick-3m/Kast & $ 3499-10000$ & $  1.70$  & $  1.73 \pm   0.05$ & \\ 
       PTF12gty & 2012-07-27 & $         -7.1$ & Palomar-5m/DBSP & $ 3002-10295$ & $  1.72$  & $  1.61 \pm   0.29$ & \\ 
       PTF12hni & 2012-08-09 & $         -6.6$ &    Lick-3m/Kast & $ 3500- 9999$ &  \nodata  & $  1.52 \pm   0.11$ & \\ 
       PTF12hni & 2012-08-19 & $         +2.5$ &     Keck-I/LRIS & $ 3408-10250$ &  \nodata  & $  1.63 \pm   0.12$ & \\ 
       PTF12hni & 2013-07-12 & $       +298.3$ &  Keck-II/DEIMOS & $ 4905-10119$ &  \nodata  & $  1.91 \pm   0.23$ & \\ 
       PTF12mxx & 2012-12-18 & $         -8.4$ &     Keck-I/LRIS & $ 3299-10199$ & $ -0.71$  & $ -0.53 \pm   0.14$ & \\ 
       PTF12mxx & 2013-01-08 & $         +7.4$ &  Keck-II/DEIMOS & $ 4499- 9635$ & $  0.14$  & $  0.22 \pm   0.08$ & \\ 
\hline
\end{longtable*}

%% file: groups.table.tex
\begin{table}
\centering
\caption{SLSN-I Spectroscopic Groups}
\label{table:groups}
\begin{tabular}{|l|c|c|l|}
  \hline
  \multicolumn{1}{|c|}{} &
  \multicolumn{2}{|c|}{\bfseries In Top 5 With} &
  \multicolumn{1}{|c|}{\bfseries Feature} \\
  \multicolumn{1}{|c}{\bfseries Name} &
  \multicolumn{1}{|c}{\bfseries SN~2011ke} &
  \multicolumn{1}{|c|}{\bfseries PTF12dam} &
  \multicolumn{1}{|c|}{\bfseries Group} \\
  \hline

      SN\,2005ap &   0 &   2 &   12dam-like     \\
     SCP06F6 &   0 &   0 &   12dam-like$^a$ \\
      SN\,2006oz &   0 &   1 &   12dam-like     \\
      SN\,2007bi &   3 &   9 &   12dam-like     \\
      PTF09q &   0 &   0 &    11ke-like$^a$ \\
     PTF09as &   0 &   0 &    11ke-like$^a$ \\
    PTF09atu &  11 &  26 &   12dam-like     \\
    PTF09cwl &   8 &  20 &   12dam-like     \\
    PTF09cnd &   8 &  34 &   12dam-like     \\
    PTF10bfz &  10 &   3 &    11ke-like     \\
    PTF10bjp &   3 &   7 &   12dam-like     \\
      SN\,2010gx &  16 &  14 &    11ke-like     \\
    PTF10hgi &   5 &   3 &    11ke-like     \\
    PTF10gvb &   0 &   0 &    11ke-like$^a$ \\
    PTF10nmn &   5 &  16 &   12dam-like     \\
    PS1-10ky &   0 &   0 &   12dam-like$^a$ \\
    PTF10uhf &   6 &   4 &    11ke-like     \\
    PTF10vqv &   1 &   4 &   12dam-like     \\
      SN\,2010hy &   8 &   8 &    11ke-like     \\
   PTF10aagc &   4 &   5 &    11ke-like     \\
   PS1-10awh &   0 &   0 &   12dam-like$^a$ \\
      SN\,2011ke &  56 &  10 &    11ke-like     \\
    PTF11mnb &   0 &   0 &   12dam-like$^a$ \\
    PTF11hrq &   3 &   9 &   12dam-like     \\
      SN\,2011kf &   4 &   0 &    11ke-like     \\
    PTF11rks &   6 &   1 &    11ke-like     \\
      SN\,2012il &   7 &   4 &    11ke-like     \\
    PTF12dam &  10 &  81 &   12dam-like     \\
    PTF12gty &   1 &  14 &   12dam-like     \\
    LSQ12dlf &  20 &   6 &    11ke-like     \\
   SSS120810 &  17 &   4 &    11ke-like     \\
    PTF12hni &   0 &   0 &   12dam-like$^a$ \\
    PTF12mxx &   5 &   7 &    11ke-like     \\
    PTF13ajg &   0 &   0 &   12dam-like$^a$ \\
      SN\,2013dg &   0 &   2 &   12dam-like     \\

\hline
\end{tabular}

$^a$Top spectral matches do not include PTF12dam or SN\,2011ke, or the
redshift is too high to match against these objects. In these cases
the subgroup is assigned based on the subgroups of the best-matching
templates.

\end{table}

%% file: PTF12dam.fe2.vels.tex
\begin{table*}
\centering
\caption{PTF12dam \ion{Fe}{2} Absorption Minimum Velocities}
\label{table:PTF12dam_fe2}
\begin{tabular}{rccccc}
  \hline
  LC Phase (d) & \ion{Fe}{2} $\lambda$4923 & \ion{Fe}{2} $\lambda$5018 & \ion{Fe}{2} $\lambda$5169 & Model & Cross-Correlation\\
  \hline

$ -23.8$ & $-11.16$  & $-11.64$  & $-11.25$  & \nodata   & $-12.15$ $\pm 0.04$ \\
$ -22.9$ & \nodata   & \nodata   & \nodata   & $-13.10$  & $-12.15$ $\pm 0.43$ \\
$ -22.0$ & $-11.05$  & \nodata   & \nodata   & $-12.54$  & $-12.78$ $\pm 0.10$ \\
$ -21.1$ & $-11.80$  & $-12.16$  & \nodata   & $-11.91$  & $-11.85$ $\pm 0.65$ \\
$ -20.2$ & \nodata   & \nodata   & \nodata   & $-11.69$  & $-11.88$ $\pm 0.70$ \\
$ -19.3$ & \nodata   & \nodata   & \nodata   & \nodata   & $-11.14$ $\pm 0.26$ \\
$ -19.3$ & \nodata   & \nodata   & \nodata   & $-11.49$  & $-12.03$ $\pm 0.07$ \\
$ -18.4$ & \nodata   & \nodata   & \nodata   & \nodata   & $-12.93$ $\pm 0.40$ \\
$ -12.9$ & \nodata   & $-12.28$  & $-11.83$  & $-12.10$  & $-11.55$ $\pm 0.18$ \\
$  -6.6$ & $-11.06$  & $-11.37$  & $-10.48$  & $-11.27$  & $-11.02$ $\pm 0.19$ \\
$  -1.2$ & \nodata   & \nodata   & \nodata   & $-12.03$  & $-12.63$ $\pm 0.22$ \\
$  +2.4$ & \nodata   & $-11.67$  & $-11.76$  & $-11.76$  & $-11.76$ $\pm 0.11$ \\
$  +5.1$ & $-11.04$  & $-11.34$  & \nodata   & $-10.79$  & $-11.05$ $\pm 0.31$ \\
$  +8.7$ & \nodata   & \nodata   & \nodata   & \nodata   & $-11.11$ $\pm 0.13$ \\
$ +12.3$ & \nodata   & \nodata   & \nodata   & $ -9.96$  & $-10.57$ $\pm 0.38$ \\
$ +18.7$ & \nodata   & \nodata   & \nodata   & \nodata   & \nodata             \\
$ +23.2$ & \nodata   & \nodata   & \nodata   & \nodata   & $-11.55$ $\pm 0.31$ \\
$ +24.1$ & \nodata   & \nodata   & \nodata   & \nodata   & \nodata             \\
$ +28.6$ & \nodata   & $-11.34$  & \nodata   & $-10.88$  & $ -9.94$ $\pm 0.28$ \\
$ +49.4$ & \nodata   & $-11.29$  & \nodata   & $-10.02$  & \nodata             \\
$ +59.3$ & \nodata   & \nodata   & \nodata   & \nodata   & \nodata             \\
$ +59.3$ & $-10.70$  & $-10.21$  & \nodata   & \nodata   & \nodata             \\

\hline
\end{tabular}

All velocities are in units of $10^3$\,km\,s$^{-1}$. \ion{Fe}{2} 
$\lambda\lambda$4923, 5018, 5169 velocities are measured from second-order 
polynomial fits to
the local absorption minimum. Model velocities are from a simultaneous
fit to a simple \ion{Fe}{2} absorption model (see text for
details). Cross-correlation velocities are relative to the $+2.4$\,d
spectrum, which is assumed to have a blueshift of 11,760\,km\,s$^{-1}$
based on the model fit. Uncertainties in the cross-correlation
velocities are from Monte-Carlo tests and represent statistical
uncertainties only.

\end{table*}

%% file: PTF12dam.o2.vels.tex
\begin{table*}
\centering
\caption{PTF12dam \ion{O}{2} Absorption Minimum Velocities}
\label{table:PTF12dam_o2}
\begin{tabular}{rccccccc}
  \hline
  LC Phase (d) & \ion{O}{2} E & \ion{O}{2} D & \ion{O}{2} C & \ion{O}{2} B & \ion{O}{2} A & Model & Cross-Correlation\\
  \hline

$ -23.8$ & $-10.72$  & $-11.01$  & $-10.76$  & $-11.97$  & $-11.37$  & $-11.10$  & $-11.10$ $\pm 0.02$ \\
$ -22.9$ & $-10.11$  & $-10.41$  & $-10.77$  & $-11.91$  & $-11.53$  & $-10.96$  & $-10.89$ $\pm 0.22$ \\
$ -22.0$ & $-10.39$  & $-10.93$  & $-10.36$  & $-11.66$  & $-11.38$  & $-10.83$  & $-10.86$ $\pm 0.03$ \\
$ -21.1$ & $-10.49$  & \nodata    & $-9.96$  & $-11.68$  & $-11.30$  & $-10.37$  & $-10.83$ $\pm 0.28$ \\
$ -20.2$  & $-9.73$  & $-10.40$  & \nodata   & $-11.42$  & $-11.08$  & $-10.31$  & $-10.68$ $\pm 0.31$ \\
$ -19.3$  & $-9.69$  & $-11.21$  & $-10.14$  & $-11.49$  & $-11.17$  & $-10.76$  & $-10.62$ $\pm 0.04$ \\
$ -19.3$ & $-11.07$  & $-10.85$  & \nodata   & $-11.72$  & $-11.33$  & $-10.50$  & $-10.35$ $\pm 0.09$ \\
$ -18.4$ & $-12.20$  & $-10.45$   & $-9.88$   & $-9.71$   & $-9.10$  & $-10.04$  & $-10.20$ $\pm 0.20$ \\
$ -12.9$  & $-9.54$  & $-10.49$   & $-8.53$  & $-10.87$  & $-10.92$  & $-10.21$  & $-10.08$ $\pm 0.14$ \\
$  -6.6$  & $-8.30$  & $-10.41$   & $-8.00$   & $-7.73$  & \nodata    & $-8.04$   & $-8.68$ $\pm 0.21$ \\
$  -1.2$  & $-7.07$  & $-10.29$   & $-8.38$   & $-7.52$   & $-7.34$   & $-6.98$   & $-8.27$ $\pm 0.29$ \\
$  +2.4$ & \nodata    & $-7.97$   & $-7.60$   & $-7.48$   & $-6.88$   & $-6.84$   & $-7.76$ $\pm 0.17$ \\
$  +5.1$  & $-8.59$  & \nodata    & $-7.52$  & \nodata   & \nodata    & $-5.88$   & $-8.59$ $\pm 0.41$ \\
$  +8.7$  & $-6.44$  & \nodata    & $-7.45$  & \nodata    & $-6.77$  & \nodata    & $-7.17$ $\pm 0.10$ \\
$ +12.3$  & $-8.62$  & \nodata    & $-7.49$  & \nodata    & $-6.11$  & \nodata    & $-8.74$ $\pm 1.26$ \\

\hline
\end{tabular}

All velocities are in units of $10^3$\,km\,s$^{-1}$. Velocities for \ion{O}{2}
E, D, C, B, and A are measured from second-order polynomial fits to
the local absorption minimum assuming rest wavelengths of
3737.59\,\AA, 3959.83\,\AA, 4115.17\,\AA, 4357.97\,\AA, and
4650.71\,\AA, respectively. Model velocities are from a simple
absorption model to the \ion{O}{2} B feature (see text for
details). Cross-correlation velocities are relative to the $-23.8$\,d
spectrum, which is assumed to have a blueshift of 11,060\,km\,s$^{-1}$
based on the model fit. Uncertainties in the cross-correlation
velocities are from Monte-Carlo tests and represent statistical
uncertainties only.

\end{table*}

%% file: paper.bbl
\begin{thebibliography}{}
\expandafter\ifx\csname natexlab\endcsname\relax\def\natexlab#1{#1}\fi

\bibitem[{{Arcavi} {et~al.}(2010){Arcavi}, {Gal-Yam}, {Kasliwal}, {Quimby},
  {Ofek}, {Kulkarni}, {Nugent}, {Cenko}, {Bloom}, {Sullivan}, {Howell},
  {Poznanski}, {Filippenko}, {Law}, {Hook}, {J{\"o}nsson}, {Blake}, {Cooke},
  {Dekany}, {Rahmer}, {Hale}, {Smith}, {Zolkower}, {Velur}, {Walters},
  {Henning}, {Bui}, {McKenna}, \& {Jacobsen}}]{arcavi2010}
{Arcavi}, I., {Gal-Yam}, A., {Kasliwal}, M.~M., {et~al.} 2010, \apj, 721, 777

\bibitem[{{Barbary} {et~al.}(2009){Barbary}, {Dawson}, {Tokita}, {Aldering},
  {Amanullah}, {Connolly}, {Doi}, {Faccioli}, {Fadeyev}, {Fruchter},
  {Goldhaber}, {Goobar}, {Gude}, {Huang}, {Ihara}, {Konishi}, {Kowalski},
  {Lidman}, {Meyers}, {Morokuma}, {Nugent}, {Perlmutter}, {Rubin}, {Schlegel},
  {Spadafora}, {Suzuki}, {Swift}, {Takanashi}, {Thomas}, \&
  {Yasuda}}]{barbary2009}
{Barbary}, K., {Dawson}, K.~S., {Tokita}, K., {et~al.} 2009, \apj, 690, 1358

\bibitem[{{Barbon} {et~al.}(1995){Barbon}, {Benetti}, {Cappellaro}, {Patat},
  {Turatto}, \& {Iijima}}]{barbon1995}
{Barbon}, R., {Benetti}, S., {Cappellaro}, E., {et~al.} 1995, \aaps, 110, 513

\bibitem[{{Barkat} {et~al.}(1967){Barkat}, {Rakavy}, \& {Sack}}]{barkat1967}
{Barkat}, Z., {Rakavy}, G., \& {Sack}, N. 1967, Physical Review Letters, 18,
  379

\bibitem[{{Bayless} {et~al.}(2013){Bayless}, {Pritchard}, {Roming}, {Kuin},
  {Brown}, {Botticella}, {Dall'Ora}, {Frey}, {Even}, {Fryer}, {Maund}, \&
  {Fraser}}]{bayless2013}
{Bayless}, A.~J., {Pritchard}, T.~A., {Roming}, P.~W.~A., {et~al.} 2013, \apjl,
  764, L13

\bibitem[{{Benetti} {et~al.}(2014){Benetti}, {Nicholl}, {Cappellaro},
  {Pastorello}, {Smartt}, {Elias-Rosa}, {Drake}, {Tomasella}, {Turatto},
  {Harutyunyan}, {Taubenberger}, {Hachinger}, {Morales-Garoffolo}, {Chen},
  {Djorgovski}, {Fraser}, {Gal-Yam}, {Inserra}, {Mazzali}, {Pumo}, {Sollerman},
  {Valenti}, {Young}, {Dennefeld}, {Le Guillou}, {Fleury}, \&
  {L{\'e}get}}]{benetti2014}
{Benetti}, S., {Nicholl}, M., {Cappellaro}, E., {et~al.} 2014, \mnras, 441, 289

\bibitem[{{Berger} {et~al.}(2012){Berger}, {Chornock}, {Lunnan}, {Foley},
  {Czekala}, {Rest}, {Leibler}, {Soderberg}, {Roth}, {Narayan}, {Huber},
  {Milisavljevic}, {Sanders}, {Drout}, {Margutti}, {Kirshner}, {Marion},
  {Challis}, {Riess}, {Smartt}, {Burgett}, {Hodapp}, {Heasley}, {Kaiser},
  {Kudritzki}, {Magnier}, {McCrum}, {Price}, {Smith}, {Tonry}, \&
  {Wainscoat}}]{berger2012}
{Berger}, E., {Chornock}, R., {Lunnan}, R., {et~al.} 2012, \apjl, 755, L29

\bibitem[{{Blondin} \& {Tonry}(2007)}]{blondin2007}
{Blondin}, S., \& {Tonry}, J.~L. 2007, \apj, 666, 1024

\bibitem[{{Blondin} {et~al.}(2012){Blondin}, {Matheson}, {Kirshner}, {Mandel},
  {Berlind}, {Calkins}, {Challis}, {Garnavich}, {Jha}, {Modjaz}, {Riess}, \&
  {Schmidt}}]{blondin2012}
{Blondin}, S., {Matheson}, T., {Kirshner}, R.~P., {et~al.} 2012, \aj, 143, 126

\bibitem[{{Bose} {et~al.}(2017){Bose}, {Dong}, {Pastorello}, {Filippenko},
  {Kochanek}, {Mauerhan}, {Romero-Canizales}, {Brink}, {Chen}, {Prieto},
  {Post}, {Ashall}, {Grupe}, {Tomasella}, {Benetti}, {Shappee}, {Stanek},
  {Cai}, {Falco}, {Lundqvist}, {Mattila}, {Mutel}, {Ochner}, {Pooley},
  {Stritzinger}, {Villanueva}, {Zheng}, {Beswick}, {Brown}, {Cappellaro},
  {Davis}, {de Jaeger}, {Elias-Rosa}, {Gall}, {Gaudi}, {Herczeg}, {Hestenes},
  {Holoien}, {Hosseinzadeh}, {Hsiao}, {Hu}, {Jaejin}, {Jeffers}, {Koff},
  {Kumar}, {Kurtenkov}, {Lau}, {Prentice}, {Rudy}, {Shahbandeh}, {Somero},
  {Stassun}, {Thompson}, {Valenti}, {Woo}, \& {Yunus}}]{bose2017}
{Bose}, S., {Dong}, S., {Pastorello}, A., {et~al.} 2017, ArXiv e-prints,
  arXiv:1708.00864

\bibitem[{{Branch} {et~al.}(2002){Branch}, {Benetti}, {Kasen}, {Baron},
  {Jeffery}, {Hatano}, {Stathakis}, {Filippenko}, {Matheson}, {Pastorello},
  {Altavilla}, {Cappellaro}, {Rizzi}, {Turatto}, {Li}, {Leonard}, \&
  {Shields}}]{branch2002}
{Branch}, D., {Benetti}, S., {Kasen}, D., {et~al.} 2002, \apj, 566, 1005

\bibitem[{{Branch} {et~al.}(2008){Branch}, {Jeffery}, {Parrent}, {Baron},
  {Troxel}, {Stanishev}, {Keithley}, {Harrison}, \& {Bruner}}]{branch2008}
{Branch}, D., {Jeffery}, D.~J., {Parrent}, J., {et~al.} 2008, \pasp, 120, 135

\bibitem[{{Bufano} {et~al.}(2009){Bufano}, {Immler}, {Turatto}, {Landsman},
  {Brown}, {Benetti}, {Cappellaro}, {Holland}, {Mazzali}, {Milne}, {Panagia},
  {Pian}, {Roming}, {Zampieri}, {Breeveld}, \& {Gehrels}}]{bufano2009}
{Bufano}, F., {Immler}, S., {Turatto}, M., {et~al.} 2009, \apj, 700, 1456

\bibitem[{{Cardelli} {et~al.}(1989){Cardelli}, {Clayton}, \&
  {Mathis}}]{cardelli1989}
{Cardelli}, J.~A., {Clayton}, G.~C., \& {Mathis}, J.~S. 1989, \apj, 345, 245

\bibitem[{{Chatzopoulos} \& {Wheeler}(2012)}]{chatzopoulos_wheeler2012}
{Chatzopoulos}, E., \& {Wheeler}, J.~C. 2012, \apj, 748, 42

\bibitem[{{Chatzopoulos} {et~al.}(2013){Chatzopoulos}, {Wheeler}, {Vinko},
  {Horvath}, \& {Nagy}}]{chatzopoulos2013}
{Chatzopoulos}, E., {Wheeler}, J.~C., {Vinko}, J., {Horvath}, Z.~L., \& {Nagy},
  A. 2013, \apj, 773, 76

\bibitem[{{Chatzopoulos} {et~al.}(2011){Chatzopoulos}, {Wheeler}, {Vinko},
  {Quimby}, {Robinson}, {Miller}, {Foley}, {Perley}, {Yuan}, {Akerlof}, \&
  {Bloom}}]{chatzopoulos2011}
{Chatzopoulos}, E., {Wheeler}, J.~C., {Vinko}, J., {et~al.} 2011, \apj, 729,
  143

\bibitem[{{Chen} {et~al.}(2013){Chen}, {Smartt}, {Bresolin}, {Pastorello},
  {Kudritzki}, {Kotak}, {McCrum}, {Fraser}, \& {Valenti}}]{chen2013}
{Chen}, T.-W., {Smartt}, S.~J., {Bresolin}, F., {et~al.} 2013, \apjl, 763, L28

\bibitem[{{Chen} {et~al.}(2017){Chen}, {Nicholl}, {Smartt}, {Mazzali}, {Yates},
  {Moriya}, {Inserra}, {Langer}, {Kr{\"u}hler}, {Pan}, {Kotak}, {Galbany},
  {Schady}, {Wiseman}, {Greiner}, {Schulze}, {Man}, {Jerkstrand}, {Smith},
  {Dennefeld}, {Baltay}, {Bolmer}, {Kankare}, {Knust}, {Maguire}, {Rabinowitz},
  {Rostami}, {Sullivan}, \& {Young}}]{chen2017}
{Chen}, T.-W., {Nicholl}, M., {Smartt}, S.~J., {et~al.} 2017, \aap, 602, A9

\bibitem[{{Chevalier} \& {Irwin}(2011)}]{chevalier_irwin2011}
{Chevalier}, R.~A., \& {Irwin}, C.~M. 2011, \apjl, 729, L6

\bibitem[{{Chomiuk} {et~al.}(2011){Chomiuk}, {Chornock}, {Soderberg}, {Berger},
  {Chevalier}, {Foley}, {Huber}, {Narayan}, {Rest}, {Gezari}, {Kirshner},
  {Riess}, {Rodney}, {Smartt}, {Stubbs}, {Tonry}, {Wood-Vasey}, {Burgett},
  {Chambers}, {Czekala}, {Flewelling}, {Forster}, {Kaiser}, {Kudritzki},
  {Magnier}, {Martin}, {Morgan}, {Neill}, {Price}, {Roth}, {Sanders}, \&
  {Wainscoat}}]{chomiuk2011}
{Chomiuk}, L., {Chornock}, R., {Soderberg}, A.~M., {et~al.} 2011, \apj, 743,
  114

\bibitem[{{Chornock} {et~al.}(2013){Chornock}, {Berger}, {Rest},
  {Milisavljevic}, {Lunnan}, {Foley}, {Soderberg}, {Smartt}, {Burgasser},
  {Challis}, {Chomiuk}, {Czekala}, {Drout}, {Fong}, {Huber}, {Kirshner},
  {Leibler}, {McLeod}, {Marion}, {Narayan}, {Riess}, {Roth}, {Sanders},
  {Scolnic}, {Smith}, {Stubbs}, {Tonry}, {Valenti}, {Burgett}, {Chambers},
  {Hodapp}, {Kaiser}, {Kudritzki}, {Magnier}, \& {Price}}]{chornock2013}
{Chornock}, R., {Berger}, E., {Rest}, A., {et~al.} 2013, \apj, 767, 162

\bibitem[{{Cikota} {et~al.}(2017){Cikota}, {De Cia}, {Schulze}, {Vreeswijk},
  {Leloudas}, {Gal-Yam}, {Perley}, {Cikota}, {Kim}, {Patat}, {Lunnan},
  {Quimby}, {Yaron}, {Yan}, \& {Mazzali}}]{cikota2017}
{Cikota}, A., {De Cia}, A., {Schulze}, S., {et~al.} 2017, \mnras, 469, 4705

\bibitem[{{Crowther} {et~al.}(2016){Crowther}, {Caballero-Nieves}, {Bostroem},
  {Ma{\'{\i}}z Apell{\'a}niz}, {Schneider}, {Walborn}, {Angus}, {Brott},
  {Bonanos}, {de Koter}, {de Mink}, {Evans}, {Gr{\"a}fener}, {Herrero},
  {Howarth}, {Langer}, {Lennon}, {Puls}, {Sana}, \& {Vink}}]{crowther2016}
{Crowther}, P.~A., {Caballero-Nieves}, S.~M., {Bostroem}, K.~A., {et~al.} 2016,
  \mnras, 458, 624

\bibitem[{{Dall'Ora} {et~al.}(2014){Dall'Ora}, {Botticella}, {Pumo},
  {Zampieri}, {Tomasella}, {Pignata}, {Bayless}, {Pritchard}, {Taubenberger},
  {Kotak}, {Inserra}, {Della Valle}, {Cappellaro}, {Benetti}, {Benitez},
  {Bufano}, {Elias-Rosa}, {Fraser}, {Haislip}, {Harutyunyan}, {Howell},
  {Hsiao}, {Iijima}, {Kankare}, {Kuin}, {Maund}, {Morales-Garoffolo},
  {Morrell}, {Munari}, {Ochner}, {Pastorello}, {Patat}, {Phillips}, {Reichart},
  {Roming}, {Siviero}, {Smartt}, {Sollerman}, {Taddia}, {Valenti}, \&
  {Wright}}]{dallora2014}
{Dall'Ora}, M., {Botticella}, M.~T., {Pumo}, M.~L., {et~al.} 2014, \apj, 787,
  139

\bibitem[{{De Cia} {et~al.}(2017){De Cia}, {Gal-Yam}, {Rubin}, {Leloudas},
  {Vreeswijk}, {Perley}, {Quimby}, {Yan}, {Sullivan}, {Fl{\"o}rs}, {Sollerman},
  {Bersier}, {Cenko}, {Gal-Yam}, {Maguire}, {Ofek}, {Prentice}, {Schulze},
  {Spyromilio}, {Valenti}, {Arcavi}, {Corsi}, {Howell}, {Mazzali}, {Kasliwal},
  {Taddia}, \& {Yaron}}]{decia2017}
{De Cia}, A., {Gal-Yam}, A., {Rubin}, A., {et~al.} 2017, ArXiv e-prints,
  arXiv:1708.01623

\bibitem[{{Dessart} \& {Hillier}(2005)}]{dessart_hillier2005}
{Dessart}, L., \& {Hillier}, D.~J. 2005, \aap, 439, 671

\bibitem[{{Dessart} {et~al.}(2012){Dessart}, {Hillier}, {Waldman}, {Livne}, \&
  {Blondin}}]{dessart2012}
{Dessart}, L., {Hillier}, D.~J., {Waldman}, R., {Livne}, E., \& {Blondin}, S.
  2012, \mnras, 426, L76

\bibitem[{{Drake} {et~al.}(2009{\natexlab{a}}){Drake}, {Mahabal}, {Djorgovski},
  {Graham}, {Williams}, {Catelan}, {Beshore}, {Larson}, {Gibbs}, {Kowalski}, \&
  {Christensen}}]{drake2009_atel1}
{Drake}, A.~J., {Mahabal}, A.~A., {Djorgovski}, S.~G., {et~al.}
  2009{\natexlab{a}}, The Astronomer's Telegram, 1980

\bibitem[{{Drake} {et~al.}(2009{\natexlab{b}}){Drake}, {Mahabal}, {Djorgovski},
  {Williams}, {Graham}, {Hsiao}, {Graham}, {Pritchet}, {Balam}, {Moskvitin},
  {Catelan}, {Beshore}, {Larson}, \& {Christensen}}]{drake2009_cbet1}
{Drake}, A.~J., {Mahabal}, A., {Djorgovski}, S.~G., {et~al.}
  2009{\natexlab{b}}, Central Bureau Electronic Telegrams, 1752

\bibitem[{{Fabricant} {et~al.}(1998){Fabricant}, {Cheimets}, {Caldwell}, \&
  {Geary}}]{fabricant1998}
{Fabricant}, D., {Cheimets}, P., {Caldwell}, N., \& {Geary}, J. 1998, \pasp,
  110, 79

\bibitem[{{Fassia} {et~al.}(2001){Fassia}, {Meikle}, {Chugai}, {Geballe},
  {Lundqvist}, {Walton}, {Pollacco}, {Veilleux}, {Wright}, {Pettini}, {Kerr},
  {Puchnarewicz}, {Puxley}, {Irwin}, {Packham}, {Smartt}, \&
  {Harmer}}]{fassia2001}
{Fassia}, A., {Meikle}, W.~P.~S., {Chugai}, N., {et~al.} 2001, \mnras, 325, 907

\bibitem[{{Filippenko}(1982)}]{filippenko1982}
{Filippenko}, A.~V. 1982, \pasp, 94, 715

\bibitem[{{Filippenko}(1997)}]{filippenko1997}
---. 1997, \araa, 35, 309

\bibitem[{{Filippenko} {et~al.}(2001){Filippenko}, {Li}, {Treffers}, \&
  {Modjaz}}]{filippenko2001}
{Filippenko}, A.~V., {Li}, W.~D., {Treffers}, R.~R., \& {Modjaz}, M. 2001, in
  Astronomical Society of the Pacific Conference Series, Vol. 246, IAU Colloq.
  183: Small Telescope Astronomy on Global Scales, ed. B.~{Paczynski}, W.-P.
  {Chen}, \& C.~{Lemme}, 121

\bibitem[{{Filippenko} {et~al.}(1992){Filippenko}, {Richmond}, {Branch},
  {Gaskell}, {Herbst}, {Ford}, {Treffers}, {Matheson}, {Ho}, {Dey}, {Sargent},
  {Small}, \& {van Breugel}}]{filippenko1992}
{Filippenko}, A.~V., {Richmond}, M.~W., {Branch}, D., {et~al.} 1992, \aj, 104,
  1543

\bibitem[{{Filippenko} {et~al.}(1995){Filippenko}, {Barth}, {Matheson},
  {Armus}, {Brown}, {Espey}, {Fan}, {Goodrich}, {Ho}, {Junkkarinen}, {Koo},
  {Lehnert}, {Martel}, {Mazzarella}, {Miller}, {Smith}, {Tytler}, \&
  {Wirth}}]{filippenko1995}
{Filippenko}, A.~V., {Barth}, A.~J., {Matheson}, T., {et~al.} 1995, \apjl, 450,
  L11

\bibitem[{{Folatelli} {et~al.}(2006){Folatelli}, {Contreras}, {Phillips},
  {Woosley}, {Blinnikov}, {Morrell}, {Suntzeff}, {Lee}, {Hamuy},
  {Gonz{\'a}lez}, {Krzeminski}, {Roth}, {Li}, {Filippenko}, {Foley},
  {Freedman}, {Madore}, {Persson}, {Murphy}, {Boissier}, {Galaz},
  {Gonz{\'a}lez}, {McCarthy}, {McWilliam}, \& {Pych}}]{folatelli2006}
{Folatelli}, G., {Contreras}, C., {Phillips}, M.~M., {et~al.} 2006, \apj, 641,
  1039

\bibitem[{{Foley} {et~al.}(2003){Foley}, {Papenkova}, {Swift}, {Filippenko},
  {Li}, {Mazzali}, {Chornock}, {Leonard}, \& {Van Dyk}}]{foley2003}
{Foley}, R.~J., {Papenkova}, M.~S., {Swift}, B.~J., {et~al.} 2003, \pasp, 115,
  1220

\bibitem[{{Fowler} \& {Hoyle}(1964)}]{fowler_hoyle1964}
{Fowler}, W.~A., \& {Hoyle}, F. 1964, \apjs, 9, 201

\bibitem[{{Fransson} {et~al.}(1996){Fransson}, {Lundqvist}, \&
  {Chevalier}}]{fransson1996}
{Fransson}, C., {Lundqvist}, P., \& {Chevalier}, R.~A. 1996, \apj, 461, 993

\bibitem[{{Fransson} {et~al.}(2005){Fransson}, {Challis}, {Chevalier},
  {Filippenko}, {Kirshner}, {Kozma}, {Leonard}, {Matheson}, {Baron},
  {Garnavich}, {Jha}, {Leibundgut}, {Lundqvist}, {Pun}, {Wang}, \&
  {Wheeler}}]{fransson2005}
{Fransson}, C., {Challis}, P.~M., {Chevalier}, R.~A., {et~al.} 2005, \apj, 622,
  991

\bibitem[{{Gal-Yam}(2012)}]{galyam2012}
{Gal-Yam}, A. 2012, Science, 337, 927

\bibitem[{{Gal-Yam}(2016)}]{galyam2016}
---. 2016, ArXiv e-prints, arXiv:1611.09353

\bibitem[{{Gal-Yam} {et~al.}(2009){Gal-Yam}, {Mazzali}, {Ofek}, {Nugent},
  {Kulkarni}, {Kasliwal}, {Quimby}, {Filippenko}, {Cenko}, {Chornock},
  {Waldman}, {Kasen}, {Sullivan}, {Beshore}, {Drake}, {Thomas}, {Bloom},
  {Poznanski}, {Miller}, {Foley}, {Silverman}, {Arcavi}, {Ellis}, \&
  {Deng}}]{galyam2009}
{Gal-Yam}, A., {Mazzali}, P., {Ofek}, E.~O., {et~al.} 2009, \nat, 462, 624

\bibitem[{{Gal-Yam} {et~al.}(2011){Gal-Yam}, {Kasliwal}, {Arcavi}, {Green},
  {Yaron}, {Ben-Ami}, {Xu}, {Sternberg}, {Quimby}, {Kulkarni}, {Ofek},
  {Walters}, {Nugent}, {Poznanski}, {Bloom}, {Cenko}, {Filippenko}, {Li},
  {Silverman}, {Walker}, {Sullivan}, {Maguire}, {Howell}, {Mazzali}, {Frail},
  {Bersier}, {James}, {Akerlof}, {Yuan}, {Law}, {Fox}, \&
  {Gehrels}}]{galyam2011}
{Gal-Yam}, A., {Kasliwal}, M.~M., {Arcavi}, I., {et~al.} 2011, \apj, 736, 159

\bibitem[{{Gezari} {et~al.}(2009){Gezari}, {Halpern}, {Grupe}, {Yuan},
  {Quimby}, {McKay}, {Chamarro}, {Sisson}, {Akerlof}, {Wheeler}, {Brown},
  {Cenko}, {Rau}, {Djordjevic}, \& {Terndrup}}]{gezari2009}
{Gezari}, S., {Halpern}, J.~P., {Grupe}, D., {et~al.} 2009, \apj, 690, 1313

\bibitem[{{Gonz{\'a}lez-Gait{\'a}n} {et~al.}(2011){Gonz{\'a}lez-Gait{\'a}n},
  {Perrett}, {Sullivan}, {Conley}, {Howell}, {Carlberg}, {Astier}, {Balam},
  {Balland}, {Basa}, {Fouchez}, {Guy}, {Hardin}, {Hook.}, {Pain}, {Pritchet},
  {Regnault}, {Rich}, \& {Lidman}}]{gonzalez-gaitan2011}
{Gonz{\'a}lez-Gait{\'a}n}, S., {Perrett}, K., {Sullivan}, M., {et~al.} 2011,
  \apj, 727, 107

\bibitem[{{Guillochon} {et~al.}(2017){Guillochon}, {Parrent}, {Kelley}, \&
  {Margutti}}]{guillochon2017}
{Guillochon}, J., {Parrent}, J., {Kelley}, L.~Z., \& {Margutti}, R. 2017, \apj,
  835, 64

\bibitem[{{Harutyunyan} {et~al.}(2008){Harutyunyan}, {Pfahler}, {Pastorello},
  {Taubenberger}, {Turatto}, {Cappellaro}, {Benetti}, {Elias-Rosa},
  {Navasardyan}, {Valenti}, {Stanishev}, {Patat}, {Riello}, {Pignata}, \&
  {Hillebrandt}}]{harutyunyan2008}
{Harutyunyan}, A.~H., {Pfahler}, P., {Pastorello}, A., {et~al.} 2008, \aap,
  488, 383

\bibitem[{{Hatano} {et~al.}(2001){Hatano}, {Branch}, {Nomoto}, {Deng}, {Maeda},
  {Nugent}, \& {Aldering}}]{hatano2001}
{Hatano}, K., {Branch}, D., {Nomoto}, K., {et~al.} 2001, in Bulletin of the
  American Astronomical Society, Vol.~33, American Astronomical Society Meeting
  Abstracts \#198, 838

\bibitem[{{Howell} {et~al.}(2006){Howell}, {Sullivan}, {Nugent}, {Ellis},
  {Conley}, {Le Borgne}, {Carlberg}, {Guy}, {Balam}, {Basa}, {Fouchez}, {Hook},
  {Hsiao}, {Neill}, {Pain}, {Perrett}, \& {Pritchet}}]{howell2006}
{Howell}, D.~A., {Sullivan}, M., {Nugent}, P.~E., {et~al.} 2006, \nat, 443, 308

\bibitem[{{Howell} {et~al.}(2013){Howell}, {Kasen}, {Lidman}, {Sullivan},
  {Conley}, {Astier}, {Balland}, {Carlberg}, {Fouchez}, {Guy}, {Hardin},
  {Pain}, {Palanque-Delabrouille}, {Perrett}, {Pritchet}, {Regnault}, {Rich},
  \& {Ruhlmann-Kleider}}]{howell2013}
{Howell}, D.~A., {Kasen}, D., {Lidman}, C., {et~al.} 2013, \apj, 779, 98

\bibitem[{{Inserra} {et~al.}(2016{\natexlab{a}}){Inserra}, {Bulla}, {Sim}, \&
  {Smartt}}]{inserra2016b}
{Inserra}, C., {Bulla}, M., {Sim}, S.~A., \& {Smartt}, S.~J.
  2016{\natexlab{a}}, \apj, 831, 79

\bibitem[{{Inserra} \& {Smartt}(2014)}]{inserra_smart2014}
{Inserra}, C., \& {Smartt}, S.~J. 2014, \apj, 796, 87

\bibitem[{{Inserra} {et~al.}(2013){Inserra}, {Smartt}, {Jerkstrand}, {Valenti},
  {Fraser}, {Wright}, {Smith}, {Chen}, {Kotak}, {Pastorello}, {Nicholl},
  {Bresolin}, {Kudritzki}, {Benetti}, {Botticella}, {Burgett}, {Chambers},
  {Ergon}, {Flewelling}, {Fynbo}, {Geier}, {Hodapp}, {Howell}, {Huber},
  {Kaiser}, {Leloudas}, {Magill}, {Magnier}, {McCrum}, {Metcalfe}, {Price},
  {Rest}, {Sollerman}, {Sweeney}, {Taddia}, {Taubenberger}, {Tonry},
  {Wainscoat}, {Waters}, \& {Young}}]{inserra2013}
{Inserra}, C., {Smartt}, S.~J., {Jerkstrand}, A., {et~al.} 2013, \apj, 770, 128

\bibitem[{{Inserra} {et~al.}(2016{\natexlab{b}}){Inserra}, {Smartt}, {Gall},
  {Leloudas}, {Chen}, {Schulze}, {Jerkstarnd}, {Nicholl}, {Anderson}, {Arcavi},
  {Benetti}, {Cartier}, {Childress}, {Della Valle}, {Flewelling}, {Fraser},
  {Gal-Yam}, {Gutierrez}, {Hosseinzadeh}, {Howell}, {Huber}, {Kankare},
  {Magnier}, {Maguire}, {McCully}, {Prajs}, {Primak}, {Scalzo}, {Schmidt},
  {Smith}, {Tucker}, {Valenti}, {Wilman}, {Young}, \& {Yuan}}]{inserra2016}
{Inserra}, C., {Smartt}, S.~J., {Gall}, E.~E.~E., {et~al.} 2016{\natexlab{b}},
  ArXiv e-prints, arXiv:1604.01226

\bibitem[{{Jeffery} \& {Branch}(1990)}]{jeffery_branch1990}
{Jeffery}, D.~J., \& {Branch}, D. 1990, in Supernovae, Jerusalem Winter School
  for Theoretical Physics, ed. J.~C. {Wheeler}, T.~{Piran}, \& S.~{Weinberg},
  149

\bibitem[{{Jerkstrand} {et~al.}(2014){Jerkstrand}, {Smartt}, {Fraser},
  {Fransson}, {Sollerman}, {Taddia}, \& {Kotak}}]{jerkstrand2014}
{Jerkstrand}, A., {Smartt}, S.~J., {Fraser}, M., {et~al.} 2014, \mnras, 439,
  3694

\bibitem[{{Jerkstrand} {et~al.}(2017){Jerkstrand}, {Smartt}, {Inserra},
  {Nicholl}, {Chen}, {Kr{\"u}hler}, {Sollerman}, {Taubenberger}, {Gal-Yam},
  {Kankare}, {Maguire}, {Fraser}, {Valenti}, {Sullivan}, {Cartier}, \&
  {Young}}]{jerkstrand2017}
{Jerkstrand}, A., {Smartt}, S.~J., {Inserra}, C., {et~al.} 2017, \apj, 835, 13

\bibitem[{{Kasen} \& {Bildsten}(2010)}]{kasen_bildsten2010}
{Kasen}, D., \& {Bildsten}, L. 2010, \apj, 717, 245

\bibitem[{{Kasliwal} {et~al.}(2009{\natexlab{a}}){Kasliwal}, {Quimby},
  {Nugent}, {Ellis}, {Howell}, {Kulkarni}, {Law}, {Ofek}, {Poznanski}, \&
  {Thomas}}]{kasliwal2009_atel1}
{Kasliwal}, M.~M., {Quimby}, R., {Nugent}, P., {et~al.} 2009{\natexlab{a}}, The
  Astronomer's Telegram, 1983

\bibitem[{{Kasliwal} {et~al.}(2009{\natexlab{b}}){Kasliwal}, {Quimby},
  {Nugent}, {Ellis}, {Howell}, {Kulkarni}, {Law}, {Ofek}, {Poznanski}, \&
  {Thomas}}]{kasliwal2009_cbet1}
---. 2009{\natexlab{b}}, Central Bureau Electronic Telegrams, 1732

\bibitem[{{Kelson}(2003)}]{kelson2003}
{Kelson}, D.~D. 2003, \pasp, 115, 688

\bibitem[{{Kodros} {et~al.}(2010){Kodros}, {Cenko}, {Li}, {Filippenko},
  {Kandrashoff}, \& {Silverman}}]{kodros2010}
{Kodros}, J., {Cenko}, S.~B., {Li}, W., {et~al.} 2010, Central Bureau
  Electronic Telegrams, 2461

\bibitem[{{Kozyreva} {et~al.}(2017){Kozyreva}, {Gilmer}, {Hirschi},
  {Fr{\"o}hlich}, {Blinnikov}, {Wollaeger}, {Noebauer}, {van Rossum}, {Heger},
  {Even}, {Waldman}, {Tolstov}, {Chatzopoulos}, \& {Sorokina}}]{kozyreva2017}
{Kozyreva}, A., {Gilmer}, M., {Hirschi}, R., {et~al.} 2017, \mnras, 464, 2854

\bibitem[{{K{\"u}mmel} {et~al.}(2009){K{\"u}mmel}, {Walsh}, {Pirzkal},
  {Kuntschner}, \& {Pasquali}}]{kummel2009}
{K{\"u}mmel}, M., {Walsh}, J.~R., {Pirzkal}, N., {Kuntschner}, H., \&
  {Pasquali}, A. 2009, \pasp, 121, 59

\bibitem[{{Law} {et~al.}(2009){Law}, {Kulkarni}, {Dekany}, {Ofek}, {Quimby},
  {Nugent}, {Surace}, {Grillmair}, {Bloom}, {Kasliwal}, {Bildsten}, {Brown},
  {Cenko}, {Ciardi}, {Croner}, {Djorgovski}, {van Eyken}, {Filippenko}, {Fox},
  {Gal-Yam}, {Hale}, {Hamam}, {Helou}, {Henning}, {Howell}, {Jacobsen},
  {Laher}, {Mattingly}, {McKenna}, {Pickles}, {Poznanski}, {Rahmer}, {Rau},
  {Rosing}, {Shara}, {Smith}, {Starr}, {Sullivan}, {Velur}, {Walters}, \&
  {Zolkower}}]{law2009}
{Law}, N.~M., {Kulkarni}, S.~R., {Dekany}, R.~G., {et~al.} 2009, \pasp, 121,
  1395

\bibitem[{{Leloudas} {et~al.}(2012){Leloudas}, {Chatzopoulos}, {Dilday},
  {Gorosabel}, {Vinko}, {Gallazzi}, {Wheeler}, {Bassett}, {Fischer}, {Frieman},
  {Fynbo}, {Goobar}, {Jel{\'{\i}}nek}, {Malesani}, {Nichol}, {Nordin},
  {{\"O}stman}, {Sako}, {Schneider}, {Smith}, {Sollerman}, {Stritzinger},
  {Th{\"o}ne}, \& {de Ugarte Postigo}}]{leloudas2012}
{Leloudas}, G., {Chatzopoulos}, E., {Dilday}, B., {et~al.} 2012, \aap, 541,
  A129

\bibitem[{{Leloudas} {et~al.}(2015{\natexlab{a}}){Leloudas}, {Patat}, {Maund},
  {Hsiao}, {Malesani}, {Schulze}, {Contreras}, {de Ugarte Postigo},
  {Sollerman}, {Stritzinger}, {Taddia}, {Wheeler}, \&
  {Gorosabel}}]{leloudas2015c}
{Leloudas}, G., {Patat}, F., {Maund}, J.~R., {et~al.} 2015{\natexlab{a}},
  \apjl, 815, L10

\bibitem[{{Leloudas} {et~al.}(2015{\natexlab{b}}){Leloudas}, {Schulze},
  {Kr{\"u}hler}, {Gorosabel}, {Christensen}, {Mehner}, {de Ugarte Postigo},
  {Amor{\'{\i}}n}, {Th{\"o}ne}, {Anderson}, {Bauer}, {Gallazzi},
  {He{\l}miniak}, {Hjorth}, {Ibar}, {Malesani}, {Morell}, {Vinko}, \&
  {Wheeler}}]{leloudas2015}
{Leloudas}, G., {Schulze}, S., {Kr{\"u}hler}, T., {et~al.} 2015{\natexlab{b}},
  \mnras, 449, 917

\bibitem[{{Leonard} {et~al.}(2000){Leonard}, {Filippenko}, {Barth}, \&
  {Matheson}}]{leonard2000}
{Leonard}, D.~C., {Filippenko}, A.~V., {Barth}, A.~J., \& {Matheson}, T. 2000,
  \apj, 536, 239

\bibitem[{{Leonard} {et~al.}(2002){Leonard}, {Filippenko}, {Li}, {Matheson},
  {Kirshner}, {Chornock}, {Van Dyk}, {Berlind}, {Calkins}, {Challis},
  {Garnavich}, {Jha}, \& {Mahdavi}}]{leonard2002}
{Leonard}, D.~C., {Filippenko}, A.~V., {Li}, W., {et~al.} 2002, \aj, 124, 2490

\bibitem[{{Li} {et~al.}(2011){Li}, {Leaman}, {Chornock}, {Filippenko},
  {Poznanski}, {Ganeshalingam}, {Wang}, {Modjaz}, {Jha}, {Foley}, \&
  {Smith}}]{li2011a}
{Li}, W., {Leaman}, J., {Chornock}, R., {et~al.} 2011, \mnras, 412, 1441

\bibitem[{{Liu} {et~al.}(2017){Liu}, {Wang}, {Wang}, {Dai}, {Yu}, \&
  {Peng}}]{liu2017}
{Liu}, L.-D., {Wang}, S.-Q., {Wang}, L.-J., {et~al.} 2017, \apj, 842, 26

\bibitem[{{Liu} {et~al.}(2016){Liu}, {Modjaz}, \& {Bianco}}]{liu2016}
{Liu}, Y.-Q., {Modjaz}, M., \& {Bianco}, F.~B. 2016, ArXiv e-prints,
  arXiv:1612.07321

\bibitem[{{Lunnan} {et~al.}(2016){Lunnan}, {Chornock}, {Berger},
  {Milisavljevic}, {Jones}, {Rest}, {Fong}, {Fransson}, {Margutti}, {Drout},
  {Blanchard}, {Challis}, {Cowperthwaite}, {Foley}, {Kirshner}, {Morrell},
  {Riess}, {Roth}, {Scolnic}, {Smartt}, {Smith}, {Villar}, {Chambers},
  {Draper}, {Huber}, {Kaiser}, {Kudritzki}, {Magnier}, {Metcalfe}, \&
  {Waters}}]{lunnan2016}
{Lunnan}, R., {Chornock}, R., {Berger}, E., {et~al.} 2016, \apj, 831, 144

\bibitem[{{Lunnan} {et~al.}(2017){Lunnan}, {Chornock}, {Berger}, {Jones},
  {Rest}, {Czekala}, {Dittmann}, {Drout}, {Foley}, {Fong}, {Kirshner},
  {Laskar}, {Leibler}, {Margutti}, {Milisavljevic}, {Narayan}, {Pan}, {Riess},
  {Roth}, {Sanders}, {Scolnic}, {Smartt}, {Smith}, {Chambers}, {Draper},
  {Flewelling}, {Huber}, {Kaiser}, {Kudritzki}, {Magnier}, {Metcalfe},
  {Wainscoat}, {Waters}, \& {Willman}}]{lunnan2017}
---. 2017, ArXiv e-prints, arXiv:1708.01619

\bibitem[{{Maguire} {et~al.}(2014){Maguire}, {Sullivan}, {Pan}, {Gal-Yam},
  {Hook}, {Howell}, {Nugent}, {Mazzali}, {Chotard}, {Clubb}, {Filippenko},
  {Kasliwal}, {Kandrashoff}, {Poznanski}, {Saunders}, {Silverman}, {Walker}, \&
  {Xu}}]{maguire2014}
{Maguire}, K., {Sullivan}, M., {Pan}, Y.-C., {et~al.} 2014, \mnras, 444, 3258

\bibitem[{{Malesani} {et~al.}(2009){Malesani}, {Fynbo}, {Hjorth}, {Leloudas},
  {Sollerman}, {Stritzinger}, {Vreeswijk}, {Watson}, {Gorosabel},
  {Micha{\l}owski}, \& {Th{\"o}ne}}]{malesani2009}
{Malesani}, D., {Fynbo}, J.~P.~U., {Hjorth}, J., {et~al.} 2009, in American
  Institute of Physics Conference Series, Vol. 1111, American Institute of
  Physics Conference Series, ed. G.~{Giobbi}, A.~{Tornambe}, G.~{Raimondo},
  M.~{Limongi}, L.~A. {Antonelli}, N.~{Menci}, \& E.~{Brocato}, 627--628

\bibitem[{{Marion} {et~al.}(2009){Marion}, {H{\"o}flich}, {Gerardy}, {Vacca},
  {Wheeler}, \& {Robinson}}]{marion2009}
{Marion}, G.~H., {H{\"o}flich}, P., {Gerardy}, C.~L., {et~al.} 2009, \aj, 138,
  727

\bibitem[{{Matheson} {et~al.}(2001){Matheson}, {Filippenko}, {Li}, {Leonard},
  \& {Shields}}]{matheson2001}
{Matheson}, T., {Filippenko}, A.~V., {Li}, W., {Leonard}, D.~C., \& {Shields},
  J.~C. 2001, \aj, 121, 1648

\bibitem[{{Mazzali} {et~al.}(2016){Mazzali}, {Sullivan}, {Pian}, {Greiner}, \&
  {Kann}}]{mazzali2016}
{Mazzali}, P.~A., {Sullivan}, M., {Pian}, E., {Greiner}, J., \& {Kann}, D.~A.
  2016, \mnras, 458, 3455

\bibitem[{{Mazzali} {et~al.}(2008){Mazzali}, {Valenti}, {Della Valle},
  {Chincarini}, {Sauer}, {Benetti}, {Pian}, {Piran}, {D'Elia}, {Elias-Rosa},
  {Margutti}, {Pasotti}, {Antonelli}, {Bufano}, {Campana}, {Cappellaro},
  {Covino}, {D'Avanzo}, {Fiore}, {Fugazza}, {Gilmozzi}, {Hunter}, {Maguire},
  {Maiorano}, {Marziani}, {Masetti}, {Mirabel}, {Navasardyan}, {Nomoto},
  {Palazzi}, {Pastorello}, {Panagia}, {Pellizza}, {Sari}, {Smartt},
  {Tagliaferri}, {Tanaka}, {Taubenberger}, {Tominaga}, {Trundle}, \&
  {Turatto}}]{mazzali2008}
{Mazzali}, P.~A., {Valenti}, S., {Della Valle}, M., {et~al.} 2008, Science,
  321, 1185

\bibitem[{{Metzger} {et~al.}(2014){Metzger}, {Vurm}, {Hasco{\"e}t}, \&
  {Beloborodov}}]{metzger2014}
{Metzger}, B.~D., {Vurm}, I., {Hasco{\"e}t}, R., \& {Beloborodov}, A.~M. 2014,
  \mnras, 437, 703

\bibitem[{{Miller} {et~al.}(2009){Miller}, {Chornock}, {Perley},
  {Ganeshalingam}, {Li}, {Butler}, {Bloom}, {Smith}, {Modjaz}, {Poznanski},
  {Filippenko}, {Griffith}, {Shiode}, \& {Silverman}}]{miller2009}
{Miller}, A.~A., {Chornock}, R., {Perley}, D.~A., {et~al.} 2009, \apj, 690,
  1303

\bibitem[{{Modjaz} {et~al.}(2009){Modjaz}, {Li}, {Butler}, {Chornock},
  {Perley}, {Blondin}, {Bloom}, {Filippenko}, {Kirshner}, {Kocevski},
  {Poznanski}, {Hicken}, {Foley}, {Stringfellow}, {Berlind}, {Barrado y
  Navascues}, {Blake}, {Bouy}, {Brown}, {Challis}, {Chen}, {de Vries},
  {Dufour}, {Falco}, {Friedman}, {Ganeshalingam}, {Garnavich}, {Holden},
  {Illingworth}, {Lee}, {Liebert}, {Marion}, {Olivier}, {Prochaska},
  {Silverman}, {Smith}, {Starr}, {Steele}, {Stockton}, {Williams}, \&
  {Wood-Vasey}}]{modjaz2009}
{Modjaz}, M., {Li}, W., {Butler}, N., {et~al.} 2009, \apj, 702, 226

\bibitem[{{Modjaz} {et~al.}(2014){Modjaz}, {Blondin}, {Kirshner}, {Matheson},
  {Berlind}, {Bianco}, {Calkins}, {Challis}, {Garnavich}, {Hicken}, {Jha},
  {Liu}, \& {Marion}}]{modjaz2014}
{Modjaz}, M., {Blondin}, S., {Kirshner}, R.~P., {et~al.} 2014, \aj, 147, 99

\bibitem[{{Nicholl} {et~al.}(2017{\natexlab{a}}){Nicholl}, {Berger},
  {Margutti}, {Blanchard}, {Guillochon}, {Leja}, \& {Chornock}}]{nicholl2017b}
{Nicholl}, M., {Berger}, E., {Margutti}, R., {et~al.} 2017{\natexlab{a}},
  \apjl, 845, L8

\bibitem[{{Nicholl} {et~al.}(2017{\natexlab{b}}){Nicholl}, {Guillochon}, \&
  {Berger}}]{nicholl2017}
{Nicholl}, M., {Guillochon}, J., \& {Berger}, E. 2017{\natexlab{b}}, ArXiv
  e-prints, arXiv:1706.00825

\bibitem[{{Nicholl} {et~al.}(2013){Nicholl}, {Smartt}, {Jerkstrand}, {Inserra},
  {McCrum}, {Kotak}, {Fraser}, {Wright}, {Chen}, {Smith}, {Young}, {Sim},
  {Valenti}, {Howell}, {Bresolin}, {Kudritzki}, {Tonry}, {Huber}, {Rest},
  {Pastorello}, {Tomasella}, {Cappellaro}, {Benetti}, {Mattila}, {Kankare},
  {Kangas}, {Leloudas}, {Sollerman}, {Taddia}, {Berger}, {Chornock}, {Narayan},
  {Stubbs}, {Foley}, {Lunnan}, {Soderberg}, {Sanders}, {Milisavljevic},
  {Margutti}, {Kirshner}, {Elias-Rosa}, {Morales-Garoffolo}, {Taubenberger},
  {Botticella}, {Gezari}, {Urata}, {Rodney}, {Riess}, {Scolnic}, {Wood-Vasey},
  {Burgett}, {Chambers}, {Flewelling}, {Magnier}, {Kaiser}, {Metcalfe},
  {Morgan}, {Price}, {Sweeney}, \& {Waters}}]{nicholl2013}
{Nicholl}, M., {Smartt}, S.~J., {Jerkstrand}, A., {et~al.} 2013, \nat, 502, 346

\bibitem[{{Nicholl} {et~al.}(2014){Nicholl}, {Smartt}, {Jerkstrand}, {Inserra},
  {Anderson}, {Baltay}, {Benetti}, {Chen}, {Elias-Rosa}, {Feindt}, {Fraser},
  {Gal-Yam}, {Hadjiyska}, {Howell}, {Kotak}, {Lawrence}, {Leloudas},
  {Margheim}, {Mattila}, {McCrum}, {McKinnon}, {Mead}, {Nugent}, {Rabinowitz},
  {Rest}, {Smith}, {Sollerman}, {Sullivan}, {Taddia}, {Valenti}, {Walker}, \&
  {Young}}]{nicholl2014}
---. 2014, \mnras, 444, 2096

\bibitem[{{Nicholl} {et~al.}(2015){Nicholl}, {Smartt}, {Jerkstrand}, {Inserra},
  {Sim}, {Chen}, {Benetti}, {Fraser}, {Gal-Yam}, {Kankare}, {Maguire}, {Smith},
  {Sullivan}, {Valenti}, {Young}, {Baltay}, {Bauer}, {Baumont}, {Bersier},
  {Botticella}, {Childress}, {Dennefeld}, {Della Valle}, {Elias-Rosa},
  {Feindt}, {Galbany}, {Hadjiyska}, {Le Guillou}, {Leloudas}, {Mazzali},
  {McKinnon}, {Polshaw}, {Rabinowitz}, {Rostami}, {Scalzo}, {Schmidt},
  {Schulze}, {Sollerman}, {Taddia}, \& {Yuan}}]{nicholl2015}
---. 2015, \mnras, 452, 3869

\bibitem[{{Nicholl} {et~al.}(2016){Nicholl}, {Berger}, {Smartt}, {Margutti},
  {Kamble}, {Alexander}, {Chen}, {Inserra}, {Arcavi}, {Blanchard}, {Cartier},
  {Chambers}, {Childress}, {Chornock}, {Cowperthwaite}, {Drout}, {Flewelling},
  {Fraser}, {Gal-Yam}, {Galbany}, {Harmanen}, {Holoien}, {Hosseinzadeh},
  {Howell}, {Huber}, {Jerkstrand}, {Kankare}, {Kochanek}, {Lin}, {Lunnan},
  {Magnier}, {Maguire}, {McCully}, {McDonald}, {Metzger}, {Milisavljevic},
  {Mitra}, {Reynolds}, {Saario}, {Shappee}, {Smith}, {Valenti}, {Villar},
  {Waters}, \& {Young}}]{nicholl2016}
{Nicholl}, M., {Berger}, E., {Smartt}, S.~J., {et~al.} 2016, \apj, 826, 39

\bibitem[{{Pastorello} {et~al.}(2004){Pastorello}, {Zampieri}, {Turatto},
  {Cappellaro}, {Meikle}, {Benetti}, {Branch}, {Baron}, {Patat}, {Armstrong},
  {Altavilla}, {Salvo}, \& {Riello}}]{pastorello2004}
{Pastorello}, A., {Zampieri}, L., {Turatto}, M., {et~al.} 2004, \mnras, 347, 74

\bibitem[{{Pastorello} {et~al.}(2006){Pastorello}, {Sauer}, {Taubenberger},
  {Mazzali}, {Nomoto}, {Kawabata}, {Benetti}, {Elias-Rosa}, {Harutyunyan},
  {Navasardyan}, {Zampieri}, {Iijima}, {Botticella}, {di Rico}, {Del Principe},
  {Dolci}, {Gagliardi}, {Ragni}, \& {Valentini}}]{pastorello2006}
{Pastorello}, A., {Sauer}, D., {Taubenberger}, S., {et~al.} 2006, \mnras, 370,
  1752

\bibitem[{{Pastorello} {et~al.}(2007){Pastorello}, {Smartt}, {Mattila},
  {Eldridge}, {Young}, {Itagaki}, {Yamaoka}, {Navasardyan}, {Valenti}, {Patat},
  {Agnoletto}, {Augusteijn}, {Benetti}, {Cappellaro}, {Boles}, {Bonnet-Bidaud},
  {Botticella}, {Bufano}, {Cao}, {Deng}, {Dennefeld}, {Elias-Rosa},
  {Harutyunyan}, {Keenan}, {Iijima}, {Lorenzi}, {Mazzali}, {Meng}, {Nakano},
  {Nielsen}, {Smoker}, {Stanishev}, {Turatto}, {Xu}, \&
  {Zampieri}}]{pastorello2007}
{Pastorello}, A., {Smartt}, S.~J., {Mattila}, S., {et~al.} 2007, \nat, 447, 829

\bibitem[{{Pastorello} {et~al.}(2008){Pastorello}, {Quimby}, {Smartt},
  {Mattila}, {Navasardyan}, {Crockett}, {Elias-Rosa}, {Mondol}, {Wheeler}, \&
  {Young}}]{pastroello2008}
{Pastorello}, A., {Quimby}, R.~M., {Smartt}, S.~J., {et~al.} 2008, \mnras, 389,
  131

\bibitem[{{Pastorello} {et~al.}(2009){Pastorello}, {Valenti}, {Zampieri},
  {Navasardyan}, {Taubenberger}, {Smartt}, {Arkharov}, {B{\"a}rnbantner},
  {Barwig}, {Benetti}, {Birtwhistle}, {Botticella}, {Cappellaro}, {Del
  Principe}, {di Mille}, {di Rico}, {Dolci}, {Elias-Rosa}, {Efimova},
  {Fiedler}, {Harutyunyan}, {H{\"o}flich}, {Kloehr}, {Larionov}, {Lorenzi},
  {Maund}, {Napoleone}, {Ragni}, {Richmond}, {Ries}, {Spiro}, {Temporin},
  {Turatto}, \& {Wheeler}}]{pastorello2009}
{Pastorello}, A., {Valenti}, S., {Zampieri}, L., {et~al.} 2009, \mnras, 394,
  2266

\bibitem[{{Pastorello} {et~al.}(2010){Pastorello}, {Smartt}, {Botticella},
  {Maguire}, {Fraser}, {Smith}, {Kotak}, {Magill}, {Valenti}, {Young},
  {Gezari}, {Bresolin}, {Kudritzki}, {Howell}, {Rest}, {Metcalfe}, {Mattila},
  {Kankare}, {Huang}, {Urata}, {Burgett}, {Chambers}, {Dombeck}, {Flewelling},
  {Grav}, {Heasley}, {Hodapp}, {Kaiser}, {Luppino}, {Lupton}, {Magnier},
  {Monet}, {Morgan}, {Onaka}, {Price}, {Rhoads}, {Siegmund}, {Stubbs},
  {Sweeney}, {Tonry}, {Wainscoat}, {Waterson}, {Waters}, \&
  {Wynn-Williams}}]{pastorello2010}
{Pastorello}, A., {Smartt}, S.~J., {Botticella}, M.~T., {et~al.} 2010, \apjl,
  724, L16

\bibitem[{{Pastorello} {et~al.}(2015){Pastorello}, {Wyrzykowski}, {Valenti},
  {Prieto}, {Koz{\l}owski}, {Udalski}, {Elias-Rosa}, {Morales-Garoffolo},
  {Anderson}, {Benetti}, {Bersten}, {Botticella}, {Cappellaro}, {Fasano},
  {Fraser}, {Gal-Yam}, {Gillone}, {Graham}, {Greiner}, {Hachinger}, {Howell},
  {Inserra}, {Parrent}, {Rau}, {Schulze}, {Smartt}, {Smith}, {Turatto},
  {Yaron}, {Young}, {Kubiak}, {Szyma{\'n}ski}, {Pietrzy{\'n}ski},
  {Soszy{\'n}ski}, {Ulaczyk}, {Poleski}, {Pietrukowicz}, {Skowron}, \&
  {Mr{\'o}z}}]{pastorello2015}
{Pastorello}, A., {Wyrzykowski}, {\L}., {Valenti}, S., {et~al.} 2015, \mnras,
  449, 1941

\bibitem[{{Patat} {et~al.}(2001){Patat}, {Cappellaro}, {Danziger}, {Mazzali},
  {Sollerman}, {Augusteijn}, {Brewer}, {Doublier}, {Gonzalez}, {Hainaut},
  {Lidman}, {Leibundgut}, {Nomoto}, {Nakamura}, {Spyromilio}, {Rizzi},
  {Turatto}, {Walsh}, {Galama}, {van Paradijs}, {Kouveliotou}, {Vreeswijk},
  {Frontera}, {Masetti}, {Palazzi}, \& {Pian}}]{patat2001}
{Patat}, F., {Cappellaro}, E., {Danziger}, J., {et~al.} 2001, \apj, 555, 900

\bibitem[{{Perley} {et~al.}(2016){Perley}, {Quimby}, {Yan}, {Vreeswijk}, {De
  Cia}, {Lunnan}, {Gal-Yam}, {Yaron}, {Filippenko}, {Graham}, {Laher}, \&
  {Nugent}}]{perley2016}
{Perley}, D.~A., {Quimby}, R.~M., {Yan}, L., {et~al.} 2016, \apj, 830, 13

\bibitem[{{Prajs} {et~al.}(2017){Prajs}, {Sullivan}, {Smith}, {Levan},
  {Karpenka}, {Edwards}, {Walker}, {Wolf}, {Balland}, {Carlberg}, {Howell},
  {Lidman}, {Pain}, {Pritchet}, \& {Ruhlmann-Kleider}}]{prajs2017}
{Prajs}, S., {Sullivan}, M., {Smith}, M., {et~al.} 2017, \mnras, 464, 3568

\bibitem[{{Prentice} {et~al.}(2016){Prentice}, {Mazzali}, {Pian}, {Gal-Yam},
  {Kulkarni}, {Rubin}, {Corsi}, {Fremling}, {Sollerman}, {Yaron}, {Arcavi},
  {Zheng}, {Kasliwal}, {Filippenko}, {Cenko}, {Cao}, \&
  {Nugent}}]{prentice2016}
{Prentice}, S.~J., {Mazzali}, P.~A., {Pian}, E., {et~al.} 2016, \mnras, 458,
  2973

\bibitem[{{Pun} {et~al.}(1995){Pun}, {Kirshner}, {Sonneborn}, {Challis},
  {Nassiopoulos}, {Arquilla}, {Crenshaw}, {Shrader}, {Teays}, {Cassatella},
  {Gilmozzi}, {Talavera}, {Wamsteker}, {Fransson}, \& {Panagia}}]{pun1995}
{Pun}, C.~S.~J., {Kirshner}, R.~P., {Sonneborn}, G., {et~al.} 1995, \apjs, 99,
  223

\bibitem[{{Quimby} {et~al.}(2009{\natexlab{a}}){Quimby}, {Kasliwal}, {Cenko},
  {Fox}, {Gal-Yam}, {Howell}, {Kulkarni}, {Law}, {Levitan}, {Mahabal},
  {Nugent}, {Ofek}, {Poznanski}, \& {Thomas}}]{quimby2009_atel1}
{Quimby}, R., {Kasliwal}, M.~M., {Cenko}, S.~B., {et~al.} 2009{\natexlab{a}},
  The Astronomer's Telegram, 2005

\bibitem[{{Quimby} {et~al.}(2009{\natexlab{b}}){Quimby}, {Kasliwal}, {Cenko},
  {Fox}, {Gal-Yam}, {Howell}, {Kulkarni}, {Law}, {Levitan}, {Mahabal},
  {Nugent}, {Ofek}, {Poznanski}, \& {Thomas}}]{quimby2009_cbet1}
---. 2009{\natexlab{b}}, Central Bureau Electronic Telegrams, 1754

\bibitem[{{Quimby} {et~al.}(2007{\natexlab{a}}){Quimby}, {Aldering}, {Wheeler},
  {H{\"o}flich}, {Akerlof}, \& {Rykoff}}]{quimby2007c}
{Quimby}, R.~M., {Aldering}, G., {Wheeler}, J.~C., {et~al.} 2007{\natexlab{a}},
  \apjl, 668, L99

\bibitem[{{Quimby} {et~al.}(2007{\natexlab{b}}){Quimby}, {Wheeler},
  {H{\"o}flich}, {Akerlof}, {Brown}, \& {Rykoff}}]{quimby2007b}
{Quimby}, R.~M., {Wheeler}, J.~C., {H{\"o}flich}, P., {et~al.}
  2007{\natexlab{b}}, \apj, 666, 1093

\bibitem[{{Quimby} {et~al.}(2013){Quimby}, {Yuan}, {Akerlof}, \&
  {Wheeler}}]{quimby2013}
{Quimby}, R.~M., {Yuan}, F., {Akerlof}, C., \& {Wheeler}, J.~C. 2013, \mnras,
  431, 912

\bibitem[{{Quimby} {et~al.}(2010){Quimby}, {Kulkarni}, {Ofek}, {Kasliwal},
  {Gal-Yam}, {Ben-Ami}, {Badenes}, {Sternberg}, {Botyanszki}, {Nugent}, \&
  {Howell}}]{quimby2010_atel2}
{Quimby}, R.~M., {Kulkarni}, S., {Ofek}, E., {et~al.} 2010, The Astronomer's
  Telegram, 2979

\bibitem[{{Quimby} {et~al.}(2011){Quimby}, {Kulkarni}, {Kasliwal}, {Gal-Yam},
  {Arcavi}, {Sullivan}, {Nugent}, {Thomas}, {Howell}, {Nakar}, {Bildsten},
  {Theissen}, {Law}, {Dekany}, {Rahmer}, {Hale}, {Smith}, {Ofek}, {Zolkower},
  {Velur}, {Walters}, {Henning}, {Bui}, {McKenna}, {Poznanski}, {Cenko}, \&
  {Levitan}}]{quimby2011}
{Quimby}, R.~M., {Kulkarni}, S.~R., {Kasliwal}, M.~M., {et~al.} 2011, \nat,
  474, 487

\bibitem[{{Rau} {et~al.}(2009){Rau}, {Kulkarni}, {Law}, {Bloom}, {Ciardi},
  {Djorgovski}, {Fox}, {Gal-Yam}, {Grillmair}, {Kasliwal}, {Nugent}, {Ofek},
  {Quimby}, {Reach}, {Shara}, {Bildsten}, {Cenko}, {Drake}, {Filippenko},
  {Helfand}, {Helou}, {Howell}, {Poznanski}, \& {Sullivan}}]{rau2009}
{Rau}, A., {Kulkarni}, S.~R., {Law}, N.~M., {et~al.} 2009, \pasp, 121, 1334

\bibitem[{{Rest} {et~al.}(2011){Rest}, {Foley}, {Gezari}, {Narayan}, {Draine},
  {Olsen}, {Huber}, {Matheson}, {Garg}, {Welch}, {Becker}, {Challis},
  {Clocchiatti}, {Cook}, {Damke}, {Meixner}, {Miknaitis}, {Minniti}, {Morelli},
  {Nikolaev}, {Pignata}, {Prieto}, {Smith}, {Stubbs}, {Suntzeff}, {Walker},
  {Wood-Vasey}, {Zenteno}, {Wyrzykowski}, {Udalski}, {Szyma{\'n}ski}, {Kubiak},
  {Pietrzy{\'n}ski}, {Soszy{\'n}ski}, {Szewczyk}, {Ulaczyk}, \&
  {Poleski}}]{rest2011}
{Rest}, A., {Foley}, R.~J., {Gezari}, S., {et~al.} 2011, \apj, 729, 88

\bibitem[{{Richardson} {et~al.}(2014){Richardson}, {Jenkins}, {Wright}, \&
  {Maddox}}]{richardson2014}
{Richardson}, D., {Jenkins}, III, R.~L., {Wright}, J., \& {Maddox}, L. 2014,
  \aj, 147, 118

\bibitem[{{Riess} {et~al.}(1997){Riess}, {Filippenko}, {Leonard}, {Schmidt},
  {Suntzeff}, {Phillips}, {Schommer}, {Clocchiatti}, {Kirshner}, {Garnavich},
  {Challis}, {Leibundgut}, {Spyromilio}, \& {Smith}}]{riess1997}
{Riess}, A.~G., {Filippenko}, A.~V., {Leonard}, D.~C., {et~al.} 1997, \aj, 114,
  722

\bibitem[{{Sahu} {et~al.}(2006){Sahu}, {Anupama}, {Srividya}, \&
  {Muneer}}]{sahu2006}
{Sahu}, D.~K., {Anupama}, G.~C., {Srividya}, S., \& {Muneer}, S. 2006, \mnras,
  372, 1315

\bibitem[{{Savitzky} \& {Golay}(1964)}]{savitky_golay1964}
{Savitzky}, A., \& {Golay}, M.~J.~E. 1964, Analytical Chemistry, 36, 1627

\bibitem[{{Scovacricchi} {et~al.}(2016){Scovacricchi}, {Nichol}, {Bacon},
  {Sullivan}, \& {Prajs}}]{scovacricchi2016}
{Scovacricchi}, D., {Nichol}, R.~C., {Bacon}, D., {Sullivan}, M., \& {Prajs},
  S. 2016, \mnras, 456, 1700

\bibitem[{{Silverman} {et~al.}(2012{\natexlab{a}}){Silverman}, {Kong}, \&
  {Filippenko}}]{silverman2012b}
{Silverman}, J.~M., {Kong}, J.~J., \& {Filippenko}, A.~V. 2012{\natexlab{a}},
  \mnras, 425, 1819

\bibitem[{{Silverman} {et~al.}(2012{\natexlab{b}}){Silverman}, {Foley},
  {Filippenko}, {Ganeshalingam}, {Barth}, {Chornock}, {Griffith}, {Kong},
  {Lee}, {Leonard}, {Matheson}, {Miller}, {Steele}, {Barris}, {Bloom}, {Cobb},
  {Coil}, {Desroches}, {Gates}, {Ho}, {Jha}, {Kandrashoff}, {Li}, {Mandel},
  {Modjaz}, {Moore}, {Mostardi}, {Papenkova}, {Park}, {Perley}, {Poznanski},
  {Reuter}, {Scala}, {Serduke}, {Shields}, {Swift}, {Tonry}, {Van Dyk}, {Wang},
  \& {Wong}}]{silverman2012}
{Silverman}, J.~M., {Foley}, R.~J., {Filippenko}, A.~V., {et~al.}
  2012{\natexlab{b}}, \mnras, 425, 1789

\bibitem[{{Smith} {et~al.}(2010){Smith}, {Chornock}, {Silverman}, {Filippenko},
  \& {Foley}}]{smith2010}
{Smith}, N., {Chornock}, R., {Silverman}, J.~M., {Filippenko}, A.~V., \&
  {Foley}, R.~J. 2010, \apj, 709, 856

\bibitem[{{Smith} {et~al.}(2007){Smith}, {Li}, {Foley}, {Wheeler}, {Pooley},
  {Chornock}, {Filippenko}, {Silverman}, {Quimby}, {Bloom}, \&
  {Hansen}}]{smith2007}
{Smith}, N., {Li}, W., {Foley}, R.~J., {et~al.} 2007, \apj, 666, 1116

\bibitem[{{Soderberg} {et~al.}(2008){Soderberg}, {Berger}, {Page}, {Schady},
  {Parrent}, {Pooley}, {Wang}, {Ofek}, {Cucchiara}, {Rau}, {Waxman}, {Simon},
  {Bock}, {Milne}, {Page}, {Barentine}, {Barthelmy}, {Beardmore}, {Bietenholz},
  {Brown}, {Burrows}, {Burrows}, {Byrngelson}, {Cenko}, {Chandra}, {Cummings},
  {Fox}, {Gal-Yam}, {Gehrels}, {Immler}, {Kasliwal}, {Kong}, {Krimm},
  {Kulkarni}, {Maccarone}, {M{\'e}sz{\'a}ros}, {Nakar}, {O'Brien}, {Overzier},
  {de Pasquale}, {Racusin}, {Rea}, \& {York}}]{soderberg2008}
{Soderberg}, A.~M., {Berger}, E., {Page}, K.~L., {et~al.} 2008, \nat, 453, 469

\bibitem[{{Spiro} {et~al.}(2014){Spiro}, {Pastorello}, {Pumo}, {Zampieri},
  {Turatto}, {Smartt}, {Benetti}, {Cappellaro}, {Valenti}, {Agnoletto},
  {Altavilla}, {Aoki}, {Brocato}, {Corsini}, {Di Cianno}, {Elias-Rosa},
  {Hamuy}, {Enya}, {Fiaschi}, {Folatelli}, {Desidera}, {Harutyunyan}, {Howell},
  {Kawka}, {Kobayashi}, {Leibundgut}, {Minezaki}, {Navasardyan}, {Nomoto},
  {Mattila}, {Pietrinferni}, {Pignata}, {Raimondo}, {Salvo}, {Schmidt},
  {Sollerman}, {Spyromilio}, {Taubenberger}, {Valentini}, {Vennes}, \&
  {Yoshii}}]{spiro2014}
{Spiro}, S., {Pastorello}, A., {Pumo}, M.~L., {et~al.} 2014, \mnras, 439, 2873

\bibitem[{{Stritzinger} {et~al.}(2012){Stritzinger}, {Taddia}, {Fransson},
  {Fox}, {Morrell}, {Phillips}, {Sollerman}, {Anderson}, {Boldt}, {Brown},
  {Campillay}, {Castellon}, {Contreras}, {Folatelli}, {Habergham}, {Hamuy},
  {Hjorth}, {James}, {Krzeminski}, {Mattila}, {Persson}, \&
  {Roth}}]{stritzinger2012}
{Stritzinger}, M., {Taddia}, F., {Fransson}, C., {et~al.} 2012, \apj, 756, 173

\bibitem[{{Sun} \& {Gal-Yam}(2017)}]{sun_galyam2017}
{Sun}, F., \& {Gal-Yam}, A. 2017, ArXiv e-prints, arXiv:1707.02543

\bibitem[{{Taddia} {et~al.}(2013){Taddia}, {Stritzinger}, {Sollerman},
  {Phillips}, {Anderson}, {Boldt}, {Campillay}, {Castell{\'o}n}, {Contreras},
  {Folatelli}, {Hamuy}, {Heinrich-Josties}, {Krzeminski}, {Morrell}, {Burns},
  {Freedman}, {Madore}, {Persson}, \& {Suntzeff}}]{taddia2013}
{Taddia}, F., {Stritzinger}, M.~D., {Sollerman}, J., {et~al.} 2013, \aap, 555,
  A10

\bibitem[{{Tak{\'a}ts} {et~al.}(2014){Tak{\'a}ts}, {Pumo}, {Elias-Rosa},
  {Pastorello}, {Pignata}, {Paillas}, {Zampieri}, {Anderson}, {Vink{\'o}},
  {Benetti}, {Botticella}, {Bufano}, {Campillay}, {Cartier}, {Ergon},
  {Folatelli}, {Foley}, {F{\"o}rster}, {Hamuy}, {Hentunen}, {Kankare},
  {Leloudas}, {Morrell}, {Nissinen}, {Phillips}, {Smartt}, {Stritzinger},
  {Taubenberger}, {Valenti}, {Van Dyk}, {Haislip}, {LaCluyze}, {Moore}, \&
  {Reichart}}]{takats2014}
{Tak{\'a}ts}, K., {Pumo}, M.~L., {Elias-Rosa}, N., {et~al.} 2014, \mnras, 438,
  368

\bibitem[{{Tanaka} {et~al.}(2012){Tanaka}, {Moriya}, {Yoshida}, \&
  {Nomoto}}]{tanaka2012}
{Tanaka}, M., {Moriya}, T.~J., {Yoshida}, N., \& {Nomoto}, K. 2012, \mnras,
  422, 2675

\bibitem[{{Taubenberger} {et~al.}(2006){Taubenberger}, {Pastorello}, {Mazzali},
  {Valenti}, {Pignata}, {Sauer}, {Arbey}, {B{\"a}rnbantner}, {Benetti}, {Della
  Valle}, {Deng}, {Elias-Rosa}, {Filippenko}, {Foley}, {Goobar}, {Kotak}, {Li},
  {Meikle}, {Mendez}, {Patat}, {Pian}, {Ries}, {Ruiz-Lapuente}, {Salvo},
  {Stanishev}, {Turatto}, \& {Hillebrandt}}]{taubenberger2006}
{Taubenberger}, S., {Pastorello}, A., {Mazzali}, P.~A., {et~al.} 2006, \mnras,
  371, 1459

\bibitem[{{Taubenberger} {et~al.}(2009){Taubenberger}, {Valenti}, {Benetti},
  {Cappellaro}, {Della Valle}, {Elias-Rosa}, {Hachinger}, {Hillebrandt},
  {Maeda}, {Mazzali}, {Pastorello}, {Patat}, {Sim}, \&
  {Turatto}}]{taubenberger2009}
{Taubenberger}, S., {Valenti}, S., {Benetti}, S., {et~al.} 2009, \mnras, 397,
  677

\bibitem[{{Taubenberger} {et~al.}(2011){Taubenberger}, {Navasardyan}, {Maurer},
  {Zampieri}, {Chugai}, {Benetti}, {Agnoletto}, {Bufano}, {Elias-Rosa},
  {Turatto}, {Patat}, {Cappellaro}, {Mazzali}, {Iijima}, {Valenti},
  {Harutyunyan}, {Claudi}, \& {Dolci}}]{taubenberger2011}
{Taubenberger}, S., {Navasardyan}, H., {Maurer}, J.~I., {et~al.} 2011, \mnras,
  413, 2140

\bibitem[{{Thomas} {et~al.}(2011){Thomas}, {Nugent}, \& {Meza}}]{thomas2011}
{Thomas}, R.~C., {Nugent}, P.~E., \& {Meza}, J.~C. 2011, \pasp, 123, 237

\bibitem[{{Tody}(1986)}]{tody1986}
{Tody}, D. 1986, in \procspie, Vol. 627, Instrumentation in astronomy VI, ed.
  D.~L. {Crawford}, 733

\bibitem[{{Tody}(1993)}]{tody1993}
{Tody}, D. 1993, in Astronomical Society of the Pacific Conference Series,
  Vol.~52, Astronomical Data Analysis Software and Systems II, ed. R.~J.
  {Hanisch}, R.~J.~V. {Brissenden}, \& J.~{Barnes}, 173

\bibitem[{{Tolstov} {et~al.}(2017{\natexlab{a}}){Tolstov}, {Nomoto},
  {Blinnikov}, {Sorokina}, {Quimby}, \& {Baklanov}}]{tolstov2017}
{Tolstov}, A., {Nomoto}, K., {Blinnikov}, S., {et~al.} 2017{\natexlab{a}},
  \apj, 835, 266

\bibitem[{{Tolstov} {et~al.}(2017{\natexlab{b}}){Tolstov}, {Zhiglo}, {Nomoto},
  {Sorokina}, {Kozyreva}, \& {Blinnikov}}]{tolstov2017b}
{Tolstov}, A., {Zhiglo}, A., {Nomoto}, K., {et~al.} 2017{\natexlab{b}}, \apjl,
  845, L2

\bibitem[{{Valenti} {et~al.}(2008){Valenti}, {Elias-Rosa}, {Taubenberger},
  {Stanishev}, {Agnoletto}, {Sauer}, {Cappellaro}, {Pastorello}, {Benetti},
  {Riffeser}, {Hopp}, {Navasardyan}, {Tsvetkov}, {Lorenzi}, {Patat}, {Turatto},
  {Barbon}, {Ciroi}, {Di Mille}, {Frandsen}, {Fynbo}, {Laursen}, \&
  {Mazzali}}]{valenti2008}
{Valenti}, S., {Elias-Rosa}, N., {Taubenberger}, S., {et~al.} 2008, \apjl, 673,
  L155

\bibitem[{Veres \& Wiese(1996)}]{veres_wiese1996}
Veres, G., \& Wiese, W.~L. 1996, Phys. Rev. A, 54, 1999

\bibitem[{{Vink{\'o}} {et~al.}(2006){Vink{\'o}}, {Tak{\'a}ts}, {S{\'a}rneczky},
  {Szab{\'o}}, {M{\'e}sz{\'a}ros}, {Csorv{\'a}si}, {Szalai}, {G{\'a}sp{\'a}r},
  {P{\'a}l}, {Csizmadia}, {K{\'o}sp{\'a}l}, {R{\'a}cz}, {Kun}, {Cs{\'a}k},
  {F{\"u}r{\'e}sz}, {DeBond}, {Grunhut}, {Thomson}, {Mochnacki}, \&
  {Koktay}}]{vinko2006}
{Vink{\'o}}, J., {Tak{\'a}ts}, K., {S{\'a}rneczky}, K., {et~al.} 2006, \mnras,
  369, 1780

\bibitem[{{Vreeswijk} {et~al.}(2014){Vreeswijk}, {Savaglio}, {Gal-Yam}, {De
  Cia}, {Quimby}, {Sullivan}, {Cenko}, {Perley}, {Filippenko}, {Clubb},
  {Taddia}, {Sollerman}, {Leloudas}, {Arcavi}, {Rubin}, {Kasliwal}, {Cao},
  {Yaron}, {Tal}, {Ofek}, {Capone}, {Kutyrev}, {Toy}, {Nugent}, {Laher},
  {Surace}, \& {Kulkarni}}]{vreeswijk2014}
{Vreeswijk}, P.~M., {Savaglio}, S., {Gal-Yam}, A., {et~al.} 2014, \apj, 797, 24

\bibitem[{{Vreeswijk} {et~al.}(2017){Vreeswijk}, {Leloudas}, {Gal-Yam}, {De
  Cia}, {Perley}, {Quimby}, {Waldman}, {Sullivan}, {Yan}, {Ofek}, {Fremling},
  {Taddia}, {Sollerman}, {Valenti}, {Arcavi}, {Howell}, {Filippenko}, {Cenko},
  {Yaron}, {Kasliwal}, {Cao}, {Ben-Ami}, {Horesh}, {Rubin}, {Lunnan}, {Nugent},
  {Laher}, {Rebbapragada}, {Wo{\'z}niak}, \& {Kulkarni}}]{vreeswijk2017}
{Vreeswijk}, P.~M., {Leloudas}, G., {Gal-Yam}, A., {et~al.} 2017, \apj, 835, 58

\bibitem[{{Whalen} {et~al.}(2013){Whalen}, {Even}, {Frey}, {Smidt}, {Johnson},
  {Lovekin}, {Fryer}, {Stiavelli}, {Holz}, {Heger}, {Woosley}, \&
  {Hungerford}}]{whalen2013}
{Whalen}, D.~J., {Even}, W., {Frey}, L.~H., {et~al.} 2013, \apj, 777, 110

\bibitem[{{Whalen} {et~al.}(2014){Whalen}, {Smidt}, {Heger}, {Hirschi},
  {Yusof}, {Even}, {Fryer}, {Stiavelli}, {Chen}, \& {Joggerst}}]{whalen2014}
{Whalen}, D.~J., {Smidt}, J., {Heger}, A., {et~al.} 2014, \apj, 797, 9

\bibitem[{{Wheeler} {et~al.}(2017){Wheeler}, {Chatzopoulos}, {Vink{\'o}}, \&
  {Tuminello}}]{wheeler2017}
{Wheeler}, J.~C., {Chatzopoulos}, E., {Vink{\'o}}, J., \& {Tuminello}, R. 2017,
  \apjl, 851, L14

\bibitem[{{Woosley}(2010)}]{woosley2010}
{Woosley}, S.~E. 2010, \apjl, 719, L204

\bibitem[{{Woosley}(2017)}]{woosley2017}
---. 2017, \apj, 836, 244

\bibitem[{{Woosley} \& {Heger}(2006)}]{woosley_heger2006}
{Woosley}, S.~E., \& {Heger}, A. 2006, \apj, 637, 914

\bibitem[{{Yan} {et~al.}(2015){Yan}, {Quimby}, {Ofek}, {Gal-Yam}, {Mazzali},
  {Perley}, {Vreeswijk}, {Leloudas}, {de Cia}, {Masci}, {Cenko}, {Cao},
  {Kulkarni}, {Nugent}, {Rebbapragada}, {Wo{\'z}niak}, \& {Yaron}}]{yan2015}
{Yan}, L., {Quimby}, R., {Ofek}, E., {et~al.} 2015, \apj, 814, 108

\bibitem[{{Yan} {et~al.}(2017{\natexlab{a}}){Yan}, {Quimby}, {Gal-Yam},
  {Brown}, {Blagorodnova}, {Ofek}, {Lunnan}, {Cooke}, {Cenko}, {Jencson}, \&
  {Kasliwal}}]{yan2017}
{Yan}, L., {Quimby}, R., {Gal-Yam}, A., {et~al.} 2017{\natexlab{a}}, \apj, 840,
  57

\bibitem[{{Yan} {et~al.}(2017{\natexlab{b}}){Yan}, {Lunnan}, {Perley},
  {Gal-Yam}, {Yaron}, {Roy}, {Quimby}, {Sollerman}, {Fremling}, {Leloudas},
  {Cenko}, {Vreeswijk}, {De Cia}, {Ofek}, {Kulkarni}, {Masci}, {Rebbapragada},
  \& {Wo{\'z}niak}}]{yan2017b}
{Yan}, L., {Lunnan}, R., {Perley}, D., {et~al.} 2017{\natexlab{b}}, ArXiv
  e-prints, arXiv:1704.05061

\bibitem[{{Yaron} \& {Gal-Yam}(2012)}]{yaron_galyam2012}
{Yaron}, O., \& {Gal-Yam}, A. 2012, \pasp, 124, 668

\bibitem[{{Yoon} \& {Langer}(2005)}]{yoon_langer2005}
{Yoon}, S.-C., \& {Langer}, N. 2005, \aap, 443, 643

\bibitem[{{Yusof} {et~al.}(2013){Yusof}, {Hirschi}, {Meynet}, {Crowther},
  {Ekstr{\"o}m}, {Frischknecht}, {Georgy}, {Abu Kassim}, \&
  {Schnurr}}]{yusof2013}
{Yusof}, N., {Hirschi}, R., {Meynet}, G., {et~al.} 2013, \mnras, 433, 1114

\bibitem[{{Zhang} {et~al.}(2014){Zhang}, {Wang}, {Mazzali}, {Bai}, {Zhang},
  {Bersier}, {Huang}, {Fan}, {Mo}, {Wang}, {Yi}, {Wang}, {Xin}, {Liangchang},
  {Zhang}, {Lun}, {Wang}, {He}, \& {Walker}}]{zhang2014}
{Zhang}, J., {Wang}, X., {Mazzali}, P.~A., {et~al.} 2014, \apj, 797, 5

\end{thebibliography}
